\documentclass[11pt
]{llncs}

\usepackage{geometry}
\usepackage{amssymb}

\usepackage{amsmath}
\allowdisplaybreaks % allow breaks among pages for align

\usepackage{breakcites}
\usepackage{caption}
\usepackage{subcaption}
\usepackage{nicefrac}
\usepackage[para,online,flushleft]{threeparttable}

\usepackage{soul}

\usepackage{ifdraft}

\usepackage{anyfontsize}
\usepackage{color}
\usepackage[dvipsnames]{xcolor}

\usepackage{graphicx} \setkeys{Gin}{draft=false}
\usepackage{mathtools}
 
\usepackage{physics}
\let\originalket\ket
\renewcommand{\ket}[1]{\originalket*{#1}} % redefine \ket to \ket*. The original \ket is still available as \originalket.
\let\originalbra\bra
\renewcommand{\bra}[1]{\originalbra*{#1}} % redefine \bra to \bra*. The original \bra is still available as \originalbra.

\usepackage{xspace}
\usepackage[utf8]{inputenc}

\usepackage{amsthm}

\usepackage{cases} 

\usepackage{enumitem}
\setitemize{leftmargin=*,topsep=0.5em,itemsep=0.5em}
\setenumerate{leftmargin=*,topsep=0.5em,itemsep=0.5em}

\usepackage{scalerel}

\usepackage{algorithm,algpseudocode}

\usepackage[pagebackref=true,draft=false]{hyperref}
\hypersetup{
    linktoc=all, 
	hidelinks, 
	colorlinks=true,
	allcolors=blue
}

\ifdraft{
  \usepackage[multiuser,inline,nomargin,marginclue,draft]{fixme}
}{
  \usepackage[multiuser,inline,nomargin,marginclue]{fixme}
}

\fxusetheme{color}
\colorlet{fxtarget}{Red}

\FXRegisterAuthor{x}{ex}{\color{Plum} Xiao}
\newcommand{\xiao}[1]{\xnote{#1}}

\FXRegisterAuthor{cf}{ecf}{\color{Violet} Chak Fai}
\newcommand{\chakfai}[1]{\cfnote{#1}}

\FXRegisterAuthor{a}{ea}{\color{Red} Takashi}

\usepackage{zref-clever}
\zcsetup{
  cap = true, % always capitalize the initila letter
  nameinlink = true
}

\zcRefTypeSetup{Claim}{
Name-sg = {Claim},
name-sg = {claim},
Name-pl = {Claims},
name-pl = {claims},
refbounds = {, , , }% {preref*, preref, postref, postref*}
}

\zcRefTypeSetup{Expression}{
Name-sg = {Expression},
name-sg = {expression},
Name-pl = {Expressions},
name-pl = {expressions},
refbounds = {, (, ), }% {preref*, preref, postref, postref*}
}

\zcRefTypeSetup{Inequality}{
Name-sg = {Inequality},
name-sg = {inequality},
Name-pl = {Inequalities},
name-pl = {inequalities},
refbounds = {, (, ), }% {preref*, preref, postref, postref*}
}

\zcRefTypeSetup{Property}{
Name-sg = {Property},
name-sg = {property},
Name-pl = {Properties},
name-pl = {properties},
refbounds = {, , , }% {preref*, preref, postref, postref*}
}

\zcRefTypeSetup{Step}{
Name-sg = {Step},
name-sg = {step},
Name-pl = {Steps},
name-pl = {steps},
refbounds = {, , , }% {preref*, preref, postref, postref*}
}

\usepackage{nameref}
\makeatletter
\let\orgdescriptionlabel\descriptionlabel
\renewcommand*{\descriptionlabel}[1]{%
  \let\orglabel\label
  \let\label\@gobble
  \phantomsection
  \edef\@currentlabel{#1}%
  \let\label\orglabel
  \orgdescriptionlabel{#1}%
}
\makeatother

\usepackage[nameinlink]{cleveref}
\Crefname{Algorithm}{Algorithm}{Algorithms}
\Crefname{appendix}{Appx.}{Appx.}
\Crefname{Case}{Case}{Cases}
\Crefname{Condition}{Condition}{Conditions}
\Crefname{definition}{Def.}{Def.}
\Crefname{equation}{Eq.}{Eq.}
\Crefname{Equation}{Eq.}{Eq.}
\Crefname{Execution}{Execution}{Executions}
\Crefname{Expression}{Expression}{Expressions}
\creflabelformat{Expression}{#2(#1)#3}
\Crefname{figure}{Fig.}{Fig.}
\Crefname{Inequality}{Inequality}{Inequalities}
\creflabelformat{Inequality}{#2(#1)#3}
\Crefname{Issue}{Issue}{Issues}
\Crefname{lemma}{Lem.}{Lem.}
\Crefname{Language}{Language}{Languages}
\creflabelformat{Language}{#2(#1)#3}
\Crefname{obs}{Observation}{Observations}
\Crefname{Path}{Path}{Paths}
\Crefname{Problem}{Problem}{Problems}
\Crefname{Property}{Property}{Properties}
\Crefname{remark}{Rmk.}{Rmk.}
\Crefname{Round}{Round}{Round}
\Crefname{section}{Sec.}{Sec.}
\Crefname{Stage}{Stage}{Stages}
\Crefname{Step}{Step}{Steps}
\Crefname{theorem}{Thm.}{Thm.}
\Crefname{Trick}{Trick}{Tricks}
\Crefname{Type}{Type}{Types}

\usepackage[most]{tcolorbox}
\newtcolorbox
[auto counter,
Crefname={Constr.}{Constr.},
]{ConstructionBox}[2][]
{
  title    = Construction~\thetcbcounter: #2,
  fonttitle=\small\bfseries,
  enforce breakable, % breakable across pages
  colframe = black,
  coltitle = black,
  colbacktitle = white, 
  coltext  = black, 
  colback  = white,
  arc=0pt,
  outer arc=0pt,
  boxrule=0.5pt,
  left=0pt, % left margin
  right=0pt, % right margin
   top=0pt,
  bottom=0pt,
  #1
}

\newtcolorbox
[auto counter,
Crefname={Prot.}{Prot.}
]{ProtocolBox}[2][]
{
  title    = Protocol~\thetcbcounter: #2,
  fonttitle=\small\bfseries,
  enforce breakable, % breakable across pages
  colframe = black,
  coltitle = black,
  colbacktitle = white, 
  coltext  = black, 
  colback  = white,
  arc=0pt,
  outer arc=0pt,
  boxrule=0.5pt,
  toprule at break = 0pt,
  bottomrule at break = 0pt,
  left=0pt, % left margin
  right=0pt, % right margin
   top=0pt,
  bottom=0pt,
  #1
}

\newtcolorbox
[auto counter,
Crefname={Fig.}{Fig.},
]{FigureBox}[2][]
{
  title    = Figure~\thetcbcounter: #2,
  fonttitle=\small\bfseries,
  enforce breakable, % breakable across pages
  colframe = black,
  coltitle = black,
  colbacktitle = white, 
  coltext  = black, 
  colback  = white,
  arc=0pt,
  outer arc=0pt,
  boxrule=0.5pt,
  toprule at break = 0pt,
  bottomrule at break = 0pt,
  left=0pt, % left margin
  right=0pt, % right margin
   top=0pt,
  bottom=0pt,
  #1
}

\newtcolorbox[auto counter]{CommentBox}[2][]
{
  breakable,
  colframe = gray!10,
  colback  = gray!2,
  coltitle = red, 
  coltext  = red, 
  title    = #2,
  arc=0pt,
  outer arc=0pt,
  boxrule=1.5pt,
  left=0mm,
  right=0pt,
  top=0pt,
  bottom=0pt,
}

\newtcolorbox[
auto counter,
number within=section, 
Crefname={Algo.}{Algo.}
]{AlgorithmBox}[2][]
{
  title    = {Algorithm~\thetcbcounter: #2},
  fonttitle=\small\bfseries,
  enforce breakable, % breakable across pages
  colframe = black,
  coltitle = black,
  colbacktitle = white, 
  coltext  = black, 
  colback  = white,
  arc = 0pt,
  outer arc = 0pt,
  boxrule = 0.5pt,
  toprule at break = 0pt,
  bottomrule at break = 0pt,
  left = 0pt, % left margin
  right = 0pt, % right margin
  top=0pt,
  bottom=0pt,
  #1
}

\newtcolorbox
[auto counter,
Crefname={Expr.}{Expr.}
]{ExperimentBox}[2][]
{
  title    = Experiment~\thetcbcounter: #2,
  fonttitle=\small\bfseries,
  enforce breakable, % breakable across pages
  colframe = black,
  coltitle = black,
  colbacktitle = white, 
  coltext  = black, 
  colback  = white,
  arc=0pt,
  outer arc=0pt,
  boxrule=0.5pt,
  toprule at break = 0pt,
  bottomrule at break = 0pt,
  left=0pt, % left margin
  right=0pt, % right margin
   top=0pt,
  bottom=0pt,
  #1
}

\newtcolorbox
[auto counter,
Crefname={Box}{Boxes}
]{BoxBox}[2][]
{
  title    = {Box~\thetcbcounter: #2},
  fonttitle=\small\bfseries,
  enforce breakable, % breakable across pages
  colframe = black,
  coltitle = black,
  colbacktitle = white, 
  coltext  = black, 
  colback  = white,
  arc=0pt,
  outer arc=0pt,
  boxrule=0.5pt,
  toprule at break = 0pt,
  bottomrule at break = 0pt,
  left=0pt, % left margin
  right=0pt, % right margin
   top=0pt,
  bottom=0pt,
  #1
}

\spnewtheorem{construction}{Construction}{\bfseries}{\itshape}
\crefname{construction}{construction}{constructions}
\Crefname{construction}{Construction}{Constructions}

\spnewtheorem*{MyComment}{Comment}{\bfseries}{\itshape}

\newcounter{myclaimcounter}
\newtheorem{MyClaim}[myclaimcounter]{Claim}
\Crefname{MyClaim}{Claim}{Claims}
\creflabelformat{Claim}{#2(#1)#3}

\makeatletter
\renewcommand\subsubsection{\@startsection{subsubsection}{3}{\z@}%
                       {-18\p@ \@plus -4\p@ \@minus -4\p@}%
                       {0.5em \@plus 0.22em \@minus 0.1em}%
                       {\normalfont\normalsize\bfseries\boldmath}}
\makeatother
\Crefname{subsection}{Sec.}{Sec.}
\Crefname{APPXsubsubsection}{Appx.}{Appx.}

\newcommand{\para}[1]{\vspace{0.5em}\noindent {\bf #1}}
\newcommand{\subpara}[1]{\vspace{0.5em}\noindent\underline{#1}}

\usepackage[normalem]{ulem}
\newcommand{\dlt}[1][0.5ex]{\bgroup\markoverwith{\textcolor{red}{\rule[#1]{1pt}{0.9pt}}}\ULon}

\newcommand{\msf}{\mathsf}
\newcommand{\mcal}{\mathcal}

\newcommand{\BQP}{\mathsf{BQP}}

\newcommand{\coNP}{\ensuremath{\mathsf{coNP}}\xspace}

\newcommand{\NP}{\ensuremath{\mathsf{NP}}\xspace}

\renewcommand{\P}{\ensuremath{\mathsf{P}}\xspace}

\newcommand{\PromiseNP}{\ensuremath{\mathsf{Promise\text{-}NP}}\xspace}
\newcommand{\PromisecoNP}{\ensuremath{\mathsf{Promise\text{-}coNP}}\xspace}

\newcommand{\RP}{\ensuremath{\mathsf{RP}}\xspace}

\newcommand{\UP}{\ensuremath{\mathsf{UP}}\xspace}

\newcommand{\Adv}{\ensuremath{\mathcal{A}}\xspace}

\renewcommand{\bar}{\overline}

\newcommand{\bits}{\ensuremath{\{0,1\}}\xspace}

\newcommand{\E}{\ensuremath{\mathbb{E}}\xspace}

\renewcommand{\epsilon}{\varepsilon}

\renewcommand{\hat}{\widehat}

\newcommand{\idind}{\ensuremath{\stackrel{\text{i.d.}}{=\mathrel{\mkern-3mu}=}}\xspace}

\newcommand{\iO}{\ensuremath{\mathit{i}O}\xspace}
\newcommand{\iObf}{\ensuremath{\mathit{i}\mathcal{O}}\xspace}

\newcommand{\Lang}{\ensuremath{\mathcal{L}}\xspace}

\newcommand{\Naturals}{\ensuremath{\mathbb{N}}\xspace}

\newcommand{\negl}{\ensuremath{\mathsf{negl}}\xspace}

\newcommand{\Ora}{\ensuremath{\mathcal{O}}\xspace}

\renewcommand{\paragraph}{\para}

\newcommand{\pick}{\ensuremath{\xleftarrow{\$}}\xspace}
\newcommand{\picks}{\pick}

\newcommand{\poly}{\ensuremath{\mathsf{poly}}\xspace}

\newcommand{\qcind}{\ensuremath{\stackrel{\text{c}_{\text{q}}}{\approx}}\xspace}

\newcommand{\red}[1]{{\color{red} #1}}

\newcommand{\Set}[1]{\ensuremath{\{#1\}}\xspace}

\newcommand{\sind}{\ensuremath{\stackrel{\text{s}}{\approx}}\xspace}

\renewcommand{\tilde}{\widetilde}

\newcommand{\rel}{R}

\newcommand{\bit}{\{0,1\}}

\newenvironment{boxfig}[2]{\begin{figure}[#1]\fbox{\begin{minipage}{0.97\linewidth}
                        \vspace{0.2em}
                        \makebox[0.025\linewidth]{}
                        \begin{minipage}{0.95\linewidth}
            {{
                        #2 }}
                        \end{minipage}
                        \vspace{0.2em}
                        \end{minipage}}}{\end{figure}}

\usepackage{bbm}
\usepackage{bm}

\makeatletter
\def\@fnsymbol#1{\ensuremath{\ifcase#1\or *\or \dagger\or \ddagger\or
   \mathsection\or \mathparagraph\or \|\or **\or \dagger\dagger
   \or \ddagger\ddagger \else\@ctrerr\fi}}
\makeatother

\begin{document}

%%%%%%%%%%%%%%%%%%%%%%%%%%%%%%%%%%%%%%%%
% The following two commands is to speed-up the latex compiling. Should remove it in the final version
%%%%%%%%%%%%%%%%%%%%%%%%%%%%%%%%%%%%%%%%
\ifdraft{
	 % stop compressing the final PDF
	 % stop compressing the object file 
}{}

\ifdraft{
    \listoffixmes
    \addcontentsline{toc}{section}{List of Corrections}
    \newpage
}{}

% Change the title as you like
\title{Structured Quantum Hardness from Unstructured Assumptions}
% \title{}
% \titlerunning{}
% \toctitle{}

\author{
Xiao Liang\inst{1}\thanks{The work described in this paper was partially supported by a grant from the Research Grants Council of the Hong Kong Special Administrative  Region, China [Project No.: CUHK 24208025].}
%\orcidID{0000-0003-0858-9289} 
\and 
Takashi Yamakawa\inst{2}
\and 
Chak Fai Yung\inst{3}
}
\institute{
The Chinese University of Hong Kong, Hong Kong\\ \email{xiaoliang@cuhk.edu.hk}
\and
NTT Social Informatics Laboratories, Japan\\ \email{takashi.yamakawa@ntt.com}
\and
The Chinese University of Hong Kong, Hong Kong\\ \email{cfyung2@cse.cuhk.edu.hk
}
}

% -- The wrap-up around "\maketitle" is used to remove paper title from ToC
\let\oldaddcontentsline\addcontentsline
\def\addcontentsline#1#2#3{}
\maketitle
\def\addcontentsline#1#2#3{\oldaddcontentsline{#1}{#2}{#3}}

\begin{abstract}

    Whether useful structured hardness can arise from unstructured assumptions is a fundamental question in complexity and cryptography. The longstanding Bennett--Gill question asks whether $\NP\cap\coNP$ remains hard relative to a random oracle; \cite[STOC'23]{STOC:GIKKLS23} recently gave a conditional affirmative answer in the classical setting. We establish the quantum analogue: if $\UP\not\subseteq\BQP$, then
\(
    \NP^{O}\cap\coNP^{O}
    \not\subseteq \BQP^{O}
\)
with probability~$1$ over a random oracle~$O$.\\[-0.7em]

Beyond extending the classical result, this separation has a distinctly quantum significance: Shor's algorithms defeat canonical number-theoretic hardness candidates, while lattice assumptions and other leading post-quantum cryptographic assumptions do not directly supply hard languages with two-sided classical certification on every input. Our conditional oracle separation provides evidence that such structured quantum hardness can nevertheless arise from unstructured sources.\\[-0.7em]

Complementing this result, we rule out fully black-box quantum reductions of  worst-case quantum hardness in $\NP\cap\coNP$ to post-quantum injective one-way functions and indistinguishability obfuscation. This barrier excludes a broad class of black-box approaches to deriving structured quantum hardness from generic cryptography primitives, providing partial evidence that the non-black-box techniques in our first result are essential.

\keywords{Quantum Complexity \and Random Oracle \and Non-Interactive Zero Knowledge \and Pseudorandom Hash functions \and Black-Box Impossibility}
\end{abstract}
%-- Starting numbering pages with Roman letters --------
	\pagenumbering{roman}
% -- Creat ToC
\tableofcontents
\addcontentsline{toc}{section}{Table of Contents}
\clearpage

%-- Starting numbering pages with arabic --------
\pagenumbering{arabic}

\newcommand{\QOra}{\ket{\mathcal{O}}}
\newcommand{\NIZK}{\ket{\msf{NIZK}}}

\section{Introduction}
   
Understanding how the structure of a computational problem governs its complexity is a central theme in theoretical computer science. Additional structure may enable efficient algorithms, but it may also endow a problem with useful properties without making it tractable. Distinguishing these possibilities is essential to understanding which forms of computational hardness can exist and what they can accomplish. This tension is particularly pronounced in cryptography, where the algebraic structure underlying a hard problem often enables powerful applications, while simultaneously providing a foothold for algorithmic attacks. A fundamental question is therefore whether useful forms of structured hardness inherently require structured assumptions, or whether they can instead arise from generic, unstructured sources of computational difficulty.

A concrete setting for studying this question is the class
$\NP\cap\coNP$, whose languages admit polynomial-size,
efficiently verifiable certificates for both membership and nonmembership.
This two-sided certification requirement provides a complexity-theoretic
notion of structure. Whether such a requirement is compatible with intractability
is captured by the classical conjecture $\P\neq\NP\cap\coNP$. However, compared with $\P\neq\NP$, this conjecture is
supported by a much narrower collection of natural candidate problems.
Among the most prominent are decision problems associated with integer
factorization and discrete logarithms. Their presumed hardness
has motivated extensive study and underlies important cryptographic constructions, but their considerable mathematical structure leaves open
the possibility of unexpected algorithmic advances.  {For example, number-field-sieve algorithms achieve
heuristic subexponential running times for integer factorization
\cite{BLP93} and for discrete logarithms in prime fields
\cite{Gordon93}.}

A longstanding line of research asks whether hardness in
$\NP\cap\coNP$ can instead be obtained from unstructured
sources
\cite{BG81,FOCS:BluImp87,Rudich88,Tardos89,
      C:BitDegVai17,STOC:GIKKLS23}.
Random oracles provide a natural framework for investigating this
question. A uniformly random function supplies computational difficulty
without imposing a particular algebraic structure; a separation relative
to such an oracle can therefore provide evidence that hardness need not
depend on special properties of an explicit mathematical construction. The seminal work of
Bennett and Gill~\cite{BG81} established 
$\P^{\Ora}\neq\NP^{\Ora}$ with
probability~$1$ over a random oracle~$\mathcal O$, and asked whether
$\P^{\Ora}\neq
\NP^{\Ora}\cap\coNP^{\Ora}$ holds as well.
They proved the latter separation when $\Ora$ is a random \emph{permutation} (a structured source of hardness),
but left the random-function case open. 

The same distinction arises when constructing hard languages in
$\NP\cap\coNP$ from cryptographic primitives.
One-way permutations (OWPs) yield such languages through black-box
constructions. Injective one-way functions (inj-OWFs), by contrast, are viewed as
an unstructured source of hardness in this context: a sufficiently
length-expanding random function is both injective and one-way except
with negligible probability~\cite{STOC:GIKKLS23}.
Here, ``unstructured'' does not mean the absence of every mathematical
property, but rather the absence of additional constraints that cannot
be obtained from a random function.
In this regard,,
results of Blum and Impagliazzo~\cite{FOCS:BluImp87} and
Rudich~\cite{Rudich88} rule out black-box constructions of hard languages
in $\NP\cap\coNP$ from OWFs;
Bitansky, Degwekar, and Vaikuntanathan
\cite{C:BitDegVai17} strengthened this barrier to
inj-OWFs, even when additionally allowing black-box
access to \emph{indistinguishability obfuscation} ($\iO$).
These results expose a substantial obstacle to deriving two-sided
certification together with hardness from generic cryptographic
primitives.

Recently,  \cite{STOC:GIKKLS23} overcame this obstacle through a
non-black-box approach. Assuming
$\UP \not\subseteq \RP$ (an assumption implied by the
existence of inj-OWFs), they established $
    \NP^{\Ora}\cap\coNP^{\Ora}\nsubseteq \P^{\Ora}
    $  with
probability~$1$ over a random oracle~$\Ora$. Their result gives a \emph{conditional} affirmative answer to 
Bennett--Gill question.
%  \xiao{I removeed the follwoing part.}
% A central ingredient is the use of non-interactive zero-knowledge
% proofs to convert an appropriate hard promise problem into a hard
% language. This distinction is essential: a promise problem need only
% admit the required certificates on inputs satisfying its promise,
% whereas a language in $\NP\cap\coNP$ must admit
% certificates of the correct answer on \emph{every} input.

\para{Structured Quantum Hardness?}
In contrast, the picture in the quantum regime is much less understood.
In fact, \emph{this is not merely asking the quantum version of a classical
question}: Shor's algorithms~\cite{FOCS:Shor94} solve integer
factorization and discrete logarithms in quantum polynomial time,
removing two of the principal sources of evidence for classical
hardness in $\NP\cap\coNP$.
These algorithms sharply illustrate the underlying concern:
the structure supporting useful certification and cryptographic
applications can also enable an efficient quantum algorithm.
Thus, the quantum setting asks whether two-sided classical
certification can coexist with quantum hardness, \emph{even after
allowing the algorithmic capabilities that defeat canonical problems
serving as evidence for the classical Bennett--Gill question}.

Moreover, the leading sources of conjectured \emph{post-quantum} hardness
makes the situation more fascinating: 
Lattice problems, including learning with errors (LWE), short integer
solution (SIS), and the approximate shortest vector problem
($\mathrm{GapSVP}$), are widely conjectured to resist quantum
algorithms at appropriate parameters and underpin much of
post-quantum cryptography~\cite{EPRINT:Peikert15}.
Nevertheless, these conjectures do not directly furnish hard
languages in $\NP\cap\coNP$.
The distinction here is subtle: the obstacle is not simply the
absence of efficiently checkable no-certificates.
Indeed, Aharonov and Regev~\cite{FOCS:AhaReg04} established
two-sided classical certification for suitable gap versions of
lattice problems, placing them in $\PromiseNP \cap  \PromisecoNP$.  
These guarantees, however, apply only to inputs satisfying the
promise; they do not provide certificates of the correct answer
on \emph{every} input.
Likewise, standard formulations of SIS and LWE do not directly supply the
required language with two-sided certification either.
 {It is also worth mentioning that other approaches to post-quantum cryptography draw on problems
in coding theory, multivariate polynomial systems, and isogenies
of elliptic curves \cite{NISTPQC16}. These approaches raise the
same complexity-theoretic issue: conjectured hardness of a search
task or a promise problem does not, by itself, provide a language
with efficiently verifiable certificates for both answers on every
input. Thus, the availability of post-quantum cryptographic candidates therefore
does not by itself resolve the question considered here.}  
This gives the quantum Bennett--Gill question a distinctive motivation
beyond its classical version.

There is also an important distinction between assuming quantum
hardness for an already structured primitive and deriving structured
quantum hardness from an unstructured source. For example, a
one-way permutation secure against quantum polynomial-time
algorithms yields a language in $\NP\cap\coNP$ outside $\BQP$:
its unique inverse provides certificates for either value of an
inverse bit, while an efficient quantum algorithm for deciding
these bits would allow inversion, using error amplification.
Here, however, the structure needed for two-sided certification
is already supplied by the assumption: every output has exactly
one preimage. Likewise, the random-permutation separation of \cite{BBBV97}
provides relativized evidence for quantum hardness in
$\NP\cap\coNP$, but starts from an oracle with this same
structural guarantee. Our aim is instead to obtain two-sided
certification together with quantum hardness without assuming
such permutation structure at the outset.

\begin{quote}
  \centering
    \textbf{Question}: 
    \emph{Can structured \ul{quantum} hardness be derived from unstructured sources?}
\end{quote}

    \subsection{Our Results} 
    \label{sec:intro:results}

 We obtain both positive and negative results concerning this question. 
We summarize them below.

    \para{Unstructured Quantum Hardness.}    Our first result gives a conditional affirmative answer to the quantum
Bennett--Gill question. Specifically, assuming that $\UP$ contains a
language outside $\BQP$, we show that a random oracle almost surely
yields a language with two-sided classical certification that remains
hard even for quantum algorithms. Note that both the assumption $\UP \nsubseteq \BQP$ and a random oracle
are regarded as sources of unstructured hardness.
Thus, \Cref{thm:informal:main} provides evidence toward an affirmative
answer to the question posed above.
(Here and throughout, $\BQP^{\Ora}$ denotes bounded-error
quantum polynomial time with \emph{quantum} query access to $\Ora$.)  
    \begin{theorem}\label{thm:informal:main}
      If $\UP \nsubseteq \BQP$, then it holds that  $\NP^{\Ora} \cap \coNP^{\Ora} \nsubseteq \BQP^{\Ora}$ with probability 1 over the choice of a random oracle $\Ora$. 
    \end{theorem}

Importantly, the proof of \Cref{thm:informal:main} makes non-black-box use of a language
$\Lang \in \UP \setminus \BQP$ to construct the desired quantum-hard
language.\footnote{As will become clear in the following sections,
non-black-box use here refers to generating a \emph{non-interactive zero-knowledge proof} for a statement
involving the definition of $\Lang$.}
If this use of $\Lang$ could instead be made black-box, one could hope
to use the random oracle itself to construct an oracle language
$\Lang^{\Ora}$ in place of $\Lang$.\footnote{A sufficiently
length-expanding restriction of a random oracle is injective and
post-quantum one-way with overwhelming probability (see \Cref{lem:injOWF-Oracle}). This yields an
oracle language in $\UP^{\Ora}$ outside $\BQP^{\Ora}$,
rather than requiring the separate assumption $\UP \nsubseteq \BQP$.}
Such an approach, if carried out, would establish the conclusion of \Cref{thm:informal:main} 
without the additional assumption $\UP \nsubseteq \BQP$, thereby resolving the quantum
Bennett--Gill question. However, replacing this non-black-box use with a black-box construction
appears challenging. Our investigation of this obstacle leads to the
impossibility result below, which provides
partial evidence that the non-black-box nature of \Cref{thm:informal:main} seems
essential.

\para{Black-Box Impossibilities for Cryptographic Attempts.} For lower bounds, an ideal result would be to rule out every
construction of a quantum-hard language in
$\NP^{\Ora}\cap\coNP^{\Ora}$ based solely on a 
random oracle. 
However, such a result would seem too ambitious at present.   Instead, we consider a cryptographic analogue: can we establish an
impossibility result when the quantum random oracle is replaced by a
cryptographic primitive capturing a comparable form of unstructured
hardness?
Such a result would still provide useful insight into the limitations of black-box
approaches. 

Injective one-way functions are a natural candidate.
As discussed earlier, inj-OWFs can be viewed as an
unstructured source of hardness.
Moreover, the analogous question has been studied in the classical
setting: Bitansky, Degwekar, and Vaikuntanathan~\cite{C:BitDegVai17}
ruled out fully black-box constructions of hard languages in
$\NP\cap\coNP$ from inj-OWFs, even with additional access
to $\iO$.  
We establish a corresponding impossibility result in the quantum
setting. 
    \begin{theorem}\label{thm:informa:impossibility}
      There is no fully black-box\footnote{Strictly speaking, what we rule out is slightly more general than the traditional definition of fully black-box constructions, due to the non-black-box use of obfuscating circuits with injOWF-gates. This is similar to \cite{C:BitDegVai17}.} quantum reduction of any (even worst-case) quantum hard problem in $\NP \cap \coNP$ to post-quantum $\iO$ and post-quantum inj-OWFs. {Consequently, there is no such reduction to the following post-quantum cryptographic primitives: public-key encryption, oblivious transfer, deniable encryption, and functional encryption.}  
    \end{theorem}

\para{Strong PRHF and a Unified Reduction for QROM NIZK.} An important technical ingredient for \Cref{thm:informal:main} is a
non-interactive zero-knowledge (NIZK) proof for $\NP$ in the quantum random oracle model (QROM),
without additional computational assumptions.
As discussed in the classical setting~\cite{STOC:GIKKLS23}, obtaining such a proof system
is challenging: standard approaches either rely on structured assumptions
beyond random oracles or provide soundness only against polynomial-time
provers.\footnote{A system with soundness only against computationally
bounded provers is called a NIZK \emph{argument}, rather than a NIZK
proof.}
Constructing NIZK proofs for $\NP$ in the (Q)ROM without additional assumptions remains an important open
problem. \cite{STOC:GIKKLS23} bypasses this obstacle using an unbounded honest prover. 
We obtain in \Cref{thm:informal:nizk} its post-quantum analogue.
However, as explained below, our construction relies on new technical
ideas rather than a direct quantization of their techniques.
\begin{theorem}\label{thm:informal:nizk}
      There exists an (unbounded-prover) post-quantum NIZK proof for $\NP$ in the quantum random oracle model.
    \end{theorem}

\para{Strong $\delta$-Dense PRHFs.}
The approach underlying \Cref{thm:informal:nizk} constitutes one of
our main technical contributions.
\cite{STOC:GIKKLS23}'s results are obtained
by introducing a cryptographic primitive called a
\emph{$\delta$-dense pseudorandom hash function} (PRHF).
Their approach has two steps:
(i) constructing the desired NIZK from a PRHF, and
(ii) showing that a random oracle satisfies PRHF security. There is, however, a subtle mismatch between these components.
The step-(i) reduction from NIZK to PRHF requires a non-uniform PRHFs secure against ``function-dependent''
adversaries.
In contrast, their step-(ii) proof of a random oracle's PRHF properties  provides
only a uniform guarantee.
Consequently, directly substituting the random oracle for the PRHF
in the NIZK construction can introduce inefficient ``oracle-dependent'' preprocessing
that render the security reduction inefficient. Therefore,
the two steps do not compose directly.
To address this issue, \cite{STOC:GIKKLS23} provides a separate
security proof for the NIZK construction in the random oracle model,
in addition to its analysis in the PRHF setting.
Indeed, \cite[Remark~7.6]{STOC:GIKKLS23} leaves open whether a
random oracle satisfies PRHF security against oracle-dependent
adversaries.

Rather than attacking this open question, we take a different approach.
Our starting point is a change in perspective:
instead of treating the mismatch solely as a limitation of existing
techniques for handling oracle-dependent adversaries, we view it as
an issue with the interface provided by the PRHF definition.
This suggests seeking a strengthened notion that both admits a
random-oracle instantiation and supports a NIZK security reduction
without the problematic oracle-dependent preprocessing. Following this perspective, we introduce
\emph{strong $\delta$-dense PRHFs}.
We show that a random oracle is a strong PRHF and that a strong PRHF
suffices to construct the desired NIZK through a unified security
reduction,  independently of how the strong
PRHF is instantiated. Moreover, we establish both components in the quantum setting,
yielding \Cref{thm:informal:nizk} and demonstrating the robustness
of the strengthened PRHF notion. More details are provided in \Cref{sec:tech-overview}. We believe that the strong PRHF is a cleaner and more natural primitive that will find further applications in cryptographic constructions.

\subsection{Related Work}

It is also worth noting that there do exist other candidates for quantum hardness in $\NP \cap \coNP$, beyond the number-theoretic problems mentioned earlier.
Parity games, simple stochastic games, and related graph-game
problems are important representatives
\cite{Condon92,STOC:CJKLS17}.
These problems have motivated substantial algorithmic research;
notably, parity games admit a quasipolynomial-time
algorithm~\cite{STOC:CJKLS17}.
Nevertheless, no polynomial-time classical or quantum algorithms
are known for these general game problems.
Their unresolved complexity supplies candidates for hardness, but
a different kind of evidence from the
number-theoretic assumptions \emph{traditionally used in cryptography}.
In particular, proposing that one of these games is quantum-hard
does not explain whether such hardness follows from a generic
post-quantum cryptographic assumption.
The evidence for quantum hardness in
$\NP\cap\coNP$ is consequently narrower than the
classical picture might initially suggest.

Another route to quantum hardness in $\NP\cap\coNP$ proceeds through classical derandomization. Under suitable circuit-hardness assumptions, $\mathsf{AM}=\NP$ and hence $\mathsf{AM}\cap\mathsf{coAM}=\NP\cap\coNP$~\cite{MV05}. Since $\mathsf{SZK}\subseteq\mathsf{AM}\cap\mathsf{coAM}$~\cite{STOC:AhaTaS03}, any language in $\mathsf{SZK}\setminus\BQP$, together with such a derandomization assumption, would yield a language in $(\NP\cap\coNP)\setminus\BQP$. Graph non-isomorphism is a natural candidate. Other candidate languages in $\mathsf{SZK}$ include the decision versions of tensor isomorphism and code equivalence over finite fields, problems whose search variants have been studied for  post-quantum cryptography purpose~\cite{GGPS23,DMR11}. 

Combined with classical derandomization, these assumptions therefore provide another route to quantum hardness in $\NP\cap\coNP$. This route, however, starts from specific algebraic problems already admitting statistical zero-knowledge proofs and invokes an additional derandomization assumption. Our focus is instead on deriving two-sided classical certification together with quantum hardness from unstructured sources.

\section{Technical Overview}
\label{sec:tech-overview}

Our starting point is the work of \cite{STOC:GIKKLS23}: 
if $\UP \nsubseteq \RP$, then $\NP^O \cap \coNP^O  \nsubseteq \P^O$ with probability~$1$ over a random oracle
$O$.
We first recall the main ingredients of their construction, emphasizing
the aspects relevant to extending their approach to the quantum setting.

\para{From a hard promise problem to a hard language.}
Fix a language $\Lang_{\msf{unique}} \in \UP \setminus \RP$, and let
$R_{\msf{unique}}$ be its polynomial-time witness relation. Thus, every
$x \in \Lang_{\msf{unique}}$ has a unique polynomial-length witness
$w_x$, whereas every $x \notin \Lang_{\msf{unique}}$ has no witness.
For simplicity, we use a fixed polynomial witness length, padding
witnesses uniquely if necessary. Consider the language
\[
    \Lang_{\msf{almost}}
    \coloneqq
    \bigl\{(x,i) \,\big|\,
        \exists w:\ R_{\msf{unique}}(x,w)=1 \wedge w_i=1
    \bigr\}.
\]
Clearly, $\Lang_{\msf{almost}} \in \NP$. Moreover, a polynomial-time
decider for $\Lang_{\msf{almost}}$ would recover the unique witness
bit by bit whenever one exists; checking the recovered witness would
then decide $\Lang_{\msf{unique}}$. Hence,
$\Lang_{\msf{almost}} \notin \P$.

The difficulty is establishing membership in $\coNP$. For a NO-instance
$(x,i)$, there are two cases:
\begin{itemize}
    \item \emph{Good case: $x \in \Lang_{\msf{unique}}$.}
    The unique witness $w_x$ certifies nonmembership by satisfying
    $R_{\msf{unique}}(x,w_x)=1$ and $(w_x)_i=0$.
    Uniqueness rules out a different witness whose $i$th bit is~$1$.

    \item \emph{Problematic case: $x \notin \Lang_{\msf{unique}}$.}
    No witness exists, and there need not be an efficiently verifiable
    certificate of this fact.
\end{itemize}
Thus, the witness-bit problem has both YES- and NO-certificates under
the promise $x \in \Lang_{\msf{unique}}$, but extending this guarantee
to all inputs is nontrivial.

The idea of \cite{STOC:GIKKLS23} is to enforce this promise by appending
an efficiently verifiable proof $\pi$ of
$x \in \Lang_{\msf{unique}}$. Given a deterministic polynomial-time
verifier $V$, define
\[
    \Lang_{\msf{proof},V}
    \coloneqq
    \bigl\{(x,i,\pi) \,\big|\,
        (V(x,\pi)=1)
        \land
        (\exists w~s.t.~ R_{\msf{unique}}(x,w)=1 \wedge w_i=1)
    \bigr\}.
\]
Here, $V$ is fixed for the language, and proofs have polynomially
bounded length. For intuition, suppose that the proof system has
perfect completeness and perfect soundness. An $\NP$ certificate is
simply $w_x$. To certify nonmembership, either the verifier rejects
$\pi$, which can be checked directly, or $w_x$ certifies that its
$i$th bit is~$0$. Soundness ensures that an accepting proof cannot
occur in the problematic case. Consequently,
$\Lang_{\msf{proof},V} \in \NP \cap \coNP$.

Preserving hardness is subtler. If $\pi$ simply reveals $w_x$, then
the construction satisfies completeness and soundness but yields an
easy language. The proof must therefore enforce the promise without
revealing the witness. This motivates using a
\emph{non-interactive zero-knowledge} (NIZK) proof: its efficient
simulator generates proofs without knowing $w_x$.

\para{Why zero knowledge preserves hardness.}
A NIZK construction in the random oracle model is required to have a  simulator, which can  produce both a
simulated proof and simulated oracle responses. Informally, a decider
for $\Lang_{\msf{proof},V}^O$, applied to the simulated proof for each
index $i$, recovers the witness bit by bit. Zero knowledge transfers
the correctness of this recovery from genuine proofs to simulated
ones. Checking the recovered witness gives a one-sided-error
algorithm for $\Lang_{\msf{unique}}$: it always rejects NO-instances
and accepts YES-instances with high probability.

Making this argument rigorous requires more than the usual
simulation argument. The desired conclusion quantifies over random
oracles and all polynomial-time oracle machines. The proof in
\cite{STOC:GIKKLS23} uses a Bennett--Gill lemma~\cite{BG81} to pass
from positive-probability decidability to a suitable fixed oracle
machine, to which zero knowledge can be applied. (We remark that extending this
hardness argument to quantum algorithms is a separate issue from
constructing a quantum-secure NIZK; we return to it in \Cref{sec:tech-overview:separation}.)

There is also a soundness qualification. The construction provides
\emph{adaptive statistical soundness}: except with negligible
probability over the oracle, no false statement of a given length
has an accepting proof. Since these failure probabilities are
summable, the Borel--Cantelli lemma implies that almost every oracle
has only finitely many exceptional statement lengths. For each such
oracle, these lengths can be handled separately when establishing
membership in $\NP^O \cap \coNP^O$. This is why statistical
\emph{proofs}, rather than merely computationally sound arguments,
are essential.

\para{NIZK from unstructured assumptions.}
The remaining challenge is to construct the required NIZK without
structured assumptions. Previously known NIZK proofs in the uniform
reference string model relied on structured primitives, whereas
standard random-oracle constructions such as Fiat--Shamir generally
provide only arguments. Importantly, this application does not
require an efficient honest prover: it requires only an efficient
verifier and simulator. The main technical contribution of
\cite{STOC:GIKKLS23} is an unbounded-prover NIZK proof system for
$\NP$ in the classical random-oracle model.

\subsection{Building NIZKs with an Unbounded Prover:
the Classical Starting Point}

\para{Z-tamperable hidden bits.}
 \cite{STOC:GIKKLS23}   proceeds through the
\emph{Z-tamperable-hidden-bits} model. The prover receives a hidden
string $r=(r_1,\ldots,r_m)$ of independent, possibly biased bits and
may reveal selected positions to the verifier. A malicious prover
may additionally change any $0$ to a $1$, but cannot change a $1$
to a $0$.

To construct NIZK in this model, \cite{STOC:GIKKLS23} adapts the
classical hidden-bits construction for graph Hamiltonicity. The
hidden bits encode many sparse matrices, some of which contain an
$n$-cycle as a submatrix and zeros elsewhere. For each such matrix,
the prover shows that the cycle is contained in a permutation of the
input graph, revealing only entries corresponding to nonedges.
The remaining matrices are fully revealed and checked not to be
of this form.

The new difficulty is that one-sided tampering lets a cheating
prover destroy every cycle matrix by adding a $1$, thereby making
it eligible for disclosure and disposal. The construction counters
this attack with a statistical check on the number of revealed
matrices containing at least $n+1$ ones. Destroying cycle matrices
increases this count. Moreover, the matrices are sufficiently sparse
that almost all matrices already containing more than $n$ ones
occupy more than $n$ rows or columns, and hence cannot be hidden
inside an $n \times n$ submatrix. Thus, the prover cannot offset
the increase by withholding enough other matrices. Concentration
bounds then yield statistical soundness.

\para{Implementing the hidden bits.}
Let $H:\bits^n \to \bits^n$ have image density approximately
$\delta$, for an efficiently approximable constant
$\delta \in (0,1)$. For a uniformly random public value $y$, define
\[
    r(y)=
    \begin{cases}
        0, & y \in \msf{Image}(H) \\
        1, & y \notin \msf{Image}(H)
    \end{cases}.
\]
An unbounded prover can determine $r(y)$ and, when $r(y)=0$, sample
a uniformly random preimage $v$ of $y$. To reveal a zero, it provides
$v$, which the verifier checks by evaluating $H(v)=y$. To reveal a
one, it provides the symbol $\bot$, without proving nonmembership.

This implements precisely the allowed direction of tampering:
a prover may withhold an existing preimage and claim a one, but
cannot produce a preimage of a value outside the image.
Furthermore, independent uniform values $y_1,\ldots,y_m$ induce
independent hidden bits with bias approximately $1-\delta$.
Density and one-sided verifiability therefore allow the soundness
of the hidden-bits protocol to carry over.

\para{The role of pseudorandomness.}
Zero knowledge requires additional properties. In
\cite[Definition~4.6]{STOC:GIKKLS23}, a
\emph{$\delta$-dense pseudorandom hash function} (PRHF) satisfies,
in addition to image density, the following:
\begin{itemize}
    \item \emph{Output pseudorandomness:}
    $H(U_n)$ is computationally indistinguishable from $U_n$.

    \item \emph{Preimage pseudorandomness:}
    the following distributions are computationally indistinguishable:
    \[
        (v,H(v)),\quad v \gets \bits^n,
        \qquad\text{and}\qquad
        (v,y),\quad
        y \gets \msf{Image}(H),\
        v \gets H^{-1}(y).
    \]
\end{itemize}
Both properties are required against non-uniform polynomial-size
adversaries. The second property is not automatic from the first:
sampling a uniform input weights each output according to its
number of preimages, whereas sampling a uniform image element does
not.

These properties enable a particularly simple simulator in the
uniform reference string (URS) model. First, run the hidden-bits simulator
to obtain the revealed bits and the associated proof. Independently
sample uniform $v_1,\ldots,v_m$ and set $y_i=H(v_i)$. For each revealed
zero, provide $v_i$; for each revealed one, provide $\bot$. Thus,
the simulator knows a preimage for every public value, including
those presented as having no preimage.

Why is this indistinguishable from an honest execution? For a
revealed zero, preimage pseudorandomness justifies replacing a
uniform image element and its uniform preimage with
$(H(v_i),v_i)$. For a one, the honest public value is uniform
outside the image. Output pseudorandomness and preimage
pseudorandomness together imply that this distribution is also
indistinguishable from $H(U_n)$: the uniform distribution is a
constant-weight mixture of the uniform image and nonimage
distributions, and the uniform image distribution is already
indistinguishable from $H(U_n)$.
After these replacements, the public values are generated
independently of the hidden bits, allowing the hidden-bits
zero-knowledge property to complete the argument.

\para{PRHF  does not directly give a
random-oracle NIZK.}
A length-preserving random function has image density approximately
$1-e^{-1}$ and satisfies the required pseudorandomness properties
against polynomial-query adversaries; see
\cite[Appendix~B]{STOC:GIKKLS23}.
One might therefore hope simply to instantiate the PRHF with a
random oracle. However, the security guarantees needed by the
above reduction do not match those established for the oracle.

The PRHF-to-NIZK proof uses non-uniform hybrid reductions to
hardwire intermediate values that need not be efficiently
computable, including hidden-bit transcripts and samples
conditioned on image membership. For a fixed explicit function
$H$, this is permitted by the PRHF definition. If $H$ is replaced
by a random oracle, however, the corresponding advice may depend
on the sampled oracle itself. The reduction would therefore
require preimage pseudorandomness against adversaries equipped
with oracle-dependent advice, potentially obtained by unbounded
preprocessing of the oracle. \cite{STOC:GIKKLS23} does not establish this
stronger guarantee. \emph{The distinction is oracle-dependent advice,
not merely whether an adversary is uniform or non-uniform}.

As explained in \cite[Remark~7.6]{STOC:GIKKLS23}, this mismatch
prevents a direct composition of the two results. Instead, the
authors prove zero knowledge separately in the classical
random-oracle model. Their simulator samples public values,
programs preimages for the simulated zeros, and answers subsequent
oracle queries by lazy sampling. A direct hybrid argument analyzes
the resulting joint distribution without invoking PRHF security
with oracle-dependent advice. This proves security against
classical polynomial-query adversaries, but does not by itself
establish security for quantum superposition queries. Consequently,
both this simulation step and the subsequent hardness argument
must be addressed when extending the approach to the quantum
setting.

\subsection{Strong $\delta$-Dense PRHFs and Their Application to NIZK}

Our starting point is a change in perspective: rather than overcoming
oracle-dependent non-uniformity by proving security against stronger
adversaries, we redesign the interface provided by the PRHF definition.
The original reduction introduces oracle-dependent auxiliary information
when it replaces individual samples while fixing the remaining
transcript. We instead formulate a security experiment that exposes
the entire polynomial-size collection of samples from the outset.
This leads to \emph{strong $\delta$-dense PRHFs}.

Our approach simultaneously removes the need for oracle-dependent
advice and extends the primitive to the post-quantum setting.
We establish two complementary results: strong post-quantum PRHFs
suffice for the NIZK construction through a uniform, black-box
reduction, and a quantum random oracle instantiates this strengthened
notion. Consequently, the random-oracle instantiation follows from
the same abstract reduction, without a separate random-oracle-specific
proof of the zero-knowledge property.

\para{Strong preimage pseudorandomness.}\label{para:strong-preimage-pseudorandomness}
Our new notion of strong $\delta$-dense PRHFs retains the density and pseudorandomness requirements:
the image of $H$ occupies approximately a $\delta$ fraction of its
codomain, and $H(v)$ for uniform $v$ is pseudorandom against quantum
adversaries with oracle access to $H$.
The principal change is the third requirement,
\emph{strong preimage pseudorandomness}.

Let $H:\bits^n\to\bits^{\ell(n)}$, let $p(n)$ be a polynomial, and set
$\gamma=1-\delta$. Consider two distributions that both expose
\[
    T \coloneqq \left(
        r = (r_1,\ldots,r_{p(n)}),~\{y_i\}_{i\in[p(n)]},~
        \{v_i\}_{i\in\mcal{I}_0}
    \right),
\] where $\mcal{I}_0\coloneqq \Set{i ~\vert~ i \in [p(n)]~s.t.~ r_i=0}$.  
In both distributions, the bits $r_i$ are sampled independently
with probability $\gamma$ of being one.
The distributions differ in how they generate the remaining samples:
\begin{itemize}
    \item \textbf{Backward sampling ($\mathcal{D}_0^H$).}
    If $r_i=0$, sample $y_i$ uniformly from $\msf{Image}(H)$
    and then sample $v_i$ uniformly from $\msf{PreImage}(y_i)$.
    If $r_i=1$, sample $y_i$ uniformly from
    $\bits^{\ell(n)}\setminus\msf{Image}(H)$.\footnote{If the set $\bits^{\ell(n)} \setminus \msf{Image}(H)$ is empty, we simply set $y_i$ to a fixed value (e.g., $0^{\ell(n)}$ or $\bot$). We adopt this convention whenever sampling from an empty set. This does not affect any of our results, since the empty-set case occurs with only negligible probability throughout the paper.} 
     This distribution need not be efficiently
    sampleable.

    \item \textbf{Forward sampling ($\mathcal{D}_1^H$).}
    Independently sample $v_i\picks\bits^n$ and set $y_i=H(v_i)$
    for every $i$. 
\end{itemize}
Let $\mcal{H}_n$ be the family of $H$ with security parameter $n$. Strong preimage pseudorandomness requires
\[
    \left|
        \Pr[\Adv^{\ket{H}}(T)=1~:~
            \begin{array}{l}
                H \gets \mcal{H}_n, \\
                T\gets\mathcal{D}^H_0
            \end{array}]
        -
        \Pr[\Adv^{\ket{H}}(T)=1~:~
            \begin{array}{l}
                H \gets \mcal{H}_n, \\
                T\gets\mathcal{D}^H_1
            \end{array}]
    \right|
    \leq\negl(n),
\]
where $\ket{H}$ indicates that $\Adv$ has quantum access to $H$ as an oracle.\footnote{Here, we model $H$ as an oracle because it is ultimately instantiated as a random oracle in our main theorem. Alternatively, $H$ may be treated as a plain-model primitive that could potentially be instantiated under algebraic assumptions in future work. In that case, the description of $H$ should be provided to $\Adv$ in the security definition.}    
We require security against  polynomial-query
quantum adversaries $\Adv$, allowing oracle-independent auxiliary input. 
Their computation between oracle queries need not be efficient;
this distinction is essential because our NIZK application invokes an
unbounded prover.

The property replaces the entire collection at once, even though
the adversary sees all hidden bits, receives preimages at every
zero position, and can query $H$ in superposition.
In particular, it accommodates subsequent computations depending
on the full hidden-bit string without fixing their outputs as
oracle-dependent advice.

\para{From strong PRHFs to NIZK with URS.}
We retain the construction of \cite{STOC:GIKKLS23}, but give a
different zero-knowledge reduction.
Recall that a Z-tamperable hidden-bits (ZHB) proof allows the prover
to reveal selected hidden bits and to change a zero into a one,
but not conversely.
To implement these bits, parse the uniform reference string as
$y_1,\ldots,y_{p_{\msf{hbs}}(n)}$ and define
\[
    r_i=
    \begin{cases}
        0 & \text{if }y_i\in\msf{Image}(H),\\
        1 & \text{otherwise}.
    \end{cases}
\]
The unbounded prover computes $r$ and runs
$\msf{ZHB.P}_{\gamma}(r,x)$, obtaining an index set $I$ and a
ZHB proof. It opens a zero by providing a uniformly sampled
preimage of $y_i$, and opens a one by providing $\bot$.
The verifier checks the supplied preimages and invokes the
ZHB verifier on the resulting bits.
A prover can always replace a preimage by $\bot$, but cannot
provide a preimage of a point outside the image. This realizes
exactly the permitted zero-to-one tampering.
Completeness and soundness then follow from the corresponding
ZHB properties and the density guarantee.

For zero knowledge, first rewrite the real experiment by sampling
the hidden bits before the public values. Conditioned on $H$,
the probability that $r_i=1$ is
$1-|\msf{Image}(H)|/2^{\ell(n)}$.
The density guarantee allows us to replace this probability by
$1- \delta$  at negligible statistical cost.
The resulting samples are distributed as $\mathcal{D}_0$ (i.e., the \textbf{Backward sampling} defined above).

We now apply strong preimage pseudorandomness \emph{once}, to
replace $\mathcal{D}_0$ by $\mathcal{D}_1$.
The reduction receives the complete challenge transcript,
runs $\msf{ZHB.P}_{\gamma}(r,x)$, retains the openings selected
by its output, and passes the resulting URS transcript to the
distinguisher. Although the ZHB prover may be computationally
unbounded, it makes no queries to $H$; hence the reduction
remains polynomial-query-bounded.
Unlike the coordinate-by-coordinate argument of
\cite{STOC:GIKKLS23}, this reduction neither fixes the ZHB proof
nor supplies the remaining samples as oracle-dependent advice.

After the replacement, all pairs $(v_i,H(v_i))$ are generated
independently of $r$. The final transcript therefore depends on
the ZHB execution only through its revealed bits and proof.
Statistical zero knowledge of the ZHB system now allows us to
replace these by the output of $\msf{ZHB.Sim}_{\gamma}(x)$.
This gives an efficient simulator: generate the simulated ZHB
transcript, sample all pairs forward, and reveal the preimages
selected by its zero bits.

The reduction is genuinely black-box in $H$, independently
of its implementation.
In \Cref{sec:tech-overview:RO-PRHF}, we establish
that a random oracle is a strong $(1-e^{-1})$-dense post-quantum
PRHF, which represents our another technical contribution.
 Together, these results yield a URS-model NIZK relative to a
quantum random oracle. 

\para{Eliminating the URS.}
To obtain the eventual QROM-only NIZK, we need to remove the URS. Our strategy is to
derive the URS from an
independent oracle $O_{\msf{URS}}$, while using
$O_{\msf{PRHF}}$ for the PRHF.
The simulator must answer superposition queries consistently
with the simulated URS and its correlations with remaining entries of the oracle(s); and classical lazy sampling does not suffice in the quantum setting.
We formalize a \emph{compressed partially-fixed oracle},
$\operatorname{Com}(H)$, which perfectly emulates a random oracle
conditioned on a prescribed partial function $H$.
It stores the fixed entries explicitly and compresses the
remaining entries using the compressed-oracle technique
\cite{C:Zhandry19}.
Our simulator runs the URS simulator using a persistent
compressed PRHF oracle, then initializes the URS oracle
conditioned on the resulting URS.
Retaining both oracle states preserves the required correlations
throughout subsequent quantum interaction, with polynomial
overhead. This step implements the simulated experiment
efficiently.

\para{Combining the oracles.}
Finally, we combine the two independent oracles into one by domain
separation:
$O_{\msf{URS}}(z)=O(0\|z)$ and
$O_{\msf{PRHF}}(z)=O(1\|z)$.
A quantum adversary may query a superposition of the two
prefixes, so separate oracle access does not immediately
give the required combined interface.
We simulate each combined query exactly using four separate
queries: compute each oracle's answer into an ancilla,
conditionally XOR it into the answer register, and uncompute.
This works even when the prefix is entangled with the
adversary's workspace.
Thus, the two-oracle security argument transfers to a single
quantum random oracle, completing the construction.

\subsection{Completing the Separation}
\label{sec:tech-overview:separation}

With our QROM NIZK in hand, we complete the separation by adapting
the argument of \cite{STOC:GIKKLS23}.
We first instantiate the verifier in
$\Lang_{\msf{proof},V}^{\Ora}$ with our NIZK verifier.
The certificate argument remains classical: adaptive statistical
soundness, together with the Borel--Cantelli lemma, ensures that
$\Lang_{\msf{proof},V}^{\Ora}\in\NP^{\Ora}\cap\coNP^{\Ora}$
for almost every oracle.
The remaining task is to rule out quantum polynomial-time deciders.
Suppose a fixed quantum oracle machine decides this language with
bounded error on every input for at least a $2/3$ fraction of oracles.
Given an accepting proof for $x\in\Lang_{\msf{unique}}$, we can
use this machine to recover the unique witness bit by bit.
Replacing the genuine proof and oracle access by our efficient
NIZK simulator preserves each bit's recovery probability by
quantum zero knowledge.
Repeating with independent simulator instances and taking majority
votes therefore recovers the entire witness with high probability.
Checking the recovered witness ensures that NO-instances are always
rejected, yielding an oracle-free $\BQP$ algorithm for
$\Lang_{\msf{unique}}$. This contradicts
$\Lang_{\msf{unique}}\notin\BQP$.

To pass from this high-probability bound to an almost-sure
separation, we give a quantum formulation of \cite[Lemma 1]{BG81}.
Its structural requirements hold here: bounded witness enumeration
gives a \emph{total}\footnote{
An oracle machine is total if it halts on every input for every
oracle. No polynomial-time bound is required.} oracle decider, and the verifier's input-length-separated
oracle access ensures that changing finitely many oracle bits affects
only finitely many language instances.
The standard finite-prefix hardwiring and finite-input correction
argument from \cite{BG81} thus applies to quantum machines as well, showing that
each fixed machine decides the language only on a measure-zero set
of oracles.
A countable union over machines completes the proof that, assuming
$\UP\nsubseteq\BQP$, we have
$\NP^{\Ora}\cap\coNP^{\Ora}\not\subseteq\BQP^{\Ora}$
with probability~$1$ over $\Ora$.

\subsection{A Random Oracle is a Strong PRHF}
\label{sec:tech-overview:RO-PRHF}

We show that a uniformly random function
\(H:\bits^n\to\bits^n\) is a strong post-quantum
\(\delta\)-dense PRHF for \(\delta=1-e^{-1}\). The \(\delta\)-density property follows from
\cite{STOC:GIKKLS23}, while  image pseudorandomness
against quantum queries follows from a standard one-way-to-hiding (O2H)  argument \cite{C:AmbHamUnr19}. We focus on the new strong preimage
pseudorandomness (recall the definition on \Cpageref{para:strong-preimage-pseudorandomness}): for every polynomial \(p(n)\),
backward sampling \(\mathcal{D}_0\) and forward sampling
\(\mathcal{D}_1\) produce  indistinguishable
collections 
\[
    T \coloneqq \left(
        r = (r_1,\ldots,r_{p(n)}),~\{y_i\}_{i\in[p(n)]},~
        \{v_i\}_{i\in\mcal{I}_0}
    \right).
\]
even when the distinguisher makes polynomially-many \emph{quantum} queries to \(H\).

Recall that backward sampling chooses \(y_i\) uniformly from
\(\msf{Image}(H)\) and then a uniform preimage \(v_i\) when
\(r_i=0\), and chooses \(y_i\) outside the image when \(r_i=1\).
Forward sampling instead chooses a uniform \(v_i\) and sets
\(y_i=H(v_i)\) at every position, revealing \(v_i\) only when
\(r_i=0\). We fix \(r\) throughout the overview, and let $\mcal{I}_0$ and $\mcal{I}_1$ denote the sets of indices where $r_i=0$ and $r_i=1$, respectively; (note that $\mcal{I}_0 \cup \mcal{I}_1 = [p(n)]$). All bounds are
uniform over its choice and therefore also hold after averaging
over \(r\). We suppress negligible collision and exceptional-size
events, which our formal proof in \Cref{sec:random-oracle-is-strong-dense-prhf} handles explicitly.

The difficulty is that \(T\) and the oracle available to the
adversary depend on the same random function. Replacing one sample
while fixing the others risks introducing oracle-dependent advice.
We instead transform the entire collection and the queried oracle
jointly. This is not a direct application of quantum tools, e.g.,  the O2H lemma~\cite{C:AmbHamUnr19}.
Although such tools can bound the effect of changing an oracle on a set of
inputs, its usefulness depends on bounding the adversary's
probability of querying that set. Here, the information in \(T\)
is correlated with the locations we wish to change. Non-trivial ideas and efforts are required to bound the query-hitting probability.
% The
% proof needs to first change the joint distribution so that this
% query-hitting probability can actually be bounded.

We build a sequence of hybrids, where we gradually switch from the backward-sampling distribution to the forward-sampling distribution. In the following, we provide a high-level overview of the main steps in the proof, while deferring the technical details to \Cref{sec:random-oracle-is-strong-dense-prhf}.

\para{Generating outputs outside the image by modifying the oracle.} We start from the backward-sampling distribution $\mcal{D}_0$.
We first change how the outputs at positions \(i \in \mcal{I}_1\) are
generated. Rather than choosing \(y_i\) outside the image, we sample \(v_i\picks \bits^n\), set \(y_i=H(v_i)\), and keep
\(v_i\) hidden. Then, we modify every entry of the oracle $H$ whose value
is one of these selected outputs, replacing its value with fresh
randomness outside the selected outputs. Consequently, these
\(y_i\)'s are absent from the image of the modified oracle, as
required by backward sampling.  {To distinguish it from the orignial oracle \(H\), we denote this modified oracle as \(H^*\); also note that $H^*$ is now the oracle supplied to the adversary.} 

To see why this modification is safe, notice that the number of outputs missing from a random function's image is
sharply concentrated. Thus, choosing a random function and then
uniform missing outputs is statistically close to choosing the
outputs first and then a random function avoiding them.
Meanwhile, polynomially many forward evaluations produce almost
uniform distinct outputs. Conditional on the locations being
modified, the other entries already avoid those outputs;
resampling the modified entries yields the corresponding
distribution of an oracle avoiding them.

{We must also account for the $(v_i, y_i)$ pairs for each $i \in \mcal{I}_0$:
these pairs were sampled backward from the original oracle,
but the adversary now receives the modified oracle.
Their sampling distribution need not coincide with backward
sampling from that modified oracle, since the modification
can change both its image and its preimage sets.
We show that this discrepancy has negligible statistical cost.
Only a small number of image elements and preimage sets are
affected, and a uniformly selected image element encounters
these changes with negligible probability.
This establishes closeness of the joint distributions of \(T\)
and the modified oracle, rather than merely closeness of the
selected outputs.}
After this step, every \(y_i\) has a designated input \(v_i\)
in the original oracle, but those with \(i\in \mcal{I}_1\) remain hidden.

\para{Canceling the global bias introduced by backward sampling.} {Next, for all $x$ satysfying $H(x)\notin\{y_i\}_{i\in[p(n)]}$, we modify the value of \(H^*(x)\) to an independent uniform value
outside \(\{y_i\}_{i\in[p(n)]}\). 
That is, we resample entries of the constructed oracle $H^*$, but
determine their locations using the original oracle \(H\).}  {A tempting argument is that, after fixing the selected outputs
and all their preimage sets under \(H\), the remaining entries
of \(H\) are independent and uniform outside those outputs.
Unfortunately, backward sampling prevents this argument from being correct.}

Indeed, conditional on the original oracle being \(H=h\),
the probability of sampling any specified  pairs
\(\{(y_i,v_i)\}_{i\in\mcal I_0}\) consistent with \(h\) is
\[
    |\msf{Image}(h)|^{-|\mcal I_0|}\cdot \prod_{i\in\mcal I_0}
    \frac{1}
    {
     |\msf{PreImage}^{h}(y_i)|}.
\]
Here the image and preimage sets are those of the original
oracle $H=h$, before any modification. Now, consider the distribution of the remaining entries of $h$ after we fix the selected outputs and their complete preimage sets under \(h\):   the preimage-size factors $\prod_{i\in\mcal I_0}
    \frac{1}
    {
     |\msf{PreImage}^{h}(y_i)|}$ are fixed, but
the factor
\(
    |\msf{Image}(h)|^{-|\mcal I_0|}
\)
can still vary across completions of the remaining oracle
entries. These completions therefore need not have equal
conditional probabilities: backward sampling introduces a
global bias into the unexposed part of the original table.

We cancel this bias by temporarily changing the distribution
of the original oracle.
We introduce a new oracle \(\widetilde H\) that plays the same role as \(H\), but have the following probability density (and all subsequent sampling and oracle-construction steps remain
unchanged, using \(\widetilde H\) in place of \(H\)):
\[
    \Pr[\widetilde H=h]
    =
    \frac{|\msf{Image}(h)|^{|\mcal{I}_0|}}
    {\displaystyle
        \sum_{f:\bits^n\to\bits^n}
        |\msf{Image}(f)|^{|\mcal{I}_0|}}.
\]
The additional weight cancels the image-size factor from
backward sampling exactly. Once the selected outputs, their
designated inputs, and all entries mapping to them are fixed,
the factors arising from choosing uniform preimages are also
fixed. The remaining entries are now independent uniform values
outside the selected outputs, so they can be replaced by fresh
randomness.

We show that this change of distribution has negligible
statistical cost. The image size is sharply concentrated
around \((1-e^{-1})2^n\), and its polynomial power varies
by only a negligible relative amount on typical tables. Notice that conditioned on
\[
    \{(v_i,y_i)\}_{i\in[p(n)]}
    \quad\text{and}\quad
    \msf{PreImage}^{\widetilde H}(\Set{y_i}_{i\in[p(n)]}),
\]
the entries \(\widetilde H(x)\) at inputs
\(
    x\notin \msf{PreImage}^{\widetilde H}(\Set{y_i}_{i \in [p(n)]})
\)
are then independent and uniform over
\(\bits^n\setminus\{y_i:i\in[p(n)]\}\).
At these inputs, the constructed oracle \(H^*\) still agrees
with \(\widetilde H\).
Replacing these entries of \(H^*\) by fresh independent
values from the same distribution therefore leaves the
joint distribution of \(T\) and \(H^*\) unchanged.
We can then return to the  original uniform oracle $H$
at negligible statistical cost. That is, the $\tilde{H}$ is only an intermediate tool helping us to analyze the statistical closeness. 
In the remainder of the overview, \(H\) again denotes this
uniformly sampled original oracle, while \(H^*\) denotes
the constructed oracle supplied to the adversary.

\para{Making the additional preimage locations hidden.} Having refreshed the entries at inputs \(x\) satisfying
\(H(x)\notin\{y_i:i\in[p(n)]\}\), we next consider the
preimages of \(y_1,\ldots,y_{p(n)}\) under \(H\).
For each \(i\), the input \(v_i\) is one preimage of \(y_i\),
but \(y_i\) may also have other preimages.
We now modify \(H^*\) at all such preimages except
\(v_1,\ldots,v_{p(n)}\); that is, at inputs in
\begin{equation}\label{eq:H*differing}
        S = \msf{PreImage}^{H}(\Set{y_i}_{i\in[p(n)]})
    \setminus\{v_1,\ldots,v_{p(n)}\}.
\end{equation}
Previously, for an input \(x\) in this set with \(H(x)=y_i\),
\(H^*(x)\) was \(y_i\) for \(i\in\mcal{I}_0\), and an independent
uniform value in
\(\bits^n\setminus\{y_j:j\in\mcal I_1\}\) for  \(i \in \mcal{I}_1\).
We now replace \(H^*(x)\) in both cases by an independent
uniform value in
\(\bits^n\setminus\{y_j:j\in[p(n)]\}\). We leave the entries at \(v_1,\ldots,v_{p(n)}\) unchanged:
\(H^*(v_i)=y_i\) for $i\in \mcal{I}_0$, while \(H^*(v_i)\)
remains a previously sampled uniform value in
\(\bits^n\setminus\{y_j:j\in\mcal I_1\}\) for $i\in \mcal{I}_1$.

Thus, the versions of \(H^*\) before and after this
modification can differ only at inputs in the set $S$ defined in \Cref{eq:H*differing}. We plan to use O2H to argue that the adversary cannot distinguish these two versions. To do so, we need to bound the probability that a query hits \(S\).

Let \(\mcal{B}\) be the \emph{good} event that (i) the selected outputs $\Set{y_i}_{i\in[p(n)]}$ are
distinct, (ii) every output of \(H\) has at most \(n\) preimages,
and (iii) \(|\msf{Image}(H)|\geq 2^{n-1}\).
It follows from standard collision bounds and  random-function
concentration bounds that \(\Pr[\mcal{B}]= 1- \negl(n)\).
Conditioned on event \(\mcal{B}\), it hold that
\(|S|\leq p(n)n\). But this size bound alone is
insufficient for O2H: a small set can still be easy to query
if the adversary's input reveals its location. We instead
prove that \(S\) is hidden in the particular experiment used
to bound the query-hitting probability.

{Consider the new experiment with the modified $H^*$.} For analysis, condition
on  the good event \(\mcal{B}\) and all pairs \(\{(v_i,y_i)\}_{i\in[p(n)]}\), including the inputs
with \(i\in \mcal{I}_1\) that are not revealed to the adversary.
Once these pairs are fixed, the $H^*$ depends only
on fresh randomness, whereas \(S\) depends on the original
$H$. The adversary's interaction with $H^*$
is therefore conditionally independent of \(S\).

Moreover, the original sampling procedure is invariant under
permutations of the domain fixing all designated inputs $\Set{v_i}_{i\in[p(n)]}$.
Conditional on the designated pairs, the good event $\mcal{B}$, and
\(|S|=s\), the set \(S\) is consequently a uniform \(s\)-element
subset of the remaining domain. Hence a measured query in
this experiment hits \(S\) with probability at most
\[
    \Pr[\text{measured query lies in }S]
    \leq \frac{p(n)n}{2^n-p(n)}.
\]
O2H now bounds the distinguishing advantage of a \(q\)-query
adversary by
\[
    2q\sqrt{\frac{p(n)n}{2^n-p(n)}}+\negl(n)
    =\negl(n).
\]

The order of these arguments is essential. We first change
the joint distribution to obtain conditional independence,
then use symmetry to bound the query-hitting probability,
and only then apply O2H. We do not infer that a hidden-set
hitting probability is preserved merely because two
distributions of the adversary's visible information are close.
Also, conditioning on the designated pairs is only an
analytical tool: no reduction receives them as
oracle-dependent advice.

\para{Identifying backward and forward sampling after resampling.}
The constructed oracle $H^*$ now depends only on the designated
pairs and fresh randomness. It no longer exposes information
about additional preimages in the original table.

{At this point, we change the sampling only for 
\(i\in \mcal{I}_0\): instead of choosing \(y_i\) uniformly from
\(\msf{Image}(H)\) and then a uniform preimage \(v_i\),
we choose \(v_i\) uniformly from \(\bits^n\) and set
\(y_i=H(v_i)\).
The positions with \(i \in \mcal{I}_1\) already use forward sampling
and remain unchanged.
Symmetry shows that these two ways of generating the
designated pairs become statistically interchangeable
after the preceding oracle modifications.} 
Conditional on distinct designated inputs
and distinct selected outputs, permutations of the domain
and range act transitively on the possible tuples
\(\{(v_i,y_i)\}_{i\in[p(n)]}\). 
Both the mixed sampling procedure before this change and
the all-forward procedure after it are invariant under
these permutations; thus, they induce the same conditional distribution on the tuples.  
Given such a tuple, the constructed oracle \(H^*\) supplied
to the adversary is generated by the same rule in both
experiments. 
Consequently, the two experiments have the same conditional
joint distribution of \(T\) and \(H^*\).
The excluded collision events contribute only negligible
statistical error.

This argument does not identify the original joint
distributions of \(T\) and \(H\).
Rather, it establishes statistical closeness of the joint
distributions of \(T\) and \(H^*\) after the preceding
resampling steps have removed the dependence of \(H^*\)
on the original preimage sets beyond the information
specified by \(\{(v_i,y_i)\}_{i\in[p(n)]}\).

\para{Restoring the original random oracle.}
Now that every pair is generated by forward sampling, we can
undo the remaining oracle modifications using quantum
collision-finding and preimage-search bounds, together with O2H.

First, restore the entries at the undisclosed inputs mapping
to selected outputs. Querying such an input produces a second
input with the same original output as a designated input,
and hence a collision in the original random function.
The reduction generates the designated pairs with polynomially
many queries and simulates the relevant intermediate oracle
by explicitly reprogramming the designated inputs with
\(r_i=1\). Quantum collision-finding bounds make the resulting
query-hitting probability negligible.

Finally, restore the entries at those hidden designated inputs.
Hitting one of them entails finding a preimage of one of the
forward-sampled outputs with \(r_i=1\).
A quantum multi-target preimage-search bound applies to this
distribution. The pairs revealed at positions \(r_i=0\) are
generated by additional oracle queries within the same
reduction, rather than supplied as oracle-dependent advice.
O2H then shows that restoring these last entries is
indistinguishable. The resulting experiment is exactly
\(\mathcal{D}_1\) with access to the unmodified random oracle.

We remark that the main work of this proof is not the final invocation of O2H or
of quantum collision and preimage bounds. It is the
distributional preparation that makes these quantum tools
useful: canceling the global bias of image sampling,
refreshing the unexposed oracle entries, and proving that
the remaining disagreement locations are hidden.
This allows the proof to handle the entire collection
jointly, with bounds depending only on the number of
oracle queries and not on the adversary's intervening
computation.

\subsection{Quantum Black-Box Impossibility via a Universal Decision Oracle}
\label{sec:overview:quantum-impossibility}

We construct an oracle world in which post-quantum injOWFs and
$\iO$ exist, but every $\NP\cap\coNP$ problem produced
by a fully black-box construction is easy.
Besides the \emph{source oracle}, which implements the two
primitives, this world contains a \emph{decision oracle}:
an additional interface that solves the decision problems
arising from such constructions.
Both the adversary and the security reduction may query this
interface, including in superposition.
The challenge is therefore to make it powerful enough to defeat
every proposed construction of $\NP\cap\coNP$ languages without enabling quantum attacks
on the source primitives, i.e.,  injOWFs and $\iO$. 

\paragraph{A powerful decision oracle with short certificates.}
An $\NP\cap\coNP$ problem has two verifiers: one certifies
membership, and the other certifies nonmembership.
For each input, exactly one verifier has an accepting witness.
Following~\cite{C:BitDegVai17}, our oracle $\mathsf{Decide}$
takes such a verifier pair and an input, and returns the side
with an accepting witness.

There is an essential restriction on the pairs it accepts.
Call a source implementation \emph{admissible} if it satisfies
the primitives' functional requirements, such as injectivity
and correct evaluation of obfuscated programs; it need not
satisfy their security requirements.
A verifier pair is valid only if, for every input and
\emph{every admissible source implementation}, exactly one
side has an accepting witness.
We call this property \emph{universal complementarity}.
Invalid pairs receive $\bot$.
The universality matches the functional guarantee of a fully
black-box construction: its output must define an
$\NP\cap\coNP$ problem for every functionally correct
implementation of its sources.
Thus, $\mathsf{Decide}$ solves the constructed problem, but
need not solve arbitrary verifier pairs that happen to be
complementary for the particular sampled oracle.

This restriction also protects the sources.
Fix a valid decision query and one source oracle, called the
\emph{reference oracle}, against which we compare later
modifications.
If the verifiers make at most $r$ source queries, an accepting
witness under the reference oracle uses at most $r$ oracle
answers.
Any modification preserving those answers preserves
acceptance. Universal complementarity prevents the opposite
side from also accepting, so the decision bit cannot change.
Each decision answer therefore has a small \emph{critical set},
even though finding its witness may require unbounded computation.
This stability property is inherited from~\cite{C:BitDegVai17}.

\paragraph{From local stability to quantum security.}
Quantum queries may place amplitude on many inputs simultaneously,
and their amplitudes depend on earlier oracle interactions.
Consequently, an unconditional bound on the probability that
a fixed answer changes is insufficient. To resolve this difficulty, we establish a conditional-coupling lemma, which is similar to the  
quantum hybrid from \cite{EC:Unruh14,FOCS:AmbRosUnr14}.
For coupled oracles $(F_1,F_2)$, suppose that
\[
    \Pr[F_1(u)\neq F_2(u)\mid F_2]\leq\gamma
    \qquad\text{for every reference realization $F_2$
    and input $u$.}
\]
Then, for a $T$-query quantum algorithm whose initial state is
independent of the sampled oracle pair, the expected trace
distance between its final states under $F_1$ and $F_2$ is
at most $2T\sqrt{\gamma}$. 
Conditioning on the entire reference oracle accommodates the
dependence of its query states on that oracle.
We apply this bound to the \emph{combined source and decision
interfaces}; critical sets control disagreement at decision
queries.

As a warm-up, let $f:\{0,1\}^n\to\{0,1\}^{n+1}$ be a random
injection. Sample $y$ uniformly outside its image and plant $y$ at an
independent uniform input $x$, obtaining $g$.
The joint distribution of $(g,x,y)$ is exactly that obtained
by sampling a uniform injection $g$ and an independent
uniform input $x$, and setting $y=g(x)$. 
In contrast, under the reference function $f$, the same
challenge has no preimage.
Conditioned on $f$ and $y$, a fixed decision answer changes
only if $x$ hits its critical set, with probability at most
$r/2^n$ for verifiers making at
most $r$ source queries. 
Checking an inverter's output with one additional query
therefore gives
\[
    \Pr[\text{invert}]
    \leq 2(T+1)\sqrt{r/2^n}.
\]
Thus, the unbounded decision interface does not destroy
one-wayness, even when queried coherently.

\paragraph{Obfuscation with adaptive quantum side information.}
Following~\cite{C:BitDegVai17}, we implement obfuscation by an
injective encoding table mapping a circuit and randomness
$(C,r)$ to a handle $\widehat C$.
An evaluation oracle executes the circuit encoded by that
handle; the obfuscated program is a wrapper calling this
interface. We must also specify evaluation on handles outside the table's
image. A \emph{backup map} $\varphi$ assigns a circuit description
$\varphi(\widehat C)$ to each (possibly malformed) handle $\widehat C$.
If $\widehat C$ is outside the image, the evaluation oracle
answers a query $(\widehat C,z)$ according to
$\varphi(\widehat C)$; otherwise, it evaluates the circuit
encoded by the table.
The admissible oracle family allows every such backup map. This freedom lets us move a challenge handle out of the
encoding table while preserving its visible functionality.

The principal difficulty is adaptive challenge selection:
the adversary chooses $C_0,C_1$ after oracle access and retains
a quantum state correlated with the encoding table.
Reprogramming only the eventual challenge entry does not
establish the required independence.
We use the row-reprogramming architecture
of~\cite{C:BitDegVai17}, analyzing the full joint distribution
of the oracle, challenge, and residual quantum state.

First, we sample the uniform tape $r\in\{0,1\}^n$ that will
be used to generate the challenge handle.
For the pre-challenge execution only, we replace the encoding
table on the entire row $\{(C,r):C\}$ by an ordered tuple $D$
of distinct fresh handles outside the original table's image.
The challenge and post-challenge oracle still use the original
table. 
The coupling hides this modification: each fixed source
query changes with probability $O(2^{-n})$, and critical sets
give an $O(q2^{-n})$ bound for decision queries whose verifiers
make at most $q$ source queries. 
Conditioned on $r$, $D$, and the modified pre-challenge oracle,
the adversary's classical outputs and residual quantum state
are independent of the erased original row.
We can therefore sample the challenge handle $\widehat C$
uniformly outside the original table's image and $D$, and
reprogram the post-challenge table to map $(C_b,r)$ to
$\widehat C$.
This replacement preserves \emph{exactly} the joint distribution
of the challenge bit, pre-challenge output state, challenge
handle, and post-challenge oracle.

Next, we restore only the original pre-challenge oracle,
retaining both the fresh challenge handle and the single-entry
post-challenge reprogramming. 
Closeness of averaged states would not suffice here, because
the subsequent computation depends on the reprogramming
randomness.
Instead, our coupling bound controls the \emph{expected trace
distance of paired executions}. 
More precisely, for each fixed original oracle, let
$\rho_{r,D}$ and $\rho_{\mathrm{ref}}$ be the complete
pre-challenge output states under the modified and original
oracles, respectively.
The bound controls
$\mathbb E_{r,D}[
\operatorname{TD}(\rho_{r,D},\rho_{\mathrm{ref}})]$,
rather than merely the distance between the averaged states.
For each fixed $(r,D,b)$, challenge generation and the
post-challenge execution form the same channel in the two
experiments, so contractivity applies before averaging.
 With the original pre-challenge oracle restored, the
pre-challenge output state is independent of $r$, and the
selected circuits again satisfy $C_0^f\equiv C_1^f$.

{Finally, we undo the single-entry post-challenge reprogramming,
restoring the original table value at $(C_b,r)$, and set
$\varphi(\widehat C)=C_0$.
The fresh handle $\widehat C$ is now outside the table's image,
so it is evaluated through the backup map.
Evaluation at this publicly known handle is unchanged because
$C_b^f\equiv C_0^f$.
The source interfaces can differ only at the encoding query
$(C_b,r)$ and at evaluation queries using its original handle.
These locations are hidden by the uniform tape $r$, allowing
another application of critical sets and the coupling bound.} 
The final experiment is independent of the challenge bit.
Throughout these comparisons, validity is tested against the
same family containing all backup maps, so every modification
remains admissible and the stability argument continues to apply.

\paragraph{Preserving one-wayness and concluding.}
To preserve one-wayness after adding obfuscation, we fix the
encoding and backup tables and coherently simulate their
interfaces using the injOWF-only oracle.
One subtlety is that fixing these tables can make a previously
invalid verifier pair complementary.
The simulation must therefore retain the \emph{original}
universal-validity test before translating decision queries.
This predicate is independent of the sampled function, so its
finite truth table can be hardwired in the query-bounded
simulation.

Our bounds restrict evaluation lengths and verifier resources
in addition to external query counts.
Every efficient construction and reduction meets these
restrictions for some polynomial bound.
The two source primitives thus remain quantum-secure, whereas
one decision query solves the constructed language exactly.
A fully black-box security reduction would turn this solver
into an attack on a source primitive, giving the contradiction.

\section{Preliminaries} \label{prelim}

\para{Notation.}     Let PPT denotes for a probabilistic polynomial time, and QPT denotes for quantum polynomial time. 
     We say that a function $f(n)$ is \emph{negligible} in $n$ if $f(n) = n^{-\omega(1)}$,    
    and we denote this by $f(n)=\negl(n)$.     We say that a function $g(n)$ is \emph{polynomial} in $n$ if
    $g(n)=p(n)$ for some fixed polynomial $p$, and we denote this by
    $g(n)=\poly(n)$.

    When we consider a non-uniform QPT adversary, we often specify it by a
sequence of polynomial-size quantum circuits with quantum advice
$\{\mathcal{A}_\lambda, \rho_\lambda\}_{\lambda \in \mathbb{N}}$.
In an execution with the security parameter $\lambda$, $\mathcal{A}$ runs
$\mathcal{A}_\lambda$ taking $\rho_\lambda$ as the advice.
For simplicity, we often omit the index $\lambda$ and just write
$\mathcal{A}_\rho$ to mean a non-uniform QPT algorithm specified by
$\{\mathcal{A}_\lambda, \rho_\lambda\}_{\lambda \in \mathbb{N}}$.

        If $R$ is a random variable (or distribution), then
       $ r \gets R$ 
    denotes sampling $r$ according to $R$. If $T$ is a finite set, then
        $i \pick T$
    denotes sampling $i$ uniformly at random from $T$. Throughout this work, we take the convention that once we need to sample from an empty set, the output is defined to be some special symbol (e.g., $\bot$ or $0^n$).

        For $t\in\mathbb{N}$, we use $[t]$ to denote the set
        $[t] := \{1,\ldots,t\}$.

 For $x\in\{0,1\}^*$, we write $|x|$ for its bit length.
 For $i\in\mathbb{N}$, we write $\langle i\rangle$ for the standard
    binary representation of $i$ without leading zeros.
     For
    $i\in\{0,\ldots,2^\ell-1\}$,we write $\langle i\rangle_\ell$ denotes its
    $\ell$-bit binary representation, padded with leading zeros when necessary.

    We collect more basic preliminaries in \Cref{sec:prelim-basic}.

\subsection{Random Oracles}\label{Ora}

Let $\mathcal{O}$ denote the distribution over functions
    $O:\bits^*\to\bits$
obtained by independently sampling
    $O(x)\pick\bits$
for every $x\in\bits^*$. We refer to
    $O\gets\mathcal{O}$
as a \emph{random oracle}. Thus, $\mathcal{O}$ denotes the distribution
of random oracles, whereas $O$ denotes a particular oracle sampled from
this distribution.

As in \cite{STOC:GIKKLS23}, a single random oracle
$O:\bits^*\to\bits$ can be viewed as simultaneously providing
independent random functions of arbitrary polynomially bounded input and
output lengths.  The following lemma justifies this interpretation by setting $m(n)=n$. 

\begin{lemma}[{\cite[Lemma 3.7]{STOC:GIKKLS23}}]
\label{lem:random-oracle-decomposition}
    For any function $m:\Naturals \to \Naturals$, a random oracle $O\gets\mathcal{O}$ can be interpreted as a
    countably infinite family of mutually independent random functions
    \[
        \{O_{n,m}:\bits^n\to\bits^{m(n)}\}_{n\in\mathbb{N}}.
    \]
    In particular, for every $n\in\mathbb{N}$, we write $\Ora_n$ as the uniform distribution over all functions from $\bits^n$ to $\bits^n$, where each instance is denoted by $O_n$. 
\end{lemma}
Throughout this paper, we will often need to sample oracles with proper input and output lengths as $O_{n,m} \gets \Ora_{n,m}$. When the input and output lengths are clear from the context, we will simply write $O \gets \Ora$, omitting the subscripts $n$ and $m$.

\para{Oracle access.}
For an oracle algorithm $\mathcal{A}$, we write $\mathcal{A}^{\ket{O}}$ to
indicate that $\mathcal{A}$ is given quantum oracle access to $O$. The quantum access to $O_{n,m}$ is implemented through the
standard unitary
\[
    U_{O_{n,m}} :
    \ket{x}\ket{y}
    \longmapsto
    \ket{x}\ket{y\oplus O_{n,m}(x)},
\]
for $x\in\bits^n$ and $y\in\bits^{m(n)}$. We refer to one application of $U_{O_{n,m}}$ as one
\emph{quantum query} to $O_{n,m}$. Classical oracle access corresponds to
querying this unitary on computational-basis states.

When we write
\[
    \Pr_{O\gets\mathcal{O}}
    \left[
        b=1 :
        b\gets\mathcal{A}^{\ket{O}}(x)
    \right],
\]
the probability is taken over the sampling of $O\gets\mathcal{O}$ as
well as all internal randomness and measurement outcomes of
$\mathcal{A}$.

\subsection{Useful Lemmas}

\begin{lemma}[{\cite[Corollary 4.13]{EC:YamZha21}}]
    For any $n,m,q\in\mathbb{N}$, any fixed $y\in\bits^n$,
    and any quantum algorithm $\mathcal{A}$ making at most $q$
    quantum queries to a random oracle
    $O_n:\bits^m\to\bits^n$, we have
    \[
        \Pr_{O\gets\Ora}\left[
            O_n(x)=y
            \;:\;
            x\gets\mathcal{A}^{\ket{O}}(y)
        \right]
        \leq \frac{(2q+1)^2}{2^n}.
    \]
\end{lemma}
Taking the probability over $y\pick \bits^n$ in the above lemma, we have the following corollary.
\begin{corollary}\label{preimage-uniform}
    For any $n,m,q\in\mathbb{N}$ and any quantum algorithm
    $\mathcal{A}$ making at most $q$ quantum queries to a  random
    oracle $O_n:\bits^m\to \bits^n$, we have
    \[
        \Pr_{O\gets \Ora,y\pick \bits^n}\left[O_n(x)=y\; : x \gets \mathcal{A}^{\ket{O}}(y)\right] \leq \frac{(2q+1)^2}{2^n}.
    \]
\end{corollary}

\begin{lemma}[Preimage Resistance]\label{preimage_search}
    For any $n,m,q\in\mathbb{N}$ and any quantum algorithm
    $\mathcal{A}$ making at most $q$ quantum queries to a  random
    oracle $O_n:\bits^{n}\to \bits^n$, we have
    \[
        \Pr_{O\gets \Ora,x^{'}\pick \bits^n}\left[O_n(x)=O_n(x')\; : x \gets \mathcal{A}^{\ket{O}}(O_n(x'))\right] \leq  {\frac{2n(2q+1)^2+1}{2^n}}.
    \]
\end{lemma}
    
    \begin{proof}
        We first show that, over the random choice of $O\gets\Ora$,
the following event occurs with probability at most $2^{-n}$:
\begin{itemize}
    \item $E$: $\displaystyle
        \max_{y\in\bits^n}
        \left|\msf{PreImage}_{O_n}(y)\right|>2n$
\end{itemize}
Then,
\begin{align*}
    \Pr_{O\gets\Ora}[E]
    &\leq
    \sum_{y\in\bits^n}
    \sum_{\substack{S\subseteq\bits^n\\|S|=2n+1}}
    \Pr_{O\gets\Ora}\left[
        O_n(z)=y\text{ for every }z\in S
    \right]
    \tag{Union Bound}\\
    &=
    \sum_{y\in\bits^n}
    \sum_{\substack{S\subseteq\bits^n\\|S|=2n+1}}
    \prod_{z\in S}\frac{1}{2^n}\\
    &=
    2^n\binom{2^n}{2n+1}2^{-n(2n+1)}\\
    &\leq
    2^n\frac{(2^n)^{2n+1}}{(2n+1)!}
    2^{-n(2n+1)}
    =
    \frac{2^n}{(2n+1)!}\\
    &=
    \frac{2^n}{\prod_{i=1}^{n}(2i)(2i+1)}
    \leq
    \frac{2^n}{4^n}
    =
    \frac{1}{2^n}.\label{eq:fixed-preimage-size-bound}
\end{align*}

    We then observe that, for each fixed oracle $O$ and each
$y\in\bits^n$,
\begin{equation}\label{eq:oracle-image-distribution}
    \Pr_{x'\pick\bits^n}\left[O_n(x')=y\right]
    =
    \frac{\left|\{z\in\bits^n:O_n(z)=y\}\right|}{2^n}.
\end{equation}
    We can now bound the success probability using
\Cref{eq:oracle-image-distribution}:
\begin{align*}
    &\Pr_{O\gets\Ora,\;x'\pick\bits^n}\left[
        O_n(x)=O_n(x')
        \;:\;
        x\gets\mathcal{A}^{\ket O}(O_n(x'))
    \right]\\
    &=
    \mathbb{E}_{O\gets\Ora}\left[
        \sum_{y\in\bits^n}
        \Pr_{x'\pick\bits^n}\left[O_n(x')=y\right]
        \Pr\left[
            O_n(x)=y
            \;:\;
            x\gets\mathcal{A}^{\ket O}(y)
        \right]
    \right]\\
    &\leq
    \frac{2n}{2^n}
    \mathbb{E}_{O\gets\Ora}\left[
        \sum_{y\in\bits^n}
        \Pr\left[
            O_n(x)=y
            \;:\;
            x\gets\mathcal{A}^{\ket O}(y)
        \right]
    \right]
    +\Pr_{O\gets\Ora}[E]
    \tag{\Cref{eq:oracle-image-distribution}+ conditional on $\neg E$}\\
    &=
    2n\,
    \Pr_{O\gets\Ora,\;y\pick\bits^n}\left[
        O_n(x)=y
        \;:\;
        x\gets\mathcal{A}^{\ket O}(y)
    \right]
    +\Pr_{O\gets\Ora}[E]\\
    &\leq
    \frac{2n(2q+1)^2}{2^n}
    +\Pr_{O\gets\Ora}[E]
    \tag{\Cref{preimage-uniform}}\\
    &\leq
    \frac{2n(2q+1)^2+1}{2^n}.
    \tag{\(\Pr_{O\gets\Ora}[E]\leq 2^{-n}\)}
\end{align*}
    This completes the proof of \Cref{preimage_search}.
\end{proof}

\begin{corollary}[Multi-Target Preimage Resistance]\label{multi_preimage_search}
    For any $n,m,k,q\in\mathbb{N}$ and any quantum algorithm
    $\mathcal{A}$ making at most $q$ quantum queries to a  random
    oracle $O_n:\bits^{n}\to \bits^n$, we have
        \begin{align*}
        \Pr\left[
            \exists i\in[k],\; O_n(x)=O_n(x_i)
            \;:\;
            \begin{array}{l}
                O\gets\Ora, ~~ 
                x_1,\ldots,x_k\pick\bits^n, \\
                x\gets\mathcal{A}^{\ket O}
            (O_n(x_1),\ldots,O_n(x_k))
            \end{array}
        \right] \leq
        k\cdot
        \frac{2n\bigl(2(q+k)-1\bigr)^2+1}{2^n}.
    \end{align*}
\end{corollary}
\begin{proof}

    The main idea is to reduce the multi-target preimage search problem to a single-target preimage search problem.
    Assume that there exist a quantum algorithm $\Adv$ that can find a preimage of one of the $k$ targets with high probability ,i.e..
    \begin{align*}
        \Pr\left[
            \exists i\in[k],\; O_n(x)=O_n(x_i)
            \;:\;
            \begin{array}{l}
                O\gets\Ora, ~~ 
                x_1,\ldots,x_k\pick\bits^n, \\
                x\gets\mathcal{A}^{\ket O}
            (O_n(x_1),\ldots,O_n(x_k))
            \end{array}
        \right] >
        k\cdot
        \frac{2n\bigl(2(q+k)-1\bigr)^2+1}{2^n}.
    \end{align*}
    We can construct a new quantum algorithm $\mcal{B}$ making at most $q+k-1$ quantum queries to the oracle $O_n$ that breaks \Cref{preimage_search} as follows:
    \begin{AlgorithmBox}[label=alg:multi_preimage_search]{ Reduction $\mcal{B}$}
        $\mcal{B}(y^*)$
        \begin{enumerate}
            \item Sample index $j\pick[k]$.
            \item Set $y_j=y^*$ , sample $x_i\pick \bits^n$ and set $y_i=O_n(x_i)$ for all $i\in[k]\setminus\{j\}$.
            \item Output $\Adv^{\ket{O}}(y_1,\ldots,y_k)$ 
        \end{enumerate}
    \end{AlgorithmBox}
    Define the index set $\mathcal{I}(x)=\{i: O_n(x) = y_i\}$ which corresponds to the set of indices of the targets that $x$ is a preimage of. 
    Notice that.
    \[\Pr[\Adv \text{ succeeds}]\leq \Pr[\mathcal{I}(x) \neq \emptyset]\]
    We have that since $j$ is sampled uniformly at random from $[k]$ and $\mcal{B} \text{ succeeds}$ if and only if $j\in \mathcal{I}(x)$, we have
    \begin{align*}
        \Pr[\mcal{B} \text{ succeeds}]
        &\geq \Pr[j\in \mathcal{I}(x)| \mathcal{I}(x)\neq \emptyset]\Pr[\mathcal{I}(x)\neq \emptyset]\\
        &\geq \frac{1}{k}\cdot\Pr[\mcal{A} \text{ succeeds}]\\
        &> {\frac{2n\bigl(2(q+k-1)+1\bigr)^2+1}{2^n}}\\
        &> {\frac{2n\bigl(2(q+k)-1\bigr)^2+1}{2^n}},
    \end{align*}
    which contradicts \Cref{preimage_search}. 
    
\end{proof}

\begin{lemma}[{\cite[Theorem 3.1]{QIC:Zhandry15}}]\label{collision_finding} 
    For any $n,m,q\in\mathbb{N}$ and any quantum algorithm
    $\mathcal{A}$ making at most $q$ quantum queries to a  random
    oracle $O_n:\bits^m\to \bits^n$, we have
    \[
        \Pr_{O\gets \Ora}\left[O_n(x)=O_n(x')\; \wedge\; x \neq x' : (x,x') \gets \mathcal{A}^{\lvert O \rangle}\right] \leq O(\frac{q^3}{2^n}).
    \]
\end{lemma}

The following lemma, usually referred to as the one-way-to-hiding lemma, serves as an important tool in the proof of \Cref{lem:strong-preimg-pseudorandomenss}.

\begin{lemma}[{\cite[Theorem 3]{C:AmbHamUnr19}}]
\label{O2H}
    Let $S$ be a random subset of $X$. Let 
    \(
        G,H:X\to Y
    \)
    be arbitrary random functions satisfying
    \(
        G(x)=H(x) 
    \) for all $x\notin S$. 
     Let \(z\) be an arbitrary, possibly random string. ($S$, $G$, $H$, and $z$ may have arbitrary joint distribution.)  

    Let $\mathcal{A}$ be a quantum oracle
    algorithm making at most $q$ quantum queries to its oracle. 
    Define $\mathcal{B}^{H}$ to be the oracle algorithm that, on input
    $z$, proceeds as follows:
    \begin{enumerate}
        \item sample $i\pick[q]$;
        \item run $\mathcal{A}^{H}(z)$ until immediately before its
        $i$-th query;
        \item measure all query-input registers in the computational
        basis and output the set $T$ of measurement outcomes.
    \end{enumerate}

    Define
    \[
        P_{\mathrm{left}}
        :=
        \Pr\!\left[
            b=1 :
            b\gets\mathcal{A}^{H}(z)
        \right],
    \]
    \[
        P_{\mathrm{right}}
        :=
        \Pr\!\left[
            b=1 :
            b\gets\mathcal{A}^{G}(z)
        \right],
    \]
    and
    \[
        P_{\mathrm{guess}}
        :=
        \Pr\!\left[
            S\cap T\neq\varnothing :
            T\gets\mathcal{B}^{H}(z)
        \right].
    \]
 Here, the probabilities are taken over the joint distribution of
    $(S,G,H,z)$, the internal randomness and measurement outcomes of the
    corresponding algorithms.
    
    Then, for any joint distribution of $G,H,S,z$, it holds that
    \[
        \left|P_{\mathrm{left}}-P_{\mathrm{right}}\right|
        \leq
        2q\sqrt{P_{\mathrm{guess}}}.
    \]
\end{lemma}

\subsection{Simulating Two Oracles using One}
\label{sec:merge-oracles}
 
In this subsection, we recall that joint quantum queries to two oracles can be
simulated exactly using separate access, with constant query
overhead and linear additional gate cost per query.
Consequently, security with respect to two independent quantum-accessible random
oracles is preserved under the following implementation
\[
    O_0(x)=O(0\|x),
    \qquad
    O_1(x)=O(1\|x),
\]
where $O$ is a single quantum-accessible random oracle.

\para{Separate and joint oracle access.}
Let $O_0,O_1:\{0,1\}^n\to\{0,1\}^n$ be functions.
Separate quantum access to these functions provides the query
unitaries
\[
    U_{O_b}:\ket{x}\ket{y}
    \longmapsto
    \ket{x}\ket{y\oplus O_b(x)}
    \qquad (\forall b\in\{0,1\}).
\]
Each oracle call applies one of these two unitaries, while the
algorithm may perform arbitrary quantum operations between calls.
In particular, its registers may be arbitrarily entangled.

Joint access additionally allows the oracle index to be queried
in superposition through the unitary
\[
    U_{O_0,O_1}:\ket{b}\ket{x}\ket{y}
    \longmapsto
    \ket{b}\ket{x}\ket{y\oplus O_b(x)}.
\]
The following lemma shows that this joint interface can be
implemented using separate access.

\begin{lemma}[Simulation of the Joint Oracle]
\label{lem:combining-quantum-oracles}
Given separate access to $O_0$ and $O_1$, the unitary
$U_{O_0,O_1}$ can be implemented exactly using two queries
to each oracle and $O(n)$ additional gates.
The simulation uses an $n$-qubit ancilla initialized and
returned to $\ket{0^n}$.
\end{lemma}

\begin{proof}
Introduce an $n$-qubit ancilla register $t$, initialized to
$\ket{0^n}$. On a computational-basis input
$\ket{b}\ket{x}\ket{y}\ket{0^n}$, perform the following steps.

\begin{enumerate}
    \item Apply $U_{O_0}$ to $(x,t)$ to compute $O_0(x)$
    into $t$, obtaining
    \[
        \ket{b}\ket{x}\ket{y}\ket{O_0(x)}.
    \]

    \item XOR $t$ into $y$ conditioned on $b=0$, obtaining
    \[
        \ket{b}\ket{x}
        \ket{y\oplus(1-b)O_0(x)}
        \ket{O_0(x)}.
    \]

    \item Apply $U_{O_0}$ again to $(x,t)$ to uncompute $t$.
    Since $U_{O_0}$ is an XOR-query unitary, the resulting
    state is
    \[
        \ket{b}\ket{x}
        \ket{y\oplus(1-b)O_0(x)}
        \ket{0^n}.
    \]

    \item Repeat the above procedure
    using $O_1$, with the XOR conditioned on $b=1$.
    This gives
    \[
        \ket{b}\ket{x}
        \ket{y\oplus(1-b)O_0(x)\oplus bO_1(x)}
        \ket{0^n}.
    \]
\end{enumerate}

Since
\[
    y\oplus(1-b)O_0(x)\oplus bO_1(x)
    =
    y\oplus O_b(x),
\]
the procedure implements $U_{O_0,O_1}$ and returns the ancilla
to $\ket{0^n}$.
By linearity, the simulation is exact for arbitrary inputs,
including those entangled with other registers.

The procedure makes two queries to each oracle.
Each controlled-XOR operation requires $O(n)$ gates, yielding
the claimed gate complexity.
\end{proof}

\subsection{Non-Interactive Zero-Knowledge Proof}

We begin by recalling the definitions necessary for formalizing NIZK proof systems in our setting.

\begin{definition}[Polynomial-Query-Bounded Quantum Oracle Algorithm]
\label{def:poly-query-bounded}
    We say that a quantum oracle algorithm $\mathcal{A}^{(\cdot)}$ is
    \emph{polynomial-query-bounded} if there exists a polynomial
    $p:\mathbb{N}\to\mathbb{N}$ such that, for every input
    $x\in\{0,1\}^*$ and every oracle $H$, the execution of
    $\mathcal{A}^{\ket{H}}$ makes at most $p(|x|)$ quantum queries
    to $H$.
\end{definition}
\begin{definition}[Efficiently-Approximable Constant]\label{effapp}
    We say that a constant (real number) $\gamma \in (0, 1)$  is $\msf{Efficiently}$-$\msf{Approximable}$ if there exists a polynomial-time deterministic algorithm $\mathsf{Approx}_\gamma$ such that for all $n \in \mathbb{N}$,
    \begin{itemize}
    \item $\mathsf{Approx}_\gamma(1^n)$ outputs a rational number $\hat{\gamma} = a/b$, where $a, b$ are integers in binary representation.  
    \item For all $n \in \mathbb{N}$, $|\gamma - \mathsf{Approx}_\gamma(1^n)| \leq \frac{1}{2^n}$.
    \end{itemize}
\end{definition}

Then, we recall the definition of NIZK proof systems in the $\msf{Z}$-$\msf{Tamperable}$-$\msf{Hidden}$-$\msf{Bits}$  model, which is a variant of the standard NIZK proof system that allows for tampering with hidden bits.
 
\begin{definition}[{NIZK Proof in the ZHB Model \cite[Definition 4.2]{STOC:GIKKLS23}}]\label{def:zhb-nizk}
    An (unbounded-prover) non-interactive zero-knowledge (NIZK) proof system in the $\msf{Z}$-$\msf{Tamperable}$-$\msf{Hidden}$-$\msf{Bits}$$(\msf{ZHB})$ model for a language $L$ is a family of algorithms
     $\{\Pi_\gamma = (\mathsf{Gen}_\gamma, \mathsf{P}_\gamma, \mathsf{V}_\gamma)\}$ parameterized by $\msf{Efficiently}$-$\msf{Approximable}$ constants $\gamma \in (0, 1)$, where:
    
    \begin{itemize}
    \item $\mathsf{Gen}_\gamma(1^n)$ is a PPT   hidden bit string generator that given an input length $n$ outputs a hidden bit string $r \in \{0, 1\}^{p_{\gamma,\mathsf{hbs}}(n)}$ for some (fixed) polynomial $p_{\gamma,\mathsf{hbs}}$ where each bit $r_i$ of $r$ is sampled independently from a Bernoulli distribution with $\Pr[r_i = 1] = \hat{\gamma}$ where $\hat{\gamma} = \mathsf{Approx}_\gamma(1^n)$.
    \item $\mathsf{P}_\gamma(r, x)$ is an (unbounded, randomized) honest prover algorithm that takes as input a hidden bit string $r \in \{0, 1\}^{p_{\gamma,\mathsf{hbs}}(|x|)}$ and $x \in L$, and outputs a set of indices $I \subseteq [p_{\gamma,\mathsf{hbs}}(|x|)]$ and a proof $\pi$.
    \item $\mathsf{V}_\gamma(x, I, \{r_i\}_{i \in I}, \pi)$ is a polynomial-time deterministic verifier that takes as input $x$, an index set $I \subseteq [p_{\gamma,\mathsf{hbs}}(|x|)]$, the bits $\{r_i\}_{i \in I}$ of the hidden bit string $r$ corresponding to $I$, and a proof $\pi$, and outputs either 1 (accept) or 0 (reject).
    \end{itemize}
    
    We require $\Pi$ to satisfy the following completeness, soundness, and zero knowledge requirements:
    
    \begin{itemize}
    \item \textbf{Perfect Completeness:} For all Efficiently-Approximable constants $\gamma \in (0, 1)$ and all $x \in L$,

    \[
    \Pr\left[\mathsf{V}_\gamma(x, I, \{r_i\}_{i \in I}, \pi) = 1 : r \leftarrow \mathsf{Gen}_\gamma(1^{|x|}), (I, \pi) \leftarrow \mathsf{P}_\gamma(r, x)\right] = 1
    \]
    
    where the probability is over the randomness of $\mathsf{Gen}_\gamma$ and $\mathsf{P}_\gamma$.
    
    \item \textbf{(Adaptive) Statistical Soundness:} For all Efficiently-Approximable constants $\gamma \in (0, 1)$, there exists a negligible function $\negl(\cdot)$ such that,
    
    $$
    \Pr_{r \leftarrow \mathsf{Gen}_\gamma(1^n)} \left[ \exists (x, I, f, \pi) \text{ such that } x \notin L,|x|=n,\mathsf{V}_\gamma(x, I, \{\tilde{r}_i = \mathsf{ZeroFlip}(r_i, f_i)\}_{i \in I}, \pi) = 1 \right] \leq \negl(|x|)
    $$
    
    where $\mathsf{ZeroFlip}(r_i, f_i) = (r_i \oplus f_i$ if $r_i = 0$; $r_i$ if $r_i = 1)$.
    
    \item \textbf{Statistical Zero Knowledge:} For all efficiently-approximable constants $\gamma \in (0, 1)$ (as per \Cref{effapp}), there exists a PPT simulator $\mathsf{Sim}_\gamma$ , for all adversaries $A$, there exists a negligible function $\negl(\cdot)$ such that all $x \in L$ , 
    \begin{align*}
        &|\Pr[A(x, I, \{r_i\}_{i \in I}, \pi) = 1 : r \leftarrow \mathsf{Gen}_\gamma(1^{|x|}), (I, \pi) \leftarrow \mathsf{P}_\gamma(r, x)] \\
        &- \Pr[A(x, I^*, \{r^*_i\}_{i \in I^*}, \pi^*) = 1 : (I^*, \{r^*_i\}_{i \in I^*}, \pi^*) \leftarrow \mathsf{Sim}_\gamma(x)] |\leq \negl(|x|)
    \end{align*}

    where the probability is over the randomness of $\mathsf{Gen}_\gamma$, $\mathsf{P}_\gamma$, $\mathsf{Sim}_\gamma$, and $A$.
    \end{itemize}
\end{definition}

\begin{definition}[NIZK Proof in the QROM]  
    An (unbounded-prover) non-interactive zero knowledge (NIZK) proof system in the quantum random oracle model (QROM) for a language $\mathcal{L}$ is a tuple of algorithms $\Pi = (P^{(\cdot)},V^{(\cdot)})$ with access to a random oracle $\mathcal{O}$.
    
    \begin{itemize}
        \item $P^{O}(x)$ is an (unbounded, randomized) honest prover algorithm that is given oracle access to a random oracle $\mathcal{O}$, takes as input $x\in \mathcal{L}$, and outputs a proof $\pi$.
        
        \item $V^{O}(x,\pi)$ is a polynomial-time deterministic verifier that is given oracle access to a random oracle $\mathcal{O}$, takes as input $x$ and a proof $\pi$, and outputs either 1 (accept) or 0 (reject).
    \end{itemize}
    
    We require $\Pi$ to satisfy the following completeness, soundness, and zero knowledge requirements:
    
    \begin{itemize}
        \item \textbf{Perfect Completeness:} For all $x\in \mathcal{L}$, it holds that
        \[
        \Pr_{O\gets \Ora}\left[V^{O}(x,\pi) = 1:\pi \leftarrow P^{O}(x)\right] = 1.
        \] 
        
        \item \textbf{(Adaptive) Statistical Soundness:} It holds that
        \[
        \Pr_{O\gets \Ora}[\exists (x,\pi)\mathrm{~such~that~}x\notin \mathcal{L},|x| = n,\mathrm{~and~}V^{O}(x,\pi) = 1]\leq \negl(|x|).
        \]
        
        \item \textbf{Zero Knowledge:} There exists a stateful QPT simulator $\operatorname{Sim} = (\operatorname{SimProof},\operatorname{SimRO})$ such that  for all $x\in \mathcal{L}$ andall polynomial-query-bounded quantum oracle algorithm $\mathcal{A}^{(\cdot)}$ (\Cref{def:poly-query-bounded}), 
        \begin{align*}
        & \Big|\operatorname{Pr}\left[\mathcal{A}^{\ket{O}}(x,\pi) = 1:\pi \leftarrow P^{O}(x)\right] \\
        & \hspace{6em} - \operatorname{Pr}\left[\mathcal{A}^{\ket{\mathrm{SimRO}(\mathrm{st},\cdot)}}(x,\pi^{*}) = 1:(\mathrm{st},\pi^{*})\leftarrow \mathrm{SimProof}(x)\right]\Big|\leq \negl(|x|).
        \end{align*} 
    \end{itemize}
    \end{definition}

    \subsection{Compressed Partially-Fixed Oracle}

    We introduced compressed partially-fixed oracles to simulate a random
oracle conditioned on a prescribed classical table. This
formulation adapts the compressed-oracle technique
of~\cite{C:Zhandry19}.

Our zero-knowledge simulator initializes the oracle with a prescribed classical table and retains its private state throughout subsequent quantum interaction. We therefore spell out this fixed-table specialization and its efficient sparse implementation.

We first formalize the conditional oracle distribution.
\begin{definition}[Conditional Distribution of a Random Oracle]\label{def:condRO}  
    Let $n,m \in \mathbb{N}$ be parameters.
    A subset $Q\subseteq\{0,1\}^n\times\{0,1\}^m$ is a \emph{partial function} if $\forall (x,y), (x,y')\in Q$, it holds that $y=y'$. We write
    \[
    \operatorname{dom}(Q):=\{x:\exists y\text{ such that }(x,y)\in Q\}
    \]
    and write $Q(x)=y$ when $(x,y)\in Q$. 
    
    Given a partial function $Q$, the \emph{conditional distribution of random oracle}, denoted as  $O|_Q$, is the uniform distribution over all functions $F:\{0,1\}^n\rightarrow\{0,1\}^m$ consistent with $Q$. Equivalently,
    $$O|_Q(x) \coloneqq 
    \begin{cases} Q(x) & x \in \operatorname{dom}(Q),\\ 
        y \picks \{0,1\}^m & x\in \bits^n \setminus \operatorname{dom}(Q) 
    \end{cases}.
    $$
\end{definition}

In the above definition, the unspecified values are sampled independently once and then
retained, so repeated queries to the same input return the same
value. We next describe an efficient stateful simulation of
quantum access to this distribution.

For a stateful simulator $\mathcal{C}\gets\operatorname{Com}(H)$,
access to $\ket{\mathcal{C}}$ means coherent access to its query
procedure, with its private state retained between queries.
We also write $\ket{\operatorname{Com}(H)}$ when initialization
is implicit; all queries then use the same initialized simulator.
Thus, $\Adv^{\ket{\operatorname{Com}(H)}}$ denotes $\Adv$ interacting 
coherently with a single initialized simulator whose private state is retained and updated between queries.

The following lemma shows that quantum access to $\Ora|_H$ can be simulated exactly by a stateful oracle, with time and space polynomial in $n,m,|H|$, and the number of queries.

\begin{lemma}[Compressed Partially-Fixed Oracle] \label{ComO}
    Let $n,m \in \mathbb{N}$ be parameters. For a given partial function $H\subseteq\{0,1\}^n\times\{0,1\}^m$, there exists a (stateful) oracle $\msf{Com(H)}$ called the \emph{compressed partially-fixed oracle} satisfying the following properties:
    \begin{itemize}
        \item {\bf Perfect Indistinguishability:} For any (potentially unbounded) quantum algorithm $\mathcal{A}$, 
           \[
        \left|\Pr\left[\mathcal{A}^{\ket{O|_H}}=1\right]
        -\Pr\left[\mathcal{A}^{\ket{\msf{Com}(H)}}=1\right]\right|=0, 
    \]
        where the first probability includes sampling $O\gets\Ora|_H$, and both include the internal randomness and measurement outcomes of the interacting algorithms.
     
    \item {\bf Efficient constructability:}  Initializing $\operatorname{Com}(H)$ and answering $q$ queries require time and space polynomial in $n,m,|H|$, and $q$.
    \end{itemize}
    
\end{lemma}
\begin{proof}
We derive the lemma from Zhandry's compressed-oracle
simulation~\cite{C:Zhandry19}.
Let $S=\operatorname{dom}(H)$, and let
$F:\{0,1\}^n\to\{0,1\}^m$ be a uniformly random function.
Define
\[
    G_{H,F}(x)=
    \begin{cases}
        H(x), & x\in S,\\
        F(x), & x\notin S.
    \end{cases}
\]
Then $G_{H,F}$ is distributed uniformly among the functions
extending $H$.

Under the standard XOR-query convention, a quantum query to
$G_{H,F}$ can be implemented using two queries to $F$.
Specifically, first compute $F(x)$ into a fresh scratch
register. Then XOR $H(x)$ into the response register when
$x\in S$, and XOR the scratch register into the response
register otherwise. Finally, use a second query to $F$ to
uncompute the scratch register. All computations involving
$H$ are performed coherently, with $H$ hardwired.
The resulting transformation is exactly
\[
    \ket{x,z,0^m}
    \longmapsto
    \ket{x,z\oplus G_{H,F}(x),0^m}.
\]

Define $\msf{Com}(H)$ by replacing access to $F$ in this
implementation with access to Zhandry's ordinary compressed
oracle. For any quantum algorithm $\mathcal A$, its
composition with the above implementation is itself a
quantum oracle algorithm. Zhandry's perfect simulation
guarantee therefore implies that replacing the uniformly
random oracle $F$ by its compressed implementation preserves
$\mathcal A$'s acceptance probability exactly.
Since $G_{H,F}$ is distributed as $\Ora|_H$, the claimed
equality follows.

\end{proof}

\section{Strong \texorpdfstring{$\delta$}{delta}-Dense PRHF: Definition and Construction}

\subsection{Strong  \texorpdfstring{$\delta$}{delta}-Dense PRHF}
     Our definition of the strong $\delta$-dense PRHF considers polynomially many
correlated samples and only requires strong pre-image
pseudorandomness against non-unifrom QPT quantum
adversaries.

In the zero-knowledge
proof of \cite{STOC:GIKKLS23}, single-instance pre-image
pseudorandomness is lifted to polynomially many samples through a
coordinate-by-coordinate hybrid argument. The reduction for each
adjacent pair non-uniformly fixes the hidden-bit string, the
corresponding ZHB proof, and all samples outside the distinguished
coordinate. Because this auxiliary information may depend on $H$,
instantiating $H$ with a random oracle would require security against
adversaries whose advice depends on the sampled oracle.

Formally, we introduce our notion of a Strong PRHF as follows.
\begin{BoxBox}[label={box:dist-D0-D1},
    sidebyside, 
    sidebyside align=top,
    float, 
    floatplacement=!t, 
    fontupper=\small, fontlower=\small
    ]{Distributions $D_0$ and $D_1$}
    \vspace{0.5em}
    \uline{\textbf{Distribution $\mathcal{D}_0(n, H_n, p(n), \vb{r})$:}}
        \begin{enumerate}
             \item According to $\vb{r} = (r_1, \ldots, r_{p(n)})$, define the index sets:
            \begin{align*}
            & \mathcal{I}_0\coloneqq\{i\in [p(n)]~|~ r_i=0\} \\
            & \mathcal{I}_1\coloneqq\{i\in [p(n)]~|~ r_i=1\}  \quad \left(= [p(n)]\setminus \mcal{I}_0\right).
            \end{align*}
                
            \item {Define preimage and image sets:}
            \begin{align*}
                &\msf{Image} 
                 \coloneqq \{y\in \bits^{\ell(n)} ~|~ \exists z \in \bits^n : H(z) = y\} \\
                 & \msf{PreImage}(y)
                \coloneqq\{z\in \bits^n ~|~ H(z) = y\}.
            \end{align*}

            \item For all $i\in [p(n)]$:
                \begin{enumerate}
                    \item If $r_i=0$, then sample $y_i\pick \msf{Image}$ and $v_i\pick \msf{PreImage}(y_i)$;
                    \item If $r_i=1$, then sample $y_i\pick \bits^\ell(n)\setminus \msf{Image}$. 
                \end{enumerate}
            \item Output:
             $$T =\left(\{r_i\}_{i\in[p(n)]},~ \{y_i\}_{i\in[p(n)]},~    \{v_i\}_{i\in\mcal{I}_0}\right).$$
        \end{enumerate}

    \tcblower

    \vspace{0.5em}
    \uline{\textbf{Distribution $\mathcal{D}_1(n, H_n, p(n), \vb{r})$:}}
        \begin{enumerate}

            \item According to $\vb{r} = (r_1, \ldots, r_{p(n)})$, define the index sets:
            \begin{align*}
                & \mathcal{I}_0\coloneqq\{i\in [p(n)]~|~ r_i=0\}\\
                & \mathcal{I}_1\coloneqq\{i\in [p(n)]~|~ r_i=1\} \quad \left(= [p(n)]\setminus \mcal{I}_0\right).
            \end{align*}
           
            \item For all $i\in [p(n)]$, sample $v_i\pick \bits^n$ and define $y_i\coloneqq H(v_i)$.
            \item Output: 
            $$T =\left(\{r_i\}_{i\in[p(n)]},~ \{y_i\}_{i\in[p(n)]},~    \{v_i\}_{i\in\mcal{I}_0}\right).$$
        \end{enumerate}  
\end{BoxBox}

    \begin{definition}[Strong $\delta$-Dense PRHF]
    \label{def:strongPRHF} 
        Let $\delta\in (0,1)$ be a $\msf{Efficiently}$-$\msf{Approximable}$ constant as per  \Cref{effapp} and $n\in \Naturals$. 
        Let $\ell: \Naturals \to \Naturals$ be a polynomial, and let $\{\mathcal{H}_n\}_{n\in\Naturals}$ be an ensemble of distributions over functions $H:\{0,1\}^n \to \{0,1\}^{\ell(n)}$. We call this ensemble a strong $\delta$-dense post-quantum PRHF if  the following conditions hold:
        \begin{enumerate}
            \item\label[Property]{delta-dense}($\delta$-Density) For any function $H_n:\{0,1\}^n \to \{0,1\}^{\ell(n)}$, let $p_{H_n}$ denote the probability that a random $y\pick\bits^{\ell(n)}$ is in the range of $H_n$. It holds that
            \[
                \Pr\left[
                    p_{H_n}\in[\delta-\negl(n),\delta+\negl(n)]~:~ H_n\gets\mcal H_n
                \right]
                \geq 1-\negl(n),
            \] 
           
            \item (Pseudorandomness) For all non-uniform QPT adversary $\Adv$, it holds that
            \[
                \left|\Pr[\Adv(1^n, H_n, H_n(x))=1~:~ 
                \begin{array}{l}
                    x\pick \bits^n,\\ 
                     H_n\gets \mcal{H}_n
                \end{array}] 
                - \Pr[\Adv(1^n, H_n, y)=1 ~:~ 
                \begin{array}{l}
                    y\pick \bits^{\ell(n)},\\ 
                     H_n\gets \mcal{H}_n
                \end{array}
                ]\right|\leq \negl(n).
            \]
            
            \item \zcsetup{reftype=Property}\label{def:strong-PRHF:strongPIP} (Strong Pre-Image Pseudorandomness)
                For any non-uniform QPT adversary $\Adv$, any polynomial $p(n)$, any bit-vector $\vb{r} = (r_1, \ldots, r_{p(n)})$, it holds that 
                \begin{align*}
                    \left| \Pr[\mathcal{A}(1^n, H_n, T) = 1 ~:~ 
                    \begin{array}{l}
                        H_n\gets \mcal{H}_n; \\
                        T \gets \mathcal{D}_0
                    \end{array}
                    ]
                    - \Pr[\mathcal{A}(1^n, H_n, T) = 1 ~:~ 
                    \begin{array}{l}
                        H_n\gets \mcal{H}_n; \\
                        T \gets \mathcal{D}_1
                    \end{array}
                    ] \right| \leq \negl(n),
                \end{align*}
            where $\mcal{D}_0$ and $\mcal{D}_1$ are the  $\mathcal{D}_0(n, H_n, p(n), \vb{r})$ and $\mathcal{D}_1(n, H_n, p(n), \vb{r})$, respectively,  defined in \Cref{box:dist-D0-D1}. 

        \end{enumerate}
    \end{definition}

\subsection{Random Oracle is a Strong \texorpdfstring{$(1-\frac{1}{e})$}{1-1/e}-Dense PRHF}\label{sec:random-oracle-is-strong-dense-prhf}

In this subsection, we show that a random oracle is a strong $(1-\frac{1}{e})$-dense PRHF as per \Cref{def:strongPRHF}. 

\begin{remark}[Random Oracles as $\delta$-dense PRHFs]
    \label{rmk:random-oracle-strong-dense-prhf}
We first need to clarify an important modeling subtlety: in \Cref{def:strongPRHF}, the QPT adversary $\Adv$ receives the full description of $H$ as input. However, this is impossible when we talk about an oracle construction, since the QPT adversary cannot receive the full description of a random oracle. Thus, when we say ``a random oracle is a strong $(1-\frac{1}{e})$-dense PRHF,'' it actually means a variant of as per \Cref{def:strongPRHF} where the QPT adversary is modeled as a QPT oracle machine and receives \emph{quantum} oracle access to $H$ (which we will instantiate as a random oracle) instead of the full description of $H$. We take this modeling  convention throughout the paper, whenever we instantiate  the strong $\delta$-dense PRHF with a random oracle.
\end{remark}

The first two requirements of our strengthened $\delta$-dense PRHF
notion—$\delta$-density and post-quantum pseudorandomness—follow from
established analyses of random oracles in
\cite{STOC:GIKKLS23} and \cite{AC:HhaXagYam19}, respectively. For completeness, we list them as \Cref{lem:RO-density-previous} and \Cref{pq-pseudoramdom} below.
\begin{lemma}[{\cite[Lemma B.3]{STOC:GIKKLS23}}]\label{lem:RO-density-previous}
    Fixed any $n\in\Naturals$, let $\Ora_n:\bits^n \to \bits^n $ be a random oracle. There exists negligible function $\negl(\cdot)$ such that

\[
    \Pr_{O_n\gets\Ora_n}\!\left[
        p_{O_n}\in
        \left[
            \left(1-\frac1e\right)-\negl(n),
            \left(1-\frac1e\right)+\negl(n)
        \right]
    \right]
    \geq 1-\negl(n),
\]
where
\[
    p_{O_n}\coloneqq
    \Pr_{y\pick\bits^n}\!\left[
        \exists x\in\bits^n:\ O_n(x)=y
    \right].
\]
\end{lemma}
\begin{lemma}[{\cite[Theorem 2]{AC:HhaXagYam19}}]\label{pq-pseudoramdom}
   For all QPT oracle-aided adversary $\Adv$, it holds that
    \[
                \left|\Pr[\Adv^{\ket{O_n}}(1^n, O_n(x))=1~:~ 
                \begin{array}{l}
                    x\pick \bits^n,\\ 
                     O_n\gets \Ora_n
                \end{array}] 
                - \Pr[\Adv^{\ket{O_n}}(1^n, y)=1 ~:~ 
                \begin{array}{l}
                    y\pick \bits^{n},\\ 
                     O_n\gets \Ora_n
                \end{array}
                ]\right|\leq \negl(n).
            \]
\end{lemma}
\para{Explanation of \Cref{pq-pseudoramdom}.} 
The result of \cite[Theorem~2]{AC:HhaXagYam19} is actually more general: it
considers a salted random oracle
$O\colon [K]\times[N]\rightarrow[M]$ for some $K,N,M \in \Naturals$ and permits the adversary to receive
an $S$-bit classical advice string that may depend on $O$. To obtain the
lemma above, set $K=1$, $N=M=2^n$, use no auxiliary advice, and let
$T=q(n)$ be the adversary's polynomial query bound. The resulting
distinguishing advantage is at most
\[
    \widetilde{O}\!\left(
        \left(\frac{q(n)^4}{2^n}\right)^{1/6}
    \right)
    =
    \widetilde{O}\!\left(
        \frac{q(n)^{2/3}}{2^{n/6}}
    \right),
\]
which is negligible in $n$. Identifying $[2^n]$ with $\bits^n$ through binary representation and
omitting the trivial salt coordinate yields precisely the statement
above.

\para{Strong Pre-Image Pseudo-randomness.} To establish strong pre-image pseudo-randomness, we depart from the
proof of \cite{STOC:GIKKLS23}. We state the lemma below and prove it in \Cref{sec:proof-strong-preimg-pseudorandomenss-for-ro}.

\begin{lemma}
\label{lem:strong-preimg-pseudorandomenss}
For any non-uniform QPT oracle-aided adversary $\Adv$, any polynomial $p(n)$, and any bit-vector $\vb{r} = (r_1, \ldots, r_{p(n)})$, it holds that
    \begin{align*}
                    \left| \Pr[\mathcal{A}^{\ket{O_n}}(1^n, T) = 1 ~:~ 
                    \begin{array}{l}
                        O_n\gets \Ora_n; \\
                        T \gets \mathcal{D}_0
                    \end{array}
                    ]
                    - \Pr[\mathcal{A}^{\ket{O_n}}(1^n, T) = 1 ~:~ 
                    \begin{array}{l}
                        O_n\gets \Ora_n; \\
                        T \gets \mathcal{D}_1
                    \end{array}
                    ] \right| \leq \negl(n),
                \end{align*}
            where $\mcal{D}_0$ and $\mcal{D}_1$ are the  $\mathcal{D}_0(n, O_n, p(n), \vb{r})$ and $\mathcal{D}_1(n, O_n, p(n), \vb{r})$,  respectively,  defined in \Cref{box:dist-D0-D1}. 
\end{lemma}

\subsection{Proof of \texorpdfstring{\Cref{lem:strong-preimg-pseudorandomenss}}{Strong Pre-Image Pseudorandomness of Random Oracles}}
\label{sec:proof-strong-preimg-pseudorandomenss-for-ro}

We proceed to prove \Cref{lem:strong-preimg-pseudorandomenss}. Our proof actually works for any (potentially unbounded) $\Adv$ as long as she makes only a polynomial number of quantum oracle queries.   

\para{Notation.} We will prove the lemma via a sequence of hybrids $H_0, H_1, \ldots, H_6$. Each hybrid $H_i$ depends on the same parameters including the security parameter $n$, the oracle distribution $\Ora_n$,  the polynomial $p(n)$,  the bit-vector $\vb{r} = (r_1, \ldots, r_{p(n)})$. For succinctness, we suppress them from the notation and simply write $H_i$, instead of $H_i(n, \Ora_n, p(n), \vb{r})$.

Throughout the proof, we use the convention that once we need to sample from an empty set, the output is defined to be some special symbol (e.g., $\bot$ or $0^n$).

It is straightforward from the description that the output of $H_0$ (\Cref{hybrid:h0}) is identically distributed to that of $\Adv^{\ket{O_n}}(1^n, T)$ when $O_n \gets \Ora_n$ and $T \gets \mathcal{D}_0(n, O_n, p(n), \vb{r})$. Similarly, the output of $H_6$ (\Cref{hybrid:h6}) is identically distributed to that of $\Adv^{\ket{O_n}}(1^n, T)$ when $O_n \gets \Ora_n$ and $T \gets \mathcal{D}_1(n, O_n, p(n), \vb{r})$. Therefore, our goal is to show that the output distributions of $H_1$ and $H_6$ are negligibly close. 

For two hybrid experiments $H_i$ and $H_j$, we write $H_i \idind H_j$ if $\Pr[H_i=1]=\Pr[H_j=1]$ for every adversary $\Adv$, $H_i \sind H_j$ if $|\Pr[H_i=1]-\Pr[H_j=1]|$ is negligible for all possibly inefficient adversaries $\Adv$, and $H_i \qcind H_j$ if $|\Pr[H_i=1]-\Pr[H_j=1]|$ is negligible for every quantum adversary $\Adv$ making polynomially many oracle queries. 

\subsubsection{Hybrids $H_0$ and $H_1$}

\para{Hybrid $H_0$:} This is the game corresponding to $\mathcal{D}_0$. We formally describe it in \Cref{hybrid:h0}.
\begin{ExperimentBox}[label={hybrid:h0}]{Hybrid $H_0$}
    
\begin{enumerate}
    \item According to the  input $\vb{r} = (r_1, \ldots, r_{p(n)})$, define
     $\begin{cases}
    \mathcal{I}_0  \coloneqq\{i\in [p(n)]~|~ r_i=0\} \\ 
    \mathcal{I}_1  \coloneqq\{i\in [p(n)]~|~ r_i=1\}
    \end{cases}$.
    \item
    Sample $O_{n} \gets \Ora_n$;    
    \item Define 
        $\begin{cases}
            \msf{Image}^{O_{n}}  \coloneqq \{y\in \bits^n : \exists z \in \bits^n \text{ such that } O_{n}(z) = y\} \\ 
            \msf{PreImage}^{O_n}(y)  \coloneqq\{z\in \bits^n : O_{n}(z) = y\}
        \end{cases}$

    \item $\forall i\in [p(n)]$, sample $\Set{y_i}_{i \in [p(n)]}$ and $\Set{v_i}_{i\in \mcal{I}_0}$ as follows:
        \begin{itemize}
            \item If $r_i=0$, then set $y_i\pick \msf{Image}^{O_{n}}, v_i\pick \msf{PreImage}^{O_n}(y_i)$,
            \item If $r_i=1$, then set $y_i\pick \bits^n\setminus \msf{Image}^{O_{n}}$. 
        \end{itemize}
    \item Invoke the adversary $\Adv^{\ket{O_n}}(\{r_i\}_{i\in[p(n)]},\{y_i\}_{i\in[p(n)]},\{v_i\}_{i\in\mcal{I}_0})$ 
    
\end{enumerate}
\end{ExperimentBox}

\para{Hybrid $H_1$:} We present $H_1$ in \Cref{hybrid:h1} and highlight its difference from $H_0$ in \red{red color}. In this hybrid, we change the sampling method for $y_i$ when $r_i=1$ and redirect the points to form a new oracle $\Ora^{*}_n$.

\begin{ExperimentBox}[label={hybrid:h1}]{Hybrid $H_1$}
    
\begin{enumerate}
    \item According to the  input $\vb{r} = (r_1, \ldots, r_{p(n)})$, define
     $\begin{cases}
    \mathcal{I}_0  \coloneqq\{i\in [p(n)]~|~ r_i=0\} \\ 
    \mathcal{I}_1  \coloneqq\{i\in [p(n)]~|~ r_i=1\}
    \end{cases}$.
    \item
    Sample $O_{n} \gets \Ora_n$;    
    \item Define 
        $\begin{cases}
            \msf{Image}^{O_{n}}  \coloneqq \{y\in \bits^n : \exists z \in \bits^n \text{ such that } O_{n}(z) = y\} \\ 
            \msf{PreImage}^{O_n}(y)  \coloneqq\{z\in \bits^n : O_{n}(z) = y\}
        \end{cases}$
    %\item $\forall i\in [p(n)]$, sample $r_i\in \{0,1\}$ where $\Pr[r_i=1]=\msf{Approx}_{\frac{1}{e}}(1^n)$;
    \item Define
     $\begin{cases}
    \mathcal{I}_0  \coloneqq\{i\in [p(n)]~|~ r_i=0\} \\ 
    \mathcal{I}_1  \coloneqq\{i\in [p(n)]~|~ r_i=1\}
    \end{cases}$
    \item $\forall i\in [p(n)]$, sample $\Set{y_i}_{i \in [p(n)]}$ and $\Set{v_i}_{i\in \mcal{I}_0}$ as follows:
        \begin{itemize}
            \item If $r_i=0$, then set $y_i\pick \msf{Image}^{O_{n}}, v_i\pick \msf{PreImage}^{O_n}(y_i)$,
            \red{\item If $r_i=1$, then set $\delta_i \pick \bits^n, y_i=O_n(\delta_i)$ }
        \end{itemize}
    \red{\item
        $O^*_n(z) \coloneqq \begin{cases}
       \picks \bits^n\setminus\{y_i\}_{i\in \mathcal{I}_1 } & \text{if } z \in \bigcup_{i\in \mathcal{I}_1} \msf{PreImage}^{O_n}(y_i) \\ 
       O_n(z) & \text{otherwise}
   \end{cases}$}
    \item Invoke the adversary $\Adv^{\red{\ket{O^*_n}}}(\{r_i\}_{i\in[p(n)]},\{y_i\}_{i\in[p(n)]},\{v_i\}_{i\in\mcal{I}_0})$ 
    
\end{enumerate}
\end{ExperimentBox}

\begin{MyClaim}\zcsetup{reftype=Claim}\label{claim:main-hyb-0-1}
    It holds that $H_0\sind H_1$.
\end{MyClaim} 

The proof of \zcref{claim:main-hyb-0-1} is one of the most technically involved parts. We present it to \Cref{sec:ind:hyb-0-1}.

\subsubsection{Hybrid $H_2$}

We present $H_2$ in \Cref{hybrid:h2} and highlight its difference from $H_1$ in \red{red color}.
 In this hybrid, we sample a fresh oracle $O^{\msf{fresh}}_n$ while preserving all the points in $\bigcup_{i\in \mcal{I}_0 \cup \mcal{I}_1}\msf{PreImage}(y_i)$ and redirect those ``bad'' points (some x not in $O^{\msf{fresh}}_n$ while preserving all the points in $\bigcup_{i\in \mcal{I}_0 \cup \mcal{I}_1}\msf{PreImage}(y_i)$ but $x\mapsto y_i$ in $\Ora^{\msf{fresh}}$) to form a new oracle $\Ora^{*}_n$.

\begin{ExperimentBox}[label={hybrid:h2}]{Hybrid $H_2$}
    
\begin{enumerate}
    \item According to the  input $\vb{r} = (r_1, \ldots, r_{p(n)})$, define
     $\begin{cases}
    \mathcal{I}_0  \coloneqq\{i\in [p(n)]~|~ r_i=0\} \\ 
    \mathcal{I}_1  \coloneqq\{i\in [p(n)]~|~ r_i=1\}
    \end{cases}$.
    \item
    Sample $O_{n} \gets \Ora_n$
    \item Define 
        $\begin{cases}
            \msf{Image}^{O_{n}}  \coloneqq \{y\in \bits^n : \exists z \in \bits^n \text{ such that } O_{n}(z) = y\} \\ 
            \msf{PreImage}^{O_n}(y)  \coloneqq\{z\in \bits^n : O_{n}(z) = y\}
        \end{cases}$
    %\item $\forall i\in [p(n)]$, sample $r_i\in \{0,1\}$ where $\Pr[r_i=1]=\msf{Approx}_{\frac{1}{e}}(1^n)$;
    \item Define
     $\begin{cases}
    \mathcal{I}_0  \coloneqq\{i\in [p(n)]~|~ r_i=0\} \\ 
    \mathcal{I}_1  \coloneqq\{i\in [p(n)]~|~ r_i=1\}
    \end{cases}$
    \item $\forall i\in [p(n)]$ :
        \begin{itemize}
            \item If $r_i=0$, then set $y_i\pick \msf{Image}^{O_n}, v_i\pick \msf{PreImage}^{O_n}(y_i)$
            \item If $r_i=1$, then set $\delta_i \pick \bits^n, y_i=O_n(\delta_i)$\\ 
        \end{itemize}
    \item
        $O^*_n(z) \coloneqq \begin{cases}
        \picks \bits^n\setminus\{y_i\}_{i\in \mathcal{I}_{1}} & \text{if } z \in \bigcup_{i\in \mathcal{I}_{1}} \msf{PreImage}^{O_n}(y_i) \\ 
        \red{O_n(z)} & \red{\text{if } z \in \bigcup_{i\in \mathcal{I}_{0}} \msf{PreImage}^{O_n}(y_i)}\\
        \red{\picks \bits^n\setminus\{y_i\}_{i\in \mathcal{I}_{0}\cup \mathcal{I}_{1}} } & \text{otherwise}
   \end{cases}$ 
    \item Invoke the adversary $\Adv^{\ket{O^{*}_n}}(\{r_i\}_{i\in[p(n)]},\{y_i\}_{i\in[p(n)]},\{v_i\}_{i\in\mcal{I}_0})$ 
\end{enumerate}
\end{ExperimentBox}

\begin{MyClaim}\zcsetup{reftype=Claim}\label{claim:first-hyb-1-2}
    It holds that $H_1\sind H_2$.
\end{MyClaim}
The proof of \zcref{claim:first-hyb-1-2} is another technically involved part of the proof.  We present it in \Cref{sec:ind:hyb-1-2}.

\subsubsection{Hybrid $H_{2.5}$}
We introduce an intermediate hybrid $H_{2.5}$. The only difference is that $H_{2.5}$ handles the $i\in \mcal{I}_0$ case in the same manner as $H_2$ (i.e., $y_i$ uniformly from the image of the random oracle, and then sample $\delta_i$ uniformly from its preimage), but everything else is the same as in $H_3$. We present the full description of $H_{2.5}$ in \Cref{hybrid:h25}, highlighting its differences from $H_2$ in \red{red color}. %and highlighting its differences from $H_2$ in \green{green color}.

\begin{ExperimentBox}[label={hybrid:h25}]{Hybrid $H_{2.5}$}
\begin{enumerate}
    \item According to the  input $\vb{r} = (r_1, \ldots, r_{p(n)})$, define
     $\begin{cases}
    \mathcal{I}_0  \coloneqq\{i\in [p(n)]~|~ r_i=0\} \\ 
    \mathcal{I}_1  \coloneqq\{i\in [p(n)]~|~ r_i=1\}
    \end{cases}$.
    \item
    Sample $O_{n} \gets \Ora_n$
    %\item $\forall i\in [p(n)]$, sample $r_i\in \{0,1\}$ where $\Pr[r_i=1]=\msf{Approx}_{\frac{1}{e}}(1^n)$;
    \item Define
     $\begin{cases}
    \mathcal{I}_0  \coloneqq\{i\in [p(n)]~|~ r_i=0\} \\ 
    \mathcal{I}_1  \coloneqq\{i\in [p(n)]~|~ r_i=1\}
    \end{cases}$
    \item $\forall i\in [p(n)]$ :
        \begin{itemize}
            \item If $r_i=0$, then sample $y_i \pick \msf{Image}^{\Ora_n}$ and $\red{\delta_i} \pick \msf{PreImage}^{\Ora_n}(y_i)$, \red{and define $v_i \coloneqq \delta_i$}.
            \item If $r_i=1$, then set $\delta_i \pick \bits^n, y_i=O_n(\delta_i)$ 
        \end{itemize}
    \item
        $O^*_n(z) \coloneqq \begin{cases}
        \picks \bits^n\setminus\{y_i\}_{i\in \mathcal{I}_{1}} & \text{if } z \in \bigcup_{i\in \mathcal{I}_{1}} \red{\{\delta_i\}} \\ 
        O_n(z) & \text{if } z \in \bigcup_{i\in \mathcal{I}_{0}} \red{\{\delta_i\}}\\
        \picks \bits^n\setminus\{y_i\}_{i\in \mathcal{I}_{0}\cup \mathcal{I}_{1}}  & \text{otherwise}
   \end{cases}$ 
    \item Invoke the adversary $\Adv^{\ket{O^{*}_n}}(\{r_i\}_{i\in[p(n)]},\{y_i\}_{i\in[p(n)]},\{v_i\}_{i\in\mcal{I}_0})$ 
\end{enumerate}
\end{ExperimentBox} 
\begin{MyClaim}\zcsetup{reftype=Claim}\label{claim:hyb-2-3}
    It holds that $H_2\qcind H_{2.5}$. 
\end{MyClaim}

The proof of \zcref{claim:hyb-2-3} is another technically involved part of the proof.  We present it in \Cref{sec:proof-H2-H3}.

\subsubsection{Hybrid $H_3$} 

We present $H_3$ in \Cref{hybrid:h3} and highlight its difference from $H_2$ in \red{red color}.
In this hybrid, we try to redirect less points and show the indistinguishability of the two oracles towards polynomial-query bound quantum adversary through one-way to hiding lemma.

\begin{ExperimentBox}[label={hybrid:h3}]{Hybrid $H_3$}
    
\begin{enumerate}
    \item According to the  input $\vb{r} = (r_1, \ldots, r_{p(n)})$, define
     $\begin{cases}
    \mathcal{I}_0  \coloneqq\{i\in [p(n)]~|~ r_i=0\} \\ 
    \mathcal{I}_1  \coloneqq\{i\in [p(n)]~|~ r_i=1\}
    \end{cases}$.
    \item
    Sample $O_{n} \gets \Ora_n$
    \item $\forall i\in [p(n)]$ :
        \begin{itemize}
            \red{\item If $r_i=0$,  then sample $\delta_i \pick \bits^n$ and define $y_i \coloneqq O_{n}(\delta_i)$ and $v_i \coloneqq \delta_i$.}
            \item If $r_i=1$, then set $\delta_i \pick \bits^n, y_i=O_n(\delta_i)$\\ 
        \end{itemize}
    \item
        $O^*_n(z) \coloneqq \begin{cases}
        \picks \bits^n\setminus\{y_i\}_{i\in \mathcal{I}_{1}} & \text{if } z \in \bigcup_{i\in \mathcal{I}_{1}} \{\delta_i\} \\ 
        O_n(z) & \text{if } z \in \bigcup_{i\in \mathcal{I}_{0}} \{\delta_i\}\\
        \picks \bits^n\setminus\{y_i\}_{i\in \mathcal{I}_{0}\cup \mathcal{I}_{1}}  & \text{otherwise}
   \end{cases}$ 
    \item Invoke the adversary $\Adv^{\ket{O^{*}_n}}(\{r_i\}_{i\in[p(n)]},\{y_i\}_{i\in[p(n)]},\{v_i\}_{i\in\mcal{I}_0})$ 
\end{enumerate}
\end{ExperimentBox}

\begin{MyClaim}\zcsetup{reftype=Claim} \label{claim:H25-H3}
    It holds that $H_{2.5}\sind H_3$. 
\end{MyClaim}
\begin{proof}    Let $\mathbf r=(r_1,\ldots,r_{p(n)})$ be the corresponding input vector.

Define the following event:
\begin{itemize}
    \item \textbf{Event $E$}: for every distinct $i,j\in [p(n)]$, it holds that $y_i \neq y_j$.
\end{itemize}
Since $O_n(\delta_i)=y_i$, event $E$ also implies that all the
$\delta_i$'s are distinct.

 Conditional on the event $E$, the tuple $\Set{(\delta_i, y_i)}_{i\in [p(n)]}$ has exactly the same distribution in both games: it is uniform over
pairs of ordered lists with distinct entries within each list. To see this, both sampling procedures are invariant under arbitrary,
independent relabelings of the domain and codomain. Those relabelings
can map any such collision-free tuple to any other.

Furthermore, it holds that once this tuple $\Set{(\delta_i, y_i)}_{i\in [p(n)]}$ is fixed, the construction
of $O_n^*$ is identical in both hybrids: the only retained original-oracle
values are
\[
O_n(\delta_i)=y_i
\qquad (i\in\mathcal{I}_0),
\]
and all other entries are freshly sampled according to the same
rules described in \Cref{hybrid:h25}.

Thus, the adversary's views conditioned on $E$
are identical.

            Next, we proceed to bound $\Pr[\neg E]$ in both hybrids.

            In $H_{2.5}$, consider any pair $\Set{y_i, y_j}$: if the index $i$ belongs
            to $\mcal I_0$, conditional on $O_n$, then $y_i$ is
            uniform over $\msf{Image}(O_n)$ and is independent of
             $y_j$.
            Hence, for such an $i\in \mcal{I}_0$, it holds that
            \begin{align*}
                \Pr_{H_{2.5}}[y_i=y_j]
                &=
                \mathop{\mathbb{E}}\limits_{H_{2.5}}\!\left[
                    \frac1{|\msf{Image}(O_n)|}
                \right]\\
                &\leq
                2^{-n/2}
                +\Pr_{H_{2.5}}[|\msf{Image}(O_n)|\leq2^{n/2}]\\
                &=2^{-n/2}+\negl(n). \tag{\Cref{lem:random-oracle-image-size}}
            \end{align*}
            Otherwise (also in $H_{2.5}$), both $i$ and $j$ belong to $\mcal I_1$, it holds that
            \[
                \Pr[y_i=y_j]
                =
                2^{-n}+(1-2^{-n})2^{-n}
                \leq 2^{1-n}
                \leq 2^{-n/2}
            \]
            for sufficiently large $n$.

            In  $H_3$, all  $\delta_i$'s are sampled independently and uniformly from  $\bits^n$. Consequently, this bound $\Pr[y_i=y_j] \leq  2^{-n/2}$ holds for all pairs $\Set{y_i, y_j}_{i, j\in [p(n)]}$ in $H_3$ as well.

            Therefore, for either $b\in\{2.5,3\}$, a union bound gives
            \begin{align*}
                \Pr_{H_b}[\neg E]
                &=
                \Pr_{H_b}[
                    \exists\,1\leq i<j\leq p(n):\;y_i=y_j
                ]\\
                &\leq
                \sum_{1\leq i<j\leq p(n)}
                \Pr_{H_b}[y_i=y_j]\\
                &\leq
                \binom{p(n)}2
                \left(2^{-n/2}+\negl(n)\right) =\negl(n).
            \end{align*}

            We then have
\begin{align*}
\Delta\!\left(
    \msf{distr}(V_{2.5}),
    \msf{distr}(V_3)
\right)
&~\leq
\Pr_{H_{2.5}}[\neg E]
+
\Pr_{H_3}[\neg E]\\
&~\leq 
2\binom{p(n)}2
\left(2^{-n/2}+\negl(n)\right)
=\negl(n),
\end{align*}
where $V_b$ denotes the adversary's view in hybrid $H_b$.
The negligible bound above is uniform over $\boldsymbol\rho$.

This completes the proof of \zcref{claim:H25-H3}.

\end{proof}

\subsubsection{Hybrid $H_4$} 

We present $H_4$ in \Cref{hybrid:h4} and highlight its difference from $H_3$ in \red{red color}.

\begin{ExperimentBox}[label={hybrid:h4}]{Hybrid $H_4$}
\begin{enumerate}
    \item According to the  input $\vb{r} = (r_1, \ldots, r_{p(n)})$, define
     $\begin{cases}
    \mathcal{I}_0  \coloneqq\{i\in [p(n)]~|~ r_i=0\} \\ 
    \mathcal{I}_1  \coloneqq\{i\in [p(n)]~|~ r_i=1\}
    \end{cases}$.
    \item
    Sample $O_{n} \gets \Ora_n$
    %\item $\forall i\in [p(n)]$, sample $r_i\in \{0,1\}$ where $\Pr[r_i=1]=\msf{Approx}_{\frac{1}{e}}(1^n)$;
    \item Define
     $\begin{cases}
    \mathcal{I}_0  \coloneqq\{i\in [p(n)]~|~ r_i=0\} \\ 
    \mathcal{I}_1  \coloneqq\{i\in [p(n)]~|~ r_i=1\}
    \end{cases}$
    \item $\forall i\in [p(n)]$ :
        \begin{itemize}
            \item If $r_i=0$,  then sample $\delta_i \pick \bits^n$ and define $y_i \coloneqq O_{n}(\delta_i)$ and $v_i \coloneqq \delta_i$.
            \item If $r_i=1$, then set $\delta_i \pick \bits^n, y_i=O_n(\delta_i)$
        \end{itemize}
    \item
        $O^*_n(z) \coloneqq \begin{cases}
        \picks \bits^n\setminus\{y_i\}_{i\in \mathcal{I}_{1}} & \text{if } z \in \bigcup_{i\in \mathcal{I}_{1}} \{\delta_i\} \\ 
        %{O_n(z)} & \text{if } z \in \bigcup_{i\in \mathcal{I}_{0}} \{\delta_i\} \\ 
        \red{\picks \bits^n\setminus\{y_i\}_{i\in \mathcal{I}_{0}\cup \mathcal{I}_{1}}}  & \text{if } z \in \red{\bigcup_{i\in \mathcal{I}_{0} \cup \mathcal{I}_{1}}\msf{PreImage}^{O_n}(y_i) \setminus\{\delta_i\}_{i\in \mathcal{I}_{0} \cup \mathcal{I}_{1}}}\\
        \red{O_n(z)} & \text{otherwise}
   \end{cases}$ 
    \item Invoke the adversary $\Adv^{\ket{O^{*}_n}}(\{r_i\}_{i\in[p(n)]},\{y_i\}_{i\in[p(n)]},\{v_i\}_{i\in\mcal{I}_0})$ 
\end{enumerate}
\end{ExperimentBox}
\begin{remark}
Note that the conditions in each piecewise oracle definition are
checked sequentially, from top to bottom. Each condition
is considered only if all preceding conditions are false;
the first satisfied condition determines the applicable
branch. The ``otherwise'' branch applies only if every
preceding condition is false.
This convention applies to both $H_3$ and $H_4$.
In particular, if $\delta_i=\delta_j$ for some
$i\in\mathcal{I}_0$ and $j\in\mathcal{I}_1$, their common
input is handled by the first branch in both hybrids.
Thus, overlapping conditions are resolved consistently,
including when designated inputs collide.
\end{remark}
\begin{MyClaim}\zcsetup{reftype=Claim}\label{claim:hyb-3-4}
    It holds that $H_3 \idind H_4$. 
\end{MyClaim}

\begin{proof}
      First, notice that the joint distribution of the  following variables is identical in both hybrids:
    \begin{equation}\zcsetup{reftype=Expression}\label{eq:joint-dist-h3-h4}
    \Set{(\delta_i,y_i)}_{i \in[p(n)]}
    % ,~ \cup_{i\in \mathcal{I}_{0}} \msf{PreImage}^{O_n}(y_i),~ \cup_{i\in \mathcal{I}_{1}} \msf{PreImage}^{O_n}(y_i)
    \end{equation}
Then, to show $H_3\idind H_4$, it suffices to show that conditioned on these variables, the distribution of the oracle table of $O^*_n$ is also identical in both hybrids. 

Toward that, we consider the distribution of $O^*_n(z)$ for any $z\in \bits^n$ conditioned on the above variables. We have the following cases:
\begin{enumerate}
    \item For $z\in \bigcup_{i\in \mathcal{I}_{0} }\{\delta_i\}$, both of the oracle return the same value $y_i$
    \item For $z\in \bigcup_{i\in \mathcal{I}_{1} }\{\delta_i\}$, both of the oracle return a uniformly random value in $\bits^n\setminus \{y_i\}_{i\in \mathcal{I}_{1}}$
    \item For the remaining $z$, conditioning on \zcref{eq:joint-dist-h3-h4} leaves $O_n(z)$ independent and uniform over $\bits^n$.
In $H_3$, $O_n^*(z)$ is sampled uniformly from $\bits^n\setminus\{y_i\}_{i\in\mathcal{I}_0\cup\mathcal{I}_1}$.
In $H_4$, $O_n(z)$ is retained if it lies outside $\{y_i\}_{i\in\mathcal{I}_0\cup\mathcal{I}_1}$; otherwise, it is resampled uniformly from $\bits^n\setminus\{y_i\}_{i\in\mathcal{I}_0\cup\mathcal{I}_1}$.
Thus, the remaining entries are independently uniform over $\bits^n\setminus\{y_i\}_{i\in\mathcal{I}_0\cup\mathcal{I}_1}$ in both hybrids.
\end{enumerate}
Thus, the two final oracles are identically distributed and we can conclude that $H_3 \idind H_4$.

This completes the proof of \zcref{claim:hyb-3-4}.

\end{proof}

\subsubsection{Hybrid $H_5$}

We present $H_5$ in \Cref{hybrid:h5} and highlight its difference from $H_4$ in \red{red color}.

\begin{ExperimentBox}[label={hybrid:h5}]{Hybrid $H_5$}
\begin{enumerate}
    \item According to the  input $\vb{r} = (r_1, \ldots, r_{p(n)})$, define
     $\begin{cases}
    \mathcal{I}_0  \coloneqq\{i\in [p(n)]~|~ r_i=0\} \\ 
    \mathcal{I}_1  \coloneqq\{i\in [p(n)]~|~ r_i=1\}
    \end{cases}$.
    \item
    Sample $O_{n} \gets \Ora_n$
    %\item $\forall i\in [p(n)]$, sample $r_i\in \{0,1\}$ where $\Pr[r_i=1]=\msf{Approx}_{\frac{1}{e}}(1^n)$;
    \item Define
     $\begin{cases}
    \mathcal{I}_0  \coloneqq\{i\in [p(n)]~|~ r_i=0\} \\ 
    \mathcal{I}_1  \coloneqq\{i\in [p(n)]~|~ r_i=1\}
    \end{cases}$
    \item $\forall i\in [p(n)]$ :
        \begin{itemize}
            \item If $r_i=0$,  then sample $\delta_i \pick \bits^n$ and define $y_i \coloneqq O_{n}(\delta_i)$ and $v_i \coloneqq \delta_i$.
            \item If $r_i=1$, then set $\delta_i \pick \bits^n, y_i=O_n(\delta_i)$ 
        \end{itemize}
    \item
        $O^*_n(z) \coloneqq \begin{cases}
        \picks \bits^n\setminus\{y_i\}_{i\in \mathcal{I}_{1}} & \text{if } z \in \bigcup_{i\in \mathcal{I}_{1}} \{\delta_i\} \\ 
        \red{O_n(z)}  & \text{if } z \in \bigcup_{i\in \mathcal{I}_{0} \cup \mathcal{I}_{1}}\msf{PreImage}^{O_n}(y_i) \setminus\{\delta_i\}_{i\in \mathcal{I}_{0} \cup \mathcal{I}_{1}}\\
        O_n(z) & \text{otherwise}
   \end{cases}$ 
    \item Invoke the adversary $\Adv^{\ket{O^{*}_n}}(\{r_i\}_{i\in[p(n)]},\{y_i\}_{i\in[p(n)]},\{v_i\}_{i\in\mcal{I}_0})$ 
\end{enumerate}
\end{ExperimentBox}

\begin{MyClaim}\zcsetup{reftype=Claim}\label{claim:hyb-4-5}
    It holds that $H_4 \qcind H_5$, where $\qcind$ denotes indistinguishability against polynomial-query quantum adversary.
\end{MyClaim}

\begin{proof}
 We use one-way-to-hiding lemma to reduce indistinguishability to collision resistance.

Apply \Cref{O2H} with auxiliary input $z=(\{r_i\}_{i\in[p(n)]},\{y_i\}_{i\in[p(n)]},\{v_i\}_{i\in\mcal{I}_0})$,  $G$ and $H$
equal to the final oracle $O^*_n$ in $H_4$ and $H_{5}$ respectively,
and 
$S=\bigcup_{i\in \mathcal{I}_{0} \cup \mathcal{I}_{1}\msf{PreImage}^{O_n}(y_i)}\setminus\{\delta_i\}_{i\in \mathcal{I}_{0} \cup \mathcal{I}_{1}}$. It is obvious that the two
oracles agree outside $S$. Let $q(n)$ be the query count
of the fixed adversary and let $B$ be the measured-query
extractor defined in \Cref{O2H}. Then its distinguishing
advantage is at most $2q(n)\sqrt{P_{\mathrm{guess}}}$, where
\begin{equation}\zcsetup{reftype=Expression}\label{eq:guessing-bound2}
P_{\mathrm{guess}}=\Pr_{H_5}[T\cap S\neq\emptyset~:~ T\gets B^{\ket{O^*_n}}(z)].
\end{equation}
Then, to finish the proof of \zcref{claim:hyb-4-5}, it suffices to show that $P_{\msf{guess}}$  defined in \eqref{eq:guessing-bound2} is negligible for any $q(n)$-query adversary. We proceed to establish that  below.

    Assume for contradiction that there is a polynomial $\poly(n)$ such that  $P_{\msf{guess}}\ge \frac{1}{\poly(n)}$ for infinitely many $n$. 
    We then construct an oracle-aided quantum algorithm \textbf{CollisionFinder} that can find a collision pair for oracle $O_n$ with non-negligible probability as follows: \chakfai{Changed}

    \para{CollisionFinder}$^{\ket{O_n}}$:
        \begin{enumerate}
            \item Repeat the steps to produce $z \coloneqq (\{y_i\}_{i\in[p(n)]},\{v_i\}_{i\in\mathcal{I}_0})$ as in hybrid $H_{5}$; In particular, record the values $\Set{\delta_i}_{i\in \mathcal{I}_0 \cup \mathcal{I}_1}$ for later use.
            \item Invoke the oracle algorithm $T \gets B^{O^*_n}(z)$ where the oracle $O^*_n$ is defined in the same manner as in the hybrid $H_{5}$.
            % \item Parse the output set $T$ from the oracle algorithm $B^{O^{(5)}_n}$, denote the $j$-th element of $T$ by $t_j$ and query all the corresponding images $\Ora_n(t_j)$.
            \item Check whether there exists a $t \in T$ such that $O_n(t)=y_i$ and $t\neq \delta_i$ for some $i \in \mcal{I}_0 \cup \mcal{I}_1$.
            \item If such a pair exists, then output the corresponding pair $(\delta_i,t)$ as the collision pair and halt. 
        \end{enumerate}
    Although $\textbf{CollisionFinder}$ is given oracle access only to
    $O_n$, it can coherently simulate $O_n^*$ using recorded value in Step 1.
    It can be easily seen that whenever $T \cap S \ne \emptyset$, this \textbf{CollisionFinder} algorithm returns a collision for $\ket{O_n}$; hence the success
    probability of collision finding is at least $P_{\mathrm{guess}}$, which is non-negligible. However, this contradicts \Cref{collision_finding}.

This completes the proof of \zcref{claim:hyb-4-5}.

\end{proof}

\subsubsection{Hybrid $H_6$} 

We present $H_6$ in \Cref{hybrid:h6} and highlight its difference from $H_5$ in \red{red color}.

\begin{ExperimentBox}[label={hybrid:h6}]{Hybrid $H_6$}
    
\begin{enumerate}
    \item According to the  input $\vb{r} = (r_1, \ldots, r_{p(n)})$, define
     $\begin{cases}
    \mathcal{I}_0  \coloneqq\{i\in [p(n)]~|~ r_i=0\} \\ 
    \mathcal{I}_1  \coloneqq\{i\in [p(n)]~|~ r_i=1\}
    \end{cases}$.
    \item
    Sample $O_{n} \gets \Ora_n$
    \item Define
     $\begin{cases}
    \mathcal{I}_0  \coloneqq\{i\in [p(n)]~|~ r_i=0\} \\ 
    \mathcal{I}_1  \coloneqq\{i\in [p(n)]~|~ r_i=1\}
    \end{cases}$
    \item $\forall i\in [p(n)]$ :
        \begin{itemize}
            \item If $r_i=0$,  then sample $\delta_i \pick \bits^n$ and define $y_i \coloneqq O_{n}(\delta_i)$ and $v_i \coloneqq \delta_i$.
            \item If $r_i=1$, then set $\delta_i \pick \bits^n, y_i=O_n(\delta_i)$ 
        \end{itemize}

    \item \red{Delete the definition of $O^*_n$.}    
    \item Invoke the adversary $\Adv^{\red{\ket{O_n}}}(\{r_i\}_{i\in[p(n)]},\{y_i\}_{i\in[p(n)]},\{v_i\}_{i\in\mcal{I}_0})$ 
\end{enumerate}
\end{ExperimentBox}

\begin{MyClaim}\zcsetup{reftype=Claim}\label{claim:hyb-5-6}
    It holds that $H_5 \qcind H_6$, where $\qcind$ denotes indistinguishability against polynomial-query quantum adversary.
\end{MyClaim}

\begin{proof}
We use one-way-to-hiding lemma to reduce indistinguishability to preimage finding.

Apply \Cref{O2H} with auxiliary input $z=(\{y_i\}_{i\in[p(n)]},\{v_i\}_{i\in\mcal{I}_0})$,  $G$ and $H$
equal to the final oracle in $H_5$ and $H_{6}$ respectively,
and 
$S=\bigcup_{i\in\mcal{I}_1}\{\delta_i\}$. It is obvious that the two
oracles agree outside $S$. Let $q(n)$ be the query count
of the fixed adversary and let $B$ be the measured-query
extractor defined in \Cref{O2H}. Then its distinguishing
advantage is at most $2q(n)\sqrt{P_{\mathrm{guess}}}$, where
\begin{equation}\zcsetup{reftype=Expression}\label{eq:guessing-bound3}
P_{\mathrm{guess}}=\Pr_{H_6}[T\cap S\neq\emptyset~:~ T\gets B^{\ket{O_n}}(z)].
\end{equation}
Then, to finish the proof of \zcref{claim:hyb-5-6}, it suffices to show that $P_{\msf{guess}}$  defined in \eqref{eq:guessing-bound3} is negligible for any $q(n)$-query adversary. We proceed to establish that  below.

    Assume for contradiction that there is a polynomial $\poly(n)$ such that  $P_{\msf{guess}}\ge \frac{1}{\poly(n)}$ for infinitely many $n$. 
    We then construct an oracle-aided quantum algorithm
    $\mathbf{PreImageFinder}$ that, given challenge values
    $z_1,\ldots,z_{p(n)}$ sampled as in
    \Cref{multi_preimage_search}, finds a preimage of one of these
    values under $O_n$ with non-negligible probability, as follows:
    
    \para{PreImageFinder}$^{\ket{O_n}}(z_1,z_2,...,z_{p(n)})$:
        \begin{enumerate}
            \item Repeat the steps to produce $z \coloneqq (\{r_i\}_{i\in[p(n)]},\{y_i\}_{i\in[p(n)]},\{v_i\}_{i\in\mathcal{I}_0})$ as in hybrid $H_{5}$; except for the following change in step 4:
                \begin{itemize}
                    \item If $r_i=1$, set $y_i=z_i$
                \end{itemize}
            \item Invoke the oracle algorithm $T \gets B^{\ket{O_n}}(z)$

            \item Check whether there exists a $t \in T$ such that such that $O_n(t)=y_i$ for some $i \in \mcal{I}_1$.
            \item If such a pair exists, then output the corresponding point $t$ as the targeted pre-image and halt. 
        \end{enumerate}
        The substitution $y_i=z_i$ for $i\in\mathcal{I}_1$ in Step~1
        preserves the joint distribution of the oracle and the transcript
        $\pi$ in $H_6$, since the challenge values in
        \Cref{multi_preimage_search} are sampled in exactly the same
        manner as the corresponding $y_i$.
        Whenever $T\cap S\neq\emptyset$, $\mathbf{PreImageFinder}$
        returns a point $t$ satisfying $O_n(t)=y_i=z_i$ for some
        $i\in\mathcal{I}_1$.
        Hence, its success probability is at least $P_{\mathrm{guess}}$,
        which is non-negligible.
        This contradicts \Cref{multi_preimage_search}.

This completes the proof of \zcref{claim:hyb-5-6}.

\end{proof}

\subsection{\texorpdfstring{Proof of \zcref{claim:main-hyb-0-1}: $H_0\sind H_1$}{From H0 to H1}}
\label{sec:ind:hyb-0-1}

\subsubsection{Proof Strategy Overview}

We prove $H_0\sind H_1$ by establishing the stronger statement
that the joint distributions of the complete oracle table and
the indexed inputs supplied to $\Adv$ are statistically close
in the two experiments. We first introduce the notation used
throughout the proof in \Cref{sec:ind:hyb-0-1:notation}. We then
establish useful technical lemmas concerning random oracles
and their sampling distributions in \Cref{sec:H0-H1:tools}.
Finally, in \Cref{sec:h0-h1-final-proof}, we use these lemmas to
construct the hybrid argument proving $H_0\sind H_1$.

To analyze these joint distributions directly, we introduce
the joint-output experiments $\hat H_0$ and $\hat H_1$,
defined in \Cref{hybrid:h00,hybrid:h01}, respectively.
Similar as $H_0$ and $H_1$, these experiments have a input bit vector
$\boldsymbol r=(r_i)_{i\in[p(n)]}$ and output the complete
oracle table together with the inputs that would be supplied
to $\Adv$, rather than invoking $\Adv$.

% Unlike $H_0$ and $H_1$, these experiments fix the bit vector
% $\boldsymbol r=(r_i)_{i\in[p(n)]}$ and output the complete
% oracle table together with the inputs that would be supplied
% to $\Adv$, rather than invoking $\Adv$.

For convenience, we organize these inputs into two ordered
tuples: $T$ contains the labels $y_i$ at indices with $r_i=1$,
and $W$ contains the pairs $(y_i,v_i)$ at indices with $r_i=0$,
both listed in increasing index order.
These tuples and the associated sampling notation are defined
formally in the following \Cref{sec:ind:hyb-0-1:notation}.

The joint-output formulation allows us to compare distributions
without tracking the adversary's oracle queries.
It also permits convenient changes in sampling order:
conditional on the originally sampled oracle $O_n$ and the
fixed bit vector, the pairs comprising $W$ are independent
of one another, of $T$, and of the randomness used to modify
the oracle. Thus, we may sample $T$ and construct the reported
oracle before sampling $W$, without changing their joint
distribution. We then establish a chain of statistically close
intermediate experiments,
\[
    \hat H_0 \sind \cdots \sind \hat H_1,
\]
by gradually changing the sampling procedures and the reported
oracle. Each transition incurs only negligible statistical
distance, with bounds uniform over $\boldsymbol r$.

Finally, given $\boldsymbol r$ and the output of either
joint-output experiment, we can reconstruct all indexed inputs
and run $\Adv$ with the reported oracle.
For each $b\in\{0,1\}$, this common randomized transformation
recovers $H_b$ conditioned on $\boldsymbol r$.
Since randomized post-processing cannot increase statistical
distance (\Cref{lem:h01-statistical-distance}), the statistical
closeness of $\hat H_0$ and $\hat H_1$ implies that of the
corresponding conditional experiments.
Averaging over the common distribution of $\boldsymbol r$
therefore yields $H_0\sind H_1$.

\subsubsection{Notation}
\label{sec:ind:hyb-0-1:notation}

Let $N=2^n$. Since $p(n)$ is polynomially bounded, we restrict
attention to sufficiently large $n$ such that $p(n)<N$.
Write
\[
\textbf{r}=(r_i)_{i\in[p(n)]},
\qquad
\mathcal I_0=\{i:r_i=0\},
\qquad
\mathcal I_1=\{i:r_i=1\}.
\]

To simplify the presentation, define the ordered tuples
\[
    T=(y_i)_{i\in\mathcal I_1},~
    W=((y_i,v_i))_{i\in\mathcal I_0},
\]
with indices in increasing order. Given $\textbf{r}$, the pair $(T,W)$
contains precisely the labels and witnesses supplied to $\Adv$.

For any function $f:\bits^n\to\bits^n$, use the image and
preimage notation from the hybrids:
\[
\msf{Image}^{f}
=
\{y\in\bits^n:\exists z\in\bits^n,\ f(z)=y\},
\qquad
\msf{PreImage}^{f}(y)
=
\{z\in\bits^n:f(z)=y\}.
\]
Let $K_f$ denote the distribution of one positive
label--witness pair:
\begin{equation}
\label{eq:h01-positive-kernel}
K_f(y,v)
=
\begin{cases}
\displaystyle
\frac{1}{|\msf{Image}^{f}|\,|\msf{PreImage}^{f}(y)|},
    & f(v)=y,\\[2mm]
0,  & f(v)\ne y.
\end{cases}
\end{equation}
Thus, sampling from $K_f$ means sampling $y$ uniformly from
$\msf{Image}^{f}$ and then sampling $v$ uniformly from
$\msf{PreImage}^{f}(y)$. The distribution $K_f^{\otimes a}$
describes all $a$ independently sampled positive pairs for some integer $a$.
When $a=0$, it is the point mass on the empty tuple.

 Finally, we recall the convention that  once we need to sample from an empty set, the output is defined to be some special symbol (e.g., $\bot$ or $0^n$).

\subsubsection{Technical Tools regarding Random Oracles} 
\label{sec:H0-H1:tools}

In this subsection, we establish several properties of random
oracles that will be used in the proof of $H_0\sind H_1$.
These results allow us to bound the abort probabilities and
justify the changes in sampling across the subsequent hybrids.

\begin{lemma}[Image size of a random oracle]
\label{lem:random-oracle-image-size}
Fix any $n\in \Naturals$, let $O_n$ be a random oracle
from $\bits^n$ to $\bits^n$.

\[
    \Pr_{O_n\gets\Ora_n}[|\msf{Image}(O_n)|\leq2^{n/2}]
    \leq 2^{n2^{n/2}-n2^{n-1}}
    =\negl(n).
\]
\end{lemma}

\begin{proof}
Let $L=\lfloor2^{n/2}\rfloor$.
For any fixed $L$-element subset $A\subseteq\bits^n$,
independence of the oracle outputs gives
\[
    \Pr_{O_n\gets\Ora_n}[\msf{Image}(O_n)\subseteq A]
    =\left(\frac{L}{2^n}\right)^{2^n}.
\]
Thus, by a union bound,
\begin{align*}
    \Pr_{O_n\gets\Ora_n}[|\msf{Image}(O_n)|\leq2^{n/2}]
    &=
    \Pr_{O_n\gets\Ora_n}[\exists A\subseteq\bits^n:
         |A|=L,\ \msf{Image}(O_n)\subseteq A]\\
    &\leq
    \binom{2^n}{L}
\left(\frac{L}{2^n}\right)^{2^n}\\
&\leq
(2^n)^L
\left(\frac{L}{2^n}\right)^{2^n}\\
&\leq
(2^n)^{2^{n/2}}
\left(\frac{2^{n/2}}{2^n}\right)^{2^n}\\
&=
2^{n2^{n/2}-n2^{n-1}}
\end{align*}
\end{proof}
\begin{lemma}[Concentration of the Image Size]
\label{lem:ro-image-concentration}
For any $n\in\mathbb{N}$, let
$O_n:\{0,1\}^n\to\{0,1\}^n$ be sampled from $\Ora_n$, and let
\[
    \mu
    \coloneqq
    \mathbb{E}_{O_n\gets\Ora_n}
    \bigl[|\msf{Image}^{O_n}|\bigr].
\]
Then
\[
    \mu
    =
    2^n\left(1-\left(1-\frac{1}{2^n}\right)^{2^n}\right)
    \geq
    \left(1-\frac{1}{e}\right)2^n
    \geq 2^{n-1}.
\]
Moreover, for every $t\geq 0$,
\[
    \Pr_{O_n\gets\Ora_n}
    \left[
        \left||\msf{Image}^{O_n}|-\mu\right|\geq t
    \right]
    \leq
    2\exp\left(-\frac{2t^2}{2^n}\right).
\]
In particular,
\begin{equation}
\label{eq:image-concentration}
    \Pr_{O_n\gets\Ora_n}
    \left[
        \left||\msf{Image}^{O_n}|-\mu\right|
        \geq 2^{3n/4}
    \right]
    \leq
    2\exp\left(-2\cdot 2^{n/2}\right).
\end{equation}
\end{lemma}

\begin{proof}
For each $y\in\{0,1\}^n$, we have
\[
    \Pr_{O_n\gets\Ora_n}
    \bigl[y\notin\msf{Image}^{O_n}\bigr]
    =
    \left(1-\frac{1}{2^n}\right)^{2^n}.
\]
For each $y\in\{0,1\}^n$, let $I_y$ be the indicator of the
event $y\in\msf{Image}^{O_n}$. Since
\[
    |\msf{Image}^{O_n}|
    =
    \sum_{y\in\{0,1\}^n} I_y,
\]
linearity of expectation gives
\[
    \mu
    =
    \sum_{y\in\{0,1\}^n}\mathbb{E}[I_y]
    =
    \sum_{y\in\{0,1\}^n}
    \Pr_{O_n\gets\Ora_n}
    \bigl[y\in\msf{Image}^{O_n}\bigr].
\]
For any fixed $y\in\{0,1\}^n$, the event
$y\notin\msf{Image}^{O_n}$ occurs precisely when
$O_n(x)\neq y$ for every $x\in\{0,1\}^n$.
The oracle entries are independent and uniformly distributed,
so
\[
    \Pr_{O_n\gets\Ora_n}
    \bigl[y\notin\msf{Image}^{O_n}\bigr]
    =
    \prod_{x\in\{0,1\}^n}
    \Pr_{O_n\gets\Ora_n}[O_n(x)\neq y]
    =
    \left(1-\frac{1}{2^n}\right)^{2^n}.
\]
Substituting this identity into the expression for $\mu$ yields
\[
    \mu
    =
    2^n\left(
        1-\left(1-\frac{1}{2^n}\right)^{2^n}
    \right).
\]
Finally, the inequality $1-u\leq e^{-u}$ for $u\geq 0$ implies
\[
    \left(1-\frac{1}{2^n}\right)^{2^n}
    \leq
    \left(e^{-2^{-n}}\right)^{2^n}
    =
    e^{-1}.
\]
Consequently,
\[
    \mu
    \geq
    \left(1-\frac{1}{e}\right)2^n
    \geq
    2^{n-1},
\]
where the last inequality follows from $e>2$

The image size is a function of the $2^n$ independent oracle
entries $\{O_n(x)\}_{x\in\{0,1\}^n}$.
Changing any one entry changes the image size by at most one.
Thus, \cref{lem:mcdiarmid}, with each $c_i=1$, gives
\[
    \Pr_{O_n\gets\Ora_n}
    \left[
        \left||\msf{Image}^{O_n}|-\mu\right|\geq t
    \right]
    \leq
    2\exp\left(-\frac{2t^2}{2^n}\right).
\]
Setting $t=2^{3n/4}$ proves
\cref{eq:image-concentration}.
\end{proof}
\begin{lemma}[Preimage distribution for forward sampling]
\label{lem:sampled-output-preimage}
Fix $n\in\Naturals$.
Sample $O_n\gets\Ora_n$ and $\delta\picks\bits^n$
independently. Then
\[
    |\msf{PreImage}^{O_n}(O_n(\delta))| \idind 1+X,
    \qquad
    X\sim\msf{Binomial}(2^n-1,2^{-n}).
\]
\end{lemma}

\begin{proof}
Condition on $\delta=z$ and $O_n(z)=y$.
The input $z$ itself belongs to $\msf{PreImage}^{O_n}(y)$.
Each of the remaining $2^n-1$ inputs maps to $y$
independently with probability $2^{-n}$.
The claimed distribution follows, since it does not
depend on the conditioned values $z$ and $y$.

\end{proof}

\begin{corollary}[of \Cref{lem:sampled-output-preimage}, tail bound for preimage size]
\label{lem:sampled-output-preimage-tail}
Sample $O_n\gets\Ora_n$ and $\delta\picks\bits^n$
independently. Then
\[
    \Pr_{\delta\picks\bits^n,\;O_n\gets\Ora_n}\!\left[
        |\msf{PreImage}^{O_n}(O_n(\delta))|>\log n
    \right]
    =\negl(n).
\]
\end{corollary}

\begin{proof}
By \Cref{lem:sampled-output-preimage}, the preimage size
is distributed as $1+X$, where
\[
    X\sim\msf{Binomial}(2^n-1,2^{-n}),
    \qquad
    \E[X]=1-2^{-n}<1.
\]
Let $t=\lfloor\log n\rfloor$.
Since $X$ is integer-valued, $1+X>\log n$ is equivalent
to $X\geq t$.
For sufficiently large $n$, we have $t\geq1$.
Applying \Cref{chernoff_2} with $\mu=1$ gives
\begin{align*}
    \Pr\!\left[
        |\msf{PreImage}^{O_n}(O_n(\delta))|>\log n
    \right]
    &=\Pr[X\geq t]\\
    &\leq e^{-1}\left(\frac et\right)^t\\
    &=\negl(n).
\end{align*}
The final bound is negligible because
$t=\Theta(\log n)$ and
$(e/t)^t=\exp(-\Theta(\log n\log\log n))$.
\end{proof}

\begin{BoxBox}[label={box:change-sampling-order},
    sidebyside,
    sidebyside align=top,
    float,
    floatplacement=!t,
    fontupper=\small,
    fontlower=\small
]{Games $G_0$ and $G_1$: Changing the sampling order}

    \vspace{0.5em}
    \uline{\textbf{Game $G_0(\mcal{I}_1)$:}}
    \begin{enumerate}
        \item Sample $O_n\gets\Ora_n$.

        \item $\forall i\in\mcal I_1$, sample 
        \(
            y_i\pick
            \bits^n\setminus\msf{Image}^{O_n}. 
        \) \\
        (Recall our convention that $y_i$ is set to a special symbol (e.g., $0^n$) if $\bits^n\setminus\msf{Image}^{O_n} = \emptyset$.)

        \item Set
        $T\coloneqq\{y_i\}_{i\in\mcal I_1}$.

        \item Output $(O_n,T)$.
    \end{enumerate}

    \tcblower

    \vspace{0.5em}
    \uline{\textbf{Game $G_1(\mcal{I}_1)$:}}
    \begin{enumerate}
        \item  $\forall i\in\mcal I_1$,  sample
        \(
            y_i\pick\bits^n.
        \)

        \item Set
        $T\coloneqq\{y_i\}_{i\in\mcal I_1}$.

        \item Sample an oracle $O_n$ where each $O_n(x)$ is independently sampled from
        $\bits^n\setminus T$.

        \item Output $(O_n,T)$.
    \end{enumerate}

\end{BoxBox}

\begin{lemma}[Changing the sampling order] 
\label{lem:change-sampling-order}
Let $\mcal{I}_1$ be any fixed subset of $[p(n)]$ where $p(n)$ is a polynomially bounded function of $n$.
The games $G_0$ and $G_1$ defined in \Cref{box:change-sampling-order} are statistically close: $G_0 \sind G_1$.
\end{lemma}

\begin{proof}
Write $N\coloneqq 2^n$ and $m\coloneqq|\mcal I_1|=\poly(n)$.  The case $m=0$ is immediate, so assume
$1\le m<N$, as holds for all sufficiently large $n$.

In this proof, instead of $T$,  we use the ordered samples
$y=(y_1,\ldots,y_m)$ in the output; replacing $y$ by its underlying
set $T$ cannot increase statistical distance.

Let $P_0(\cdot, \cdot)$ and $P_1(\cdot, \cdot)$ denote the probability mass functions of the joint
distribution of $(O_n,y)$ in $G_0$ and $G_1$, respectively, and write
$\Pr_{G_b}$ for probability in game $G_b$.

Define $K(O_n)\coloneqq N-|\msf{Image}^{O_n}|$.
To make $G_0$ well-defined when $K(O_n)=0$, set
$y=(0^n,\ldots,0^n)$ in that case; the analysis below shows that this
convention affects only a negligible probability mass.

Define the bad event
\begin{equation}\label{eq:sampling-order-bad-event}
E_{\msf{bad}}\coloneqq
\left(
\left|K(O_n)-N/e\right|>N^{3/4}
\right) \ \lor\
\left(\exists\,1\le i<j\le m,~s.t.~ y_i=y_j
\right),
\end{equation}
and define the set of good output pairs by
\[
\mathcal G\coloneqq
\left\{
(O_n,y):
\begin{array}{l}
y_i\notin\msf{Image}^{O_n}\text{ for every }i\in[m],\\
y_i\ne y_j\text{ for every }1\le i<j\le m,\\
|K(O_n)-N/e|\le N^{3/4}
\end{array}
\right\}.
\]
For all sufficiently large $n$, in either game the output belongs to
$\mathcal G$ exactly when $E_{\msf{bad}}$ does not occur.

For a uniform random oracle, the missing-image count satisfies
\[
\mathbb E[K(O_n)]=N(1-1/N)^N=N/e+O(1)
\]
For a uniformly sampled random oracle, the missing-image count
$K(O_n)=N-|\msf{Image}^{O_n}|$ satisfies
\[
    \mathbb{E}[K(O_n)]
    =
    N\left(1-\frac{1}{N}\right)^N
    =
    \frac{N}{e}+O(1).
\]
Since $K(O_n)$ differs from the image size only by subtraction
from $N$, \cref{lem:ro-image-concentration} gives the same
concentration bound for $K(O_n)$ around its expectation.
Moreover, for all sufficiently large $N$,
\[
    \left|\mathbb{E}[K(O_n)]-\frac{N}{e}\right|
    \leq \frac{1}{2}N^{3/4}.
\]
Therefore,
\[
    \begin{aligned}
        \Pr_{O_n\gets\Ora_n}
        \left[
            \left|K(O_n)-\frac{N}{e}\right|>N^{3/4}
        \right]
        &\leq
        \Pr_{O_n\gets\Ora_n}
        \left[
            \left|K(O_n)-\mathbb{E}[K(O_n)]\right|
            >\frac{1}{2}N^{3/4}
        \right]\\
        &\leq
        2e^{-\frac{1}{2}\sqrt{N}}.
    \end{aligned}
\]
Otherwise, $K(O_n)=\Omega(N)$, and a union bound bounds the probability
of a sample collision by $\binom m2/K(O_n)=O(m^2/N)$.
For all sufficiently large $n$, the case $K(O_n)=0$ is included
in the deviation event. Hence, by \Cref{eq:sampling-order-bad-event},
\begin{equation}\label{eq:sampling-order-bad-probability}
\Pr_{G_0}[E_{\msf{bad}}]
=O(e^{-\frac{1}{2}\sqrt{N}}+m^2/N)
=\negl(n).
\end{equation}

Fix $(O_n,y)\in\mathcal G$. Since the samples are distinct,
\[
P_0(O_n,y)=N^{-N}K(O_n)^{-m},
\qquad
P_1(O_n,y)=N^{-m}(N-m)^{-N}.
\]
Set $a\coloneqq N^{-N}(e/N)^m$.
For every pair in $\mathcal G$,
$K(O_n)=(N/e)(1+O(N^{-1/4}))$, uniformly over these pairs.
Consequently,
\[
P_0(O_n,y)=a\cdot\exp\!\bigl(O(mN^{-1/4})\bigr),
\qquad
P_1(O_n,y)=a\cdot\exp\!\bigl(O(m^2/N)\bigr).
\]
Since $m=\operatorname{poly}(n)$, there exists a negligible function
$\varepsilon=\varepsilon(n)$ such that, uniformly over
$(O_n,y)\in\mathcal G$,
\begin{equation}\label{eq:sampling-order-pointwise}
|P_0(O_n,y)-a|\le\varepsilon a,
\qquad
|P_1(O_n,y)-a|\le\varepsilon a.
\end{equation}
In particular,
\[
(1-\varepsilon)|\mathcal G|a
\le \sum_{(O_n,y)\in\mathcal G}P_0(O_n,y)
\le 1.
\]
Summing \Cref{eq:sampling-order-pointwise} therefore yields
\begin{equation}\label{eq:sampling-order-good-mass}
\sum_{(O_n,y)\in\mathcal G}
|P_0(O_n,y)-P_1(O_n,y)|
\le 2\varepsilon|\mathcal G|a
\le \frac{2\varepsilon}{1-\varepsilon}
=\negl(n).
\end{equation}
By \Cref{eq:sampling-order-bad-probability,eq:sampling-order-good-mass},
\[
\begin{aligned}
\Pr_{G_1}[E_{\msf{bad}}]
&\le \Pr_{G_0}[E_{\msf{bad}}]
 +\left|
 \Pr_{G_0}[(O_n,y)\in\mathcal G]
 -\Pr_{G_1}[(O_n,y)\in\mathcal G]
 \right|\\
&=\negl(n).
\end{aligned}
\]
Finally,
\[
\Delta(P_0,P_1)
\le \frac12\left(
\sum_{(O_n,y)\in\mathcal G}|P_0(O_n,y)-P_1(O_n,y)|
+\Pr_{G_0}[E_{\msf{bad}}]
+\Pr_{G_1}[E_{\msf{bad}}]
\right)
=\negl(n).
\]

\end{proof}

\begin{lemma}[Resampling outside the preimage]
\label{lem:resample-outside-preimage}
The games $G_0$ and $G_1$ defined in
\Cref{box:resample-outside-preimage} are identically distributed:
\(
    G_0 \idind G_1.
\)
\end{lemma}

\begin{BoxBox}[label={box:resample-outside-preimage},
    sidebyside,
    sidebyside align=top,
    float,
    floatplacement=!t,
    fontupper=\small,
    fontlower=\small
]{Games $G_0$ and $G_1$: Resampling Outside the Preimage}

    \vspace{0.5em}
    \uline{\textbf{Game $G_0$:}}
    \begin{enumerate}
        \item Sample $O_n\gets\Ora_n$.

        \item For each $i\in\mcal I_1$, independently sample
        $\delta_i\pick\bits^n$ and set $y_i\coloneqq O_n(\delta_i).$

        \item Set
        $T\coloneqq\{y_i\}_{i\in\mcal I_1}$.

        \item Set $S\coloneqq \bigcup_{i\in\mcal{I}_1}\msf{PreImage}(y_i)$
        and  $\bar S\coloneqq\bits^n\setminus S$.
        \item Name the resulting function generated by $O_n$ restricted to $\bar S$ by $O_n|_{\bar S}$.
    \end{enumerate}
    Output $(S,T,O_n|_{\bar S})$.

    \tcblower

    \vspace{0.5em}
    \uline{\textbf{Game $G_1$:}}
    \begin{enumerate}
        \item Run Steps~1--4 of $G_0$ to obtain
        $(O_n,S, \bar{S}, T)$.

        \item For every $x\in\bar S$, independently
        resample
        \[
            O_n(x)\pick\bits^n\setminus T.
        \]
        \item Name the resulting function generated by $O_n$ restricted to $\bar S$ by $O_n|_{\bar S}$.
    \end{enumerate}
        Output $(S,T,O_n|_{\bar S})$.

\end{BoxBox}

\begin{proof}
Fix any attainable pair $(S,T)$.
Since $S=\bigcup_{i\in\mcal{I}_1}\msf{PreImage}(y_i)$, every oracle value outside $S$
must lie outside $T$.
All remaining constraints defining $(S,T)$ involve only
the sampled inputs and the restriction $O_n|_S$.
Consequently, conditional on $(S,T)$, the values
$\{O_n(x)\}_{x\in\bar S}$ are independent and uniformly
distributed over $\bits^n\setminus T$.

The resampling step in $G_1$ therefore samples
$O_n|_{\bar S}$ from exactly its conditional distribution
in $G_0$, preserving the joint distribution of
$(S,T,O_n|_{\bar S})$.

\end{proof}

\subsubsection{
\texorpdfstring{Proving $H_0\sind H_1$}{Indistinguishability of H0 and H1}}
\label{sec:h0-h1-final-proof}

\chakfai{Somehow the previous version hybrids magically can be written as $H_0,\hat{H}_{0.1},\cdots ,H_1$ 
without adding or removing any hybrids. It is just renaming.}

\para{Hybrid $\hat{H}_0$:}
We present $\hat{H}_0$, the joint-output experiment corresponding
to $H_0$, in \Cref{hybrid:h00}.
\begin{ExperimentBox}[label={hybrid:h00}]{Hybrid $\hat{H}_0(\{r_i\}_{i\in [p(n)]})$}
    
\begin{enumerate}
    \item
    Sample $O_{n} \gets \Ora_n$;    
    \item For $i \in \mcal{I}_1:$
        \begin{itemize}
            \item Sample $y_i\pick \bits^n\setminus \msf{Image}^{O_{n}}$.
        \end{itemize}
    \item Set $T\coloneqq \{y_i\}_{i\in\mcal{I}_1}$
    \item Sample $W\sim K_{O_n}^{\otimes |\mcal{I}_0|}$ 
    \item Output $(O_n,T,W)$
\end{enumerate}
\end{ExperimentBox}

\para{Hybrid $\hat{H}_{0.1}$:}
We present $\hat{H}_{0.1}$ in \Cref{hybrid:h0p1} and highlight its difference from $\hat{H}_0$ in \red{red color}.
In this hybrid, we first sample $y_i\pick\bits^n$ independently
for each $i\in\mcal{I}_1$ and set
$T=\{y_i\}_{i\in\mcal{I}_1}$.
We then sample each $O_n(x)$ independently and uniformly from
$\bits^n\setminus T$, ensuring that
$T\cap\msf{Image}^{O_n}=\varnothing$.
For $i\in\mcal{I}_0$, the pairs $(y_i,v_i)$ are sampled as
before, namely $W\sim K_{O_n}^{\otimes|\mcal{I}_0|}$.

\begin{ExperimentBox}[label={hybrid:h0p1}]{Hybrid $\hat{H}_{0.1}(\{r_i\}_{i\in [p(n)]})$ } 
    
\begin{enumerate} 
    \item For $i \in \mcal{I}_1:$
        \begin{itemize}
            \item Sample \red{$y_i\pick \bits^n$}
        \end{itemize}
    \item Set $T\coloneqq \{y_i\}_{i\in\mcal{I}_1}$
    \item
    Sample $O_{n}$ where each $x\in \bits^n$ , $O_n(x)$ is sampled independently and uniformly from $\bits^n\setminus T$;   
    \item Sample $W\sim K_{O_n}^{\otimes |\mcal{I}_0|}$ 
    \item Output $(O_n,T,W)$
\end{enumerate}
\end{ExperimentBox}
\begin{MyClaim}\zcsetup{reftype=Claim}\label{claim:hybvs1}
    It holds that $\hat{H}_0\sind \hat{H}_{0.1}$.
\end{MyClaim}

\begin{proof}
First, 
by \Cref{lem:change-sampling-order}, the joint distributions
of the following variables are statistically indistinguishable
in the two hybrids:
\begin{equation}\zcsetup{reftype=Expression}
\label{eq:joint-dist-h00-h0p1}
    O_n,\qquad T=\{y_i\}_{i\in\mcal{I}_1}.
\end{equation}

Next, observe that, conditioned on the variables in
\zcref{eq:joint-dist-h00-h0p1}, both hybrids sample $W$ from
the same distribution $K_{O_n}^{\otimes|\mcal{I}_0|}$.
Thus, their full outputs $(O_n,T,W)$ are obtained by applying
the same randomized transformation to their respective
distributions of the variables in
\zcref{eq:joint-dist-h00-h0p1}.
Since statistical distance cannot increase under such a
transformation by \Cref{lem:h01-statistical-distance}, the full output distributions are also
statistically indistinguishable.

This yields $\hat{H}_0\sind \hat{H}_{0.1}$,
completing the proof of \zcref{claim:hybvs1}.

\end{proof}

\para{Hybrid $\hat{H}_{0.2}$:}
We present $\hat{H}_{0.2}$ in \Cref{hybrid:h0p2} and highlight its difference from $\hat{H}_{0.1}$ in \red{red color}.
In this hybrid, we rename the sampled function $O_n$ to $F$.
Accordingly, we sample $W\sim K_F^{\otimes|\mcal{I}_0|}$
and output $(F,T,W)$.
This change is purely notational and preserves the output distribution.
\begin{ExperimentBox}[label={hybrid:h0p2}]{Hybrid $\hat{H}_{0.2}(\{r_i\}_{i\in [p(n)]})$ }
    
\begin{enumerate}  
    \item For $i \in \mcal{I}_1:$
        \begin{itemize}
            \item Sample $y_i\pick \bits^n$
        \end{itemize}
    \item Set $T\coloneqq \{y_i\}_{i\in\mcal{I}_1}$
    \item
    Sample $\red{F}$ where each $x\in \bits^n$ , $\red{F}(x)$ is sampled independently and uniformly from $\bits^n\setminus T$;   
    \item Sample $W\sim K_{\red{F}}^{\otimes |\mcal{I}_0|}$ 
    \item Output $(\red{F},T,W)$
\end{enumerate}
\end{ExperimentBox}

\begin{MyClaim}\zcsetup{reftype=Claim}\label{claim:hybvs2}
    It holds that $\hat{H}_{0.1} \idind \hat{H}_{0.2}$. 
\end{MyClaim}
\begin{proof}
    This hybrid is identical to $\hat{H}_{0.1}$ except a notational change where we rename the sampled function $O_n$ to $F$. Therefore, the two distributions are identical.
\end{proof}

\para{Hybrid $\hat{H}_{0.3}$:}
We present $\hat{H}_{0.3}$ in \Cref{hybrid:h0p3} and highlight its difference from $\hat{H}_{0.2}$ in \red{red color}.
In this hybrid, we generate $\{y_i\}_{i\in\mcal{I}_1}$
by first sampling $O_n\gets\Ora_n$ and then independently
sampling $\delta_i\pick\bits^n$ and setting $y_i=O_n(\delta_i)$
for each $i\in\mcal{I}_1$.
We set $T=\{y_i\}_{i\in\mcal{I}_1}$ and introduce
$S=\bigcup_{i\in\mcal{I}_1}\msf{PreImage}^{O_n}(y_i)$,
the set of all inputs that $O_n$ maps into $T$.
\begin{ExperimentBox}[label={hybrid:h0p3}]{Hybrid $\hat{H}_{0.3}(\{r_i\}_{i\in [p(n)]})$ }

\begin{enumerate}  
    \item \red{Sample $O_n\gets \Ora_n$;}
    \item 
    \red{ 
        For $i \in \mcal{I}_1:$
        \begin{itemize}
            \item Sample $\delta_i\pick \bits^n,y_i=O_n(\delta_i)$
        \end{itemize}
        }
    \item Set $T\coloneqq \{y_i\}_{i\in\mcal{I}_1}$
    \item \red{Define $S\coloneqq \bigcup_{i\in\mcal{I}_1}\msf{PreImage}^{O_n}(y_i)$, $\bar S\coloneqq\bits^n\setminus S$}
    \item Sample a function $F:\bits^n\to\bits^n$ where each entry independently:
    \[
F(x) =
\begin{cases}
\text{pick uniformly from } \bits^n\setminus T & \text{if } x\in S, \\
\text{pick uniformly from } \bits^n\setminus T & \text{if } x \in \bar S.
\end{cases}
\]
    \item Sample $W\sim K_{F}^{\otimes |\mcal{I}_0|}$ 
    \item Output $(F,T,W)$
\end{enumerate}
\end{ExperimentBox}
\begin{MyClaim}\zcsetup{reftype=Claim}\label{claim:hybvs3}
    It holds that $\hat{H}_{0.2} \sind \hat{H}_{0.3}$. 
\end{MyClaim}
\begin{proof}
    Write $k=|\mcal{I}_1|$.
    Let $E$ denote the event that the inputs
    $\{\delta_i\}_{i\in\mcal{I}_1}$ sampled in $\hat{H}_{0.3}$
    are pairwise distinct.

    First, conditioned on $E$, the values $y_i=O_n(\delta_i)$
    are independent and uniformly distributed over $\bits^n$.
    Thus, the distribution of $T$ in $\hat{H}_{0.3}$ conditioned on $E$
    is identical to its distribution in $\hat{H}_{0.2}$.

    Next, conditioned on $T$ in $\hat{H}_{0.2}$ and on $T$ together
    with $E$ in $\hat{H}_{0.3}$, both hybrids sample every entry of $F$
    independently and uniformly from $\bits^n\setminus T$,
    and then sample $W\sim K_F^{\otimes|\mcal{I}_0|}$.
    Consequently, we have $\hat{H}_{0.2}\idind (\hat{H}_{0.3}\mid E).$
    Therefore,
    \[
        \Delta(\hat{H}_{0.2},\hat{H}_{0.3})
        \leq \Pr[\neg E]
        \leq \Pr[\exists i\ne j\in\mcal{I}_1:\delta_i=\delta_j]
        \leq \binom{|\mcal{I}_1|}{2}2^{-n}
        =\negl(n).
    \]
\end{proof}

\para{Hybrid $\hat{H}_{0.4}$:}
We present $\hat{H}_{0.4}$ in \Cref{hybrid:h0p4} and highlight its difference from $\hat{H}_{0.3}$ in \red{red color}.
In this hybrid, we set $F(x)=O_n(x)$ for every
$x\in\bits^n\setminus S$ instead of resampling these values.
Since retaining these values has the same distribution as
resampling them, we have $\hat{H}_{0.3}\idind \hat{H}_{0.4}$.

\begin{ExperimentBox}[label={hybrid:h0p4}]{Hybrid $\hat{H}_{0.4}(\{r_i\}_{i\in [p(n)]})$ }
\begin{enumerate}  
    \item Sample $O_n\gets \Ora_n$;
    \item For $i \in \mcal{I}_1:$
        \begin{itemize}
            \item Sample $\delta_i\pick \bits^n,y_i=O_n(\delta_i)$
        \end{itemize}
    \item Set $T\coloneqq \{y_i\}_{i\in\mcal{I}_1}$
    \item Define $S\coloneqq \bigcup_{i\in\mcal{I}_1}\msf{PreImage}^{O_n}(y_i)$, and $\bar S\coloneqq\bits^n\setminus S$
    \item Sample a function $F:\bits^n\to\bits^n$ where each entry independently:
    \[
F(x) =
\begin{cases}
\text{pick uniformly from } \bits^n\setminus T & \text{if } x\in S, \\
\red{O_n(x)} & \text{if } x \in \bar S.
\end{cases}
\]
    \item Sample $W\sim K_{F}^{\otimes |\mcal{I}_0|}$ 
    \item Output $(F,T,W)$
\end{enumerate}
\end{ExperimentBox}
\begin{MyClaim}\zcsetup{reftype=Claim}\label{claim:hybvs4}
    It holds that $\hat{H}_{0.3} \idind \hat{H}_{0.4}$. 
\end{MyClaim}
\begin{proof}

    First, by \Cref{lem:resample-outside-preimage}, the following
    variables have identical joint distributions in both hybrids:
    \begin{equation}\zcsetup{reftype=Expression}
    \label{eq:joint-dist-h0p3-h0p4}
        S,~T,~F|_{\bar S},
    \end{equation}
    where $F|_{\bar S}$
denotes the restriction of $F$ to inputs in $\bar S$.
    Here, $\hat{H}_{0.3}$ independently resamples the oracle entries
    at inputs $x\in\bar S$, while $\hat{H}_{0.4}$ retains their
    original values.

    Next, conditioned on the variables in
    \zcref{eq:joint-dist-h0p3-h0p4}, both hybrids sample each
    $F(x)$ for $x\in S$ independently and uniformly from
    $\bits^n\setminus T$.
    Thus, $(S,T,F)$ has the same joint distribution in both hybrids.

    Finally, both hybrids sample
    $W\sim K_F^{\otimes|\mcal{I}_0|}$ using fresh randomness.
    Consequently, their full outputs $(F,T,W)$ are identically
    distributed, establishing $\hat{H}_{0.3}\idind \hat{H}_{0.4}$.

\end{proof}
\para{Hybrid $\hat{H}_{0.5}$:} We present $\hat{H}_{0.5}$ in \Cref{hybrid:h0p5} and highlight its difference from $\hat{H}_{0.4}$ in \red{red color}.
In this hybrid, for each $i\in\mcal{I}_0$, we sample
$y_i\pick\msf{Image}(F)$ and abort if
$\msf{PreImage}^{F}(y_i)\neq\msf{PreImage}^{O_n}(y_i)$.
Intuitively, an abort is unlikely because it can occur only
when $y_i\in F(S)$, which is small compared with
$\msf{Image}(F)$, giving $\hat{H}_{0.4}\sind \hat{H}_{0.5}$.
\begin{ExperimentBox}[label={hybrid:h0p5}]{Hybrid $\hat{H}_{0.5}(\{r_i\}_{i\in [p(n)]})$ }
\begin{enumerate}  
    \item Sample $O_n\gets \Ora_n$;
    \item For $i \in \mcal{I}_1:$
        \begin{itemize}
            \item Sample $\delta_i\pick \bits^n,y_i=O_n(\delta_i)$
        \end{itemize}
    \item Set $T\coloneqq \{y_i\}_{i\in\mcal{I}_1}$
    \item Define $S\coloneqq \bigcup_{i\in\mcal{I}_1}\msf{PreImage}^{O_n}(y_i)$, $\bar S\coloneqq\bits^n\setminus S$
    \item Sample a function $F:\bits^n\to\bits^n$ where each entry independently:
    \[
F(x) =
\begin{cases}
\text{pick uniformly from } \bits^n\setminus T & \text{if } x\in S, \\
O_n(x) & \text{if } x \in \bar S.
\end{cases}
\]
    \item Initialize $W=\{\}$.
    \item For $i \in \mcal{I}_0:$
        \begin{itemize}
            \item Sample $y_i\pick \msf{Image}(F)$
            \item \red{Abort if $\msf{PreImage}^{F}(y_i)\neq \msf{PreImage}^{O_n}(y_i)$}
            \item Sample $v_i\pick \msf{PreImage}^{\red{O_n}}(y_i)$
            \item $W=W\cup \{(y_i,v_i)\}$
        \end{itemize}
    \item Output $(F,T,W)$
\end{enumerate}
\end{ExperimentBox}
\begin{MyClaim}\zcsetup{reftype=Claim}\label{claim:hybvs5}
    It holds that $\hat{H}_{0.4} \sind \hat{H}_{0.5}$. 
\end{MyClaim}
\begin{proof}
    \chakfai{Notice that the probability and expectation are taken over different space but somehow I think it looks even more confusing if we written out everything since almost everything is taken from slightly different space. }
First, the following variables have the same joint distribution
in both hybrids:
\begin{equation}\zcsetup{reftype=Expression}
    \label{eq:joint-dist-h0p4-h0p5}
        O_n,~S,~T,~F.
    \end{equation}
    Next, conditioned on the variables in
    \zcref{eq:joint-dist-h0p4-h0p5}, whenever $\hat{H}_{0.5}$ does not
    abort, both hybrids use the same preimage sets and therefore
    sample each $v_i$ from the same distribution.
    Thus, it suffices to bound the probability that $\hat{H}_{0.5}$ aborts.

    Since $F$ agrees with $O_n$ outside $S$ and never outputs
    a value in $T$, the preimages of a label
    $y\in\msf{Image}(F)$ can differ only if $y\in F(S)$.
    Thus, a union bound over the $|\mcal I_0|$ label samples gives
    \begin{align*}
        \Pr[\msf{Abort}\mid O_n,S,T,F]
        &\leq
        \sum_{i\in\mcal I_0}
        \Pr[y_i\in F(S)\mid O_n,S,T,F]\\
        &=
        \sum_{i\in\mcal I_0}
        \frac{|F(S)|}{|\msf{Image}(F)|}\\
        &=
        \frac{|\mcal I_0|\,|F(S)|}{|\msf{Image}(F)|}\\
        &\leq
        \frac{|\mcal I_0|\,|S|}{|\msf{Image}(F)|}.
    \end{align*}
    Moreover,
    \begin{align*}
        \mathbb E[|S|]
        &=
        \mathbb E\!\left[
            \left|
            \bigcup_{i\in\mcal I_1}
            \msf{PreImage}^{O_n}(O_n(\delta_i))
            \right|
        \right]\\
        &\leq
        \mathbb E\!\left[
            \sum_{i\in\mcal I_1}
            |\msf{PreImage}^{O_n}(O_n(\delta_i))|
        \right]\\
        &=
        \sum_{i\in\mcal I_1}
        \mathbb E\!\left[
            |\msf{PreImage}^{O_n}(O_n(\delta_i))|
        \right]\\
        &=
        \sum_{i\in\mcal I_1}
        \left(1+\frac{2^n-1}{2^n}\right)\tag{\Cref{lem:sampled-output-preimage}}\\
        &=
        |\mcal I_1|\left(1+\frac{2^n-1}{2^n}\right)\\
        &\leq 2|\mcal I_1|.
    \end{align*}
    Here each preimage class contains $\delta_i$ itself,
    and each other input has the same oracle output
    with probability $2^{-n}$.

    By \Cref{lem:random-oracle-image-size},
    \[
        \Pr[|\msf{Image}(O_n)|\leq2^{n/2}]
        \leq 2^{n2^{n/2}-n2^{n-1}}.
    \]
    Outside this exceptional event, $F$ retains every
    original image value outside $T$, so
    \[
        |\msf{Image}(F)|
        \geq |\msf{Image}(O_n)|-|T|
        > 2^{n/2}-|\mcal I_1|
        \geq 2^{n/2-1},
    \]
    where the last inequality holds for sufficiently large
    $n$, since $|\mcal I_1|$ is polynomially bounded.

    Combining these bounds and averaging over the 
    randomness yields
    \begin{align*}
        \Pr[\msf{Abort}]
        &\leq
        \Pr[|\msf{Image}(O_n)|\leq2^{n/2}]
        +\Pr[\msf{Abort}\ \land \
            |\msf{Image}(O_n)|>2^{n/2}]\\
        &\leq
        2^{n2^{n/2}-n2^{n-1}}
        +\frac{|\mcal I_0|\,\mathbb E[|S|]}{2^{n/2-1}}\\
        &\leq
        2^{n2^{n/2}-n2^{n-1}}
        +\frac{4|\mcal I_0|\,|\mcal I_1|}{2^{n/2}}\\
        &=\negl(n).
    \end{align*}
    We therefore establish
    $\hat{H}_{0.4}\sind \hat{H}_{0.5}$.
\end{proof}
\para{Hybrid $\hat{H}_{0.6}$:} We present $\hat{H}_{0.6}$ in \Cref{hybrid:h0p6} and highlight its difference from $\hat{H}_{0.5}$ in \red{red color}.
In this hybrid, we remove the abort check introduced in
$\hat{H}_{0.5}$.
Since this check causes an abort only with negligible
probability, removing it gives $\hat{H}_{0.5}\sind \hat{H}_{0.6}$.
\begin{ExperimentBox}[label={hybrid:h0p6}]{Hybrid $\hat{H}_{0.6}(\{r_i\}_{i\in [p(n)]})$ }
\begin{enumerate}  
    \item Sample $O_n\gets \Ora_n$;
    \item For $i \in \mcal{I}_1:$
        \begin{itemize}
            \item Sample $\delta_i\pick \bits^n,y_i=O_n(\delta_i)$
        \end{itemize}
    \item Set $T\coloneqq \{y_i\}_{i\in\mcal{I}_1}$
    \item Define $S\coloneqq \bigcup_{i\in\mcal{I}_1}\msf{PreImage}^{O_n}(y_i)$, $\bar S\coloneqq\bits^n\setminus S$
    \item Sample a function $F:\bits^n\to\bits^n$ where each entry independently:
    \[
F(x) =
\begin{cases}
\text{pick uniformly from } \bits^n\setminus T & \text{if } x\in S, \\
O_n(x) & \text{if } x \in \bar S.
\end{cases}
\]
    \item Initialize $W=\{\}$.
    \item For $i \in \mcal{I}_0:$
        \begin{itemize}
            \item Sample $y_i\pick \msf{Image}(F)$
            \item \dlt{Abort if $\msf{PreImage}^{F}(y_i)\neq \msf{PreImage}^{O_n}(y_i)$}
            \item Sample $v_i\pick \msf{PreImage}^{O_n}(y_i)$
            \item $W=W\cup \{(y_i,v_i)\}$
        \end{itemize}
    \item Output $(F,T,W)$
\end{enumerate}
\end{ExperimentBox}
\begin{MyClaim}\zcsetup{reftype=Claim}\label{claim:hybvs6}
    It holds that $\hat{H}_{0.5} \sind \hat{H}_{0.6}$. 
\end{MyClaim}
\begin{proof}
    The two hybrids differ only in the abort check removed
in $\hat{H}_{0.6}$.
The argument for $\hat{H}_{0.4} \sind \hat{H}_{0.5}$ already shows that
this check is triggered with probability at most
$\negl(n)$, establishing $\hat{H}_{0.5} \sind \hat{H}_{0.6}$.
\end{proof}
\para{Hybrid $\hat{H}_{0.7}$:} We present $\hat{H}_{0.7}$ in \Cref{hybrid:h0p7} and highlight its difference from $\hat{H}_{0.6}$ in \red{red color}.
In this hybrid, we introduce abort checks to ensure that
$|S|$ is polynomially bounded and $|\msf{Image}(O_n)|$ is
exponentially large. These bounds let us control the effect
of modifying $O_n$ on $S$ in the subsequent hybrids.
Since the checks trigger only with negligible probability,
we have $\hat{H}_{0.6}\sind \hat{H}_{0.7}$.
\begin{ExperimentBox}[label={hybrid:h0p7}]{Hybrid $\hat{H}_{0.7}(\{r_i\}_{i\in [p(n)]})$ }
\begin{enumerate}  
    \item Sample $O_n\gets \Ora_n$;
    \item For $i \in \mcal{I}_1:$
        \begin{itemize}
            \item Sample $\delta_i\pick \bits^n,y_i=O_n(\delta_i)$
        \end{itemize}
    \item Set $T\coloneqq \{y_i\}_{i\in\mcal{I}_1}$
    \item Define $S\coloneqq \bigcup_{i\in\mcal{I}_1}\msf{PreImage}^{O_n}(y_i)$, $\bar S\coloneqq\bits^n\setminus S$
    \item Sample a function $F:\bits^n\to\bits^n$ where each entry independently:
    \[
F(x) =
\begin{cases}
\text{pick uniformly from } \bits^n\setminus T & \text{if } x\in S, \\
O_n(x) & \text{if } x \in \bar S.
\end{cases}
\]
    \item \red{Abort if $|S|\geq p(n)\log(n)$ or $|\msf{Image}(O_n)|\leq 2^{n/2}$}
    \item Initialize $W=\{\}$.
    \item For $i \in \mcal{I}_0:$
        \begin{itemize}
            \item Sample $y_i\pick \msf{Image}(F)$
            \item Sample $v_i\pick \msf{PreImage}^{O_n}(y_i)$
            \item $W=W\cup \{(y_i,v_i)\}$
        \end{itemize}
    \item Output $(F,T,W)$
\end{enumerate}
\end{ExperimentBox}
\begin{MyClaim}\zcsetup{reftype=Claim}\label{claim:hybvs7}
    It holds that $\hat{H}_{0.6} \sind \hat{H}_{0.7}$. 
\end{MyClaim}
\begin{proof}
    The two hybrids differ only in the two abort checks introduced
in $\hat{H}_{0.7}$. Thus, it suffices to bound the abort probability.

First, recall that $S$ is the union of the preimage sets of
the $k=|\mcal I_1|$ sampled oracle outputs $y_i=O_n(\delta_i)$.
Hence,
\begin{align*}
    |S|
    &=
    \left|
        \bigcup_{i\in\mcal I_1}
        \msf{PreImage}^{O_n}(O_n(\delta_i))
    \right|\\
    &\leq
    \sum_{i\in\mcal I_1}
    |\msf{PreImage}^{O_n}(O_n(\delta_i))|.
\end{align*}
Since $|\mcal I_1|\leq p(n)$, the event
$|S|\geq p(n)\log n$ implies that at least one of these
preimage sets has size at least $\log n$.

Next, 
\Cref{lem:sampled-output-preimage-tail} also bounds the
probability that each such preimage set has size at least
$\log n$ by $\negl(n)$: take $t=\lceil\log n\rceil-1$
in that calculation.
Thus, by a union bound and
\Cref{lem:random-oracle-image-size},
\begin{align*}
    \Pr[\msf{Abort}]
    &\leq
    \Pr[|S|\geq p(n)\log n]
    +\Pr[|\msf{Image}(O_n)|\leq2^{n/2}]\\
    &\leq
    \sum_{i\in\mcal I_1}
    \Pr\!\left[
        |\msf{PreImage}^{O_n}(O_n(\delta_i))|\geq\log n
    \right]
    +\Pr[|\msf{Image}(O_n)|\leq2^{n/2}]\\
    &\leq
    |\mcal I_1|\cdot\negl(n)
    +2^{n2^{n/2}-n2^{n-1}}
    \tag{\Cref{lem:sampled-output-preimage-tail}+\Cref{lem:random-oracle-image-size}}\\
    &=\negl(n).
\end{align*}
We therefore conclude that $\hat{H}_{0.6}\sind \hat{H}_{0.7}$.
\end{proof}

\para{Hybrid $\hat{H}_{0.8}$:} We present $\hat{H}_{0.8}$ in \Cref{hybrid:h0p8} and highlight its difference from $\hat{H}_{0.7}$ in \red{red color}.
In this hybrid, for each $i\in\mcal{I}_0$, we sample
$y_i$ uniformly from $\msf{Image}(O_n)$ instead of
$\msf{Image}(F)$.
When the abort checks pass, the two images differ by only
a negligible fraction of their elements, giving
$\hat{H}_{0.7}\sind \hat{H}_{0.8}$.
\begin{ExperimentBox}[label={hybrid:h0p8}]{Hybrid $\hat{H}_{0.8}(\{r_i\}_{i\in [p(n)]})$ }
\begin{enumerate}  
    \item Sample $O_n\gets \Ora_n$;
    \item For $i \in \mcal{I}_1:$
        \begin{itemize}
            \item Sample $\delta_i\pick \bits^n,y_i=O_n(\delta_i)$
        \end{itemize}
    \item Set $T\coloneqq \{y_i\}_{i\in\mcal{I}_1}$
    \item Define $S\coloneqq \bigcup_{i\in\mcal{I}_1}\msf{PreImage}^{O_n}(y_i)$, $\bar S\coloneqq\bits^n\setminus S$
    \item Sample a function $F:\bits^n\to\bits^n$ where each entry independently:
    \[
F(x) =
\begin{cases}
\text{pick uniformly from } \bits^n\setminus T & \text{if } x\in S, \\
O_n(x) & \text{if } x \in \bar S.
\end{cases}
\]
    \item \zcsetup{reftype=Step}\label{step:S-size-check}
     Abort if $|S|\geq p(n)\log(n)$ or $|\msf{Image}(O_n)|\leq 2^{n/2}$
    \item Initialize $W=\{\}$.
    \item For $i \in \mcal{I}_0:$
        \begin{itemize}
            \item Sample $y_i\pick \msf{Image}(\red{O_n})$
            \item Sample $v_i\pick \msf{PreImage}^{O_n}(y_i)$
            \item $W=W\cup \{(y_i,v_i)\}$
        \end{itemize}
    \item Output $(F,T,W)$
\end{enumerate}
\end{ExperimentBox}
\begin{MyClaim}\zcsetup{reftype=Claim}\label{claim:hybvs8}
    It holds that $\hat{H}_{0.7} \sind \hat{H}_{0.8}$. 
\end{MyClaim}
\begin{proof}
    \xiao{Go over this with Chak Fai}
First, the two hybrids have identical initial sampling and
abort checks. In particular, the following variables have
the same joint distribution:
\begin{equation}\zcsetup{reftype=Expression}
\label{eq:joint-dist-h0p7-h0p8}
    O_n,~T,~F.
\end{equation}
If either initial abort check is triggered, both hybrids abort.
Otherwise, condition on fixed values of the variables in
\zcref{eq:joint-dist-h0p7-h0p8}. Define the following event:
\begin{itemize}
    \item \textbf{Event $E$}: for all $i\in\mcal I_0$, it holds that $
        y_i\in\msf{Image}(F)\cap\msf{Image}(O_n)$.
\end{itemize}

Next, conditioned on these variables and on $E$, both hybrids
sample each $y_i$ independently and uniformly from
$\msf{Image}(F)\cap\msf{Image}(O_n)$, and then sample each $v_i$
uniformly from $\msf{PreImage}^{O_n}(y_i)$.
Their conditional output distributions are therefore identical.

It remains to bound the probability that $E$ fails
under each hybrid.
Notice that $F$ and $O_n$ differ only on $S$; then, since the difference in their range must be induced by the difference in their domain, we have
\begin{align}
|\msf{Image}(F)\setminus\msf{Image}(O_n)|  & \le |S| \le p(n)\log n \label{eq:diff1}\\
|\msf{Image}(O_n)\setminus\msf{Image}(F)|
    & \le |S| \leq  p(n)\log n \label{eq:diff2}
\end{align}
where the $|S| \leq p(n)\log n$ bound follows from the common abort check at \zcref{step:S-size-check} in $\hat{H}_{0.7}$ and $\hat{H}_{0.8}$.

Moreover, the initial abort checks also ensure that
\begin{align}
|\msf{Image}(O_n)|&>2^{n/2} \label{eq:imageO}\\
|\msf{Image}(F)|
    &\geq |\msf{Image}(O_n)|-|S|\\
    &>2^{n/2}-p(n)\log n \label{eq:imageF}
\end{align}
Thus, by a union bound,
\begin{align*}
    \Pr_{\hat{H}_{0.7}}[\neg E\mid O_n,T,F]
    &=
    \Pr_{\hat{H}_{0.7}}\!\left[
        \exists i\in\mcal I_0:\;
        y_i\in\msf{Image}(F)\setminus\msf{Image}(O_n)
        \,\middle|\, O_n,T,F
    \right]\\
    &\leq
    \frac{|\mcal I_0|\,
    |\msf{Image}(F)\setminus\msf{Image}(O_n)|}
    {|\msf{Image}(F)|}\\
    &\leq
    \frac{|\mcal I_0|p(n)\log n}
    {2^{n/2}-p(n)\log n} \tag{\Cref{eq:diff1}~and~\Cref{eq:imageF}}.
\end{align*}
Similarly,
\begin{align*}
    \Pr_{\hat{H}_{0.8}}[\neg E\mid O_n,T,F]
    &=
    \Pr_{\hat{H}_{0.8}}\!\left[
        \exists i\in\mcal I_0:\;
        y_i\in\msf{Image}(O_n)\setminus\msf{Image}(F)
        \,\middle|\, O_n,T,F
    \right]\\
    &\leq
    \frac{|\mcal I_0|\,
    |\msf{Image}(O_n)\setminus\msf{Image}(F)|}
    {|\msf{Image}(O_n)|}\\
    &\leq
    \frac{|\mcal I_0|p(n)\log n}{2^{n/2}} \tag{\Cref{eq:diff2}~and~\Cref{eq:imageO}}.
\end{align*}

For each fixed nonaborting initial sample, the
statistical distance is at most the sum of these
two failure probabilities by \Cref{lem:h01-statistical-distance}.
Averaging over the common initial randomness gives
\[
    \Delta(\hat{H}_{0.7},\hat{H}_{0.8})
    \leq
    \frac{|\mcal I_0|p(n)\log n}
    {2^{n/2}-p(n)\log n}+\frac{|\mcal I_0|p(n)\log n}{2^{n/2}}
    =\negl(n),
\]
establishing $\hat{H}_{0.7}\sind \hat{H}_{0.8}$.
\end{proof}

\para{Hybrid $\hat{H}_{0.9}$:} We present $\hat{H}_{0.9}$ in \Cref{hybrid:h0p9} and highlight its difference from $\hat{H}_{0.8}$ in \red{red color}. 
\begin{ExperimentBox}[label={hybrid:h0p9}]{Hybrid $\hat{H}_{0.9}(\{r_i\}_{i\in [p(n)]})$ }
\begin{enumerate}  
    \item Sample $O_n\gets \Ora_n$;
    \item For $i \in \mcal{I}_1:$
        \begin{itemize}
            \item Sample $\delta_i\pick \bits^n,y_i=O_n(\delta_i)$
        \end{itemize}
    \item Set $T\coloneqq \{y_i\}_{i\in\mcal{I}_1}$
    \item Define $S\coloneqq \bigcup_{i\in\mcal{I}_1}\msf{PreImage}^{O_n}(y_i)$, $\bar S\coloneqq\bits^n\setminus S$
    \item Sample a function $F:\bits^n\to\bits^n$ where each entry independently:
    \[
F(x) =
\begin{cases}
\text{pick uniformly from } \bits^n\setminus T & \text{if } x\in S, \\
O_n(x) & \text{if } x \in \bar S.
\end{cases}
\]
    \item \dlt{Abort if $|S|\geq p(n)\log(n)$ or $|\msf{Image}(O_n)|\leq 2^{n/2}$}
    \item Initialize $W=\{\}$.
    \item For $i \in \mcal{I}_0:$
        \begin{itemize}
            \item Sample $y_i\pick \msf{Image}(O_n)$
            \item Sample $v_i\pick \msf{PreImage}^{O_n}(y_i)$
            \item $W=W\cup \{(y_i,v_i)\}$
        \end{itemize}
    \item Output $(F,T,W)$
\end{enumerate}
\end{ExperimentBox}
\begin{MyClaim}\zcsetup{reftype=Claim}\label{claim:hybvs9}
    It holds that $\hat{H}_{0.8} \sind \hat{H}_{0.9}$. 
\end{MyClaim}
\begin{proof}
    The two hybrids differ only in the abort check removed
in $\hat{H}_{0.9}$.
and the proof of $\hat{H}_{0.6}\sind \hat{H}_{0.7}$ already shows that this
check triggers with negligible probability,
establishing $\hat{H}_{0.8}\sind \hat{H}_{0.9}$.
\end{proof}
\para{Hybrid $\hat{H_1}$:} We present $\hat{H_{1}}$ in \Cref{hybrid:h01} and highlight its difference from $\hat{H}_{0.9}$ in \red{red color}.
In this hybrid, we express the sampling of the pairs
$(y_i,v_i)$ for $i\in\mcal I_0$ as
$W\sim K_O^{\otimes|\mcal I_0|}$.
This is purely a notational change and preserves the output
distribution, giving $\hat{H}_{0.9}\idind\hat{H_1}$.
\begin{ExperimentBox}[label={hybrid:h01}]{Hybrid $\hat{H_1}(\{r_i\}_{i\in [p(n)]})$ }
\begin{enumerate}  
    \item Sample $O_n\gets \Ora_n$;
    \item For $i \in \mcal{I}_1:$
        \begin{itemize}
            \item Sample $\delta_i\pick \bits^n,y_i=O_n(\delta_i)$
        \end{itemize}
    \item Set $T\coloneqq \{y_i\}_{i\in\mcal{I}_1}$
    \item Define $S\coloneqq \bigcup_{i\in\mcal{I}_1}\msf{PreImage}^{O_n}(y_i)$, $\bar S\coloneqq\bits^n\setminus S$
    \item Sample a function $F:\bits^n\to\bits^n$ where each entry independently:
    \[
F(x) =
\begin{cases}
\text{pick uniformly from } \bits^n\setminus T & \text{if } x\in S, \\
O_n(x) & \text{if } x \in \bar S.
\end{cases}
\]
    \item \red{Sample $W\sim K_{O_n}^{\otimes |\mcal{I}_0|}$ }
    \item Output $(F,T,W)$
\end{enumerate}
\end{ExperimentBox}
\begin{MyClaim}\zcsetup{reftype=Claim}\label{claim:hybvs10}
    It holds that $\hat{H}_{0.9}\idind \hat{H_1}$. 
\end{MyClaim}
\begin{proof}
    This is just a notational change,where we do not explicity write out the sampling of $W$ in $\hat{H}_{0.9}$ anymore. Therefore, the two distributions are identical.
\end{proof}

\para{Completing the proof of $H_0\sind H_1$:}
By the previous hybrids indistinguishability claims and the triangle inequality, for every
$\boldsymbol r=(r_i)_{i\in[p(n)]}$, we have
\[
    \Delta(\hat{H}_0(\boldsymbol r),
           \hat{H_1}(\boldsymbol r))
    =\negl(n),
\]

These experiments only reorder independent sampling operations
and replace the invocation of $\Adv$ by the output of its
complete input and oracle table.
Thus, applying the original adversary invocation  to $\hat{H_b}$
recovers $H_b$ for
$b\in\{0,1\}$. By \Cref{lem:h01-statistical-distance}, statistical distance
cannot increase under this transformation.
Then the above hybrids indistinguishability establishing $H_0\sind H_1$.

\subsection{\texorpdfstring{Proof of $H_1\sind H_2$}
{From  H1 to H2}}
\label{sec:ind:hyb-1-2}

\para{Proof Strategy Overview.}
We prove $H_1\sind H_2$ by establishing the stronger
statement that the joint distributions of the complete
constructed oracle $O_n^*$ and the indexed inputs supplied
to $\Adv$ are statistically close.

Recall that the two hybrids have the same initial sampling.
For each $i\in\mcal I_0$, they independently sample
\[
    y_i\pick\msf{Image}^{O_n},
    \qquad
    v_i\pick\msf{PreImage}^{O_n}(y_i),
\]
and, for each $i\in\mcal I_1$, they independently sample
\[
    \delta_i\pick\bits^n,
    \qquad
    y_i\coloneqq O_n(\delta_i).
\]
Their constructions of $O_n^*$ differ only at inputs
\[
    z\notin
    \bigcup_{i\in[p(n)]}\msf{PreImage}^{O_n}(y_i).
\]
At these inputs, $H_1$ retains $O_n(z)$, whereas $H_2$
independently samples a fresh uniform value from
$\bits^n\setminus\{y_i:i\in[p(n)]\}$.

Intuitively, after fixing the sampled information and the
preimage sets of the selected labels, the remaining oracle
entries should already be uniform over the outputs other
than the selected labels.
However, sampling $y_i$ uniformly from the image for each
$i\in\mcal I_0$ introduces the factor
\[
    \frac{1}{|\msf{Image}^{O_n}|^{|\mcal I_0|}}.
\]
The image size can depend on the remaining oracle entries,
so these entries are not immediately independent and uniform.

To remove this dependence, we introduce modified experiments
$\tilde H_1$ and $\tilde H_2$ in which each initial oracle
table is sampled with probability proportional to
$|\msf{Image}^{O_n}|^{|\mcal I_0|}$.
The two factors then cancel exactly, allowing us to prove
$\tilde H_1\idind\tilde H_2$.
We next show that the modified oracle sampling is
statistically close to uniform sampling, establishing
$H_b\sind\tilde H_b$ for each $b\in\{1,2\}$.

Throughout the proof, restrict attention
to sufficiently large $n$ such that $p(n)<2^n$.
We retain the notation
\[
    \mcal I_0=\{i\in[p(n)]:r_i=0\},
    \qquad
    \mcal I_1=\{i\in[p(n)]:r_i=1\}.
\]
The bit vector has the same distribution in both original
hybrids and is independent of the initially sampled oracle.
Thus, we may sample the bit vector first without changing
either original experiment.

\para{Hybrid $\tilde H_1$.}
We present $\tilde H_1$ in \Cref{hybrid:tilde-h1}.
This hybrid differs from $H_1$ only in the distribution
of the initially sampled oracle.
The modified sampling is used only for the proof and
need not have an efficient implementation.

\begin{ExperimentBox}[label={hybrid:tilde-h1}]
{Hybrid $\tilde H_1$}
\begin{enumerate}
    \item According to the  input $\vb{r} = (r_1, \ldots, r_{p(n)})$, define
     $\begin{cases}
    \mathcal{I}_0  \coloneqq\{i\in [p(n)]~|~ r_i=0\} \\ 
    \mathcal{I}_1  \coloneqq\{i\in [p(n)]~|~ r_i=1\}
    \end{cases}$.

    \item Sample $O_n\in\Ora_n$ according to
    \[
        \Pr[O_n=f\mid\{r_i\}_{i\in[p(n)]}]
        =
        \frac{|\msf{Image}^{f}|^{|\mcal I_0|}}
        {\displaystyle
            \sum_{g\in\Ora_n}
            |\msf{Image}^{g}|^{|\mcal I_0|}}
        \qquad\text{for every }f\in\Ora_n.
    \]

    \item For each $i\in\mcal I_0$, independently sample
    \[
        y_i\pick\msf{Image}^{O_n},
        \qquad
        v_i\pick\msf{PreImage}^{O_n}(y_i).
    \]

    \item For each $i\in\mcal I_1$, independently sample
    \[
        \delta_i\pick\bits^n,
        \qquad
        y_i\coloneqq O_n(\delta_i).
    \]

    \item Construct $O_n^*$ as in $H_1$.
    In particular, set
    \[
        O_n^*(z)\coloneqq O_n(z)
        \qquad\text{if }\;
        z\notin
        \bigcup_{i\in[p(n)]}
        \msf{PreImage}^{O_n}(y_i).
    \]
    All other branches of the oracle construction remain
    unchanged.

    \item Invoke $\Adv$ with the same indexed inputs and
    oracle access as in $H_1$, and output its result.
\end{enumerate}
\end{ExperimentBox}

\para{Hybrid $\tilde H_2$.}
We present $\tilde H_2$ in \Cref{hybrid:tilde-h2}.
This hybrid uses the same modified initial sampling as
$\tilde H_1$, but resamples the oracle entries outside
the selected preimage sets, as in $H_2$.
Equivalently, it differs from $H_2$ only in the distribution
of the initially sampled oracle.

\begin{ExperimentBox}[label={hybrid:tilde-h2}]
{Hybrid $\tilde H_2$}
\begin{enumerate}
    \item Run Steps~1--4 of $\tilde H_1$ to obtain
    \[
        O_n,\quad
        \{y_i\}_{i\in[p(n)]},\quad
        \{v_i\}_{i\in\mcal I_0},\quad
        \{\delta_i\}_{i\in\mcal I_1}.
    \]

    \item Construct $O_n^*$ as in $H_2$.
    In particular, for every input
    \[
        z\notin
        \bigcup_{i\in[p(n)]}
        \msf{PreImage}^{O_n}(y_i),
    \]
    independently sample
    \[
        O_n^*(z)\pick
        \bits^n\setminus\{y_i:i\in[p(n)]\},
    \]
    using fresh randomness.
    On all remaining inputs, use the same oracle
    construction as in $\tilde H_1$.

    \item Invoke $\Adv$ with the same indexed inputs and
    oracle access as in $H_2$, and output its result.
\end{enumerate}
\end{ExperimentBox}

\begin{MyClaim}
\zcsetup{reftype=Claim}
\label{claim:tilde-h1-h2-identical}
    It holds that $\tilde H_1\idind\tilde H_2$.
\end{MyClaim}

\begin{proof}
We prove that the complete constructed oracle $O_n^*$
and the indexed inputs supplied to $\Adv$ have identical
joint distributions in the two experiments.

First, the following variables have the same joint
distribution in both experiments:
\begin{equation}
\zcsetup{reftype=Expression}
\label{eq:tilde-h1-h2-fixed-data}
    \{y_i\}_{i\in[p(n)]},\qquad
    \{v_i\}_{i\in\mcal I_0},\qquad
    \{\delta_i\}_{i\in\mcal I_1},\qquad
    \bigl(\msf{PreImage}^{O_n}(y_i)\bigr)_{i\in[p(n)]}.
\end{equation}
Condition on any attainable values of these variables.

Consider an oracle table $f$ consistent with the fixed
values in \zcref{eq:tilde-h1-h2-fixed-data}.
Conditional on $O_n=f$, the probability of sampling the
specified labels, witnesses, and inputs is
\[
    2^{-n|\mcal I_1|}
    \prod_{i\in\mcal I_0}
    \frac{1}
    {|\msf{Image}^{f}|\,
     |\msf{PreImage}^{f}(y_i)|}.
\]
Indeed, for each $i\in\mcal I_1$, sampling the specified
$\delta_i$ has probability $2^{-n}$, after which $y_i$
is determined by $f(\delta_i)$.
For each $i\in\mcal I_0$, the two denominator terms
correspond to sampling $y_i$ uniformly from the image
and then $v_i$ uniformly from its preimage set.

Multiplying by the modified probability of sampling
the table $f$ gives
\begin{align*}
    &\frac{|\msf{Image}^{f}|^{|\mcal I_0|}}
    {\displaystyle
        \sum_{g\in\Ora_n}
        |\msf{Image}^{g}|^{|\mcal I_0|}}
    \cdot
    2^{-n|\mcal I_1|}
    \prod_{i\in\mcal I_0}
    \frac{1}
    {|\msf{Image}^{f}|\,
     |\msf{PreImage}^{f}(y_i)|}
    \\
    &\qquad =
    \frac{2^{-n|\mcal I_1|}}
    {\displaystyle
        \sum_{g\in\Ora_n}
        |\msf{Image}^{g}|^{|\mcal I_0|}}
    \prod_{i\in\mcal I_0}
    \frac{1}{|\msf{PreImage}^{f}(y_i)|}.
\end{align*}
The factors involving $|\msf{Image}^{f}|$ cancel exactly.
Moreover, the preimage sizes in the remaining expression
are fixed by \zcref{eq:tilde-h1-h2-fixed-data}.
Thus, every consistent oracle table has the same
conditional probability.

Next, the fixed preimage sets determine the oracle value
at every input in
\[
    \bigcup_{i\in[p(n)]}\msf{PreImage}^{O_n}(y_i).
\]
At any other input, the oracle value must lie in
$\bits^n\setminus\{y_i:i\in[p(n)]\}$.
Conversely, any assignment of the remaining oracle entries
to values in this set preserves all the fixed information
in \zcref{eq:tilde-h1-h2-fixed-data}.

Consequently, conditional on the fixed information,
the entries
\[
    \left\{
        O_n(z):
        z\notin
        \bigcup_{i\in[p(n)]}
        \msf{PreImage}^{O_n}(y_i)
    \right\}
\]
are independent and uniformly distributed over
$\bits^n\setminus\{y_i:i\in[p(n)]\}$.
This conclusion does not require the labels $y_i$ to
be distinct.

Therefore, $\tilde H_1$ retains entries that already
have exactly the distribution from which $\tilde H_2$
resamples.
On the remaining inputs, both experiments use the same
oracle construction, which depends only on the fixed
information and fresh randomness.
Their complete constructed oracle tables and indexed
inputs therefore have identical conditional distributions.

Averaging over \zcref{eq:tilde-h1-h2-fixed-data} and then
over the common distribution of
$\{r_i\}_{i\in[p(n)]}$ preserves this identity.
Applying the common adversary invocation establish
$\tilde H_1\idind\tilde H_2$.

Notice that the identity concerns the constructed oracle
$O_n^*$ supplied to $\Adv$.
We do not additionally reveal the original table $O_n$.
\end{proof}

\begin{MyClaim}
\zcsetup{reftype=Claim}
\label{claim:hb-tilde-hb-close}
    For every $b\in\{1,2\}$, it holds that
    $H_b\sind\tilde H_b$.
\end{MyClaim}

\begin{proof}
Consider an arbitrary $b\in\{1,2\}$.
The experiments $H_b$ and $\tilde H_b$ differ only in
the distribution of the initially sampled oracle.
Thus, it suffices to bound the statistical distance
between uniform oracle sampling and the modified sampling.

Intuitively, the modified sampling assigns larger weights to oracle
tables with larger images. However, almost all uniformly sampled
tables have nearly the same image size, so their sampling weights
are nearly equal.

Let
\[
    \mu
    \coloneqq
    \mathbb{E}_{O_n\gets\Ora_n}
    \bigl[|\msf{Image}^{O_n}|\bigr].
\]
By \Cref{lem:ro-image-concentration}, we have $\mu\geq 2^{n-1}$ and
\begin{equation}
\label{eq:h12-image-concentration}
    \Pr_{O_n\gets\Ora_n}
    \left[
        \left||\msf{Image}^{O_n}|-\mu\right|>2^{3n/4}
    \right]
    \leq
    2e^{-2\cdot 2^{n/2}}.
\end{equation}
We use this concentration bound to compare the weights of the
typical tables and separately bound the total contribution of
the exceptional tables to the normalization.

For the following calculation, define the rescaled
weight of a table $f$ by
\[
    w(f)\coloneqq
    \left(
        \frac{|\msf{Image}^{f}|}{\mu}
    \right)^{|\mcal I_0|}.
\]
Dividing all weights by the same number
$\mu^{|\mcal I_0|}$ does not change the modified
sampling distribution.

First, consider a table $f$ satisfying
\[
    \left||\msf{Image}^{f}|-\mu\right|
    \leq 2^{3n/4}.
\]
Since $\mu\geq 2^{n-1}$, we have
\[
    1-2^{1-n/4}
    \leq
    \frac{|\msf{Image}^{f}|}{\mu}
    \leq
    1+2^{1-n/4}.
\]
Consequently,
\[
    (1-2^{1-n/4})^{|\mcal I_0|}
    \leq w(f)
    \leq
    (1+2^{1-n/4})^{|\mcal I_0|}.
\]
Using $|\mcal I_0|\leq p(n)$, together with
$(1-x)^k\geq1-kx$ and $(1+x)^k\leq e^{kx}$, gives
\begin{equation}
\label{eq:h12-good-weight}
    |w(f)-1|
    =
    O\bigl(p(n)2^{-n/4}\bigr).
\end{equation}
Here we use $p(n)2^{-n/4}=o(1)$.
The bound is uniform over all such tables and all choices
of the bit vector.

Next, for every oracle table, including those outside
the preceding image-size bound, we have
\[
    0\leq w(f)=\left(
        \frac{|\msf{Image}^{f}|}{\mu}
    \right)^{|\mcal I_0|}
    \leq
    \left(\frac{2^n}{\mu}\right)^{|\mcal I_0|}
    \leq 2^{p(n)}.
\]
where the last inequality uses $\mu\geq 2^{n-1}$.
Thus, the exceptional tables have very small probability,
and even their largest possible weights have negligible
total contribution.
More precisely, combining
\Cref{eq:h12-image-concentration,eq:h12-good-weight}
gives
\begin{align}
    \mathbb E_{O_n\gets\Ora_n}
    \bigl[|w(O_n)-1|\bigr]
    &\leq
    O\bigl(p(n)2^{-n/4}\bigr)
    +(2^{p(n)}+1)\cdot 2e^{-2\cdot 2^{n/2}}
    \nonumber\\
    &=\negl(n).
    \label{eq:h12-average-weight-deviation}
\end{align}

It remains to account for the normalization of the weights.
Let $\varepsilon(n)$ be a negligible upper bound on
\zcref{eq:h12-average-weight-deviation}, uniform over
the bit vector.
Then by triangle inequality, we have
\[
    \left|
        \mathbb E_{O_n\gets\Ora_n}[w(O_n)]-1
    \right|
    \leq
    \mathbb E_{O_n\gets\Ora_n}[|w(O_n)-1|]
    \leq\varepsilon(n).
\]
In particular, the average weight is also close to one.

Since $|\Ora_n|=2^{n\cdot 2^n}$, the probability of a particular
table $f$ under the modified sampling is
\[
    \frac{w(f)}{\displaystyle\sum_{g\in\Ora_n}w(g)}
    =
    \frac{1}{2^{n\cdot 2^n}}\cdot
    \frac{w(f)}
    {\mathbb E_{O_n\gets\Ora_n}[w(O_n)]}.
\]
Therefore, the statistical distance between uniform
and modified oracle sampling is
\begin{align*}
    &\frac12\sum_{f\in\Ora_n}
    \left|
        \frac{1}{2^{n\cdot 2^n}}
        -
        \frac{1}{2^{n\cdot 2^n}}\cdot
        \frac{w(f)}
        {\mathbb E_{O_n\gets\Ora_n}[w(O_n)]}
    \right|
    \\
    &\quad =
    \frac{
        \mathbb E_{O_n\gets\Ora_n}
        \left[
            \left|
                w(O_n)
                -
                \mathbb E_{O_n\gets\Ora_n}[w(O_n)]
            \right|
        \right]
    }{
        2\mathbb E_{O_n\gets\Ora_n}[w(O_n)]
    }
    \\
    &\quad \leq
    \frac{
        \mathbb E_{O_n\gets\Ora_n}[|w(O_n)-1|]
        +
        \left|
            1-\mathbb E_{O_n\gets\Ora_n}[w(O_n)]
        \right|
    }{
        2\mathbb E_{O_n\gets\Ora_n}[w(O_n)]
    }
    \\
    &\quad \leq
    \frac{\varepsilon(n)}{1-\varepsilon(n)}
    =
    \negl(n).
\end{align*}

Finally, for each fixed bit vector and each
$b\in\{1,2\}$, all subsequent steps of $H_b$ and
$\tilde H_b$ apply the same randomized transformation
to the initially sampled oracle.
Since statistical distance cannot increase under such
a transformation, the joint distributions of the
constructed oracle and the indexed inputs are
statistically close.
The same holds after invoking $\Adv$.

The bound is uniform over $\{r_i\}_{i\in[p(n)]}$.
Averaging over its common distribution therefore gives
$H_b\sind\tilde H_b$ for every $b\in\{1,2\}$.

\end{proof}

\para{Completing the proof of $H_1\sind H_2$.}
By \zcref{claim:tilde-h1-h2-identical,claim:hb-tilde-hb-close},
we have
\[
    H_1
    \sind
    \tilde H_1
    \idind
    \tilde H_2
    \sind
    H_2.
\]
The triangle inequality therefore gives
\[
    \Delta(H_1,H_2)
    \leq
    \Delta(H_1,\tilde H_1)
    +
    \Delta(\tilde H_1,\tilde H_2)
    +
    \Delta(\tilde H_2,H_2)
    =
    \negl(n),
\]
establishing $H_1\sind H_2$.

\subsection{\texorpdfstring{Proof of $H_2\qcind H_{2.5}$}
{From  H2 to H2.5}}
\label{sec:proof-H2-H3}

Below, we establishes  $H_2 \qcind H_{2.5}$.

\para{Proof strategy overview.}
We couple the two experiments so that their constructed
oracles agree outside a small set of inputs.
We then show that, in $H_{2.5}$, the locations of these
inputs remain hidden even given the complete constructed
oracle and the sampled information.
Consequently, measuring a query in an execution with the
$H_{2.5}$ oracle hits this set with only negligible
probability.
By the one-way-to-hiding (O2H) bound, a non-negligible
distinguishing advantage would contradict this query bound.

Fix a quantum polynomial-time adversary $\Adv$, and let
$q=q(n)\geq 1$ be a polynomial upper bound on its number
of oracle queries.
Write $p=p(n)$.
For each $i\in\mcal I_0$, write $\delta_i=v_i$, and define
\[
    W\coloneqq
    \bigl((y_i,\delta_i)\bigr)_{i\in[p]},
    \qquad
    D\coloneqq\{\delta_i:i\in[p]\},
    \qquad
    Y\coloneqq\{y_i:i\in[p]\}.
\]
We also write
\[
    Y_1\coloneqq\{y_i:i\in\mcal I_1\}.
\]
The variable $W$ contains all the indexed inputs supplied
to $\Adv$, as well as the designated inputs
$\delta_i$ for $i\in\mcal I_1$.
We condition on these additional inputs only for the
analysis; the experiments do not reveal them to $\Adv$.

\para{Restricting to the good event.}
For an oracle table $f$, define
\[
    K(f)\coloneqq
    \max_{y\in\bits^n}|\msf{PreImage}^{f}(y)|,
    \qquad
    M(f)\coloneqq|\msf{Image}^{f}|.
\]
Consider the event
\[
    B\coloneqq
    \bigl[\text{all }y_i\text{ are distinct}\bigr]
    \cap
    \bigl[K(O_n)\leq n\bigr]
    \cap
    \bigl[M(O_n)\geq 2^{n-1}\bigr].
\]
The distinctness bound established in the proof of
\zcref{claim:H25-H3} applies here, since $H_2$ and $H_{2.5}$
have the same initial sampling.
Combining this bound with \Cref{lem:ro-image-concentration}
and the preimage-size bound obtained from
\Cref{lem:sampled-output-preimage} and \Cref{chernoff_2}
by a union bound over all $2^n$ inputs, we obtain
\[
    \Pr[\neg B]=\negl(n).
\]
The event $B$ concerns only the common initial sampling,
so it has the same probability in both experiments.

For now, condition both experiments on $B$.
Since $O_n(\delta_i)=y_i$ for every $i\in[p]$,
distinctness of the labels implies distinctness of
the designated inputs. Thus, $|D|=p$.
Define the hidden set
\[
    S\coloneqq
    \left(
        \bigcup_{i\in[p]}
        \msf{PreImage}^{O_n}(y_i)
    \right)\setminus D.
\]
On $B$, we have
\begin{equation}
\label{eq:h2-h25-hidden-set-size}
    |S|\leq pn.
\end{equation}

\para{Coupling the constructed oracles.}
Let $O_{n,2}^*$ and $O_{n,2.5}^*$ denote the constructed
oracles in $H_2$ and $H_{2.5}$, respectively.
Use the same initial oracle $O_n$ and the same sampled
data $W$ in both experiments.
Couple the remaining oracle randomness as follows:
\begin{itemize}
    \item
    At every $\delta_i$ with $i\in\mcal I_1$, use the
    same fresh uniform output from $\bits^n\setminus Y_1$.

    \item
    At every $\delta_i$ with $i\in\mcal I_0$, both
    constructed oracles output $y_i$.

    \item
    At every input outside
    $\bigcup_{i\in[p]}\msf{PreImage}^{O_n}(y_i)$,
    use the same fresh uniform output from
    $\bits^n\setminus Y$.

    \item
    At the remaining inputs, sample the oracle values
    according to their respective prescribed
    distributions.
\end{itemize}
All fresh samples at distinct inputs are independent
within each oracle construction.
Since the labels are distinct on $B$, the designated-input
cases do not conflict.
This coupling therefore preserves both conditional
experiment distributions and satisfies
\begin{equation}
\label{eq:h2-h25-agreement}
    O_{n,2}^*(z)=O_{n,2.5}^*(z)
    \qquad\text{for every }z\notin S.
\end{equation}

The agreement outside the selected preimage sets uses
the fact that $H_2$ already resamples its background
entries uniformly from $\bits^n\setminus Y$.

\para{The hidden set remains hidden in $H_{2.5}$.}
Fix any attainable value of $W$ under the conditioning
on $B$.
The complete oracle $O_{n,2.5}^*$ can be generated from
this fixed information and independent fresh randomness:
at each designated input its distribution is determined
by $W$, and at every other input it is independently
uniform over $\bits^n\setminus Y$.
In particular, the construction does not otherwise use
the initial oracle $O_n$.

It follows that
\begin{equation}
\label{eq:h2-h25-independence}
    O_{n,2.5}^* \text{ and } S
    \text{ are conditionally independent given } (W,B).
\end{equation}
This remains true despite conditioning on $B$, since
$B$ depends only on the initial sampling, not on the
fresh randomness used to construct $O_{n,2.5}^*$.

Intuitively, after fixing $W$ and conditioning on $B$ and
$|S|=s$, every $s$-element subset of $\bits^n\setminus D$
should be equally likely to be $S$.
We justify this symmetry by reshuffling the truth table
outside $D$, preserving the sampling probabilities while
permuting the locations of $S$.

Formally, consider any permutation $\pi$ of $\bits^n$ that
fixes every point in $D$, and replace the initial oracle
table $f$ by $f\circ\pi^{-1}$.
Uniform oracle sampling is invariant under this transformation.
Moreover, it preserves the image of $f$, the cardinality of
every preimage set, and the values at all designated inputs.
Consequently, the probability of generating the fixed data
$W$ is unchanged.
Indeed, for $i\in\mcal I_0$, sampling $y_i$ uniformly from
the image and then $\delta_i$ uniformly from its preimage
has the same probability before and after the transformation.
For $i\in\mcal I_1$, the same holds for sampling
$\delta_i$ uniformly from $\bits^n$ and setting
$y_i=f(\delta_i)$.

The event $B$ is also invariant under this transformation,
whereas the hidden set $S$ is mapped to $\pi(S)$.
Consequently, for every attainable $s$,
\begin{equation}
\label{eq:h2-h25-uniform-hidden-set}
    S\mid(W,B,|S|=s)
    \quad\text{is uniform among the $s$-subsets of }
    \bits^n\setminus D.
\end{equation}
Indeed, permutations fixing $D$ act transitively on
these $s$-subsets.

Combining
\Cref{eq:h2-h25-independence,eq:h2-h25-uniform-hidden-set},
we conclude that conditioning additionally on the
complete table $O_{n,2.5}^*$ does not change this uniform
distribution.
Thus, even the complete accessible oracle does not
reveal the locations of $S$ outside $D$.

\para{Bounding the probability of a measured query.}
Consider an execution of $\Adv$ with oracle
$O_{n,2.5}^*$ and its prescribed indexed inputs.
For each $j\in[q]$, consider the experiment that runs
$\Adv$ until immediately before its $j$-th oracle query
and measures that query's input register in the
computational basis.
An execution making fewer than $j$ queries is treated
as a failure to hit $S$.

Fix attainable values of
\[
    W,\quad O_{n,2.5}^*,\quad |S|=s
\]
under the conditioning on $B$.
Let $\mu_j(z)$ denote the probability that the measured
query is $z$ under this conditioning.
The adversary's execution uses only its indexed inputs,
the oracle $O_{n,2.5}^*$, and its own randomness.
Hence its query distribution does not depend further
on $S$.

By the conditional uniformity established above,
\begin{align*}
    &\Pr\bigl[
        \text{measured query }j\in S
        \,\big|\,
        W,O_{n,2.5}^*,B,|S|=s
    \bigr]
    \\
    &\qquad =
    \sum_{z\notin D}
        \mu_j(z)\frac{s}{2^n-p}
    \leq
    \frac{s}{2^n-p}
    \leq
    \frac{pn}{2^n-p}.
\end{align*}
The last inequality follows from
\Cref{eq:h2-h25-hidden-set-size}.
Averaging over the conditioned variables preserves
the bound.

Let $P_{\mathrm{hit}}$ be the probability of obtaining
an element of $S$ when choosing $j$ uniformly from
$[q]$ and performing the preceding measurement in
the $H_{2.5}$ execution, with the initial sampling
conditioned on $B$.
Then
\begin{equation}
\label{eq:h2-h25-query-bound}
    P_{\mathrm{hit}}\leq\frac{pn}{2^n-p}.
\end{equation}

\para{Applying O2H.}
Let $\Adv[H_b]$ denote the output of $\Adv$ in
experiment $H_b$, and define the conditional
distinguishing advantage
\[
    \epsilon_B
    \coloneqq
    \left|
        \Pr[\Adv[H_2]=1\mid B]
        -
        \Pr[\Adv[H_{2.5}]=1\mid B]
    \right|.
\]
The coupled oracles agree outside $S$ by
\Cref{eq:h2-h25-agreement}.
The O2H bound, with measured queries taken from the
execution using $O_{n,2.5}^*$, therefore gives
\begin{equation}
\label{eq:h2-h25-o2h}
    \epsilon_B
    \leq
    2q\sqrt{P_{\mathrm{hit}}}.
\end{equation}
The bound permits arbitrary correlations among the
two oracle tables, the set $S$, and the auxiliary input.
The conditional independence proved above is needed
to bound $P_{\mathrm{hit}}$, not to apply O2H itself.

\para{Removing the conditioning.}
Finally, the two experiments have the same initial
sampling distribution and therefore the same
probability of $B$.
Decomposing their acceptance probabilities according
to $B$ and $\neg B$, we obtain
\begin{align*}
    &\left|
        \Pr[\Adv[H_2]=1]
        -
        \Pr[\Adv[H_{2.5}]=1]
    \right|
    \\
    &\qquad\leq
    \Pr[B]\epsilon_B+\Pr[\neg B]
    \\
    &\qquad\leq
    2q(n)\sqrt{
        \frac{p(n)n}{2^n-p(n)}
    }
    +\negl(n)
    =
    \negl(n).
\end{align*}
Since $\Adv$ was arbitrary, this establishes
$H_2\qcind H_{2.5}$.

This completes the proof of \zcref{claim:hyb-2-3}.

\section{NIZK in the URS-PRHF Model}\label{sec:URS-NIZK}

In this section, we construct a post-quantum NIZK proof from a
strong $\delta$-dense post-quantum PRHF, using a uniform random
string (URS) as setup.

This construction is not a URS NIZK in the standard sense:
both the prover and the verifier additionally require access
to an honestly generated (strong $\delta$-dense post-quantum) PRHF as per \Cref{def:strongPRHF}.
The PRHF sampling algorithm samples a function $H$ from a
family $\mcal{H}_n$.
If this algorithm is public-coin, its random coins can be
included in the URS.
However, future potential instantiations of PRHFs from concrete
assumptions may require
private-coin sampling. 
In that case, our construction does not yield a URS NIZK in
the standard sense. We therefore formalize the notion achieved by our construction
in \Cref{def:URS-PRHF-NIZK}: in addition to a standard URS, the prover and
the verifier are given access to an honestly generated PRHF.
We call this the \emph{URS-PRHF model}.

This model serves only as an intermediate step toward our
final construction in the QROM in \Cref{sec:NIZK-QROM:major}, where we
implement both the URS and the PRHF using a random oracle.
The URS-PRHF model suffices for this purpose.

\begin{definition}[NIZK Proof in the URS-PRHF Model]\label{def:URS-PRHF-NIZK}
    Let $\{\mcal{H}_n\}_{n\in\mathbb{N}}$ be a (strong $\delta$-dense post-quantum) PRHF as per \Cref{def:strongPRHF}.
    An (unbounded-prover) non-interactive zero knowledge (NIZK)
    proof system in the URS-PRHF model for a language
    $\mathcal{L}$ is a tuple of algorithms $\Pi=(P,V)$.
    The setup consists of a uniform random string
    $\sigma\picks\{0,1\}^{\ell(n)}$, where $\ell$ is a polynomial,
    and an independently sampled PRHF $H_n\gets\mcal{H}_n$,
    where $n=|x|$.
    \begin{itemize}
        \item $P(H_n,\sigma,x)$ is an (unbounded, randomized)
        honest prover algorithm that takes as input the code
        of $H_n$, the URS $\sigma$, and $x\in\mathcal{L}$,
        and outputs a proof $\pi$.

        \item $V(H_n,\sigma,x,\pi)$ is a polynomial-time
        deterministic verifier that takes as input the code
        of $H_n$, the URS $\sigma$, $x$, and a proof $\pi$,
        and outputs either 1 (accept) or 0 (reject).
    \end{itemize}

    We require $\Pi$ to satisfy the following completeness,
    soundness, and zero knowledge requirements:

    \begin{itemize}
        \item \textbf{Perfect Completeness:}
        For all $x\in\mathcal{L}$ with $|x|=n$, it holds that
        \[
        \Pr\left[
            V(H_n,\sigma,x,\pi)=1:
            \begin{array}{l}
                H_n\gets\mcal{H}_n,\\
                \sigma\picks\{0,1\}^{\ell(n)},\\
                \pi\leftarrow P(H_n,\sigma,x)
            \end{array}
        \right]=1.
        \]

        \item \textbf{Statistical Soundness:}
        It holds that
        \[
        \Pr_{\substack{
            H_n\gets\mcal{H}_n\\
            \sigma\picks\{0,1\}^{\ell(n)}
        }}\left[
            \exists (x,\pi)\mathrm{~such~that~}
            x\notin\mathcal{L},\ |x|=n,\mathrm{~and~}
            V(H_n,\sigma,x,\pi)=1
        \right]\leq\negl(n).
        \]

        \item \textbf{Zero Knowledge:}
        There exists a QPT simulator $\operatorname{Sim}$
        such that, for all $x\in\mathcal{L}$ with $|x|=n$
        and all QPT algorithms $\mathcal{A}$,
        \begin{align*}
        \Bigg|
        &\Pr\left[
            \mathcal{A}(H_n,\sigma,x,\pi)=1:
            \begin{array}{l}
                H_n\gets\mcal{H}_n,~
                \sigma\picks\{0,1\}^{\ell(n)},\\
                \pi\leftarrow P(H_n,\sigma,x)
            \end{array}
        \right]\\
        &  \hspace{3em} -
        \Pr\left[
            \mathcal{A}(H_n,\sigma^*,x,\pi^*)=1:
            \begin{array}{l}
                H_n\gets\mcal{H}_n,\\
                (\sigma^*,\pi^*)\leftarrow
                \operatorname{Sim}(H_n,x)
            \end{array}
        \right]
        \Bigg|\leq\negl(n).
        \end{align*}
        The simulator is given the code of the honestly
        sampled $H_n$ and simulates the URS and the proof,
        without modifying $H_n$.
    \end{itemize}
\end{definition}

\subsection{Construction}\label{sec:URS-NIZK-construction}

In this subsection, we present  a construction satisfying \Cref{def:URS-PRHF-NIZK}. The construction and its security proof also apply when
$H_{\msf{PRHF}}$ is instantiated by a random oracle: The proof
relativizes by replacing each evaluation of $H_{\msf{PRHF}}$
with an oracle query.  We use this oracle
formulation in the QROM construction of
\Cref{sec:QROM-final-construction}.

We first recall the following existence result.

\begin{lemma}[{\cite[Theorem~5.1]{STOC:GIKKLS23}}]
\label{lem:zhb-existence}
For every language $L\in\NP$ and every efficiently approximable
constant $\gamma\in(0,1)$, there exists a NIZK proof system
in the Z-Tamperable-Hidden-Bits model,
\[
    \msf{ZHB\text{-}NIZK}_\gamma
    =
    \bigl(
        \msf{ZHB.Gen}_\gamma,~
        \msf{ZHB.P}_\gamma,~
        \msf{ZHB.V}_\gamma
    \bigr),
\]
with simulator $\msf{ZHB.Sim}_\gamma$, satisfying the
requirements of \Cref{def:zhb-nizk}.
\end{lemma}

The construction closely follows the framework of \cite{STOC:GIKKLS23}. The main differences are that we instantiate it with our strong \(\delta\)-dense post-quantum PRHF in place of the $\delta$-dense hash function used there, and establish security against quantum adversaries.
\begin{AlgorithmBox}[label={URS-Gen}]{URS Generator $\msf{URS.Gen}(1^n)$}
    $\msf{URS.Gen}(1^n)$:
    \begin{enumerate}
         \item For $i\in [p_{\msf{hbs}}(n)]$, sample a uniformly random $y_i \in \bits^{\ell(n)}$
         \item Output $\msf{urs}=y_1\cdots y_{p_{\msf{hbs}}(n)}$
    \end{enumerate}
\end{AlgorithmBox}
\begin{AlgorithmBox}[label={URS-Prover}]{URS Prover $\msf{URS.P}$}
    $\msf{URS.P}(\msf{urs},H_{\msf{PRHF}},x)$:
    \begin{enumerate}
        \item Parse $\msf{urs}$ as $y_1, \dots, y_{p_{\msf{hbs}}(n)}$ where $n = |x|$.
        \item For $i \in [p_{\msf{hbs}}(n)]$,
          \begin{enumerate}
          \item[(a)] If $\exists z \in \{0, 1\}^n$ such that $H_{\msf{PRHF}}(z) = y_i$, sample $v_i \picks \{z \in \{0, 1\}^n : H_{\msf{PRHF}}(z) = y_i\}$, and set $r_i = 0$.
          \item[(b)] If $\nexists z \in \{0, 1\}^n$ such that $H_{\msf{PRHF}}(z) = y_i$, set $v_i = \perp$ and $r_i = 1$.
          \end{enumerate}
        \item $( \msf{ZHB.I}, \msf{ZHB.\pi} ) \leftarrow \msf{ZHB.P}_\gamma(r, x)$ where $r = r_1 \dots r_{p_{\msf{hbs}}(n)}$.
        \item Return $\pi := (\msf{ZHB.I}, \{v_i\}_{i\in \msf{ZHB.I}}, \msf{ZHB.\pi})$.
    \end{enumerate}
\end{AlgorithmBox}
\begin{AlgorithmBox}[label={URS-Verifier}]{URS Verifier $\msf{URS.V}$}
     $\msf{URS.V}(\msf{urs},H_{\msf{PRHF}}, x, \pi)$:
    \begin{enumerate}
        \item Parse $\mathsf{urs}$ as $y_1, \dots, y_{p_{\mathsf{hbs}}(n)}$ where $n = |x|$.
        \item Parse $\pi$ as $(\mathsf{ZHB.I}, \{v_i\}_{i\in \mathsf{ZHB.I}}, \mathsf{ZHB.\pi})$.
        \item For $i \in \mathsf{ZHB.I}$,
          \begin{itemize}
          \item If $v_i = \perp$, set $\tilde{r}_i = 1$. 
          \item If $v_i \neq \perp$, and $H_{\mathsf{PRHF}}(v_i) = y_i$, set $\tilde{r}_i = 0$.  
          \item If $v_i \neq \perp$, and $H_{\mathsf{PRHF}}(v_i) \neq y_i$, reject.
          \end{itemize}
        \item Return $\mathsf{ZHB.V}_\gamma(x, \mathsf{ZHB.I}, \{\tilde{r}_i\}_{i\in \mathsf{ZHB.I}}, \mathsf{ZHB.\pi})$
    \end{enumerate}
\end{AlgorithmBox}

Throughout the following analysis, let $L\in\NP$ denote the
language for which the underlying $\msf{ZHB\text{-}NIZK}_\gamma$
is defined. The constructed $\msf{URS\text{-}NIZK}$ proves
membership in the same language $L$, and we write $n=|x|$
for the length of a statement $x$.

\subsection{Security Proof}\label{URS_NIZK}

Note that the completeness and soundness of the above NIZK proof system directly follows from \cite{STOC:GIKKLS23} as the quantum-accessibility of zero-knowledge adversary will not affect the the above two properties, we will then focus on the zero-knowledge property.
\begin{lemma}
    If $H_{\mathsf{PRHF}}$ is a $\delta$-dense post-quantum PRHF for some $\msf{Efficiently\text{-}Approximable}$ constant $\delta \in (0, 1)$ and $\mathsf{ZHB.NIZK}^\gamma$ with $\gamma = 1 - \delta$ satisfies statistical zero knowledge in the Z-Tamperable-Hidden-Bits model, then there exists a QPT simulator $\mathsf{URS.Sim}$ and for all $x\in \mathcal{L}$ and all non-uniform QPT adversary $\Adv$, 
        \begin{align*}
                    & \Bigg| \Pr[\mathcal{A}(\mathrm{urs},H_{\msf{PRHF}},x,\pi) = 1 ~:~ 
                     \begin{array}{l}
                        H_{\msf{PRHF}}\gets \mcal{H}_n; \\
                        \mathrm{urs}\leftarrow \msf{URS.Gen}(1^{|x|});\\
                        \pi \leftarrow \msf{URS.P}(\mathrm{urs},H_{\msf{PRHF}},x)
                    \end{array}
                    ] - \\
                    &\hspace{4em} \Pr[\mathcal{A}(\mathrm{urs}^{*},H_{\msf{PRHF}},x,\pi^{*}) = 1 ~:~ 
                    \begin{array}{l}
                        H_{\msf{PRHF}}\gets \mcal{H}_n; \\
                        (\mathrm{urs}^{*},\pi^{*})\leftarrow \operatorname{URS.Sim}(H_{\msf{PRHF}},x);
                    \end{array}
                    ] \Bigg| \leq \negl(n).
        \end{align*}
\end{lemma}
\begin{proof}
    We complete the proof by giving the explicit construction of the simulator $\mathsf{URS.Sim}$ and proving that it is indistinguishable from the real prover $\mathsf{URS.P}$ through hybrid arguments.
\begin{AlgorithmBox}[label={URS Simulator}]{URS Simulator $\mathsf{URS.Sim}$}
     $\mathsf{URS.Sim}(H_{\msf{PRHF}},x)$:
    \begin{enumerate}
        \item Let $n = |x|$.
        \item $(\mathsf{ZHB.I}, \{ r_i \}_{i \in \mathsf{ZHB.I}}, \mathsf{ZHB.\pi}) \leftarrow \mathsf{ZHB.Sim}_\gamma(x)$.
        \item For $i \in [p_{\mathsf{hbs}}(n)]$, sample $v_i \leftarrow \{0, 1\}^n$, and set $y_i = H_{\mathsf{PRHF}}(v_i)$.
        \item For $i \in \mathsf{ZHB.I}$,
          \begin{enumerate}
          \item[(a)] If $r_i = 1$, set $v'_i = \perp$
          \item[(b)] If $r_i = 0$, set $v'_i = v_i$.
          \end{enumerate}
        \item Output $(\mathsf{urs} = (y_1 \dots y_{p_{\mathsf{hbs}}(n)}), \pi = (\mathsf{ZHB.I}, \{ v'_i \}_{i \in \mathsf{ZHB.I}}, \mathsf{ZHB.\pi}))$.
    \end{enumerate}
\end{AlgorithmBox}

\para{Hybrid $H_0$:} This is the real-world game where the adversary interacts with the real prover.We formally describe it in \Cref{hybrid:h40}.
\begin{ExperimentBox}[label={hybrid:h40}]{Hybrid $H_0$}
    
\begin{enumerate}
    \item Let $n = |x|$.
    \item For $i \in [p_{\mathsf{hbs}}(n)]$,
      \begin{enumerate}
      \item Sample $y_i \picks \{0, 1\}^{\ell(n)}$.
      \item If $\exists z \in \{0, 1\}^n$ such that $H_{\mathsf{PRHF}}(z) = y_i$, and sample $v_i \picks \{z \in \{0, 1\}^n : H_{\mathsf{PRHF}}(z) = y_i\}$, and set $r_i = 0$.
      \item If $\nexists z \in \{0, 1\}^n$ such that $H_{\mathsf{PRHF}}(z) = y_i$, set $v_i = \perp$ and $r_i = 1$.
      \end{enumerate}
    \item $(\mathsf{ZHB.\mathcal{I}}, \mathsf{ZHB.\pi}) \leftarrow \mathsf{ZHB.P}_\gamma(r, x)$ where $r = r_1 \dots r_{p_{\mathsf{hbs}}(n)}$.
    \item Output $(\mathsf{urs} = (y_1 \dots y_{p_{\mathsf{hbs}}(n)}), \pi = (\mathsf{ZHB.\mathcal{I}}, \{ v_i \}_{i \in \mathsf{ZHB.\mathcal{I}}}, \mathsf{ZHB.\pi}))$.
\end{enumerate}

\end{ExperimentBox}

\para{Hybrid: $H_1$} We present $H_1$ in \Cref{hybrid:h41} and highlight its difference from $H_0$ in \red{red color}.
In this hybrid, we just interchange the order of sampling $y_i$ and sampling $v_i$ for those $r_i=0$ without changing the distribution.

\begin{ExperimentBox}[label={hybrid:h41}]{Hybrid $H_1$}
    
\begin{enumerate}
    \item Let $n = |x|$.
    \item For $i \in [p_{\mathsf{hbs}}(n)]$,
      \begin{enumerate}
      \red{\item Sample $r_i \in \{0, 1\}$ such that $\Pr[r_i = 1] = \gamma'$ where\\
       $\gamma' = 1 - \Pr_{y \leftarrow \{0, 1\}^{\ell(n)}} [\exists z \text{ s.t. } H_{\mathsf{PRHF}}(z) = y]$} 
      \end{enumerate}
    \item $(\mathsf{ZHB.\mathcal{I}}, \mathsf{ZHB.\pi}) \leftarrow \mathsf{ZHB.P}_\gamma(r, x)$ where $r = r_1 \dots r_{p_{\mathsf{hbs}}(n)}$.
    \red{\item Define 
        $\begin{cases}
            \msf{Image} \coloneqq \{y\in \bits^{\ell(n)} : \exists z \in \bits^n \text{ such that } H_{\mathsf{PRHF}}(z) = y\} \\ 
            \msf{PreImage}(y)\coloneqq\{z\in \bits^n : H_{\mathsf{PRHF}}(z) = y\}
        \end{cases}$}
    \red{\item For $i \in [p_{\mathsf{hbs}}(n)]$,
        \begin{enumerate}
        \item If $r_i = 0$, sample $y_i \leftarrow \mathsf{Image}$ and $v_i \leftarrow \mathsf{PreImage}(y_i)$, and set $v'_i = v_i$.
        \item If $r_i = 1$, sample $y_i \leftarrow \{0, 1\}^{\ell(n)} \setminus \mathsf{Image}$, and set $v'_i = \perp$.
        \end{enumerate}
    }
    \item Output $(\mathsf{urs} = (y_1 \dots y_{p_{\mathsf{hbs}}(n)}), \pi = (\mathsf{ZHB.\mathcal{I}}, \{ \red{v'_i} \}_{i \in \mathsf{ZHB.\mathcal{I}}}, \mathsf{ZHB.\pi}))$.
\end{enumerate}

\end{ExperimentBox}
\begin{MyClaim}\zcsetup{reftype=Claim} \label{claim:lastH0-H1}
    It holds that $H_{0}\idind H_{1}$. 
\end{MyClaim}
\begin{proof}
    The above hybrid is the same as the previous hybrid, except that we have changed the way of generating the $(r_i,v_i,y_i)$.

Observe the following statistical facts in hybrid $H_1$:
\begin{itemize}
    \item The distribution of $(\msf{ZHB}.\mathcal{I}, \msf{ZHB}.\pi)$ only depends on the random $r$ and the input $x$.
    \item For any $i\neq j$, the distribution of $(r_i,v_i,y_i)$ is independently and identically distributed as $(r_j,v_j,y_j)$.
    \item For any $i$, we have $\Pr[r_i=1]=\Pr_{y \pick \bits^{\ell(n)}}[\nexists z \in \bits^n \text{ such that } H_{\msf{PRHF}}(z) = y]$.
    \item For any $i$, conditioned on $r_i=0$, $y_i$ is uniformly distributed over all elements in the image of $H_{\msf{PRHF}}$,
     and $v_i$ is uniformly distributed over all pre-images of $y_i$.
    \item For any $i$, conditioned on $r_i=1$, $y_i$ is uniformly distributed over all elements that are not in the image of $H_{\msf{PRHF}}$,
    and  $v_i$ = $\perp$.
\end{itemize}
    Therefore, it is identical for us to first sample each $r_i$ based on the distribution of $H_{\msf{PRHF}}$, 
    and then sample $y_i$ and $v_i$ based on conditional probability induced by $r_i$.
    Also, as $(\msf{ZHB}.\mathcal{I}, \msf{ZHB}.\pi)$ only depends on $r$,
    as long as the distribution of $r$ is the same, the distribution of final output for $(\msf{ZHB}.\mathcal{I}, \msf{ZHB}.\pi)$ will also be the same.
\end{proof}

\para{Hybrid: $H_2$} We present $H_2$ in \Cref{hybrid:h42} and highlight its difference from $H_1$ in \red{red color}.
In this hybrid, we replace the exact sampling probability of
$r_i$ with its approximation, changing the resulting distribution
by only a negligible statistical distance.

\begin{ExperimentBox}[label={hybrid:h42}]{Hybrid $H_2$}
    
\begin{enumerate}
    %\item $H_{\mathsf{PRHF}}\gets \mcal{H}$
    \item Let $n = |x|$.
    \item For $i \in [p_{\mathsf{hbs}}(n)]$,
      \begin{enumerate}
      \red{\item Sample $r_i \in \{0, 1\}$ such that $\Pr[r_i = 1] = 1-\msf{Approx}_\delta(1^n)$\\
        Notice that this change reflects the
$\delta$-density property in \Cref{delta-dense}, with
$\delta$ replaced by its efficiently computable approximation
$\msf{Approx}_\delta(1^n)$ from \Cref{effapp}.}
        \end{enumerate}
    \item $(\mathsf{ZHB.\mathcal{I}}, \mathsf{ZHB.\pi}) \leftarrow \mathsf{ZHB.P}_\gamma(r, x)$ where $r = r_1 \dots r_{p_{\mathsf{hbs}}(n)}$.
    \item Define 
        $\begin{cases}
            \msf{Image} \coloneqq \{y\in \bits^{\ell(n)} : \exists z \in \bits^n \text{ such that } H_{\mathsf{PRHF}}(z) = y\} \\ 
            \msf{PreImage}(y)\coloneqq\{z\in \bits^n : H_{\mathsf{PRHF}}(z) = y\}
        \end{cases}$
    \item For $i \in [p_{\mathsf{hbs}}(n)]$,
        \begin{enumerate}
        \item If $r_i = 0$, sample $y_i \leftarrow \mathsf{Image}$ and $v_i \leftarrow \mathsf{PreImage}(y_i)$, and set $v'_i = v_i$.
        \item If $r_i = 1$, sample $y_i \leftarrow \{0, 1\}^{\ell(n)} \setminus \mathsf{Image}$, and set $v'_i = \perp$.
        \end{enumerate}
    \item Output $(\mathsf{urs} = (y_1 \dots y_{p_{\mathsf{hbs}}(n)}), \pi = (\mathsf{ZHB.\mathcal{I}}, \{ v'_i \}_{i \in \mathsf{ZHB.\mathcal{I}}}, \mathsf{ZHB.\pi}))$.
\end{enumerate}

\end{ExperimentBox}
\begin{MyClaim}\zcsetup{reftype=Claim} \label{claim:lastH1-H2}
    It holds that $H_1\sind H_2$. 
\end{MyClaim}
\begin{proof}
    Let $D_{0,x}$ and $D_{1,x}$ be the distributions of binary string $r$ sampled in hybrid $H_1(x)$ and $H_2(x)$ respectively.\\
Based on the $\delta$-density property of $H_\msf{PRHF}$, we have:
\begin{align*}
    \gamma^{'}&= 1-\Pr_{y \pick \bits^{\ell(n)} }[\exists z \in \bits^n \text{ such that } H_{\msf{PRHF}}(z) = y] \\
    &=1-\delta \pm\negl(n) \tag{\Cref{delta-dense} of \Cref{def:strongPRHF}}\\
    &=1-\msf{Approx}_{\delta}(1^n) \pm\negl(n) \tag{\Cref{effapp}}
\end{align*}
As the distribution of each $r_i$ is independently and identically distributed, we can bound the statistical distance between $D_{0,x}$ and $D_{1,x}$ with \Cref{MaxCoupling} as follows:
\[\Delta(D_{0,x},D_{1,x}) \leq p_{\msf{hbs}}(n)\cdot\negl_1(n) = \negl_2(n) \]
As these two hybrids only differ in the way of sampling $r$, and thus there exist a negligible function $\negl(n)$ such that:
\[\Delta(H_1(x),H_2(x)) \leq \negl(n)\]
\end{proof}

\para{Hybrid: $H_3$} We present $H_3$ in \Cref{hybrid:h43} and highlight its difference from $H_2$ in \red{red color}.
In this hybrid, we changed the way of sampling $v_i'$ as we defined in the Strong-PreImage Pseudorandomness of the $\delta$-dense post-quantum PRHF .

\begin{ExperimentBox}[label={hybrid:h43}]{Hybrid $H_3$}
    
\begin{enumerate}
    %\item $H_{\mathsf{PRHF}}\gets \mcal{H}$
    \item Let $n = |x|$.
    \item For $i \in [p_{\mathsf{hbs}}(n)]$,
      \begin{enumerate}
      \item Sample $r_i \in \{0, 1\}$ such that $\Pr[r_i = 1] = 1-\msf{Approx}_\delta(1^n)$
        \end{enumerate}
    \item $(\mathsf{ZHB.\mathcal{I}}, \mathsf{ZHB.\pi}) \leftarrow \mathsf{ZHB.P}_\gamma(r, x)$ where $r = r_1 \dots r_{p_{\mathsf{hbs}}(n)}$.
    \item \red{Deletion of the definition of $\msf{Image}$ and $\msf{PreImage}$}
    \item For $i \in [p_{\mathsf{hbs}}(n)]$,
        \red{\begin{enumerate}
            \item If $r_i=0$, then set $v_i\pick \bits^n ,y_i=H_{\mathsf{PRHF}}(v_i)$ and set $v'_i = v_i$
            \item If $r_i=1$, then set $v_i\pick \bits^n ,y_i=H_{\mathsf{PRHF}}(v_i)$ and set $v'_i = \perp$
        \end{enumerate}}
    \item Output $(\mathsf{urs} = (y_1 \dots y_{p_{\mathsf{hbs}}(n)}), \pi = (\mathsf{ZHB.\mathcal{I}}, \{ v'_i \}_{i \in \mathsf{ZHB.\mathcal{I}}}, \mathsf{ZHB.\pi}))$.
\end{enumerate}

\end{ExperimentBox}
\begin{MyClaim}\zcsetup{reftype=Claim} \label{claim:lastH2-H3}
    It holds that $H_2\qcind H_3$. 
\end{MyClaim}
\begin{proof}
    In this argument, we try to reduce the distinguishability between hybrid $H_2$ and hybrid $H_3$ to breaking the Strong-PreImage Pseudorandomness of the $\delta$-dense post-quantum PRHF.\\
\chakfai{Note that this distinguisher is not polynomial time because of the ZHB prover\\}
Assume for contradiction that there exist a QPT quantum distinguisher $\Adv$ such that:
\[\abs{\Pr[\Adv(H_2(x))=1]-\Pr[\Adv(H_3(x))=1]} \geq \frac{1}{\poly(n)}\]
We can then construct a non-uniform algorithm $\mcal{B}_{\rho}$ that can break the Strong-PreImage Pseudorandomness of the $\delta$-dense post-quantum PRHF w.r.t.\ the game defined in \Cref{box:dist-D0-D1}. 

The algorithm $\mcal{B}_{\rho}$'s non-uniform advice $\rho$ contains the snapshot of the execution of $H_3$ at  (inclusively)  Step 3. In particular, it contains the statement $x$ and the corresponding $(\mathsf{ZHB.\mathcal{I}}, \mathsf{ZHB.\pi})$, and the bit-vector $\vb{r} = (r_1 \dots r_{p_{\mathsf{hbs}}(n)})$. Then, $\mcal{B}_{\rho}$ starts to participate the Strong-PreImage Pseudorandomness game by receiving the challenge transcript $(\vb{r}, \{y_i\}_{i\in[p_{\mathsf{hbs}}(n)]}, \{v_i\}_{i\in\mathsf{\mcal{I}_0}})$ from the challenger. (Note that we can assume without loss of generality that this $\vb{r}$ in the challenge is the same $\vb{r}$ contained in $\mcal{B}_\rho$'s non-uniform advice, because the strong pre-image pseudo-randomness defined in \zcref{def:strong-PRHF:strongPIP} holds against any bit-vector $\vb{r}$.)

\begin{AlgorithmBox}[label={A-1-Section6}]{$\mcal{B}_\rho$ that can break the Strong-PreImage Pseudorandomness}
    $\mcal{B}_{\rho}(1^n, H_{\msf{PRHF}}, \vb{r}, \{y_i\}_{i \in [p_{\mathsf{hbs}}(n)]},\{v_i\}_{i \in \mathsf{\mcal{I}_0}})$:
\begin{enumerate}
    \item Extract $x$ and the corresponding $(\mathsf{ZHB.\mathcal{I}}, \mathsf{ZHB.\pi})$ from the non-uniform advice $\rho$ of $\mcal{B}$. 

    \item For $i \in [p_{\mathsf{hbs}}(n)]$,
    \begin{enumerate}
        \item If $r_i=0$, set $v'_i = v_i$
        \item If $r_i=1$, set $v'_i = \perp$
    \end{enumerate}

    \item Set $\mathsf{urs} = (y_1 \dots y_{p_{\mathsf{hbs}}(n)}), \pi = (\mathsf{ZHB.\mathcal{I}}, \{ v'_i \}_{i \in \mathsf{ZHB.\mathcal{I}}}, \mathsf{ZHB.\pi})$
    \item Invoke $\Adv(\msf{urs},H_{\msf{PRHF}},x,\pi)$ and output whatever $\Adv$ outputs.
\end{enumerate}
\end{AlgorithmBox}

Refering to the strong pre-image pseudorandomness property in
\Cref{def:strong-PRHF:strongPIP} of \Cref{def:strongPRHF}, the challenge
transcript
\[
    \left(\{r_i\}_{i\in[p(n)]},
        \{y_i\}_{i\in[p_{\mathsf{hbs}}(n)]},
        \{v_i\}_{i\in\mathsf{\mcal{I}_0}}
    \right)
\]
is distributed exactly as in hybrid $H_2$ when sampled from
$\mathcal{D}_0$, and exactly as in hybrid $H_3$ when sampled from
$\mathcal{D}_1$. Consequently,
\[
    \Pr[\mcal{B}_\rho(\mathcal{D}_0)=1]
    =
    \Pr[\Adv(H_2(x))=1]
    \quad\text{and}\quad
    \Pr[\mcal{B}_\rho(\mathcal{D}_1)=1]
    =
    \Pr[\Adv(H_3(x))=1].
\]
Therefore, we have:
\[\abs{\Pr[\mcal{B}_\rho(D_0)=1]-\Pr[\mcal{B}_\rho(D_1)=1]} = \abs{\Pr[\Adv(H_2(x))=1]-\Pr[\Adv(H_3(x))=1]} \geq \frac{1}{\poly(n)},\]
which contradicts the Strong-PreImage Pseudorandomness of the $\delta$-dense post-quantum PRHF.\\
Thus we can conclude that $H_2$ and $H_3$ are indistinguishable towards QPT adversaries with PRHF-independent non-uniform advice.

\end{proof}
\para{Hybrid $H_4$:} We present $H_4$ in \Cref{hybrid:h44} and highlight its difference from $H_3$ in \red{red color}.
In this hybrid, we replace the real proof generated by
$\mathsf{ZHB.P}_\gamma$ with the output of the statistical zero-knowledge
simulator $\mathsf{ZHB.Sim}_\gamma$.

\begin{ExperimentBox}[label={hybrid:h44}]{Hybrid $H_4$}
    
\begin{enumerate}
    %\item $H_{\mathsf{PRHF}}\gets \mcal{H}$
    \item Let $n = |x|$.
    \item \red{
    $(\mathsf{ZHB.\mathcal{I}},
    \{r_i\}_{i\in\mathsf{ZHB.\mathcal{I}}},
    \mathsf{ZHB.\pi})
    \leftarrow \mathsf{ZHB.Sim}_\gamma(x)$.
    }

    \item \red{ 
    For every $i\in[p_{\mathsf{hbs}}(n)]$, sample
    $v_i\picks\{0,1\}^n$ and set
    $y_i=H_{\mathsf{PRHF}}(v_i)$.
    }

    \item \red{For every $i\in\mathsf{ZHB.\mathcal{I}}$,
    \begin{enumerate}
        \item If $r_i=0$, set $v'_i=v_i$.
        \item If $r_i=1$, set $v'_i=\perp$.
    \end{enumerate}
    }

    \item Output $(\mathsf{urs} = (y_1 \dots y_{p_{\mathsf{hbs}}(n)}), \pi = (\mathsf{ZHB.\mathcal{I}}, \{ v'_i \}_{i \in \mathsf{ZHB.\mathcal{I}}}, \mathsf{ZHB.\pi}))$.
\end{enumerate}

\end{ExperimentBox}
\begin{MyClaim}\zcsetup{reftype=Claim} \label{claim:lastH3-H4}
    It holds that $H_3\sind H_4$. 
\end{MyClaim}

\begin{proof}
    In this argument, we try to reduce the distinguishability between hybrid
$H_3$ and hybrid $H_4$ to breaking the statistical zero-knowledge property
of the $\mathsf{ZHB.NIZK}$.

Recall that $\gamma=1-\delta$. We choose the approximation algorithm for
$\gamma$ such that
\[
\mathsf{Approx}_\gamma(1^n)
=
1-\mathsf{Approx}_\delta(1^n).
\]
Therefore, the distribution of $r$ in $H_3$ is identical to the
distribution generated by $\mathsf{ZHB.Gen}_\gamma(1^n)$.

Assume for contradiction that there exists a quantum
distinguisher $\Adv$ such that
\[
\abs{
\Pr[\Adv(H_3(x))=1]
-
\Pr[\Adv(H_4(x))=1]
}
\geq \frac{1}{\poly(n)}.
\]
We can then construct an adversary $\Adv_2$ that breaks the statistical
zero-knowledge property of the $\mathsf{ZHB.NIZK}$ as follows:
\begin{AlgorithmBox}[label={A-2-Section6}]{$\Adv_2$ that breaks the Statistical
Zero-Knowledge Property of the $\mathsf{ZHB.NIZK}$ }
    $\Adv_2(x,\mathsf{ZHB.\mathcal{I}},
\{r_i\}_{i\in\mathsf{ZHB.\mathcal{I}}},
\mathsf{ZHB.\pi})$:
\begin{enumerate}
    \item Let $n=|x|$.
    \item Sample $H_{\mathsf{PRHF}}\picks\mcal{H}$ .
    \item For every $i\in[p_{\mathsf{hbs}}(n)]$, sample
    $v_i\picks\{0,1\}^n$ independently and set
        $y_i=H_{\mathsf{PRHF}}(v_i).$

    \item For every $i\in\mathsf{ZHB.\mathcal{I}}$,
    \begin{enumerate}
        \item If $r_i=0$, set $v'_i=v_i$.
        \item If $r_i=1$, set $v'_i=\perp$.
    \end{enumerate}

    \item Set
    $\msf{urs}=(y_1\dots y_{p_{\msf{hbs}}(n)})$
    and
    $\pi=\left(\msf{ZHB.\mathcal{I}},
        \{v'_i\}_{i\in\msf{ZHB.\mathcal{I}}},
        \msf{ZHB.\pi}\right)$

    \item Output $\Adv(\mathsf{urs},H_{\mathsf{PRHF}},x,\pi)$.
\end{enumerate}
\end{AlgorithmBox}

We first argue that $\Adv_2$ exactly simulates hybrid $H_3$ when its input
is sampled from the real-world distribution of the
$\mathsf{ZHB.NIZK}$. In this case,
\[
r\leftarrow\mathsf{ZHB.Gen}_\gamma(1^n)
\]
and
\[
(\mathsf{ZHB.\mathcal{I}},\mathsf{ZHB.\pi})
\leftarrow
\mathsf{ZHB.P}_\gamma(r,x).
\]
After receiving the revealed bits
$\{r_i\}_{i\in\mathsf{ZHB.\mathcal{I}}}$, the adversary $\Adv_2$
samples every $v_i$ independently, sets
$y_i=H_{\mathsf{PRHF}}(v_i)$, and defines $v'_i$ in exactly the same
manner as hybrid $H_3$. Therefore,
\[
\begin{aligned}
&\Pr\left[
\Adv_2
\left(
x,\mathsf{ZHB.\mathcal{I}},
\{r_i\}_{i\in\mathsf{ZHB.\mathcal{I}}},
\mathsf{ZHB.\pi}
\right)=1:
\begin{array}{l}
r\leftarrow\mathsf{ZHB.Gen}_\gamma(1^n),\\
(\mathsf{ZHB.\mathcal{I}},\mathsf{ZHB.\pi})
\leftarrow\mathsf{ZHB.P}_\gamma(r,x)
\end{array}
\right]
\\
&\hspace{35mm}
=
\Pr[\Adv(H_3(x))=1].
\end{aligned}
\]

Similarly, when the input of $\Adv_2$ is sampled from the simulated-world
distribution
\[
\left(
\mathsf{ZHB.\mathcal{I}},
\{r_i\}_{i\in\mathsf{ZHB.\mathcal{I}}},
\mathsf{ZHB.\pi}
\right)
\leftarrow
\mathsf{ZHB.Sim}_\gamma(x),
\]
the output distribution of $\Adv_2$ is identical to the output
distribution of $\Adv$ in hybrid $H_4$. Therefore,
\[
\begin{aligned}
&\Pr\left[
\Adv_2
\left(
x,\mathsf{ZHB.\mathcal{I}},
\{r_i\}_{i\in\mathsf{ZHB.\mathcal{I}}},
\mathsf{ZHB.\pi}
\right)=1:
\left(
\mathsf{ZHB.\mathcal{I}},
\{r_i\}_{i\in\mathsf{ZHB.\mathcal{I}}},
\mathsf{ZHB.\pi}
\right)
\leftarrow
\mathsf{ZHB.Sim}_\gamma(x)
\right]\\
&=
\Pr[\Adv(H_4(x))=1].
\end{aligned}
\]

By combining the above two equalities, we have
\[
\begin{aligned}
&\abs{
\Pr\left[
\Adv_2(\mathsf{Real}_{\mathsf{ZHB}}(x))=1
\right]
-
\Pr\left[
\Adv_2(\mathsf{Sim}_{\mathsf{ZHB}}(x))=1
\right]
}
\\
&\qquad =
\abs{
\Pr[\Adv(H_3(x))=1]
-
\Pr[\Adv(H_4(x))=1]
}
\geq \frac{1}{\poly(n)}.
\end{aligned}
\]
This contradicts the statistical zero-knowledge property of the
$\mathsf{ZHB.NIZK}$. Thus, we can conclude that hybrid $H_3$ and hybrid
$H_4$ are statistically indistinguishable.
\end{proof}

\para{Completes the proof of Zero-Knowledge:}
By combining the indistinguishability between hybrid $H_0$ and hybrid $H_4$, we can conclude that the real-world interaction between the adversary and the real prover is computationally indistinguishable from the output of the simulator towards any QPT quantum adversary with $H$-independent non-uniform advice.
This completes the proof of the ZK property  in the URS-PRHF model.

\end{proof}
\section{NIZK in QROM}
\label{sec:NIZK-QROM:major}

\subsection{Structure of Our Construction}
\label{sec:nizk-qrom}

We now turn our URS-PRHF-model NIZK construction into a NIZK
proof system in the QROM by realizing both the PRHF and
the reference string using a single quantum random oracle.
Our strengthened post-quantum PRHF definition allows us
to reuse the proof for the zero-knowledge property from \Cref{sec:URS-NIZK} when we instantiate
the PRHF using a random oracle. This differs from \cite{STOC:GIKKLS23}, where
the use of hash-dependent non-uniform advice prevents
the URS zero-knowledge proof from directly carrying over
to the random-oracle setting.
The remaining work is to simulate the oracle supplying
the URS and to justify combining the two oracle interfaces.
We organize the construction into the following three steps:
\begin{enumerate}
    \item Instantiate the PRHF in our URS-PRHF-model NIZK
    construction (\Cref{sec:URS-NIZK-construction}) with
    a quantum random oracle $O_\msf{PRHF}$. This step
    relies on our result that random oracles instantiate
    strong post-quantum PRHFs
    (\Cref{sec:random-oracle-is-strong-dense-prhf}),
    and yields a URS-PRHF-model NIZK proof system relative
    to $O_\msf{PRHF}$.

    \item Replace the URS with values of an independent
    random oracle $O_\msf{URS}$, obtaining a NIZK proof
    system with access to two independent quantum random
    oracles. The conditional compressed oracle enables
    efficient simulation of $O_\msf{URS}$ conditioned on
    the reference string produced by the URS simulator.

    \item Apply the two-oracle-to-one-oracle technique
    (\Cref{sec:merge-oracles}) to implement both oracles
    using a single quantum random oracle, yielding the
    desired NIZK proof system in the QROM.
\end{enumerate}

\newcounter{qromalg}[section]
\renewcommand{\theqromalg}{\thesection.\arabic{qromalg}}

\newenvironment{qrombox}[1]{%
    \refstepcounter{qromalg}%
    \par\medskip\noindent
    \begin{tabular}{|p{\dimexpr
        \linewidth-2\tabcolsep-2\arrayrulewidth\relax}|}
    \hline
    \textbf{Algorithm \theqromalg: #1}\\
    \hline
    \begin{minipage}{\linewidth}
    \smallskip
}{%
    \smallskip
    \end{minipage}\\
    \hline
    \end{tabular}
    \par\medskip
}

For the random-oracle instantiation of the URS-PRHF-model
NIZK construction (\Cref{sec:URS-NIZK-construction}),
the PRHF density parameter and the complementary
hidden-bit parameter are
\[
    \delta=1-\frac{1}{e},
    \qquad
    \gamma=1-\delta=\frac{1}{e}.
\]
The construction requires both constants to be
efficiently approximable. It suffices to establish
this property for $\gamma=1/e$: subtracting an
approximation to $\gamma$ from $1$ yields an
approximation to $\delta$ with the same error.
The proof is standard and is deferred to \Cref{sec:eff-approx-e}.

\subsection{Instantiating the PRHF}

Our analysis (\Cref{sec:random-oracle-is-strong-dense-prhf}) shows that
a quantum random oracle $O_\msf{PRHF}\gets\Ora$
instantiates a $(1-1/e)$-dense post-quantum PRHF.
Since $1-1/e$ is efficiently approximable
(\Cref{cor:efficient-approx-one-minus-inv-e}), applying
our URS-PRHF-model construction
(\Cref{sec:URS-NIZK-construction}) yields a NIZK proof
system with both URS and RO:
\[
    \Pi_\msf{URS}
    =
    \bigl(
        \msf{URS.Gen},
        \msf{URS.P}^{O_\msf{PRHF}},
        \msf{URS.V}^{O_\msf{PRHF}}
    \bigr),
\]
with simulator $\msf{URS.Sim}^{\ket{O_\msf{PRHF}}}$.

We use the zero-knowledge guarantee established in
\Cref{sec:URS-NIZK-construction} against polynomial-time quantum distinguishers
with quantum oracle access to $O_\msf{PRHF}$.
All security probabilities include the sampling of
$O_\msf{PRHF}\gets\Ora$, together with the randomness
and measurement outcomes of the relevant algorithms.
The oracle is sampled once and remains fixed throughout
each experiment.

\subsection{Replacing the URS}

\begin{lemma}[Eliminating the URS]
\label{lem:qrom-eliminate-urs}
Suppose that $\Pi_{\mathrm{URS}}$, relative to
$O_{\msf{PRHF}}$, has perfect completeness,
adaptive statistical soundness with error
$\varepsilon_{\mathrm{s}}(n)$, and zero knowledge
against polynomial-query-bounded quantum distinguishers,
with a QPT oracle simulator.

Then there exists an unbounded-prover NIZK proof system
$\Pi_\msf{TWO-QROM}$ for $L$ with separate quantum access
to two independent random oracles $O_\msf{URS}$ and
$O_\msf{PRHF}$, without an additional reference string.
The system preserves perfect completeness, adaptive
statistical soundness with error $\varepsilon_{\mathrm{s}}(n)$,
and zero knowledge against the same class of distinguishers,
with a stateful QPT simulator.
\end{lemma}

\begin{proof}
We construct $\Pi_\msf{TwoQROM}$ by deriving the reference
string from $O_\msf{URS}$ and invoking the prover and
verifier of the URS-PRHF-model construction with quantum access
to $O_\msf{PRHF}$. We then analyze its security properties
separately.

Throughout the following analysis, an integer index $i$
used as an input to $O_\msf{URS}$ is understood to denote
$\langle i\rangle_n$, its padded $n$-bit binary representation
as defined in \Cref{prelim}. The same convention applies
to integer-labelled oracle locations in sets, conditioning
events, and partial tables. For simplicity, we suppress
these encodings until the final construction in
\Cref{sec:QROM-final-construction}, where we write them explicitly.

Let $n=|x|$ and $m=p_\msf{hbs}(n)$. The reference string
in the instantiated URS system consists of $m$ independent,
uniformly distributed $n$-bit blocks. We obtain these blocks
from a random oracle $O_\msf{URS}$, sampled independently
of $O_\msf{PRHF}$, by setting
\[
    y_i:=O_\msf{URS}(i),
    \qquad i\in[m].
\]
Define
\[
    U_n(O_\msf{URS})
    :=
    O_\msf{URS}(1)\|\cdots\|O_\msf{URS}(m).
\]
The prover and verifier are defined as follows.

\begin{AlgorithmBox}[label={TwoQROM-Prover}]
{Two-Oracle QROM Prover $\msf{TwoQROM.P}$}
    $\msf{TwoQROM.P}^{O_\msf{URS},O_\msf{PRHF}}(x)$:
    \begin{enumerate}
        \item Set $n=|x|$ and $m=p_\msf{hbs}(n)$.
        \item For each $i\in[m]$, query $O_\msf{URS}$ at $i$
        to obtain $y_i=O_\msf{URS}(i)$.
        \item Set $\msf{urs}=y_1\|\cdots\|y_m$.
        \item 
        $
            \pi\gets
            \msf{URS.P}^{O_\msf{PRHF}}(\msf{urs},x)
        $
        and return $\pi$.
    \end{enumerate}
\end{AlgorithmBox}

\begin{AlgorithmBox}[label={TwoQROM-Verifier}]
{Two-Oracle QROM Verifier $\msf{TwoQROM.V}$}
    $\msf{TwoQROM.V}^{O_\msf{URS},O_\msf{PRHF}}(x,\pi)$:
    \begin{enumerate}
        \item Set $n=|x|$ and $m=p_\msf{hbs}(n)$.
        \item For each $i\in[m]$, query $O_\msf{URS}$ at $i$
        to obtain $y_i=O_\msf{URS}(i)$.
        \item Set $\msf{urs}=y_1\|\cdots\|y_m$.
        \item Return
        $
            \msf{URS.V}^{O_\msf{PRHF}}
            (\msf{urs},x,\pi).
        $
    \end{enumerate}
\end{AlgorithmBox}

Since the values of $O_\msf{URS}$ at distinct inputs are
independent and uniform, $U_n(O_\msf{URS})$ has exactly
the distribution generated by $\msf{URS.Gen}(1^n)$.
Moreover, it is independent of $O_\msf{PRHF}$. Hence,
\begin{equation}
\label{eq:qrom-urs-distribution}
    \bigl(O_\msf{PRHF},U_n(O_\msf{URS})\bigr)
    \equiv
    \bigl(O_\msf{PRHF},\msf{URS.Gen}(1^n)\bigr),
\end{equation}
where the reference string on the right is generated
independently of $O_\msf{PRHF}$.

Conditioning on the reference string fixes only the values
$O_\msf{URS}(1),\ldots,O_\msf{URS}(m)$. The remaining
values $O_\msf{URS}(z)$, for $z\notin[m]$, remain mutually
independent and uniformly distributed, and are independent
of $O_\msf{PRHF}$. For a fixed statement $x$, this property
also holds after conditioning on the honest proof $\pi$:
the prover accesses $O_\msf{URS}$ only to obtain
$\msf{urs}$, and its subsequent computation depends only
on $\msf{urs}$, $x$, its internal randomness, and
$O_\msf{PRHF}$.

More explicitly, fix $x$ and let
$z_1,\ldots,z_t\notin[m]$ be distinct oracle inputs.
For arbitrary $v_1,\ldots,v_t\in\{0,1\}^n$,
\[
    \Pr\left[
        \bigwedge_{j=1}^{t}O_\msf{URS}(z_j)=v_j
        \,\middle|\,
        O_\msf{PRHF}=F,\,
        \msf{urs}=u_0,\,
        \pi=\pi_0
    \right]
    =2^{-nt},
\]
whenever the conditioning event has positive probability.
Thus, writing $u_0=y_1\|\cdots\|y_m$, quantum access to
$O_\msf{URS}$ under this conditioning is exactly emulated
by the partially-fixed compressed oracle initialized with
the partial table
\[
    \{(i,y_i):i\in[m]\}.
\]

\subsubsection{Completeness}
\begin{lemma}[Perfect completeness]
\label{lem:qrom-completeness}
If $\Pi_\msf{URS}$ has perfect completeness, then
$\Pi_\msf{TwoQROM}$ has perfect completeness.
\end{lemma}

\begin{proof}
Fix $x\in L$ and let $n=|x|$. By construction and
the distributional identity in
\eqref{eq:qrom-urs-distribution},
\begin{align*}
&\Pr\left[
    \msf{TwoQROM.V}^{O_\msf{URS},O_\msf{PRHF}}
    (x,\pi)=1
    :
    \pi\gets
    \msf{TwoQROM.P}^{O_\msf{URS},O_\msf{PRHF}}(x)
\right]
\\
&\quad =
\Pr\left[
    \msf{URS.V}^{O_\msf{PRHF}}
    (\msf{urs},x,\pi)=1
    :
    \begin{array}{l}
        \msf{urs}\gets\msf{URS.Gen}(1^n),\\
        \pi\gets
        \msf{URS.P}^{O_\msf{PRHF}}(\msf{urs},x)
    \end{array}
\right]
\\
&\quad =1.
\end{align*}
The first probability includes the independent sampling
of $O_\msf{URS}$ and $O_\msf{PRHF}$; the second includes
the independent sampling of $O_\msf{PRHF}$ and
$\msf{urs}$. Both probabilities also include the internal
randomness of the algorithms.
The final equality follows from the perfect completeness
of $\Pi_\msf{URS}$.
\end{proof}

\subsubsection{Soundness}
\begin{lemma}[Adaptive statistical soundness]
\label{lem:qrom-soundness}
If $\Pi_\msf{URS}$ has adaptive statistical soundness
with error $\varepsilon_{\mathrm{s}}(n)$, then
$\Pi_\msf{TwoQROM}$ has adaptive statistical
soundness with the same error.
\end{lemma}

\begin{proof}
For a fixed oracle $O_\msf{PRHF}$ and reference
string $\msf{urs}$, define
\[
    \msf{Bad}_n(O_\msf{PRHF},\msf{urs})
    :=
    \left[
        \exists x\in\{0,1\}^n\setminus L,\
        \exists\pi:
        \msf{URS.V}^{O_\msf{PRHF}}
        (\msf{urs},x,\pi)=1
    \right].
\]
Every statement of length $n$ uses the same reference
string $U_n(O_\msf{URS})$. Therefore,
\begin{align*}
&\Pr_{O_\msf{URS},\,O_\msf{PRHF}}\left[
    \exists x\in\{0,1\}^n\setminus L,\
    \exists\pi:
    \msf{TwoQROM.V}^{O_\msf{URS},O_\msf{PRHF}}
    (x,\pi)=1
\right]
\\
&\quad =
\Pr_{O_\msf{URS},\,O_\msf{PRHF}}\left[
    \msf{Bad}_n
    \bigl(O_\msf{PRHF},U_n(O_\msf{URS})\bigr)
\right]
\\
&\quad =
\Pr_{\substack{
    O_\msf{PRHF}\gets\Ora,\\
    \msf{urs}\gets\msf{URS.Gen}(1^n)
}}\left[
    \msf{Bad}_n(O_\msf{PRHF},\msf{urs})
\right]
\\
&\quad \leq\varepsilon_{\mathrm{s}}(n).
\end{align*}
The second equality follows from
\eqref{eq:qrom-urs-distribution}, and the final inequality
is the adaptive statistical soundness of
$\Pi_\msf{URS}$.

The existential quantifiers range over all false
statements of length $n$ and all proofs. Thus, the
argument permits adaptive statement selection and
computationally unbounded cheating provers.
\end{proof}

\subsubsection{Zero Knowledge}

For a reference string $\msf{urs}=y_1\|\cdots\|y_m$, define
\[
    \msf{HD}_{\msf{urs}}
    :=
    \{(i,y_i):i\in[m]\}.
\]
This table defines a consistent partial function because
its inputs are distinct. By \Cref{ComO} ,
$\msf{Com}(\msf{HD}_{\msf{urs}})$ perfectly
emulates quantum access to a random oracle conditioned
on its first $m$ values being $y_1,\ldots,y_m$.
All remaining values are mutually independent,
uniformly distributed $n$-bit strings.

\begin{AlgorithmBox}[label={TwoQROM-Proof-Simulator}]
{Proof Simulator $\msf{TwoQROM.SimProof}$}
    $\msf{TwoQROM.SimProof}(x)$:
    \begin{enumerate}
        \item Set $n=|x|$ and $m=p_\msf{hbs}(n)$.

        \item Initialize the PRHF-oracle simulator:
        \(
            \mathcal{C}_\msf{PRHF}
            \gets\operatorname{Com}(\varnothing).
        \)

        \item Run:
        \(
            (\msf{urs}^{*},\pi^{*})
            \gets
            \msf{URS.Sim}^{\ket{\mathcal{C}_\msf{PRHF}}}(x).
        \)\\
        Retain the resulting state of
        $\mathcal{C}_\msf{PRHF}$.

        \item Parse
        $\msf{urs}^{*}=y_1^{*}\|\cdots\|y_m^{*}$
        and set
        \(
            \msf{HD}^{*}
            :=\{(i,y_i^{*}):i\in[m]\}.
        \)

        \item Initialize the URS-oracle simulator:
        \(
            \mathcal{C}_\msf{URS}
            \gets\operatorname{Com}(\msf{HD}^{*}).
        \)

        \item Output $(\msf{st},\pi^{*})$, where
        $\msf{st}$ contains the current joint state of
        $\mathcal{C}_\msf{URS}$ and
        $\mathcal{C}_\msf{PRHF}$.
    \end{enumerate}
\end{AlgorithmBox}

\begin{AlgorithmBox}[label={TwoQROM-Oracle-Simulator}]
{Oracle Simulator $\msf{TwoQROM.SimRO}$}
    $\msf{TwoQROM.SimRO}$ maintains $\msf{st}$
    and provides two quantum oracle interfaces:
    \begin{enumerate}
        \item Upon a query to $\ket{O_\msf{URS}}$,
        apply the query operation of
        $\mathcal{C}_\msf{URS}$.

        \item Upon a query to $\ket{O_\msf{PRHF}}$,
        apply the query operation of
        $\mathcal{C}_\msf{PRHF}$.
    \end{enumerate}
    Each operation acts coherently on the query registers
    and the corresponding compressed-oracle state.
    Return the query registers and retain the resulting
    joint simulator state for subsequent queries.
\end{AlgorithmBox}

The oracle simulation continues from the state
produced during the execution of $\msf{URS.Sim}$,
preserving its correlations with the simulated reference
string and proof. Only the URS-oracle simulator is
initialized with a nonempty conditioning table,
namely $\msf{HD}^{*}$. At initialization, conditioned on
$(\msf{urs}^{*},\pi^{*})$, the unspecified URS-oracle
values are fresh, mutually independent uniform strings,
independent of the preceding simulation.

\begin{lemma}[Zero knowledge]
\label{lem:qrom-zero-knowledge}
Suppose that $\Pi_\msf{URS}$ satisfies 
zero knowledge against polynomial-query-bounded quantum 
distinguishers with access to $\ket{O_\msf{PRHF}}$.
Then $\Pi_\msf{TwoQROM}$ satisfies zero knowledge
against polynomial-query-bounded quantum distinguishers with
separate access to $\ket{O_\msf{URS}}$ and
$\ket{O_\msf{PRHF}}$, with an efficient stateful simulator.
\end{lemma}

\begin{proof}
Fix a polynomial-query-bounded quantum distinguisher $\mathcal{A}$
and a statement $x\in L$, and let $n=|x|$.
We construct a URS-PRHF-model distinguisher
$\mathcal{B}^{\ket{O_\msf{PRHF}}}$ such that any non-negligible
distinguishing advantage of $\mathcal{A}$ contradicts the
zero-knowledge guarantee of \Cref{URS_NIZK}.

\begin{AlgorithmBox}[label={TwoQROM-URS-Distinguisher}]
{URS Distinguisher $\mathcal{B}$}
    $\mathcal{B}^{\ket{O_\msf{PRHF}}}(\msf{urs},x,\pi)$:
    \begin{enumerate}
        \item Set $n=|x|$ and $m=p_\msf{hbs}(n)$,
        and parse $\msf{urs}=y_1\|\cdots\|y_m$.

        \item Initialize
        \[
            \mathcal{C}_\msf{URS}
            \gets
            \operatorname{Com}(\msf{HD}_{\msf{urs}}).
        \]

        \item Run $\mathcal{A}(x,\pi)$, answering its
        queries to $\ket{O_\msf{URS}}$ using
        $\mathcal{C}_\msf{URS}$ and forwarding its
        queries to $\ket{O_\msf{PRHF}}$ to the supplied
        oracle. Retain the state of
        $\mathcal{C}_\msf{URS}$ throughout the interaction.

        \item Output the bit returned by $\mathcal{A}$.
    \end{enumerate}
\end{AlgorithmBox}

\paragraph{Real experiment.}
When $(\msf{urs},\pi)$ is a real URS transcript,
$\mathcal{B}$ exactly reproduces the real experiment
for $\Pi_\msf{TwoQROM}$.
Indeed, $\msf{urs}$ is uniform and independent of
$O_\msf{PRHF}$. Conditioned on $\msf{urs}$,
$O_\msf{PRHF}$, and $\pi$, the URS-oracle values outside
$[m]$ remain mutually independent and uniform, as shown
above. \Cref{ComO} therefore gives an exact emulation of
quantum access to the URS oracle under this conditioning.
Consequently,
\begin{align*}
p_\msf{real}
&:=
\Pr\left[
    \mathcal{A}^{\ket{O_\msf{URS}},\ket{O_\msf{PRHF}}}
    (x,\pi)=1
    :
    \pi\gets
    \msf{TwoQROM.P}^{O_\msf{URS},O_\msf{PRHF}}(x)
\right]
\\
&=
\Pr\left[
    \mathcal{B}^{\ket{O_\msf{PRHF}}}
    (\msf{urs},x,\pi)=1
    :
    \begin{array}{l}
        \msf{urs}\gets\msf{URS.Gen}(1^n),\\
        \pi\gets
        \msf{URS.P}^{O_\msf{PRHF}}(\msf{urs},x)
    \end{array}
\right].
\end{align*}
This equality also holds when queries to the two
oracles are interleaved. For each fixed
$O_\msf{PRHF}$, its query operations are intervening
operations on the distinguisher's registers, which
are covered by the exact simulation in \Cref{ComO}.

\paragraph{Simulated experiment.}
When $(\msf{urs}^{*},\pi^{*})$ is generated by
$\msf{URS.Sim}^{\ket{O_\msf{PRHF}}}$,
the reduction runs $\mathcal{A}(x,\pi^{*})$ with
access to the same $O_\msf{PRHF}$ and to
$\operatorname{Com}(\msf{HD}_{\msf{urs}^{*}})$
as its URS-oracle interface.

Apply \Cref{ComO} with the empty partial function to
this entire computation: the execution of
$\msf{URS.Sim}$, the initialization of the conditional
URS oracle, and the subsequent execution of
$\mathcal{A}$. Replacing $O_\msf{PRHF}$ throughout
by a single persistent
$\operatorname{Com}(\varnothing)$ preserves the
output distribution exactly.

The resulting computation is precisely the simulation
defined above. Retaining the PRHF-oracle state throughout
preserves all correlations among $\msf{urs}^{*}$,
$\pi^{*}$, and subsequent oracle responses.
In the following experiment,
$\ket{\mathcal{C}_\msf{URS}}$ and
$\ket{\mathcal{C}_\msf{PRHF}}$ denote the two stateful
interfaces maintained by $\msf{TwoQROM.SimRO}$ using
the joint state $\msf{st}$ returned by
$\msf{TwoQROM.SimProof}$. We obtain
\begin{align*}
p_\msf{sim}
&:=
\Pr\left[
    \mathcal{A}^{\ket{\mathcal{C}_\msf{URS}},
                  \ket{\mathcal{C}_\msf{PRHF}}}
    (x,\pi^{*})=1
    :
    (\msf{st},\pi^{*})
    \gets\msf{TwoQROM.SimProof}(x)
\right]
\\
&=
\Pr\left[
    \mathcal{B}^{\ket{O_\msf{PRHF}}}
    (\msf{urs}^{*},x,\pi^{*})=1
    :
    (\msf{urs}^{*},\pi^{*})
    \gets
    \msf{URS.Sim}^{\ket{O_\msf{PRHF}}}(x)
\right].
\end{align*}

\paragraph{Reduction and efficiency.}
The reduction $\mathcal{B}$ makes no additional queries
to $O_\msf{PRHF}$ beyond those made by $\mathcal{A}$.
By \Cref{ComO}, simulating the URS oracle adds only
polynomial overhead in $n$, the query count, and the
size of $\msf{HD}_{\msf{urs}}$.
Since $\mathcal{A}$ is polynomial-query-bounded,
$\mathcal{B}$ is also polynomial-query-bounded.

Applying the zero-knowledge guarantee
of $\Pi_\msf{URS}$ to $\mathcal{B}$ and using the two
identities above yields
\[
    \bigl|p_\msf{real}-p_\msf{sim}\bigr|
    \leq \operatorname{negl}(n).
\]

Finally, $\msf{URS.Sim}$ is QPT, and a QPT distinguisher
makes only polynomially many additional oracle queries.
The PRHF compressed oracle handles their combined query
sequence while retaining its state throughout.
The URS compressed oracle starts with $m$ prescribed
entries and handles polynomially many subsequent queries.
\Cref{ComO} therefore gives polynomial time and space for
both oracle simulations.

Thus, the proof simulator is QPT, the stateful oracle
simulator runs in polynomial time throughout any QPT
interaction, and the transformation introduces no
additional distinguishing error.
\end{proof}

Together,
\Cref{lem:qrom-completeness,lem:qrom-soundness,lem:qrom-zero-knowledge}
prove \Cref{lem:qrom-eliminate-urs}.

\end{proof}

\subsection{Combining the Two Oracles}

\label{cor:qrom-nizk}
There exists an unbounded-prover NIZK proof system for
NP in the QROM with perfect completeness, adaptive
statistical soundness, and zero knowledge against
quantum polynomial-time oracle-aided distinguishers.

Let $O\gets\Ora$ be a single random oracle. By
\Cref{lem:random-oracle-decomposition}, interpret $O$ as
providing mutually independent random functions
\[
    O_{n+1,n}:\bits^{n+1}\to\bits^n,
    \qquad n\in\Naturals.
\]
For each $n$ and every $z\in\bits^n$, define
\[
    O_{\msf{URS}}(z):=O_{n+1,n}(0\|z),
    \qquad
    O_{\msf{PRHF}}(z):=O_{n+1,n}(1\|z).
\]
These restrictions are independent random functions from
$\bits^n$ to $\bits^n$, since their input domains are disjoint. The resulting system
therefore inherits perfect completeness and adaptive
statistical soundness from $\Pi_\msf{TwoQROM}$.

The oracle-merging lemma
(\Cref{sec:merge-oracles}) gives an exact implementation
of each query to $\ket{O_{n+1,n}}$ using four queries to
the separate interfaces $\ket{O_\msf{URS}}$ and
$\ket{O_\msf{PRHF}}$, with polynomial-time overhead.
This implementation also applies when the prefix
register is in superposition or entangled with other
registers. Thus, any QPT distinguisher with access to
$\ket{O}$ induces a QPT distinguisher with separate
access to $\ket{O_\msf{URS}}$ and
$\ket{O_\msf{PRHF}}$.

\subsection{Final Construction}
\label{sec:QROM-final-construction}
Combining the above results and exploiting the exact construction in \Cref{sec:URS-NIZK-construction}, we state the final construction of a NIZK proof system for NP in the QROM.
\begin{AlgorithmBox}[label={RO-prover}]{The Prover $\msf{QRO.P}^{O}$}
$\msf{QRO.P}^{O}(x)$:
    \begin{enumerate}
        \item Let $n = |x|$.
        \item For $i\in [p_{\msf{hbs}}(n)]$, set $y_i = O_{n+1,n}(0||\langle i \rangle_n)$ \\
        where $\langle i \rangle_n$ is the $n$-bit binary representation of $i$ as we defined in \Cref{prelim}
        \item For $i \in [p_{\msf{hbs}}(n)]$,
          \begin{enumerate}
          \item[(a)] If $\exists z \in \{0, 1\}^n$ such that $O_{n+1,n}(1||z) = y_i$, sample $v_i \picks \{z \in \{0, 1\}^n : O_{n+1,n}(1||z) = y_i\}$, and set $r_i = 0$.
          \item[(b)] If $\nexists z \in \{0, 1\}^n$ such that $O_{n+1,n}(1||z) = y_i$, set $v_i = \perp$ and $r_i = 1$.
          \end{enumerate}
        \item $( \msf{ZHB.I}, \msf{ZHB.\pi} ) \leftarrow \msf{ZHB.P}_{e^{-1}}(r, x)$ where $r = r_1 \dots r_{p_{\msf{hbs}}(n)}$.
        \item Return $\pi := (\msf{ZHB.I}, \{v_i\}_{i\in \msf{ZHB.I}}, \msf{ZHB.\pi})$.
    \end{enumerate}
\end{AlgorithmBox}
\begin{AlgorithmBox}[label={RO-verifier}]{The Verifier $\msf{QRO.V}^{O}$}
    $\mathsf{QRO.V}^{O}(x, \pi)$:
    \begin{enumerate}
        \item Let $n = |x|$.
        \item For $i\in [p_{\msf{hbs}}(n)]$, set $y_i = O_{n+1,n}(0||\langle i \rangle_n)$
        \item Parse $\pi$ as $(\mathsf{ZHB.I}, \{v_i\}_{i\in \mathsf{ZHB.I}}, \mathsf{ZHB.\pi})$.
        \item For $i \in \mathsf{ZHB.I}$,
          \begin{itemize}
          \item If $v_i = \perp$, set $\tilde{r}_i = 1$.
          \item If $v_i \neq \perp$, and $O_{n+1,n}(1||v_i) = y_i$, set $\tilde{r}_i = 0$. 
          \item If $v_i \neq \perp$, and $O_{n+1,n}(1||v_i) \neq y_i$, reject.
          \end{itemize}
        \item Return $\mathsf{ZHB.V}_{e^{-1}}(x, \mathsf{ZHB.I}, \{\tilde{r}_i\}_{i\in \mathsf{ZHB.I}}, \mathsf{ZHB.\pi})$ 
    \end{enumerate}
\end{AlgorithmBox}
\begin{AlgorithmBox}[label={RO-ProofSim}]{Proof Simulator $\mathsf{QRO.SimProof}$}
    $\mathsf{QRO.SimProof}(x)$:
    \begin{enumerate}
        \item Let $n = |x|$.
    \item $(\mathsf{ZHB.I}, \{ r_i \}_{i \in \mathsf{ZHB.I}}, \mathsf{ZHB.\pi}) \leftarrow \mathsf{ZHB.Sim}_{\frac{1}{e}}(x)$.
    \item Create history database $\msf{HD}=\{\}$.
    \item For $i \in [p_{\mathsf{hbs}}(n)]$ 
    \begin{enumerate}
        \item[(a)] Sample $v_i\picks\{0,1\}^n$.

        \item[(b)] If $(1\|v_i,y)\in\msf{HD}$ for some $y$,
        set $y_i=y$.
        Otherwise, sample $y_i\picks\{0,1\}^n$.
        \item[(c)] Update history database $\msf{HD} = \msf{HD} \cup \{(0||\langle i \rangle_n,y_i),(1||v_i,y_i)\}$
    \end{enumerate}
    \item For $i \in \mathsf{ZHB.I}$,
      \begin{enumerate}
        \item[(a)] If $r_i = 1$, set $v'_i = \perp$
        \item[(b)] If $r_i = 0$, set $v'_i = v_i$.
      \end{enumerate}
    \item Set $ \pi = (\mathsf{ZHB.I}, \{ v'_i \}_{i \in \mathsf{ZHB.I}}, \mathsf{ZHB.\pi})$.
    \item Return $(\msf{st}, \pi)$, where $\msf{st}$ contains the current state of the compressed oracle $\operatorname{Com}(\msf{HD})$. 
    \end{enumerate}
\end{AlgorithmBox}
\begin{AlgorithmBox}[label={RO-OraSim}]{Oracle Simulator $\mathsf{QRO.SimRO}$}
    $\msf{QRO.SimRO}(\msf{st})$:
    \begin{enumerate}
        \item Answer each quantum query using the
        compressed partially fixed oracle
        with regard to the current state $\msf{st}$, retaining
        its updated state for subsequent queries.
    \end{enumerate}
\end{AlgorithmBox}

\section{Oracle Separation between \texorpdfstring{$\NP\cap\coNP$}{NP cap coNP} and \texorpdfstring{$\BQP$}{BQP}}

We now prove the random-oracle separation informally stated in
\Cref{thm:informal:main}.

\begin{theorem}\label{thm:random-oracle-separation}
If $\UP \nsubseteq \BQP$, then
\[
    \Pr_{O\gets\Ora}\!\left[
        \NP^O \cap \coNP^O \nsubseteq \BQP^O
    \right] = 1.
\]
\end{theorem}

To prove \Cref{thm:random-oracle-separation}, we adapt the language
construction of \cite{STOC:GIKKLS23} using the QROM NIZK proof
system from \Cref{sec:QROM-final-construction}. We start with a
language $\Lang_0\in\UP\setminus\BQP$ and define an oracle-aided
language $\Lang^O$ on triples $(x,i,\pi)$. The proof $\pi$
certifies that $x\in\Lang_0$, and membership in $\Lang^O$ asks
whether the $i$-th bit of the unique witness for $x$ is $1$.
Uniqueness and the statistical soundness of the proof system
give $\Lang^O\in\NP^O\cap\coNP^O$ for almost every oracle
(\Cref{lem:separation:membership}).

We then establish quantum hardness in
\Cref{thm:quantum-hardness-of-oracle-aided-lang}. If a fixed
quantum oracle machine decided $\Lang^O$ on at least a
$2/3$ fraction of oracles, we could use it together with the
NIZK simulator to recover the unique witness bit by bit
without access to $O$. Verifying the recovered witness would
give a $\BQP$ decider for $\Lang_0$. Finally, we apply the
Bennett--Gill argument in \Cref{BG81-lemma} to conclude that,
with probability~$1$ over $O$, no quantum polynomial-time
oracle machine decides $\Lang^O$.

\subsection{Technical Tool}\label{sec:separation:tools}

We use a quantum formulation of the Bennett--Gill argument
\cite[Lemma~1]{BG81} to pass from a bound for each fixed quantum
oracle machine to an almost-sure hardness result. Under the
finite-dependence and decidability conditions below, a uniform gap
from probability~$1$ implies that each fixed machine decides the
language only on a measure-zero set of oracles. Since there are
countably many such machines, their success sets together still
have measure zero.

This lemma serves as a key technical tool in
\Cref{sec:separation:non-membership}, where we use it to show
that $\Lang^O\notin\BQP^O$ for almost every oracle $O$.

\begin{lemma}[{\cite[Lemma 1]{BG81}}, Refurnished] \label{BG81-lemma}
Let $\Lang^{(\cdot)}$ be an oracle-dependent language such that each bit of the oracle affects the membership of only finitely many strings in $\Lang^{\Ora}$; that is, for every $j \in \mathbb{N}$ there exists a finite set $F_j$ such that for any two oracles $\Ora$ and $\Ora'$ differing only in their $j$-th bit, and for any $(x,i,\pi) \notin F_j$, we have
$\Lang^{\Ora}(x,i,\pi) = \Lang^{\Ora'}(x,i,\pi)$, where \[
\Lang^{\Ora}(x,i,\pi) \coloneqq
\begin{cases}
   1& (x,i,\pi)\in\Lang^{\Ora}\\
   0&\text{otherwise}
\end{cases}.
\] 
Moreover, assume that $\Lang^{\Ora}$ is decidable by an oracle Turing machine that halts on all inputs for every oracle $\Ora$.

    Let $\mathcal{D}^{(\cdot)} = \{D_1^{(\cdot)}, D_2^{(\cdot)}, \ldots\}$ be the set of all possible oracle-aided polynomial-time quantum Turing Machines. If 
    there exists a constant $\epsilon \in (0,1]$ such that for all $k \in \mathbb{N}$, 
        \begin{align*}
    \Pr_{\Ora}\left[\forall (x,i,\pi)\in\{0,1\}^*, \Pr[D_k^{\QOra}(x,i,\pi) = \Lang^{\Ora}(x,i,\pi)]\geq \frac{2}{3}\right] < 1-\epsilon,
    \end{align*}
    \if0
    \begin{align*}
    \Pr_{\Ora}[(\forall (x,i,\pi) \in \Lang^{\Ora}, (\Pr[D_k^{\QOra}(x,i,\pi) = 1]\geq \frac{2}{3})) \land 
    (\forall (x,i,\pi) \notin \Lang^{\Ora}, \Pr[D_k^{\QOra}(x,i,\pi) = 0] \geq \frac{2}{3})] < 1-\epsilon
    \end{align*}
    \fi
then
\begin{equation}\label{eq:neg-BQP}
    \Pr_{\Ora}\left[\exists k \in \mathbb{N} \text{ such that } \forall (x,i,\pi)\in\{0,1\}^*, \Pr[D_k^{\QOra}(x,i,\pi) = \Lang^{\Ora}(x,i,\pi)]\geq \frac{2}{3}\right] = 0.
\end{equation}

\if0
\[
\Pr_{\Ora}\left[\exists k \in \mathbb{N} \text{ such that } \left(\forall (x,i,\pi) \in \Lang^{\Ora}, D_k^{\Ora}(x,i,\pi) = 1\right) \land \left(\forall (x,i,\pi) \notin \Lang^{\Ora}, D_k^{\Ora}(x,i,\pi) = 0\right)\right] = 0.
\]
\fi
\end{lemma}
\begin{proof}  
Since a countable union of events of probability zero still has probability zero, it suffices to prove that, for every $k\in\mathbb{N}$,
\[
\Pr_{\Ora}\left[\forall (x,i,\pi)\in\{0,1\}^*,\ 
\Pr[D_k^{\QOra}(x,i,\pi)=\Lang^{\Ora}(x,i,\pi)]\ge \frac{2}{3}\right]=0 .
\]
Below, we fix $k$.

For each $n\in\mathbb{N}$, let $S_n$ be the set of all oracles $\Ora$ such that
\[
\forall (x,i,\pi)\in\{0,1\}^{\le n},\ 
\Pr[D_k^{\QOra}(x,i,\pi)=\Lang^{\Ora}(x,i,\pi)]\ge \frac{2}{3},
\]
where $\{0,1\}^{\le n}$ denotes the set of all bit strings of length at most $n$.

Then, it suffices to prove that
\[
\Pr_\Ora\left[\Ora\in \bigcap_{n\in\mathbb{N}} S_n\right]=0.
\]

To this end, we establish the following claim.

\begin{MyClaim}\label[MyClaim]{claim:BG81:intermediate}
For any $m\in\mathbb{N}$, there exists $m'>m$ such that
\[
\Pr_\Ora[\Ora\in S_{m'} \mid \Ora\in S_m] < 1-\epsilon .
\]
\end{MyClaim}

We first complete the proof of \Cref{BG81-lemma} assuming \Cref{claim:BG81:intermediate}, and then proceed to prove \Cref{claim:BG81:intermediate} itself.

By the claim, we may choose an increasing sequence
$m_0<m_1<m_2<\cdots$ 
such that
\[
\Pr_\Ora[\Ora\in S_{m_j}\mid \Ora\in S_{m_{j-1}}] < 1-\epsilon
\quad\text{for all } j\in\mathbb{N}.
\]
Since $\Set{S_n}_{n}$ is monotonically decreasing, we have  
\begin{align*}
    \Pr_\Ora[\Ora\in S_{m_j}] 
    & = \Pr_\Ora[\Ora\in S_{m_j} \wedge \Ora\in S_{m_{j-1}}] \\
    & = \Pr_\Ora[\Ora\in S_{m_j} ~|~ \Ora\in S_{m_{j-1}}]  \Pr_\Ora[\Ora\in S_{m_{j-1}}] \\ 
    & < 
     (1-\epsilon) \Pr_\Ora[\Ora\in S_{m_{j-1}}].
\end{align*}
By induction, we have 
\[
\Pr_\Ora[\Ora\in S_{m_j}]
< (1-\epsilon)^j \Pr_\Ora[\Ora\in S_{m_0}]
\le (1-\epsilon)^j .
\]
Since $\Set{S_n}_{n}$ is monotonically decreasing, the following holds as desired
\[
\Pr_\Ora\left[\Ora\in \bigcap_{n\in\mathbb{N}} S_n\right]
= \lim_{n\rightarrow \infty} \Pr_\Ora[\Ora\in S_{n}]
\le \lim_{j\rightarrow \infty }(1-\epsilon)^j=0
\]

This finishes the proof of \Cref{BG81-lemma}, modulo the proof of \Cref{claim:BG81:intermediate}, which we provide below.

\subpara{Proof of \Cref{claim:BG81:intermediate}:} Let $T_k(m)$ denote the maximum running time of $D_k$ on inputs of length at most $m$.
Then, on inputs of length at most $m$, the machine $D_k$ can make oracle queries of length at most $T_k(m)$, and hence can access only the first $2^{T_k(m)}$ bits of $\Ora$.  
Also, let $M$ be the oracle Turing machine guaranteed by
the assumption of the lemma, which decides $L^{\cal O}$
and halts on every input for every oracle $\cal O$.
For every $m$, there exists a finite bound $U(m)$
such that $M^{\cal O}$ halts within $U(m)$ steps on
every input of length at most $m$, uniformly over
all oracles $\cal O$.
Indeed, this follows from König's lemma, since otherwise
there would exist an oracle $\cal O$ and an input on
which $M^{\cal O}$ does not halt.

Therefore, whether $\mathcal{O} \in S_m$ depends only on
the first $2^{R_k(m)}$ bits of $\cal O$, where
$R_k(m)\coloneqq \max\{T_k(m),U(m)\}$.

Consequently, to prove the claim it suffices to show that there exists $m'$ such that, for every $s\in\{0,1\}^{2^{R_k(m)}}$, 
\[
\Pr_\Ora[\Ora\in S_{m'} \mid \Ora \text{ has prefix } s] < 1-\epsilon,
\]
or equivalently,
\[
\Pr_\Ora[s*\Ora \in S_{m'}] < 1-\epsilon,
\]
where $s*\Ora$ denotes the oracle obtained by replacing the first $2^{R_k(m)}$ bits of $\Ora$ with $s$.

Since there are only finitely many such strings $s$, and since the sets $S_n$ are monotonically decreasing in $n$, we may interchange the order of quantifiers over $s$ and $m'$.\footnote{Suppose that for each $s$ there exists $m'_s$ such that $\Pr_\Ora[s*\Ora\in S_{m'_s}]<1-\epsilon$. Let $m'\coloneqq \max_s m'_s$. Then $S_{m'}\subseteq S_{m'_s}$ for all $s$, and hence
\[
\Pr_\Ora[s*\Ora\in S_{m'}]
\le \Pr_\Ora[s*\Ora\in S_{m'_s}]
< 1-\epsilon .
\]}
Namely, it suffices to show that for every $s\in\{0,1\}^{2^{R_k(m)}}$ there exists $m'$ such that
\[
\Pr_\Ora[s*\Ora\in S_{m'}] < 1-\epsilon .
\]
Below, we fix $s$, and prove that the above holds for some $m'$. 

Since $2^{R_k(m)}$ is a finite constant, by the premise of \Cref{BG81-lemma} there exists a finite set $B_m\subseteq\{0,1\}^*$ such that for all oracles $\Ora$,
\[
\Lang^\Ora(x,i,\pi)=\Lang^{s*\Ora}(x,i,\pi)
\quad\text{for all } (x,i,\pi)\notin B_m .
\]
We define a modified oracle-aided quantum Turing machine $\tilde{D}_k^{\QOra}$ as follows:
\begin{itemize}
\item
$\tilde{D}_k^{\QOra}$:
On input $(x,i,\pi)$, if $(x,i,\pi)\in B_m$, output $\Lang^\Ora(x,i,\pi)$; otherwise, output $D_k^{\ket{s*\Ora}}(x,i,\pi)$.
\end{itemize}
Note that $\widetilde{D}_k$ is also a QPT machine,
since $s$ and $B_m$ are finite and, as shown above,
$\Lang^{\cal O}$ is computable in oracle-independent
constant time on $B_m$.  Namely, $\tilde{D}^{(\cdot)}_k$ belongs to the family $\mcal{D}^{(\cdot)}$ defined in the statement of \Cref{BG81-lemma}.

Thus, it follows from the premise of \Cref{BG81-lemma} that
\[
\Pr_{\Ora}\left[\forall (x,i,\pi)\in\{0,1\}^*,\ 
\Pr[\tilde{D}_k^{\QOra}(x,i,\pi)=\Lang^{\Ora}(x,i,\pi)]\ge \frac{2}{3}\right]
< 1-\epsilon .
\]
By construction, for all
$\Ora$ and all $(x,i,\pi)\in B_m$, it holds that
\[
\tilde{D}_k^{\QOra}(x,i,\pi)=\Lang^{\Ora}(x,i,\pi).
\]
Moreover, for for all
$\Ora$ and all $(x,i,\pi)\notin B_m$,
\[
\tilde{D}_k^{\QOra}(x,i,\pi)=D_k^{\ket{s*\Ora}}(x,i,\pi)
\quad\text{and}\quad
\Lang^{\Ora}(x,i,\pi)=\Lang^{s*\Ora}(x,i,\pi) .
\]
Therefore,
\[
\Pr_{\Ora}\left[\forall (x,i,\pi)\notin B_m,\ 
\Pr[D_k^{\ket{s*\Ora}}(x,i,\pi)=\Lang^{s*\Ora}(x,i,\pi)]\ge \frac{2}{3}\right]
< 1-\epsilon .
\]
This immediately implies
\[
\Pr_{\Ora}\left[\forall (x,i,\pi)\in\{0,1\}^*,\ 
\Pr[D_k^{\ket{s*\Ora}}(x,i,\pi)=\Lang^{s*\Ora}(x,i,\pi)]\ge \frac{2}{3}\right]
< 1-\epsilon ,
\]
or equivalently (by the definition of $S_n$),
\[
\Pr_{\Ora}\left[s*\Ora\in \bigcap_{n\in\mathbb{N}} S_n\right] < 1-\epsilon .
\]

Since the sets $S_n$ are monotonically decreasing,
\[
\lim_{n\to\infty}\Pr_{\Ora}[s*\Ora\in S_n]
= \Pr_{\Ora}\left[s*\Ora\in \bigcap_{n\in\mathbb{N}} S_n\right]
< 1-\epsilon .
\]
Hence, there exists $m'$ such that
\[
\Pr_{\Ora}[s*\Ora\in S_{m'}] < 1-\epsilon .
\]
This completes the proof of \Cref{claim:BG81:intermediate}.

\end{proof}

\subsection{Post-Quantum injOWF  Implies $\UP \nsubseteq \BQP$}\label{sec:separation:injOWF-UP-BQP}

For a better illustration of the oracle separation result between $\NP \cap \coNP$ and $\BQP$, we start by first establishing the connection between the existence of injective post-quantum one-way function and the separation between $\UP$ and $\BQP$, instead of directly starting with a vague language that lies in $\UP \setminus \BQP$.
It does not mean that the existence of injective post-quantum one-way function is necessary for the separation between $\NP \cap \coNP$ and $\BQP$, but it provides us a more clear starting point to construct such oracle separation result.
\begin{lemma}
    If there exists an injective post-quantum one-way function $f: \{0,1\}^{*} \to \{0,1\}^{*}$, then $\UP \not \subseteq \BQP$. 
\end{lemma}
    
\begin{proof}
    Consider the language built from the function $f$ as follows: 
\[
\Lang \coloneqq  \left\{
(y,i)\in\bits^{*}\times\mathbb{N} :
\exists x\in\bits^{*},\
f(x)=y \land x_i=1 \land i\leq |x| 
\right\}.
\]
Here, the index $i$ is encoded in binary.
    
    It is easy to see that $\Lang \in \NP$ since we can verify the membership of every $y$ in $\Lang$ with the corresponding $x$ as the witness.
    The uniqueness of the witness $x$ is guaranteed by the injectivity of $f$.
    Also, the $\BQP$ hardness of $\Lang$ follows from the post-quantum one-wayness of $f$, otherwise the pre-image can be inverted with bit-by-bit recovery alongside with the standard  amplification of $\BQP$.
    We can then conclude that $\Lang \in \UP \setminus \BQP$.
\end{proof}

\subsection{The Oracle-Aided Language \texorpdfstring{$\Lang^O$}{L^O}} \label{sec:separation:oracle-aided-lang}

Following \cite{STOC:GIKKLS23}, we construct an oracle-aided
language whose instances ask for a bit of a unique witness.
Fix a base language $\Lang_0\in\UP\setminus\BQP$ (Existence is guaranteed by the assumption $\UP \nsubseteq \BQP$) and a
witness relation $\rel_{\Lang_0}$. Let
$\Pi_{\msf{NIZK}}=(\mathcal{P}^{(\cdot)},\mathcal{V}^{(\cdot)})$
be the QROM NIZK proof system from
\Cref{sec:QROM-final-construction} instantiated for $\Lang_0$.

We form an oracle-aided language by attaching a proof of
$x\in\Lang_0$ to each query about a bit of its unique witness.
For a fixed oracle $O$, define
\begin{equation}
\label{eq:separation:oracle-aided-language}
    \Lang^{\Ora} \coloneqq
    \left\{(x,i,\pi) \,\middle|\,
        \bigl(\exists w:\ (x,w)\in\rel_{\Lang_0}\land w_i=1\bigr)
        \land \Pi_{\msf{NIZK}}.\mathcal{V}^{\Ora}(x,\pi)=1
    \right\}.
\end{equation}

We proceed to show that $\Lang^O$ belongs to $\NP^O\cap\coNP^O$ for
almost every oracle in \Cref{sec:separation:membership}.
We establish its quantum hardness in
\Cref{sec:separation:non-membership}.

\subsection{Membership in \texorpdfstring{$\NP^O\cap\coNP^O$}{NP^O cap coNP^O}}\label{sec:separation:membership}

The following lemma follows directly from
\cite[Lemma~8.4]{STOC:GIKKLS23}. As in their construction,
our QROM proof system has a classical prover and verifier
and satisfies perfect completeness and adaptive statistical soundness. We only use its
zero-knowledge guarantee against quantum adversaries in
\Cref{sec:separation:non-membership}.

\begin{lemma}[{\cite[Lemma~8.4]{STOC:GIKKLS23}}]\label{lem:separation:membership}
If $\UP\nsubseteq\BQP$, then with probability~$1$ over the choice of a random oracle
$O\gets\Ora$, the language $\Lang^O$ defined in \Cref{eq:separation:oracle-aided-language} belongs to
$\NP^O\cap\coNP^O$. Equivalently,
\[
    \Pr_{O\gets\Ora}\!\left[
        \Lang^O\in\NP^O\cap\coNP^O
    \right]=1.
\]
\end{lemma}

\subsection{Non-Membership in \texorpdfstring{$\BQP^O$}{BQP^O}}\label{sec:separation:non-membership}

We now show that the oracle-aided language defined in
\Cref{eq:separation:oracle-aided-language} remains hard for
quantum polynomial-time algorithms with superposition access
to the oracle. The proof uses the QROM NIZK simulator to turn
a machine that decides $\Lang^O$ on every input for at least
a $2/3$ fraction of oracles into an oracle-free $\BQP$ decider
for $\Lang_0$. We then apply \Cref{BG81-lemma} to turn the
resulting bound for each fixed machine into an almost-sure
hardness result.

\begin{lemma}
\label{thm:quantum-hardness-of-oracle-aided-lang}
If $\UP\nsubseteq\BQP$, then with probability~$1$ over the
choice of a random oracle $O\gets\Ora$, the language $\Lang^O$ defined in \Cref{eq:separation:oracle-aided-language}
lies outside $\BQP^O$. Equivalently,
\[
    \Pr_{O\gets\Ora}\!\left[\Lang^O\notin\BQP^O\right]=1.
\]
\end{lemma}
\begin{proof}
    Fix any oracle-aided QPT machine $D^{(\cdot)}$, and define the set of good oracles on which $D$ decides $\Lang^{\Ora}$ on all inputs as
    \[
    G_D\coloneqq \left\{\Ora:\forall (x,i,\pi)\in\{0,1\}^*,
    \Pr[D^{\QOra}(x,i,\pi)=\Lang^{\Ora}(x,i,\pi)]\geq \frac{2}{3}\right\}.
    \]
    To  apply \Cref{BG81-lemma} for the separation results, we first have to show that $\Pr_{\Ora}[\Ora\in G_D]<2/3$, note that $\frac{2}{3}$ here can be replaced by any real number $\epsilon$ where $\frac{1}{2}<\epsilon<1$ 
    
    Assume for contradiction that $\Pr_{\Ora}[\Ora\in G_D]\geq 2/3$
    We then build a quantum decider in $\BQP$ for the base language $\Lang_0$ with the help of the zero-knowledge simulator constructed in the previous section.
    
    In particular, the real oracle-aided decider for the $i$-th bit is defined as follows:\\
    \begin{AlgorithmBox}[label={Real-decider}]{Real-decider for $i$-th bit: $D_{(i)}^{\QOra}$}
         $D_{(i)}^{\QOra}(x,\pi)$
        \begin{enumerate}
            \item Prepare an empty multiset $\msf{Bits}=\{\}$
            \item for $j \in [|x|]$
            \begin{enumerate}
                \item Invoke $D^{\QOra}(x,i,\pi)$, add the output to $\msf{Bits}$
            \end{enumerate}
            \item Return the majority element in $\msf{Bits}$.
        \end{enumerate}
    \end{AlgorithmBox}

    \begin{lemma}\label{real-decider-correctness}
        For all $(x,w)\in R_{\Lang_0}$ and all
        $i\in[p(|x|)]$, we have
        \[
            \Pr\left[
                D_{(i)}^{\QOra}(x,\pi)=w_i
                \;:\;
                \begin{array}{l}
                    O\gets\Ora,\\
                    \pi\gets P^O(x)
                \end{array}
            \right]
            \geq \frac{2}{3}-\negl(|x|).
        \]
        Here, $P$ is the honest NIZK prover, and the decider
        uses the same sampled oracle $O$. The probability
        also includes the internal randomness and measurement
        outcomes of the algorithms.
        \end{lemma}

    \begin{proof}
        By perfect completeness, an honestly generated proof
$\pi\gets P^O(x)$ satisfies $V^O(x,\pi)=1$
with probability $1$. Hence, conditioned on
$O\in G_D$, each invocation of
$D^{\QOra}(x,i,\pi)$ outputs $w_i$ with probability
at least $2/3$.

        Let $X$ be the random variable of the count of $w_i$ in $\msf{Bits}$ and $X_j=1$ if $D^{\QOra}(x,i,\pi)=w_i$ in the $j$-th iteration,  otherwise $X_j=0$.
        Notice the event that the output of $D_{(i)}^{\QOra}(x,\pi)$ is $w_i$ is equivalent to the event that $X$ is larger than $\frac{|\msf{Bits}|}{2}$, and 
        the bad event probability is bounded by the Chernoff bound as follows:
        \begin{align*}
            \Pr[D_{(i)}^{\QOra}(x,\pi)\neq w_i]
            =&
            \Pr[X\leq\frac{|\msf{Bits}|}{2}]\\
            =& \Pr\left[\sum_{j=1}^{|x|}X_j\leq\frac{|x|}{2}\right]\\
            =& \Pr\left[\sum_{j=1}^{|x|}X_j\leq (1-\frac{1}{4})\frac{2|x|}{3}\right]\\
            \leq& \Pr\left[\sum_{j=1}^{|x|}X_j\leq (1-\frac{1}{4})\E[X]\right]\\
            \leq& \exp\left(-\E[X]\cdot(\frac{1}{4})^2\cdot\frac{1}{2}\right) \tag{Chernoff Bound}\\
            \leq& \exp\left(-\frac{2|x|}{3}\cdot \frac{1}{32}\right)\\
            =& \exp\left(-\frac{|x|}{48}\right)\\
            =& \negl(|x|) 
        \end{align*}
         Averaging over the choice of $\Ora$ , we obtain
        \begin{align*}
        \Pr[D_{(i)}^{\QOra}(x,\pi)=w_i]
        &\geq \Pr_{\Ora}[\Ora\in G_D]-\negl(|x|)\\
        &\geq \frac{2}{3}-\negl(|x|),
        \end{align*}
        as desired.
    \end{proof}
    The quantum decider $\tilde{D}$ without actual access to the oracle is defined as follows:\\
    \begin{AlgorithmBox}[label={Simulated-decider}]{Simulated-decider for $i$-th bit: $\tilde{D}(x,i)$}
          $\tilde{D}(x,i)$
        \begin{enumerate}
            \item Invoke $\msf{SimProof}(x)$ to get $(st,\pi)$
            \item Return $D_{(i)}^{\ket{\msf{SimRO}(st,\cdot)}}(x,\pi)$.
        \end{enumerate}
    \end{AlgorithmBox}
   \begin{lemma} \label{simulator-correctness}
        For all $(x,w)$ in the relation $R_{\Lang_0}$ and all $i \in [p(|x|)]$,
        \[
        \Pr[\tilde{D}(x,i)=w_i]
        \geq \frac{2}{3}-\negl(|x|).
        \]
    \end{lemma}
    \begin{proof}
        Apply the zero-knowledge property of the NIZK to $D_{(i)}^{\QOra}$ which is polynomial-bound. The probability that this bit equals $w_i$ differs by at most a negligible amount. Combining this with \Cref{real-decider-correctness} proves the lemma.
    \end{proof}
    By \Cref{simulator-correctness}, for some negligible function $\negl_{\msf{sim}}$, the success probability of $\tilde{D}(x,i)$ is at least $2/3-\negl_{\msf{sim}}(|x|)$. Since $\negl_{\msf{sim}}$ is negligible, there exists a constant $n_0$ such that, for every $n\geq n_0$, we have $\negl_{\msf{sim}}(n)\leq 1/15$. Therefore, for every $|x|\geq n_0$,
    \[
    \frac{2}{3}-\negl_{\msf{sim}}(|x|)\geq \frac{2}{3}-\frac{1}{15}=\frac{3}{5}.
    \]
    This bound is not required to hold when $|x|<n_0$. Instead, for these input lengths we can exhaustively search for the unique witness using the efficiently decidable relation $R_{\Lang_0}$ as $\Lang_0 \in \textbf{UP}$. Although this search takes exponential time as a function of $|x|$, its running time on the branch $|x|<n_0$ is bounded by $2^{p(n_0)}\cdot\poly(n_0)$, which is a fixed constant.
    
    We define the amplified quantum decider as follows:
    \begin{AlgorithmBox}[label={amplified-decider}]{Amplified decider for $i$-th bit: $\widehat{D}(x,i)$}
          $\widehat{D}(x,i)$
        \begin{enumerate}
            \item Let $n=|x|$.
            \item If $n<n_0$:
            \begin{enumerate}
                \item Enumerate every $w\in\{0,1\}^{\leq p(n)}$.
                \item If $(x,w)\in R_{\Lang_0}$, return $w_i$.
                \item If no such witness exists, return $0$.
            \end{enumerate}
            \item Prepare an empty multiset $\msf{Bits}=\{\}$.
            \item For $j\in[n]$:
            \begin{enumerate}
                \item Invoke a fresh instance of $\tilde{D}(x,i)$, and add its output to $\msf{Bits}$.
            \end{enumerate}
            \item Return the majority element in $\msf{Bits}$.
        \end{enumerate}
    \end{AlgorithmBox}
        We then show that the amplified decider recovers each witness bit with overwhelming probability.
    \begin{lemma}\label{amplified-decider-correctness}
        For all $(x,w)$ in the relation $R_{\Lang_0}$ and all $i\in[p(|x|)]$,
        \[
        \Pr[\widehat{D}(x,i)=w_i]\geq 1-\negl(|x|).
        \]
    \end{lemma}
    \begin{proof}
        For the case that $|x|<n_0$, 
        it is clearly that $\widehat{D}(x,i)$ decide the witness bit $w_i$ perfectly, which means
        \[\Pr[\widehat{D}(x,i)=w_i]=1  \]
        For the case that $|x|\geq n_0$,
        
        Let $X_j$ be the indicator random variable for the event that the $j$-th invocation of $\tilde{D}(x,i)$ outputs $w_i$, and let $X=\sum_{j=1}^{|x|}X_j$. Every invocation uses fresh randomness for the simulator and the quantum machine. By \Cref{simulator-correctness}, for all sufficiently large $|x|$ we have $\E[X]\geq 3|x|/5$. Therefore, by a Chernoff bound,
        \begin{align*}
        \Pr[\widehat{D}(x,i)\neq w_i]
        &=\Pr\left[X\leq\frac{|x|}{2}\right]\\
        &\leq\Pr\left[X\leq\left(1-\frac{1}{6}\right)\E[X]\right]\\
        &\leq\exp\left(-\frac{\E[X]}{72}\right)\\
        &\leq\exp\left(-\frac{|x|}{120}\right)\\
        &=\negl(|x|).
        \end{align*}
        which means
            \[\Pr[\widehat{D}(x,i)=w_i]\geq 1-\negl(|x|). \]
        
            This completes the proof of \Cref{amplified-decider-correctness}.
            
    \end{proof}
    We then give the final decision algorithm $D^*(x)$ for deciding the membership of the language $\Lang_0$ as follows:
     \begin{AlgorithmBox}[label={final-decider}]{Final witness-recovering decider: $D^*(x)$}
           $D^*(x)$
        \begin{enumerate}
            \item For $i \in [p(|x|)]$
                \begin{enumerate}
                   
                \item $w_i \leftarrow$ output of $\widehat{D}(x,i)$.
                \end{enumerate}
            \item If there exists a prefix $u$ of $w$ (possibly $u=w$)
      such that $(x,u)\in R_{\mathcal{L}_0}$, output $1$.
                    \\Else output $0$
        \end{enumerate}
    \end{AlgorithmBox}
   
    We claim that the above algorithm $D^*(x)$ is a $\BQP$ decider for $\Lang_0$.\\\\
    \subpara{For those $x\notin \Lang_0$}\\
        There does not exist any $w$ such that $(x,w)\in R_{\mathcal{L}_0}$ by the definition of $\textbf{UP}$, 
        thus the output of $D^*(x)$ is always $0$.\\
        Therefore, we conclude that $D^*(x)$ correctly decides membership in $\Lang_0$, which is
        \[\forall x \notin \Lang_0, \Pr[D^*(x) = 0] = 1\]
    \subpara{For those $x\in \Lang_0$}\\
        Notice that $D^*$ decides $x$ only if the corresponding uniquen witness $w$ satisfies $(x,w)\in R_{\mathcal{L}_0}$ is recovered.
        
        Formally, for any fixed $x \in L_{0}$ , we analyze the probability of the event that $D^{*}(x)$ outputs $1$: 
        \begin{align*}
            &\Pr\left[D^{*}(x) = 1\right]\\
            &\geq\Pr\left[\forall i \in [p(|x|)], \widehat{D}(x, i) = w_i\right]\\
            &= 1 - \Pr[ \exists i \in [p(|x|)], \widehat{D}(x, i) \neq w_i ]\\
            &\geq 1 - p(|x|) \Pr[ \widehat{D}(x, i) \neq w_i ] \tag{Union Bound}\\ 
            &\geq 1 - p(|x|) \cdot \negl(|x|) \tag{\Cref{amplified-decider-correctness}}\\
            &\geq 1 - \negl(|x|)
        \end{align*}
        for some negligible function $\negl(\cdot)$.
        
        Thus, we conclude that $D^{*}(x)$ correctly decides membership in $\Lang_0$.

        The above two cases imply that $D^{*}$ is a $\BQP$ decider for the language $L_{0}$, contradicting the assumption that $L_{0} \in \textbf{UP} \setminus \BQP$. Therefore, for every fixed oracle-aided QPT machine $D^{(\cdot)}$,
        \[
        \Pr_{\Ora}[\Ora\in G_D]<\frac{2}{3}=1-\frac{1}{3}.
        \]
         It remains to check that $\Lang^{\Ora}$ satisfies the two structural conditions of \Cref{BG81-lemma}. The language is decidable by an oracle Turing machine that always halts: on input $(x,i,\pi)$, the machine can enumerate the polynomially bounded witnesses for $\Lang_0$, check the relation, and run the NIZK verifier. Moreover, a fixed bit of the random oracle belongs to a fixed input-length component. The verifier on an instance $x$ accesses only this component, and the index $i$ and every well-formed proof $\pi$ have polynomially bounded length; Hence, changing one oracle bit affects only finitely many strings in $\Lang^{\Ora}$.

        We can now invoke \Cref{BG81-lemma} with $\epsilon=1/3$. It follows that, with probability $1$ over the choice of $\Ora$, no oracle-aided QPT machine decides $\Lang^{\Ora}$ on all inputs. Equivalently,
        \[
        \Pr_{\Ora}[\Lang^{\Ora}\notin\BQP^{\Ora}]=1.
        \]
        
\end{proof}

\para{Finishing the proof of \Cref{thm:random-oracle-separation}.}
Assume $\UP\nsubseteq\BQP$. By
\Cref{sec:separation:membership,sec:separation:non-membership},
the events $\Lang^O\in\NP^O\cap\coNP^O$ and
$\Lang^O\notin\BQP^O$ each have probability~$1$ over
$O\gets\Ora$.
It follows that
\[
    \Pr_{O\gets\Ora}\!\left[
        \NP^O\cap\coNP^O\nsubseteq\BQP^O
    \right]=1.
\]

\section{Black-Box Impossibility of Constructing Quantum-Hard
\texorpdfstring{$\NP\cap\coNP$}{NP intersect coNP} Problems}

We next establish our main impossibility result.
\begin{theorem}\label{thm:main:impossibility}
    There is no fully black-box quantum reduction of any
    (even worst-case) quantum-hard problem in $\NP \cap \coNP$
    to post-quantum $\iO$ and post-quantum inj-OWFs.
\end{theorem}

To prove \Cref{thm:main:impossibility}, we follow the
\emph{two-oracle paradigm} introduced by Hsiao and Reyzin
\cite{C:HsiRey04}. This paradigm rules out fully black-box
constructions of a primitive $Y$ from a primitive $X$ by
exhibiting two oracles: an oracle $F$ that realizes $X$ and
an auxiliary oracle $G$ that breaks the proposed construction of $Y$,
while $X$ remains secure even given access to both $F$ and $G$.
A fully black-box construction and its security reduction would
yield a contradiction in this oracle world.
We use the quantum formulation of this framework due to
Hosoyamada and Yamakawa \cite{AC:HosYam20}, in which quantum
algorithms may have superposition access to both oracles.

Our auxiliary oracle is a universal decision oracle for complementary
verifier pairs. Importantly, it is required to decide only those pairs
that remain complementary relative to every implementation of the
source primitive under consideration. This is precisely the structure
guaranteed by a fully black-box construction. 

We start by specifying necessary preliminaries (\Cref{sec:impossibility:preliminaries}). We then show a warm-up case of separating injOWFs only, without considering $\iO$ (\Cref{sec:quantize-injOWF}). In this
warm-up, the base oracle is a family of random injective functions.
We show that it remains one-way against quantum adversaries even in
the presence of the quantum-accessible decision oracle. We then extend
the oracle construction to incorporate $\iO$ (\Cref{sec:quantize-iO-injOWF}).

\subsection{Black-Box Framework and Technical Tools}
\label{sec:impossibility:preliminaries}
We first specify the oracle interfaces and the black-box model.
We then define the decision oracle and establish the structural
property that makes it possible to analyze its effect on source
security. Finally, we give a quantum hybrid lemma that converts this
structural property into a bound against superposition queries.

\subsubsection{ Oracle Access and the Black-Box Model}

Throughout, a sampled classical oracle is a deterministic function.
Quantum access to a classical function $h$ means access to its
unitary evaluation oracle
\[
    O_h\ket{k,v}=\ket{k,v\oplus h(k)}.
\]

When an oracle implements a randomized classical procedure, its
randomness is supplied as an explicit input. In particular, an
obfuscation oracle has the form $\iO(C,r)$, where $r$ is the
obfuscator's random tape. Any randomness used to sample an oracle
table is sampled once at the beginning of the experiment; the table
is then held fixed throughout all queries.

We use post-quantum $\iO$ for classical
oracle-aided circuits. The adversary may make quantum oracle queries
both before and after receiving the obfuscation.

\begin{definition}[Post-Quantum Indistinguishability Obfuscation]
\label{PQ-iO}
Let
\(
    \mathcal C=\{\mathcal C_n\}_{n\in\mathbb N}
\)
be a class of classical oracle-aided circuits, where every
$C\in\mathcal C_n$ has size $n$. Let
\(
    \Gamma=\{\Gamma_n\}_{n\in\mathbb N}
\)
be an oracle distribution ensemble. In the experiments below, the
oracle tuple is sampled from $\Gamma_n$ and is held fixed throughout
the experiment.

A PPT algorithm $\iObf$ is a post-quantum indistinguishability
obfuscator for $\mathcal C$ relative to $\Gamma$ if the following hold.
\begin{enumerate}
    \item \textbf{Functionality.}
    For every $n\in\mathbb N$ and every $C\in\mathcal C_n$,
    \[
        \Pr_{\Gamma,\iObf}
        \left[
            \hat C^\Gamma\equiv C^\Gamma
            \;:\;
            \hat C\leftarrow \iObf^\Gamma(1^n,C)
        \right]
        =1.
    \]

    \item \textbf{Post-Quantum Indistinguishability.}
    For every non-uniform QPT oracle-aided adversary
    $\mathcal A=(\mathcal A_1,\mathcal A_2)$ that produces
    $C_0,C_1\in\mathcal C_n$ satisfying
    $C_0^\Gamma\equiv C_1^\Gamma$ with probability one, 
    \[
        \operatorname{Adv}^{\mathsf{iO}}_
        {\Gamma,\iObf,\mathcal C,\mathcal A}(n)
        =
        \left|
            \Pr\left[
                \mathsf{Exp}^{\mathsf{iO}}_
                {\Gamma,\iObf,\mathcal C,\mathcal A}(n)=1
            \right]
            -\frac12
        \right| \leq \negl(n),
    \]
    where $\mathsf{Exp}^{\mathsf{iO}}_
                {\Gamma,\iObf,\mathcal C,\mathcal A}(n)$  is the experiment defined in \Cref{iO-Game}.
    
\end{enumerate}
\end{definition}
\begin{ExperimentBox}[label={iO-Game}]
    {Security for Post-Quantum $\iO$}
    {
    $\mathsf{Exp}^{\mathsf{iO}}_
    {\Gamma,\iObf,\mathcal C,\mathcal A}(n):$
    \begin{enumerate}
        \item Sample the oracle tuple according to $\Gamma_n$.
        \item
        $(C_0,C_1,\mathsf{st})
        \leftarrow
        \mathcal A_1^{\ket{\Gamma}}(1^n)$. (The register $\mathsf{st}$ may contain quantum side information.)
        \item
        $b\xleftarrow{\$}\bits$.
        \item
        $\hat C\leftarrow \iObf^\Gamma(1^n,C_b)$.
        \item
        $b'\leftarrow
        \mathcal A_2^{\ket{\Gamma}}(\mathsf{st},\hat C)$.
        \item Output $1$ if $b'=b$, and output $0$ otherwise.
    \end{enumerate}
    }
    \end{ExperimentBox}

We next formalize the black-box constructions ruled out by our result.
The verifiers are classical, as required for an
$\NP\cap\coNP$ language, whereas the security reduction may be quantum.
The universal structure requirement below is essential: the same
verifier pair must remain complementary for every functionally correct
implementation of the source primitives.

\begin{definition}[Quantum Fully Black-Box Construction]
\label{def:quantum-fully-black-box}
A post-quantum fully black-box construction of a hard
$\NP\cap\coNP$ problem from injOWFs and $\iO$ for a class $\mathcal C$
of classical circuits with injOWF-gates consists of two classical
polynomial-time oracle-aided verifiers $(V_0,V_1)$ and a quantum
oracle-aided reduction $\mathcal R$ satisfying the following
conditions.
In the efficient formulation, $\mathcal R$ is polynomial-time.
For the quantitative formulation, we allow an otherwise inefficient
reduction and bound relevant oracle-query complexity explicitly with parameters $q_{\mathcal R}(n)$ and $\varepsilon_{\mathcal R}(n)$ below.

\begin{enumerate}
    \item \textbf{$\NP\cap\coNP$ Structure.}
    Let $\mathfrak S$ be the family of all oracle tuples
    $(f,\iO)$ satisfying the prescribed syntax and output-length
    bounds, such that $f$ is injective and $\iO$ is a
    deterministic function of the circuit and its random tape with
    \(
        \hat C^{\,f}\equiv C^f 
    \) 
    for every $C\in\mathcal C$, every random tape $r$, and
    \(
        \hat C^{(\cdot)}\coloneqq \iO(C,r).
    \)
    For every $(f,\iO)\in\mathfrak S$ and every input $z$,
    exactly one bit $b\in\bits$ satisfies
    \[
        \exists w:\;
        V_b^{f,\iO}(z,w)=1.
    \]
    The witnesses have polynomially bounded length.
    Thus, defining membership by the verifier $V_1$, the pair
    $(V_0,V_1)$ defines a language
    \[
        L^{f,\iO}
        \in
        \NP^{f,\iO}\cap\coNP^{f,\iO}.
    \]

    \item \textbf{Quantum Black-Box Security Proof.}
    There exist functions
    $q_{\mathcal R}(\cdot)\geq 1$ and
    $\varepsilon_{\mathcal R}(\cdot)>0$ such that the following holds. (In the efficient formulation, $q_{\mathcal R}(n)$ is  polynomial and $\varepsilon_{\mathcal R}(n)$ is inverse-polynomial.) 
    Let $\Gamma$ be any distribution over oracle tuples supported on
    $\mathfrak S$. For any  quantum oracle-aided adversary
    $\mathcal A$ that decides $L^{f,\iO}$ exactly in the
    worst case, namely, for every input $z$,
    \[
        \Pr_{(f,\iO)\leftarrow\Gamma}
        \left[
            \mathcal A^{\ket f,\ket{\iO}}(z)
            = b \quad \forall z \in \bit^n, b \in \bits~s.t.~ \exists w: V^{f, \iO}_b(z, w) = 1 
        \right]
        =1,
    \]
    the reduction breaks either $f$ or $\iO$.
    More precisely, for infinitely many $n\in\mathbb N$, either
    \[
        \Pr_{\substack{
            (f,\iO)\leftarrow\Gamma\\
            x\picks\bits^n
        }}
        \left[
            \mathcal R^{\ket{\mathcal A},\ket f,\ket{\iO}}
            (1^n,f(x))
            =x
        \right]
        \geq \varepsilon_{\mathcal R}(n),
    \]
    or
    \[
        \operatorname{Adv}^{\mathsf{iO}}_
        {\Gamma,\iO,\mathcal C,
        \mathcal R^{\ket{\mathcal A}}}(n)
        \geq \varepsilon_{\mathcal R}(n).
    \]
    In both experiments, $\mathcal R$ makes at most
    $q_{\mathcal R}(n)$ queries in total to its oracle interfaces.
    Moreover, for every computational-basis input $z$ appearing in a
    query to $\mathcal A$, every $(f,\iO)\in\mathfrak S$,
    every witness $w$, and each $b\in\bits$, the execution
    \(
        V_b^{f,\iO}(z,w)
    \)
    makes at most $q_{\mathcal R}(n)$ oracle queries.
\end{enumerate}

The oracle interface supplied for a quantum adversary is part of the
black-box model; the notation $\ket{\mathcal A}$ does not assert a
canonical coherent implementation of an arbitrary quantum channel.
For a deterministic classical adversary, we use its unitary
function-evaluation interface. This is the only adversary interface
needed for the decision-oracle attack below.
\end{definition}

\begin{remark}[Role of the auxiliary quantitative formulation] 
\Cref{def:quantum-fully-black-box} includes both an
efficient and a quantitative formulation. The efficient
formulation is the standard quantum fully black-box notion and is
the notion targeted by our main separation theorem.
The quantitative formulation serves only as an auxiliary
framework for proving this separation.

Our quantitative results (e.g., \Cref{thm:main:injOWF} and \Cref{thm:full-oracle-injOWF})  concern this auxiliary
formulation \emph{under additional restrictions} on the permitted
queries, including bounds on evaluation-query lengths and
on the resources of verifiers submitted to the decision
oracle. These results exclude otherwise inefficient
reductions with certain super-polynomial query bounds $q_{\mcal{R}}$
 satisfying those restrictions. They do \emph{not} rule out
the quantitative formulation in general, nor do they
establish lower bounds on bare oracle-query complexity
for unrestricted inefficient reductions.

These restricted quantitative results nevertheless suffice
for our main goal: For every fixed efficient construction
and reduction, we show that composition with our separating
adversary satisfies all the additional resource restrictions
for some polynomial bound $p(n)$. Since this polynomial
bound is eventually dominated by the superpolynomial bounds
covered by our quantitative results, those results rule out
the standard efficient formulation (e.g., see \Cref{cor:injOWF-black-box-separation,cor:iO-injOWF-black-box-separation}). The inefficient
formulation is therefore a tool for establishing the
efficient separation, rather than a stronger notion that
we claim to rule out in full generality.
\end{remark}

The corresponding definition for constructions from injOWFs alone
is obtained by omitting the obfuscation oracle and the alternative
of breaking iO.

\subsubsection{The Universal Decision Oracle}

We now define the auxiliary oracle that breaks a proposed
$\NP\cap\coNP$ construction. Its validity test depends only on the
oracle family $\mathfrak S$, rather than on the particular realized
oracle. This universal validity requirement is what allows us to
control how its answer changes when the realized oracle is modified.

\begin{definition}[{\cite[Definition~4.4]{C:BitDegVai17}}]
\label{def:universal-decide}
For a family of classical oracles $\mathfrak S$ and an oracle
$\Psi\in\mathfrak S$, define the classical function
$\msf{Decide}_{\mathfrak S}^{\Psi}$ as follows.
Its input is a pair of classical oracle-aided verifier circuits
$(V_0,V_1)$ and an input $z$.
\begin{enumerate}
    \item Check that the circuits have compatible input formats and,
    for every $\Psi'\in\mathfrak S$ and every input $z'$ of their
    common input length, exactly one bit $b\in\bits$ satisfies
    \[
        \exists w:\;V_b^{\Psi'}(z',w)=1.
    \]
    If this condition fails, output $\bot$.
    We call a verifier pair satisfying this condition \emph{valid}
    relative to $\mathfrak S$.

    \item Output the unique bit $b\in\bits$ for which
    \[
        \exists w:\;V_b^{\Psi}(z,w)=1.
    \]
\end{enumerate}
Malformed inputs are answered by $\bot$.
The oracle is defined only for realizations $\Psi\in\mathfrak S$.
\end{definition}

The definition specifies a function and imposes no computational
bound on the validity test or the search for an accepting witness.
We give quantum algorithms access to the reversible evaluation
oracle of this function.

If a fixed verifier pair remains complementary for every
$\Psi'\in\mathfrak S$, then its associated language is decidable
using one call to $\msf{Decide}_{\mathfrak S}^{\Psi}$.
In particular, this applies to every verifier pair satisfying the
structure requirement of \Cref{def:quantum-fully-black-box}.
It does not apply to an arbitrary pair that is complementary only
for the realized oracle $\Psi$, nor does it apply to verifiers
that themselves query $\msf{Decide}_{\mathfrak S}^{\Psi}$.

To analyze source security, we distinguish the number of queries
made by an adversary from the number made by the verifiers in a
decision-oracle query.

\begin{definition}[Bounded Decision-Oracle Queries]
\label{def:bounded-decide-query}
A query $Q=(V_0,V_1,z)$ is \emph{$r$-bounded} relative to
$\mathfrak S$ if, for every $\Psi\in\mathfrak S$, every witness $w$,
and each $b\in\bits$, the execution
\(
    V_b^\Psi(z,w)
\)
makes at most $r$ queries to $\Psi$.

An adversary making at most $T$ quantum queries to the base oracle
and the decision oracle is required to have only $r$-bounded
well-formed decision queries in its computational-basis query
support, for every oracle realization under consideration. 
Malformed queries are allowed and receive the fixed answer $\bot$.
When $T=r=q$, we call such an adversary \emph{$q$-query}.
\end{definition}

Although the decision oracle itself is computationally unbounded,
its answer to a valid $r$-bounded query is certified by at most $r$
base-oracle answers. The following consequence of the
$\NP\cap\coNP$ structure is the key property used in our analysis.

\begin{lemma}[{\cite[Claim~4.6]{C:BitDegVai17}}]
\label{lem:critical-queries}
For every valid $r$-bounded query $Q=(V_0,V_1,z)$ and every
$\Psi\in\mathfrak S$, there is a set
$C=C(\Psi,V_0,V_1,z)$ of at most $r$ classical oracle queries such
that, for every $\Psi'\in\mathfrak S$,
\[
    \Psi|_C=\Psi'|_C
    \quad\Longrightarrow\quad
    \msf{Decide}^{\Psi}_{\mathfrak S}(Q)
    =
    \msf{Decide}^{\Psi'}_{\mathfrak S}(Q).
\]
\end{lemma}

\begin{proof}
We include the proof for completeness.
Let
\(
    b=\msf{Decide}^{\Psi}_{\mathfrak S}(V_0,V_1,z),
\)
and fix an accepting witness $w$ for $V_b^\Psi(z,w)$.
Let $C$ contain the oracle queries made in this verification.
Since the query is $r$-bounded, $|C|\leq r$.

If $\Psi'$ agrees with $\Psi$ on $C$, the same verification follows
the same execution and accepts relative to $\Psi'$.
Since the verifier pair remains complementary for every oracle in
$\mathfrak S$, the decision bit relative to $\Psi'$ must also be $b$.

\end{proof}

\subsubsection{A Quantum Hybrid Bound for Coupled Oracles}

The classical argument of \cite{C:BitDegVai17} bounds the probability
that an adversary encounters a query on which two oracle realizations
disagree. This reasoning does not directly apply to quantum
adversaries, whose queries may be superpositions of many inputs:
measuring a query to identify a differing point can disturb the
computation.

We instead establish a conditional-coupling lemma for the quantum
oracle hybrid argument. If the probability of disagreement at every
fixed input is at most $\gamma$, conditioned on the reference oracle,
then any $T$-query quantum adversary has distinguishing advantage at
most $2T\sqrt{\gamma}$. The proof follows the hybrid and projector
method underlying \cite[Lemmas~35 and~37]{FOCS:AmbRosUnr14}.
Combined with \Cref{lem:critical-queries}, this formulation lets us
use classical pointwise disagreement bounds even when the adversary
has superposition access to the decision oracle.

We first recall the trace-distance estimate used in the proof. Throughout this section, $\operatorname{TD}$ denotes trace distance with the
$\frac12\|\cdot\|_1$ convention.

\begin{lemma}[{\cite{EC:Unruh14,FOCS:AmbRosUnr14}}]
\label{lem:ARU14-Lemma35}
Let $\ket{\Psi_0}$ and $\ket{\Psi_1}$ be normalized quantum states
that can be written as
\[
    \ket{\Psi_b}=\ket{\Psi_b^*}+\ket{\Phi^*},
    \qquad b\in\bits,
\]
where both $\ket{\Psi_b^*}$ are orthogonal to $\ket{\Phi^*}$.
Then
\[
    \operatorname{TD}(\ket{\Psi_0},\ket{\Psi_1})
    \leq 2\left\|\ket{\Psi_1^*}\right\|.
\]
\end{lemma}

\begin{lemma}[Quantum Queries to Coupled Oracle Distributions]
\label{lem:coupled-distributions} 
Let $\gamma\in[0,1]$, and let $\Gamma_1$ and $\Gamma_2$ be
distributions over classical functions from $\bits^n$ to $\bits^m$.
Suppose there is a joint distribution $\mu$ on pairs $(f_1,f_2)$
whose marginals are $\Gamma_1$ and $\Gamma_2$, respectively, such that
for every $f'_2$ in the support of $\Gamma_2$ and every $z\in\bits^n$,
\begin{equation}\zcsetup{reftype=Inequality}\label{eq:coupling-hypothesis}
    \Pr_{(f_1,f_2)\leftarrow\mu}
    \left[
        f_1(z)\neq f_2(z)
        \,\middle|\,
        f_2=f'_2
    \right]
    \leq\gamma.
\end{equation}

Let $\mathcal A$ be a quantum oracle algorithm making at most $T$
queries, whose initial state is independent of the sampled oracle
pair. If $\rho_f$ denotes its final state with oracle $f$, then
\[
    \mathop{\mathbb E}_{(f_1,f_2)\leftarrow\mu}
    \left[
        \operatorname{TD}(\rho_{f_1},\rho_{f_2})
    \right]
    \leq 2T\sqrt{\gamma}.
\]
In particular, if $\mathcal A$ outputs a bit, then
\[
    \left|
        \Pr_{f_1\leftarrow\Gamma_1}
        [\mathcal A^{\ket{f_1}}=1]
        -
        \Pr_{f_2\leftarrow\Gamma_2}
        [\mathcal A^{\ket{f_2}}=1]
    \right|
    \leq 2T\sqrt{\gamma}.
\]
\end{lemma}

\begin{proof}
We purify the algorithm's initial state and intermediate operations,
retaining any purification registers as part of its workspace.
It suffices to prove the bound for these purified final states,
since discarding registers cannot increase trace distance.

Let $A,K,V$ denote the workspace, query-input, and query-output
registers, respectively. For a function $f$, write
\[
    O_f\ket{a,k,v}_{AKV}
    =
    \ket{a,k,v\oplus f(k)}_{AKV}.
\]
Absorbing the initial oracle-independent computation into
$\ket{\Psi^0}$, let
\[
    \ket{\Psi_f^0}=\ket{\Psi^0},
    \qquad
    \ket{\Psi_f^i}=U_iO_f\ket{\Psi_f^{i-1}},
    \quad i\in\{1,\ldots,T\},
\]
where the $U_i$ are oracle-independent unitaries.
Algorithms making fewer queries can be padded with inactive
query slots; the same argument applies to their active-query
subspaces.

\subpara{Hybrid bound.}
For fixed $(f_1,f_2)$, define
\(
    D_i^{f_1,f_2}
    \coloneqq
    \operatorname{TD}
    (\ket{\Psi_{f_1}^i},\ket{\Psi_{f_2}^i}).
\)
By the triangle inequality and unitary invariance of trace distance,
\begin{align*}
    D_i^{f_1,f_2}
    &=
    \operatorname{TD}
    \left(
        O_{f_1}\ket{\Psi_{f_1}^{i-1}},
        O_{f_2}\ket{\Psi_{f_2}^{i-1}}
    \right)\\
    &\leq
    D_{i-1}^{f_1,f_2}
    +
    \operatorname{TD}
    \left(
        O_{f_1}\ket{\Psi_{f_2}^{i-1}},
        O_{f_2}\ket{\Psi_{f_2}^{i-1}}
    \right).
\end{align*}
Since $D_0^{f_1,f_2}=0$, it follows that
\[
    D_T^{f_1,f_2}
    \leq
    \sum_{i=0}^{T-1}
    \operatorname{TD}
    \left(
        O_{f_1}\ket{\Psi_{f_2}^{i}},
        O_{f_2}\ket{\Psi_{f_2}^{i}}
    \right).
\]

\subpara{Projector bound.}
For $z\in\bits^n$, define
\(
    Q_z \coloneqq I_A\otimes\ket z\!\bra z_K\otimes I_V,
\)
and define the disagreement projector
\[
    Q^{f_1,f_2}
    =
    \sum_{z:\,f_1(z)\neq f_2(z)}Q_z.
\]
Fix $i$, and abbreviate
$Q=Q^{f_1,f_2}$ and $\ket\psi=\ket{\Psi_{f_2}^i}$.
The two states $O_{f_1}\ket\psi$ and $O_{f_2}\ket\psi$
have the common component
\[
    \ket{\Phi^*}
    =
    O_{f_1}(I-Q)\ket\psi
    =
    O_{f_2}(I-Q)\ket\psi.
\]
Their remaining components are
\[
    \ket{\Psi_b^*}=O_{f_b}Q\ket\psi,
    \qquad b\in\{1,2\}.
\]
Both oracle unitaries preserve the query-input register.
Consequently, each $\ket{\Psi_b^*}$ is orthogonal to
$\ket{\Phi^*}$ and has norm $\|Q\ket\psi\|$.
By \Cref{lem:ARU14-Lemma35},
\[
    \operatorname{TD}
    \left(
        O_{f_1}\ket{\Psi_{f_2}^i},
        O_{f_2}\ket{\Psi_{f_2}^i}
    \right)
    \leq
    2\left\|
        Q^{f_1,f_2}\ket{\Psi_{f_2}^i}
    \right\|.
\]

\subpara{Averaging over the coupling.}
Taking expectation over $(f_1,f_2)\leftarrow\mu$ gives
\begin{equation}\zcsetup{reftype=Inequality}\label{eq:expected-hybrid}
    \mathop{\mathbb E}_{\mu}[D_T^{f_1,f_2}]
    \leq
    2\sum_{i=0}^{T-1}
    \mathop{\mathbb E}_{\mu}
    \left[
        \left\|
            Q^{f_1,f_2}\ket{\Psi_{f_2}^i}
        \right\|
    \right].
\end{equation}
By Jensen's inequality,
\[
    \mathop{\mathbb E}_{\mu}
    \left[
        \left\|
            Q^{f_1,f_2}\ket{\Psi_{f_2}^i}
        \right\|
    \right]
    \leq
    \sqrt{
        \mathop{\mathbb E}_{\mu}
        \left[
            \left\|
                Q^{f_1,f_2}\ket{\Psi_{f_2}^i}
            \right\|^2
        \right]
    }.
\]
The state $\ket{\Psi_{f_2}^i}$ depends only on the reference oracle
$f_2$. Since the projectors $Q_z$ are mutually orthogonal,
\begin{align*}
    \mathop{\mathbb E}_{\mu}
    \left[
        \left\|
            Q^{f_1,f_2}\ket{\Psi_{f_2}^i}
        \right\|^2
    \right]
    &\quad=
    \mathop{\mathbb E}_{f_2\leftarrow\Gamma_2}
    \left[
        \sum_z
        \left\|Q_z\ket{\Psi_{f_2}^i}\right\|^2
        \Pr_{\mu}
        [f_1(z)\neq f_2(z)\mid f_2]
    \right]\\
    &\quad\leq
    \gamma\,
    \mathop{\mathbb E}_{f_2\leftarrow\Gamma_2}
    \left[
        \sum_z
        \left\|Q_z\ket{\Psi_{f_2}^i}\right\|^2
    \right]\\
    &\quad=\gamma,
\end{align*}
where the inequality uses \zcref{eq:coupling-hypothesis}.
Substituting into \zcref{eq:expected-hybrid}, we obtain
\[
    \mathop{\mathbb E}_{\mu}[D_T^{f_1,f_2}]
    \leq 2T\sqrt{\gamma}.
\]

Finally, for any fixed final measurement, the difference in the
probability of an outcome is at most the trace distance.
Since $\mu$ has marginals $\Gamma_1$ and $\Gamma_2$,
\begin{align*} 
    \left|
        \Pr_{f_1\leftarrow\Gamma_1}
        [\mathcal A^{\ket{f_1}}=1]
        -
        \Pr_{f_2\leftarrow\Gamma_2}
        [\mathcal A^{\ket{f_2}}=1]
    \right| \qquad\leq
    \mathop{\mathbb E}_{(f_1,f_2)\leftarrow\mu}
    \left[
        D_T^{f_1,f_2}
    \right]
    \leq 2T\sqrt{\gamma}.
\end{align*}
This completes the proof of \Cref{lem:coupled-distributions}.

\end{proof}

\subsection{Warm-Up: Separation from injOWFs}
\label{sec:quantize-injOWF}

This subsection serves as a warm-up for the \Cref{sec:quantize-iO-injOWF} by demonstrating
the main technical tool underlying our black-box separation.
We construct an injective function oracle that remains one-way
against quantum queries even in the presence of the universal
decision oracle. Since the decision oracle decides every verifier
pair arising from a fully black-box construction, this suffices
to rule out such constructions from injOWFs alone.
In the next subsection, we strengthen the result by incorporating $\iO$.

\para{Notation.}
Let $\mathbf I_n^m$ denote the uniform distribution on all injective
functions from $\bits^n$ to $\bits^m$.
In this subsection, $\mathfrak S$ denotes the family of full
oracle ensembles
\[
    f=\{f_k\}_{k\in\mathbb N},
    \qquad
    f_k:\bits^k\longrightarrow\bits^{k+1},
\]
where each $f_k$ is injective.
The random injOWF oracle is sampled by choosing independently
\[
    f_k\leftarrow\mathbf I_k^{k+1}
    \qquad\text{for every }k\in\mathbb N.
\]
We write $f(z)=f_{|z|}(z)$.

For the security proof at input length $n$, we prove a stronger
statement: all components
\[
    f_{-n}\coloneqq\{f_k:k\neq n\}
\]
may be fixed arbitrarily, and only $f_n$ is sampled at random.
This allows the adversary and the verifiers in its decision queries
to access the entire oracle family, including input lengths other
than $n$.

The following \Cref{thm:main:injOWF} is the post-quantum analogue of the injOWF security
statement in \cite[Theorem~4.8]{C:BitDegVai17}.
Its role in the separation is to establish source security in the
two-oracle construction. The other requirement follows from the
definition of $\msf{Decide}_{\mathfrak S}^{f}$: every universally
complementary verifier pair can be decided with one oracle query.
We give the resulting black-box contradiction after the proof.

\begin{theorem}\label{thm:main:injOWF}
Fix $n\in\mathbb N$, arbitrary injective components $f_{-n}$,
and integers $T\geq 0$ and $r\geq 1$.
Let $\Adv$ be an otherwise computationally unbounded quantum
oracle adversary making at most $T$ queries in total to $f$ and
$\msf{Decide}_{\mathfrak S}^{f}$, with all well-formed
decision-oracle queries $r$-bounded.
The description and initial state of $\Adv$ are independent of
the sampled component $f_n$.
Then
\[
    \Pr_{\substack{
        f_n\leftarrow\mathbf I_n^{n+1}\\
        x\picks\bits^n 
    }}
    \left[
        \Adv^{\ket f,\ket{\msf{Decide}_{\mathfrak S}^{f}}}
        (f(x))
        =x
    \right]
    \leq
        2(T+1)\sqrt{\frac{r}{2^n}}
   .
\]
In particular, if $T,r\leq q(n)=O(2^{n/6})$, the inversion
probability is $O(2^{-n/4})$.
The same bounds hold when all components of $f$ are sampled
independently as above.
\end{theorem}

\begin{proof}
Fix $n$, $f_{-n}$, and $\Adv$ as in the theorem.
The proof first samples a fresh value $y$ outside
$\operatorname{Image}(f_n)$ and plants it at a uniformly random
input $x\in\bits^n$. The planted experiment has the same
distribution as the real inversion experiment.
We then remove the planting while keeping the challenge $y$.
In the resulting experiment, $y$ has no preimage under $f_n$.
Thus, a successful inverter distinguishes the two experiments
by making one additional query to check its proposed preimage.

The key structural ingredient is \Cref{lem:critical-queries}.
For each fixed reference oracle $f$, every valid $r$-bounded
decision query has a critical set of at most $r$ inputs.
Planting at a uniformly random $n$-bit input therefore changes
the answer to any fixed decision query with probability at most
$r/2^n$.
We apply \Cref{lem:coupled-distributions} to the combined oracle
$(f,\msf{Decide}_{\mathfrak S}^{f})$ to obtain a quantum
distinguishing bound of
\[
    2(T+1)\sqrt{r/2^n}.
\]

\para{Hybrid $H_0$.}
This is the real inversion experiment.
The components $f_{-n}$ remain fixed throughout all hybrids.

\begin{ExperimentBox}[label={hybrid:h200}]{Hybrid $H_0$}
{
\begin{enumerate}
    \item $f_n\leftarrow\mathbf I_n^{n+1}$.
    \item $x\picks\bits^n$.
    \item Invoke
    \(
        \Adv^{\ket f,\ket{\msf{Decide}_{\mathfrak S}^{f}}}
        (f(x)).
    \)
\end{enumerate}
}
\end{ExperimentBox}

The adversary wins if its output equals $x$.

\para{Hybrid $H_1$.}
We sample $y$ uniformly outside $\operatorname{Image}(f_n)$ and
plant it at a uniformly random input $x$.
The resulting full oracle $g$ agrees with $f$ everywhere except
at $x$. The difference from $H_0$ is highlighted in \red{red color}.

\begin{ExperimentBox}[label={hybrid:h201}]{Hybrid $H_1$}
{
\begin{enumerate}
    \item $f_n\leftarrow\mathbf I_n^{n+1}$.
    \item $x\picks\bits^n$.
    \item
    \red{$y\picks
    \bits^{n+1}\setminus\operatorname{Image}(f_n)$.}
    \item 
    \red{Define the full oracle $g$ by
    \(
        g(z)=
        \begin{cases}
            y & \text{if }z=x,\\
            f(z) & \text{otherwise}.
        \end{cases}
    \)}
    \item Invoke
    \(
        \Adv^{\ket{\red{g}},\ket{\msf{Decide}_{\mathfrak S}^{\red{g}}}}(\red{y}).
    \)
\end{enumerate}
}
\end{ExperimentBox}

Again, the adversary wins if its output equals $x$.
Since $y\notin\operatorname{Image}(f_n)$, the modified component
$g_n$ is injective and $g\in\mathfrak S$.

\begin{MyClaim}\zcsetup{reftype=Claim}\label{claim:hyb-0-1}
The joint distributions of the base oracle, the planted point,
and the challenge satisfy
\[
    (f,x,f(x))\text{ in }H_0
    \;\idind\;
    (g,x,y)\text{ in }H_1.
\]
Consequently, the adversary's winning probabilities in $H_0$
and $H_1$ are equal.
\end{MyClaim}

\begin{proof}
For each fixed $x$, the modified function $g_n$ is uniformly
distributed over $\mathbf I_n^{n+1}$.
Indeed, for any fixed injective $g_n$, its challenge value must be
$y=g_n(x)$, and its possible predecessors $f_n$ are obtained by
replacing $g_n(x)$ with any value outside
$\operatorname{Image}(g_n)$.
There are exactly $2^n$ such predecessors, each having the same
sampling probability.

Thus, $g_n$ is uniform and independent of the uniform point $x$.
Since $y=g_n(x)$ and $g_{-n}=f_{-n}$, the stated joint
distributions are identical.
The decision oracle is determined by the full underlying oracle,
so including it preserves this equality of distributions.

This completes the proof of \zcref{claim:hyb-0-1}.

\end{proof}

\para{Hybrid $H_2$.}
We remove the planting and give the adversary access to the
reference oracle $f$, while keeping the challenge $y$ outside
$\operatorname{Image}(f_n)$. The difference from $H_1$ is highlighted in \red{red color}.

\begin{ExperimentBox}[label={hybrid:h202}]{Hybrid $H_2$}
{
\begin{enumerate}
    \item $f_n\leftarrow\mathbf I_n^{n+1}$.
    \item $x\picks\bits^n$.
    \item
    $y\picks
    \bits^{n+1}\setminus\operatorname{Image}(f_n)$.
    \item Invoke
    \(
        \Adv^{\ket{\red{f}},\ket{\msf{Decide}_{\mathfrak S}^{\red{f}}}}(y).
    \)
\end{enumerate}
}
\end{ExperimentBox}

To compare these hybrids, we turn the inverter into a bit-output
distinguisher. Given a challenge $y$ and access to an underlying
oracle $h$ and its decision oracle, define $\mathcal B$ as follows:
\begin{enumerate}
    \item Run
    $\Adv^{\ket h,\ket{\msf{Decide}_{\mathfrak S}^{h}}}(y)$
    and obtain an output $a$.
    \item If $a\notin\bits^n$, output $0$.
    \item Query $h(a)$ and output $1$ if and only if $h(a)=y$.
\end{enumerate}
The algorithm $\mathcal B$ makes at most $T+1$ oracle queries,
and its decision-oracle queries remain $r$-bounded.
In the following claim, the subscripts $H_1$ and $H_2$ specify
the sampling of the oracle and challenge in the corresponding
hybrids.

\begin{MyClaim}\zcsetup{reftype=Claim}\label{claim:hyb-1-2}
It holds that
\[
    \left|
        \Pr_{H_1}
        \left[
            \mathcal B^{\ket g,
            \ket{\msf{Decide}_{\mathfrak S}^{g}}}(y)=1
        \right]
        -
        \Pr_{H_2}
        \left[
            \mathcal B^{\ket f,
            \ket{\msf{Decide}_{\mathfrak S}^{f}}}(y)=1
        \right]
    \right|
    \leq
    2(T+1)\sqrt{\frac{r}{2^n}}.
\]
\end{MyClaim}

\begin{proof}
We apply \Cref{lem:coupled-distributions} to a tagged oracle
combining the base oracle and the decision oracle.
For every $h\in\mathfrak S$, define
\[
    \widetilde F_h(0,z)=h(z),
    \qquad
    \widetilde F_h(1,Q)
    =
    \msf{Decide}_{\mathfrak S}^{h}(Q),
\]
where $Q=(V_0,V_1,z)$ is an $r$-bounded query.
Malformed inputs receive the fixed answer $\bot$.
Encodings of non-$r$-bounded decision queries can also be assigned
$\bot$ when extending $\widetilde F_h$ to a total function, since
they never occur in the query support of the algorithms under
consideration.

For each fixed input length, the querying algorithm has finite
query registers. We pad all relevant encodings and answers to
fixed lengths. Thus $\widetilde F_h$ can be viewed as a classical
function on a finite fixed-length domain, and access to the two
oracle interfaces is simulated by access to $\widetilde F_h$
without increasing the number of queries.

\para{Conditioning on the challenge.}
Fix $y\in\bits^{n+1}$.
Since every injective $f_n$ has exactly $2^n$ values outside its
image, the conditional distribution of $f_n$ given the challenge
$y$ is
\[
    f_n\leftarrow\mathbf I_n^{n+1}
    \mid y\notin\operatorname{Image}(f_n).
\]
The point $x$ remains uniform and independent of $f$ under this
conditioning.

Let $\Gamma_{1,y}$ and $\Gamma_{2,y}$ be the corresponding
distributions of merged oracles:
\begin{align*}
    \Gamma_{1,y}:\quad&
    f_n\leftarrow\mathbf I_n^{n+1}
    \mid y\notin\operatorname{Image}(f_n),\qquad
    x\picks\bits^n,\\
    &g=f_{x\to y},
    \qquad\text{output }\widetilde F_g;\\[1mm]
    \Gamma_{2,y}:\quad&
    f_n\leftarrow\mathbf I_n^{n+1}
    \mid y\notin\operatorname{Image}(f_n),
    \qquad\text{output }\widetilde F_f.
\end{align*}
Here $f_{-n}$ is the fixed collection of components from the
theorem statement, and $f_{x\to y}$ changes only the value at $x$.
We couple these distributions by sampling the same $f$ and a
uniform $x$ and outputting
\[
    (\widetilde F_g,\widetilde F_f),
    \qquad g=f_{x\to y}.
\]

\para{Pointwise disagreement.}
Fix the full reference oracle $f$ in the support of the conditional
distribution, and fix a merged-oracle input $w$.
We claim that
\begin{equation}\zcsetup{reftype=Inequality}\label{eq:point-wise-decide-bound}
    \Pr_{x\picks\bits^n}
    \left[
        \widetilde F_g(w)\neq\widetilde F_f(w)
    \right]
    \leq \frac{r}{2^n}.
\end{equation}
There are three cases.
\begin{itemize}
    \item If $w=(0,z)$ is a direct query to the base oracle and
    $|z|=n$, then
    \[
        \Pr_x[g(z)\neq f(z)]
        =
        \Pr_x[x=z]
        =
        2^{-n}
        \leq r2^{-n}.
    \]
    If $|z|\neq n$, the disagreement probability is zero.

    \item If $w=(1,Q)$ is an $r$-bounded decision query whose
    verifier pair is invalid relative to $\mathfrak S$, both
    answers are $\bot$.
    Indeed, the validity test depends only on the verifier pair
    and the family $\mathfrak S$, not on the realized oracle.
    Malformed inputs likewise have disagreement probability zero.

    \item Suppose $w=(1,Q)$ is a valid $r$-bounded query, where
    $Q=(V_0,V_1,z)$.
    Let
    \(
        C_Q=C(f,V_0,V_1,z)
    \)
    be the critical set from \Cref{lem:critical-queries}.
    Then $|C_Q|\leq r$.
    Since $g\in\mathfrak S$ and $f$ and $g$ differ only at $x$,
    the condition $x\notin C_Q$ implies
    \(
        \msf{Decide}_{\mathfrak S}^{f}(Q)
        =
        \msf{Decide}_{\mathfrak S}^{g}(Q).
    \)
    Consequently,
    \begin{align*}
        \Pr_x
        \left[
            \msf{Decide}_{\mathfrak S}^{f}(Q)
            \neq
            \msf{Decide}_{\mathfrak S}^{g}(Q)
        \right]
        \leq
        \Pr_x[x\in C_Q]
        =
        \frac{|C_Q\cap\bits^n|}{2^n}
        \leq\frac{r}{2^n}.
    \end{align*}
\end{itemize}
Inputs outside the allowed query domain have a fixed answer in
both merged oracles and hence also have disagreement probability
zero. This proves \zcref{eq:point-wise-decide-bound}.

The estimate holds for every fixed full reference oracle $f$.
Since $\widetilde F_f$ is a deterministic function of $f$,
averaging over reference oracles with the same merged-oracle
realization gives the same estimate conditioned on
$\widetilde F_f$. Thus the coupling satisfies the hypothesis of
\Cref{lem:coupled-distributions} with $\gamma=\min\{1,r2^{-n}\}$.

\para{Applying the quantum hybrid bound.}
Let $\mathcal B_y$ be the algorithm obtained by hardwiring $y$
as the challenge input to $\mathcal B$ and routing its queries
through the merged oracle.
It makes at most $T+1$ queries.
By \Cref{lem:coupled-distributions},
\begin{equation*}
    \left|
        \Pr_{\widetilde F_g\leftarrow\Gamma_{1,y}}
        [\mathcal B_y^{\ket{\widetilde F_g}}=1]
        -
        \Pr_{\widetilde F_f\leftarrow\Gamma_{2,y}}
        [\mathcal B_y^{\ket{\widetilde F_f}}=1]
    \right| \leq
    2(T+1)\sqrt{\gamma}
    \leq
    2(T+1)\sqrt{\frac{r}{2^n}}.
\end{equation*}

Finally, the challenge has the same uniform marginal distribution
in both hybrids. For every fixed $y\in\bits^{n+1}$,
\[
    \Pr[y\text{ is the sampled challenge}]
    =
    \frac{
        \Pr_{f_n}[y\notin\operatorname{Image}(f_n)]
    }{2^n}
    =
    \frac{1/2}{2^n}
    =
    2^{-(n+1)}.
\]
Averaging the preceding bound over $y$ completes the proof of \zcref{claim:hyb-1-2}.

\end{proof}

\para{Completing the proof of \Cref{thm:main:injOWF}.}
In $H_1$, the function $g_n$ is injective and $g_n(x)=y$.
Therefore,
\[
    \Pr_{H_1}
    \left[
        \mathcal B^{\ket g,
        \ket{\msf{Decide}_{\mathfrak S}^{g}}}(y)=1
    \right]
    =
    \Pr[H_1\text{ wins}].
\]
In $H_2$, the challenge $y$ is outside
$\operatorname{Image}(f_n)$, and $\mathcal B$ rejects every output
outside $\bits^n$. Hence
\[
    \Pr_{H_2}
    \left[
        \mathcal B^{\ket f,
        \ket{\msf{Decide}_{\mathfrak S}^{f}}}(y)=1
    \right]
    =0.
\]
Combining \zcref{claim:hyb-0-1,claim:hyb-1-2}, we obtain
\[
    \Pr[H_0\text{ wins}]
    =
    \Pr[H_1\text{ wins}]
    \leq
    2(T+1)\sqrt{\frac{r}{2^n}}.
\]
The probability is also at most $1$, proving the quantitative
bound.

If $T,r\leq q(n)=O(2^{n/6})$, then
\[
    2(T+1)\sqrt{\frac{r}{2^n}}
    \leq
    2(q(n)+1)\sqrt{\frac{q(n)}{2^n}}
    =
    O(2^{-n/4}).
\]
The bound holds uniformly over the fixed components $f_{-n}$,
so averaging over independently sampled components gives the
same result for the full random oracle ensemble.

This completes the proof of \Cref{thm:main:injOWF}.

\end{proof}

\para{Consequence for black-box constructions.}
We now relate the security theorem to the goal of this subsection.

\begin{corollary}\label{cor:injOWF-black-box-separation}
There is no fully black-box quantum construction of a
worst-case quantum-hard language in $\NP\cap\coNP$ from
post-quantum injOWFs, in the sense of the injOWF-only version
of \Cref{def:quantum-fully-black-box}.
\end{corollary}

\begin{proof}
Suppose such a construction $(V_0,V_1,\mathcal R)$ exists.
Instantiate its source oracle by the random injective family $f$
defined above.

By the structure requirement, $(V_0,V_1)$ remains complementary
for every $f\in\mathfrak S$.
Thus an adversary can decide the constructed language exactly by
submitting the corresponding verifier circuits and input to
$\msf{Decide}_{\mathfrak S}^{f}$.
For each input length, the polynomial-time verifier machines are
compiled into classical oracle-aided circuits.
The resulting decision procedure has a unitary implementation
using the decision oracle, so it also supplies the coherent
interface needed by the quantum reduction.

Replace every call by $\mathcal R$ to this adversary with its
decision-oracle implementation.
The resulting inverter makes polynomially many queries to
$f$ and $\msf{Decide}_{\mathfrak S}^{f}$, and every
decision-oracle query contains verifiers making polynomially many
base-oracle queries.
By \Cref{thm:main:injOWF}, its inversion probability is
$O(2^{-n/4})$.

On the other hand, the black-box security requirement gives
inverse-polynomial inversion probability for infinitely many
input lengths, since the adversary decides the constructed
language exactly.
This is a contradiction.

This completes the proof of \Cref{cor:injOWF-black-box-separation}.

\end{proof}

\subsection{Separation from \texorpdfstring{$\iO$}{iO} and injOWFs}
\label{sec:quantize-iO-injOWF}

We now extend the warm-up construction to include $\iO$.
Following \cite{C:BitDegVai17}, we augment the injective function
oracle with an injective encoding oracle and an evaluation oracle.
We show that both source primitives remain secure against quantum
adversaries with access to the universal decision oracle.

We first specify the oracle family and the resource bounds used
in the security analysis. We then prove one-wayness by a coherent
simulation using the warm-up construction, and prove obfuscation
security by a sequence of coupled-oracle hybrids. Finally, we
show that every efficient fully black-box reduction induces an
adversary satisfying these resource bounds for some polynomial
parameter.

The quantitative bounds below concern a resource model that
controls evaluation-query lengths as well as oracle-query counts.
They do not assert a lower bound on bare oracle-query complexity
for unrestricted inefficient reductions.

\subsubsection{Oracle Construction and Resource Bounds}

\para{Oracle setting.}\label{iO-Oracle}
We use full oracle families indexed by all input lengths.
As in the warm-up, let
\[
    f=\{f_k\}_{k\in\mathbb N},
    \qquad
    f_k:\bits^k\longrightarrow\bits^{k+1},
\]
where each $f_k$ is injective. Let
\[
    \Ora=\{\Ora_k\}_{k\in\mathbb N},
    \qquad
    \Ora_k:\bits^{2k}\longrightarrow\bits^{5k},
\]
where each $\Ora_k$ is injective. We parse an input to $\Ora_k$
as a pair $(C,r)\in\bits^k\times\bits^k$.

We fix a padded circuit-description convention under which a
$k$-bit description specifies a classical circuit with $f$-gates
and at most $k$ oracle queries on every input. Malformed circuit
descriptions denote a fixed zero circuit with the prescribed
interface. Incompatible evaluation inputs receive $\bot$.
These conventions make circuit evaluation a total function and
apply equally to circuits encoded by $\Ora$ and to the backup
circuits defined below.

A \emph{backup map} is a family
\[
    \varphi=\{\varphi_k\}_{k\in\mathbb N},
    \qquad
    \varphi_k:\bits^{5k}\longrightarrow\bits^k.
\]
For $\widehat C\in\bits^{5k}$, define
\[
    \mathsf{Eval}^{f,\Ora}_{\varphi}(\widehat C,z)
    =
    \begin{cases}
        C^f(z)
        & \text{if }\Ora_k(C,r)=\widehat C
          \text{ for some }(C,r),\\[1mm]
        \bigl(\varphi_k(\widehat C)\bigr)^f(z)
        & \text{if }\widehat C\notin
          \operatorname{Image}(\Ora_k).
    \end{cases}
\]
Injectivity of $\Ora_k$ makes the first case unambiguous.
Malformed oracle inputs receive $\bot$.

Write
\[
    \Psi(f,\Ora,\varphi)
    \coloneqq
    \bigl(f,\Ora,\mathsf{Eval}^{f,\Ora}_{\varphi}\bigr),
    \qquad
    \Psi_\varphi=\Psi(f,\Ora,\varphi).
\]
The underlying oracle family is
\[
    \mathfrak S_\Psi
    \coloneqq
    \left\{
        \Psi(f,\Ora,\varphi):
        \begin{array}{l}
            f_k:\bits^k\to\bits^{k+1}
            \text{ is injective for every }k,\\
            \Ora_k:\bits^{2k}\to\bits^{5k}
            \text{ is injective for every }k,\\
            \varphi_k:\bits^{5k}\to\bits^k
            \text{ is arbitrary for every }k
        \end{array}
    \right\}.
\]
In particular, $\mathfrak S_\Psi$ ranges over all backup maps,
even when a security experiment fixes one particular $\varphi$.
We use this same family in the validity test of
$\msf{Decide}_{\mathfrak S_\Psi}$ throughout every experiment
and hybrid.

For a fixed backup map $\varphi$, the random oracle realization
is obtained by sampling independently
\[
    f_k\leftarrow\mathbf I_k^{k+1},
    \qquad
    \Ora_k\leftarrow\mathbf I_{2k}^{5k}
    \qquad\text{for every }k\in\mathbb N.
\]
All sampled tables remain fixed throughout the experiment.
Define the augmented oracle tuple
\[
    \Gamma(f,\Ora,\varphi)
    \coloneqq
    \left(
        f,\Ora,\mathsf{Eval}^{f,\Ora}_{\varphi},
        \msf{Decide}_{\mathfrak S_\Psi}^{\Psi(f,\Ora,\varphi)}
    \right),
\]
and let $\Gamma_\varphi$ denote its distribution under this
sampling.

\para{Obfuscation and the implementation interface.}
On input a circuit description $C\in\bits^n$, the obfuscator
samples $r\picks\bits^n$, computes
\(
    \widehat C=\Ora_n(C,r),
\)
and returns the canonical oracle-aided wrapper $E_{\widehat C}$
that evaluates
\(
    E_{\widehat C}^{\,\Psi_\varphi}(z)
    =
    \mathsf{Eval}^{f,\Ora}_{\varphi}(\widehat C,z).
\)
We denote this obfuscator by $\iObf$.
For every realization in $\mathfrak S_\Psi$, every circuit $C$,
and every random tape $r$,
\(
    E_{\Ora_n(C,r)}^{\,\Psi_\varphi}\equiv C^f.
\)
Thus functionality holds for every admissible implementation,
not only with high probability over the sampled tables.

The input circuit has only $f$-gates, whereas the returned
wrapper has an $\mathsf{Eval}$-gate. Accordingly, returned
programs execute relative to the ambient oracle tuple.
This is the implementation convention used when applying the
black-box framework in this subsection: calls to obfuscation
and executions of returned programs are compiled into calls to
$\Ora$ and $\mathsf{Eval}$, respectively.
It is not a claim that $E_{\widehat C}$ is an $f$-only circuit.

The wrapper has size polynomial in $n$, and its description
can be constructed from $\widehat C$ and parsed to recover
$\widehat C$. Hence giving the adversary $\widehat C$ in the
security experiment is equivalent to giving it the canonical
wrapper $E_{\widehat C}$.

\para{Reprogramming notation.}
For a fixed $r\in\bits^n$, write
\[
    T=(\ast,r)
    \coloneqq
    \{(C,r):C\in\bits^n\},
\]
ordered lexicographically by $C$.
If $D$ is an ordered tuple of $2^n$ distinct elements of
$\bits^{5n}$, then $\Ora_{T\mapsto D}$ maps each element of
$T$ to the corresponding entry of $D$ and agrees with $\Ora$
elsewhere. Sampling an ordered subset means sampling uniformly
without replacement, including a uniformly random ordering.

Similarly, $\Ora_{(C,r)\mapsto\widehat C}$ changes only the
value at $(C,r)$, and $\varphi_{\widehat C\mapsto C}$ changes
only the backup circuit assigned to $\widehat C$.
All these modifications leave components at other lengths
unchanged.

\para{Resource bounds.}
Unlike the warm-up in \Cref{sec:quantize-injOWF}, a single evaluation query can invoke a
circuit containing many $f$-gates. We therefore bound the
lengths of evaluation encodings in addition to the number of
external oracle queries.

\begin{definition}[Resource-Bounded Queries]
\label{def:full-q-query}
Let $q\geq 1$.
In this subsection, a quantum adversary with access to
$\Gamma(f,\Ora,\varphi)$ is \emph{$q$-query} if it satisfies
the following conditions.
\begin{enumerate}
    \item \textbf{Direct function queries.}
    It makes at most $q$ queries to $f$.

    \item \textbf{Encoding and evaluation queries.}
    It makes at most $q$ queries in total to $\Ora$ and
    $\mathsf{Eval}$. Every well-formed evaluation query
    $(\widehat C,z)$ has $|\widehat C|\leq 5q$.
    Thus the selected circuit, whether recovered from $\Ora$
    or supplied by $\varphi$, makes at most $q$ queries to $f$.

    \item \textbf{Decision queries.}
    It makes at most $q$ queries to
    $\msf{Decide}_{\mathfrak S_\Psi}$.
    For every well-formed query $(V_0,V_1,z)$, every
    $\Psi\in\mathfrak S_\Psi$, every witness $w$, and each
    $b\in\bits$, the execution $V_b^\Psi(z,w)$ satisfies
    the first two conditions above.
    The verifiers have access only to $\Psi$, not to the
    decision oracle.
\end{enumerate}
For quantum queries, the restrictions apply to every
computational-basis query in the support of the query state,
including in the hybrid executions used below.
Malformed queries are allowed and receive $\bot$.

For a two-stage adversary, the query bounds apply to the two
stages combined.
\end{definition}

Thus a $q$-query adversary makes at most $3q$ queries to the
augmented tuple. Each verifier in a decision query makes at
most $2q$ queries to the underlying tuple $\Psi$.
Consequently, \Cref{lem:critical-queries} supplies a critical
set of size at most $2q$ for every valid decision query.

The adversaries considered below may otherwise be
computationally unbounded. Their descriptions and initial
states are independent of the sampled oracle component,
apart from information obtained through the stated experiment.

\subsubsection{One-Wayness in the Presence of the Augmented Oracle}

We first show that access to the encoding, evaluation, and
decision interfaces does not destroy one-wayness.
The proof fixes $\Ora$ and $\varphi$ and simulates their
interfaces using the warm-up oracle.
The simulation must preserve the validity test of the full
decision oracle, including on invalid verifier pairs. 
\begin{theorem}\label{thm:full-oracle-injOWF}
Fix $n\in\mathbb N$, an integer $q\geq 1$, a backup map
$\varphi$, an injective encoding family $\Ora$, and arbitrary
injective components $f_{-n}$.
Let $\Adv$ be a $q$-query quantum adversary whose description
and initial state are independent of $f_n$.
Then
\[
    \Pr_{\substack{
        f_n\leftarrow\mathbf I_n^{n+1}\\
        x\picks\bits^n
    }}
    \left[
        \Adv^{\ket{\Gamma(f,\Ora,\varphi)}}(f_n(x))=x
    \right]
    \leq
    O\left(q^3 2^{-n/2}\right)
    .
\]
In particular, if $q(n)=O(2^{n/12})$, the inversion
probability is $O(2^{-n/4})$.
The same bounds hold when all components of $f$ and $\Ora$
are sampled independently as above, with $\varphi$ fixed.
\end{theorem}

\begin{proof}
Fix $n,q,\varphi,\Ora,f_{-n}$, and $\Adv$ as in the theorem.
Let $\mathfrak S_f$ denote the injective-function family from
\Cref{sec:quantize-injOWF}.
We construct a quantum adversary
\(
    \mathcal B^{\ket f,
    \ket{\msf{Decide}_{\mathfrak S_f}^{f}}}
\)
that perfectly simulates
$\Adv^{\ket{\Gamma(f,\Ora,\varphi)}}$.
The simulator makes $O(q^2)$ queries in total, and each
translated decision query is $O(q^2)$-bounded relative to
$\mathfrak S_f$. Applying \Cref{thm:main:injOWF} with these
two bounds gives the result.

\para{Fixed tables and oracle-independent computation.}
At the fixed security parameter, all query registers have
finite length. The relevant portions of $\Ora$, its inverse,
and $\varphi$ can therefore be hardwired into a finite
simulation circuit. This circuit may be inefficient, as
permitted by the query model, but its description is
independent of the sampled component $f_n$.

The same observation applies to the predicate
\(
    \mathsf{Valid}_{\mathfrak S_\Psi}(V_0,V_1)
\)
from \Cref{def:universal-decide}.
On the finite set of verifier descriptions that fit in the
query registers, this predicate has a finite truth table.
We hardwire that table and evaluate it reversibly without
queries to $f$. This does not require an efficient or uniform
algorithm for deciding universal validity.

\para{Simulation of the base interfaces.}
Given the challenge $y=f_n(x)$, the simulator runs $\Adv(y)$
and answers its queries as follows.
\begin{enumerate}
    \item \textbf{$f$-queries.}
    The simulator forwards the query to its own $f$ oracle.

    \item \textbf{$\Ora$-queries.}
    The simulator answers using the fixed encoding tables.
    This requires no queries to $f$.

    \item \textbf{$\mathsf{Eval}$-queries.}
    On input $(\widehat C,z)$, the simulator determines the
    circuit
    \[
        C=
        \begin{cases}
            \pi_1\bigl(\Ora^{-1}(\widehat C)\bigr)
            & \text{if }\widehat C\in
              \operatorname{Image}(\Ora),\\
            \varphi(\widehat C)
            & \text{otherwise}.
        \end{cases}
    \]
    It then evaluates $C^f(z)$.
    By the evaluation-length restriction, this circuit makes
    at most $q$ queries to $f$.

    To implement the oracle response coherently, the simulator
    reversibly selects $C$, evaluates it while retaining the
    computation history, XORs the result into the answer
    register, and reverses the evaluation and selection.
    All workspace is restored to zero. The resulting map is
    \[
        \ket{\widehat C,z,v}\ket 0
        \longmapsto
        \ket{\widehat C,z,
        v\oplus\mathsf{Eval}^{f,\Ora}_{\varphi}(\widehat C,z)}
        \ket 0.
    \]
    This costs $O(q)$ queries to $f$ per evaluation query,
    including uncomputation.
\end{enumerate}
The same construction handles superpositions of different
query inputs. Controlled routing between interfaces incurs
at most constant-factor query overhead.

\para{Simulation of the decision interface.}
For a well-formed query $Q=(V_0,V_1,z)$, the simulator first
reversibly computes
$\mathsf{Valid}_{\mathfrak S_\Psi}(V_0,V_1)$.
If the pair is invalid, it returns $\bot$.

Suppose the pair is valid. The simulator compiles each
$V_b$ into an $f$-aided verifier $D_b$ by replacing its
$\Ora$- and $\mathsf{Eval}$-queries with the classical
simulations described above. For every $f'\in\mathfrak S_f$,
every input $z'$, and every witness $w$,
\[
    D_b^{f'}(z',w)
    =
    V_b^{\Psi(f',\Ora,\varphi)}(z',w).
\]
Since $\Psi(f',\Ora,\varphi)\in\mathfrak S_\Psi$ and
$(V_0,V_1)$ is valid relative to $\mathfrak S_\Psi$,
the pair $(D_0,D_1)$ is valid relative to $\mathfrak S_f$.
Therefore,
\[
    \msf{Decide}_{\mathfrak S_f}^{f}(D_0,D_1,z)
    =
    \msf{Decide}_{\mathfrak S_\Psi}^{\Psi(f,\Ora,\varphi)}
    (V_0,V_1,z).
\]

The original validity test is essential here.
An invalid pair relative to $\mathfrak S_\Psi$ can become
valid after $\Ora$ and $\varphi$ are fixed. Querying the
translated pair without first checking its original validity
would therefore not simulate the full decision oracle.

The translation of a valid query is computed reversibly
while retaining the original query. The simulator evaluates
the translated decision oracle into the answer register and
then uncomputes the translation and validity flag.
For invalid or malformed queries, it instead returns the
fixed answer $\bot$ and cleans its workspace.
Thus this simulation implements the exact unitary
function-evaluation interface, including on superpositions
of valid and invalid queries.

\para{Query accounting.}
The simulator forwards at most $q$ direct $f$-queries and
simulates at most $q$ evaluation queries, each at cost
$O(q)$. It also makes $O(q)$ queries to
$\msf{Decide}_{\mathfrak S_f}^{f}$.
Hence its total query count satisfies
\(
    T_{\mathcal B}=O(q^2)
\). For a translated verifier $D_b$, the direct $f$-queries
contribute at most $q$, and the simulated evaluation queries
contribute at most $q^2$. Thus every translated decision
query is $r_{\mathcal B}$-bounded for
\(
    r_{\mathcal B}=O(q^2).
\)
 This bound holds for every $f'\in\mathfrak S_f$ and every
witness on the submitted input.
The verifier's internal queries are not additional queries
made by $\mathcal B$; they enter the warm-up theorem through
the separate parameter $r_{\mathcal B}$.

\para{Applying the warm-up theorem.}
The simulation is exact and its initial state is independent
of $f_n$. By \Cref{thm:main:injOWF},
\begin{align*}
    \Pr\left[
        \Adv^{\ket{\Gamma(f,\Ora,\varphi)}}(f_n(x))=x
    \right]
    &=
    \Pr\left[
        \mathcal B^{\ket f,
        \ket{\msf{Decide}_{\mathfrak S_f}^{f}}}(f_n(x))=x
    \right]\\
    &\leq
    2(T_{\mathcal B}+1)
    \sqrt{\frac{r_{\mathcal B}}{2^n}}\\
    &=
    O\left(q^3 2^{-n/2}\right).
\end{align*}
Here the probabilities average over
$f_n\leftarrow\mathbf I_n^{n+1}$ and $x\picks\bits^n$.

For $q(n)=O(2^{n/12})$, the last expression is
$O(2^{-n/4})$.
The estimate is uniform over the fixed choices of
$f_{-n}$ and $\Ora$, so averaging over their independent
sampling gives the same bound for the full random oracle
ensemble.

This completes the proof of \Cref{thm:full-oracle-injOWF}.

\end{proof}

\subsubsection{Obfuscation Security in the Presence of the Augmented Oracle}

We next prove post-quantum obfuscation security.
The proof follows the row-reprogramming hybrids of
\cite{C:BitDegVai17}, using
\Cref{lem:coupled-distributions} to control quantum queries.

The first reprogramming hides all encodings using a random
tape $r$ during the pre-challenge phase. This allows us to
replace the challenge encoding by a fresh value without
changing the joint distribution of the challenge,
post-challenge oracle, and residual quantum state.
We then restore the original pre-challenge oracle.
Finally, we replace the challenge encoding by a backup
evaluation of $C_0$. Functional equivalence of $C_0$ and
$C_1$ makes the resulting experiment independent of the
challenge bit.

\begin{theorem}\label{thm:full-oracle-iO}
Fix $n\in\mathbb N$, an integer $q\geq 1$, a backup map
$\varphi$, an injective family $f$, and injective encoding
components $\Ora_{-n}$.
Sample $\Ora_n\leftarrow\mathbf I_{2n}^{5n}$.
Let $\Adv=(\Adv_1,\Adv_2)$ be a $q$-query quantum
adversary whose description and initial state are independent
of $\Ora_n$.
In the real experiment, $\Adv_1$ outputs equal-size circuit
descriptions $C_0,C_1\in\bits^n$ satisfying
\(
    C_0^f\equiv C_1^f
\)
with probability one.
Then, it hold that
\[
    \left|
        \Pr[\mathsf{Exp}^{\mathsf{iO}}(n)=1]
        -\frac12
    \right|
    \leq
        O\left(q^{3/2}2^{-n/2}\right).
\]
The experiment uses the augmented oracle
$\Gamma(f,\Ora,\varphi)$ and the obfuscator $\iObf$.

In particular, for $q(n)=O(2^{n/6})$, the advantage is
$O(2^{-n/4})$.
Averaging over independently sampled $f$ and $\Ora_{-n}$
gives the same bound for
\(
    \operatorname{Adv}^{\mathsf{iO}}_
    {\Gamma_\varphi,\iObf,\mathcal C,\Adv}(n).
\)
\end{theorem}

\begin{proof}
Fix $n,q,\varphi,f,\Ora_{-n}$, and $\Adv$ as in the theorem.
Only $\Ora_n$ is sampled in the experiments below.
We write $\operatorname{Image}(\Ora_n)$ explicitly when
referring to the component being modified.

\para{Merged-oracle notation.}
For fixed $f$, define the tagged classical function
\begin{align*}
    \mathsf G_{\Ora,\varphi}(0,z)
    &\coloneqq f(z),\\
    \mathsf G_{\Ora,\varphi}(1,(C,r))
    &\coloneqq \Ora(C,r),\\
    \mathsf G_{\Ora,\varphi}(2,(\widehat C,z))
    &\coloneqq
    \mathsf{Eval}^{f,\Ora}_{\varphi}(\widehat C,z),\\
    \mathsf G_{\Ora,\varphi}(3,Q)
    &\coloneqq
    \msf{Decide}_{\mathfrak S_\Psi}^{\Psi(f,\Ora,\varphi)}(Q).
\end{align*}
For the fixed query-register lengths, we pad encodings and
answers to obtain a function on a finite fixed-length domain.
Malformed queries receive $\bot$.
Inputs outside the resource-bounded domain may also be assigned
$\bot$, since they never occur in the executions under
consideration.

Access to the four interfaces is thereby represented by at
most $3q$ queries to $\mathsf G_{\Ora,\varphi}$.
All comparisons below use the same family
$\mathfrak S_\Psi$ in the decision-oracle validity test.

\para{Hybrid $H_0$.}
This is the real obfuscation experiment.

\begin{ExperimentBox}[label={hybrid:h30}]{Hybrid $H_0$}
{
\begin{enumerate}
    \item $\Ora_n\leftarrow\mathbf I_{2n}^{5n}$.
    \item $b\picks\bits$ and $r\picks\bits^n$.
    \item
    $(C_0,C_1,\mathsf{st})
    \leftarrow
    \Adv_1^{\ket{\Gamma(f,\Ora,\varphi)}}(1^n)$.
    \item $\widehat C\leftarrow\Ora_n(C_b,r)$.
    \item
    $b'\leftarrow
    \Adv_2^{\ket{\Gamma(f,\Ora,\varphi)}}
    (\mathsf{st},\widehat C)$.
    \item Output $1$ if $b'=b$, and output $0$ otherwise.
\end{enumerate}
}
\end{ExperimentBox}

The register $\mathsf{st}$ may contain quantum side information.
We say that a hybrid wins when its output is $1$.

\para{Hybrid $H_1$.}
We change only the pre-challenge oracle.
For a uniformly random tape $r$, we replace the entire row
$T=(\ast,r)$ by fresh encodings $D$.
The challenge and post-challenge oracle still use the original
table $\Ora$. Its difference from $H_0$ is highlighted in \red{red color}.

\begin{ExperimentBox}[label={hybrid:h31}]{Hybrid $H_1$}
{
\begin{enumerate}
    \item $\Ora_n\leftarrow\mathbf I_{2n}^{5n}$.
    \item $b\picks\bits$ and $r\picks\bits^n$.
    \item
    \red{Sample $D$ uniformly as an ordered subset of size
    $2^n$ of
    $\bits^{5n}\setminus\operatorname{Image}(\Ora_n)$.}
    \item \red{Set $T=(\ast,r)$.}
    \item
    $(C_0,C_1,\mathsf{st})
    \leftarrow
    \Adv_1^{\ket{\Gamma(f,\red{\Ora_{T\mapsto D}},\varphi)}}
    (1^n)$.
    \item $\widehat C\leftarrow\Ora_n(C_b,r)$.
    \item
    $b'\leftarrow
    \Adv_2^{\ket{\Gamma(f,\Ora,\varphi)}}
    (\mathsf{st},\widehat C)$.
    \item Output $1$ if $b'=b$, and output $0$ otherwise.
\end{enumerate}
}
\end{ExperimentBox}

We first establish the pointwise estimate needed to compare
the pre-challenge states. This is the quantum-hybrid input
corresponding to \cite[Claim~4.18]{C:BitDegVai17}. 
\begin{lemma}[Row-Reprogramming Disagreement]
\label{point-wise-4.18}
Fix a full reference encoding family $\Ora$.
Sample $r\picks\bits^n$ and an ordered subset $D$ as in $H_1$,
and let $T=(\ast,r)$.
For every fixed allowed merged-oracle input $w$,
\[
    \Pr_{r,D}
    \left[
        \mathsf G_{\Ora_{T\mapsto D},\varphi}(w)
        \neq
        \mathsf G_{\Ora,\varphi}(w)
    \right]
    \leq 4q\,2^{-n}.
\]
\end{lemma}

\begin{proof}
Fix $w$. Since $D$ consists of distinct values outside
$\operatorname{Image}(\Ora_n)$, the modified encoding oracle
remains injective. Both underlying tuples therefore belong
to $\mathfrak S_\Psi$.

We consider the four query types.
\begin{enumerate}
    \item \textbf{Queries to $f$.}
    The function family is unchanged, so the disagreement
    probability is zero.

    \item \textbf{Queries to $\Ora$.}
    A query at a length other than $n$ is unchanged.
    At length $n$, a fixed input $(C,s)$ changes only if
    $s=r$. Hence its disagreement probability is at most
    $2^{-n}$.

    \item \textbf{Queries to $\mathsf{Eval}$.}
    An evaluation query whose encoding has length other than
    $5n$ is unchanged. At length $5n$, its answer can change
    only if its encoding $\widetilde C$ belongs to
    $\Ora_n(T)$ or to $D$.

    By injectivity, a fixed $\widetilde C$ has at most one
    preimage under $\Ora_n$, so
    \[
        \Pr_r[\widetilde C\in\Ora_n(T)]\leq 2^{-n}.
    \]
    Also,
    \[
        \Pr_D[\widetilde C\in D]
        \leq
        \frac{2^n}{2^{5n}-2^{2n}}
        \leq 2^{1-4n}.
    \]
    Thus the disagreement probability is at most
    \[
        2^{-n}+2^{1-4n}\leq 2^{1-n}.
    \]

    \item \textbf{Queries to $\msf{Decide}$.}
    If the verifier pair is invalid relative to
    $\mathfrak S_\Psi$, both answers are $\bot$.
    Otherwise, let
    \[
        C_Q
        =
        C\bigl(\Psi(f,\Ora,\varphi),V_0,V_1,z\bigr)
    \]
    be the reference critical set from
    \Cref{lem:critical-queries}.
    By \Cref{def:full-q-query}, $|C_Q|\leq 2q$.

    If the two underlying tuples agree on $C_Q$, their
    decision answers agree. A union bound using the preceding
    direct-query estimates gives
    \begin{align*}
        &\Pr_{r,D}
        \left[
            \msf{Decide}_{\mathfrak S_\Psi}^{
                \Psi(f,\Ora_{T\mapsto D},\varphi)}(Q)
            \neq
            \msf{Decide}_{\mathfrak S_\Psi}^{
                \Psi(f,\Ora,\varphi)}(Q)
        \right]\\
        &\qquad\leq
        |C_Q|\,2^{1-n}
        \leq 4q\,2^{-n}.
    \end{align*}
\end{enumerate}
Malformed inputs and inputs outside the allowed domain have
the same fixed answer in both oracles.
This completes the proof of \Cref{point-wise-4.18}.
\end{proof}

For subsequent comparisons, set
\[
    \gamma_{\mathrm{pre}}
    \coloneqq
    4q\,2^{-n},
    \qquad
    \Delta_{\mathrm{pre}}
    \coloneqq
    6q\sqrt{\gamma_{\mathrm{pre}}}.
\]

\begin{MyClaim}\zcsetup{reftype=Claim}
\label{claim:io-h0-h1}
It holds that
\[
    \left|
        \Pr[H_1\text{ wins}]
        -
        \Pr[H_0\text{ wins}]
    \right|
    \leq \Delta_{\mathrm{pre}}.
\]
\end{MyClaim}

\begin{proof}
Fix the full reference oracle $\Ora$, in addition to the
already fixed $f$ and $\varphi$.
Couple the pre-challenge oracles by sampling $(r,D)$ as in
$H_1$ and taking
\[
    (F_1,F_2)
    =
    \left(
        \mathsf G_{\Ora_{T\mapsto D},\varphi},
        \mathsf G_{\Ora,\varphi}
    \right).
\]
The reference oracle $F_2$ is fixed.
By \Cref{point-wise-4.18}, this coupling satisfies the
conditional disagreement hypothesis of
\Cref{lem:coupled-distributions} with
$\gamma=\gamma_{\mathrm{pre}}$.

Let $\rho_{r,D}$ denote the complete classical--quantum
output state of $\Adv_1$ under the modified oracle, including
$C_0,C_1$ and its residual state. Let $\rho_{\mathrm{ref}}$
denote the corresponding state under the reference oracle.
Since $\Adv_1$ makes at most $3q$ merged-oracle queries,
\Cref{lem:coupled-distributions} gives
\[
    \mathop{\mathbb E}_{r,D}
    \left[
        \operatorname{TD}(\rho_{r,D},\rho_{\mathrm{ref}})
    \right]
    \leq
    6q\sqrt{\gamma_{\mathrm{pre}}}
    =
    \Delta_{\mathrm{pre}}.
\]

For each fixed $(r,D,b)$, challenge generation and the
execution of $\Adv_2$ form the same channel in the two
experiments: the challenge is $\Ora_n(C_b,r)$ and the
post-challenge oracle is $\Gamma(f,\Ora,\varphi)$.
Contractivity of trace distance therefore bounds the
difference in winning probabilities for this continuation
by
$\operatorname{TD}(\rho_{r,D},\rho_{\mathrm{ref}})$.
Averaging over $(r,D,b)$ and then over $\Ora_n$ proves the
claim.

The expected trace-distance conclusion is important here.
The continuation depends on $r$, so closeness of only the
averaged pre-challenge states would not suffice.

This completes the proof of \zcref{claim:io-h0-h1}.
\end{proof}

\para{Hybrid $H_2$.}
We keep the reprogrammed pre-challenge oracle, but replace
the challenge by a fresh encoding and reprogram its
post-challenge entry accordingly. Its difference from $H_1$ is highlighted in \red{red color}.

\begin{ExperimentBox}[label={hybrid:h32}]{Hybrid $H_2$}
{
\begin{enumerate}
    \item $\Ora_n\leftarrow\mathbf I_{2n}^{5n}$.
    \item $b\picks\bits$ and $r\picks\bits^n$.
    \item Sample $D$ uniformly as an ordered subset of size
    $2^n$ of
    $\bits^{5n}\setminus\operatorname{Image}(\Ora_n)$.
    \item Set $T=(\ast,r)$.
    \item
    $(C_0,C_1,\mathsf{st})
    \leftarrow
    \Adv_1^{\ket{\Gamma(f,\Ora_{T\mapsto D},\varphi)}}(1^n)$.
    \item
    \red{$\widehat C\picks
    \bits^{5n}\setminus
    \bigl(\operatorname{Image}(\Ora_n)\cup D\bigr)$.}
    \item
    $b'\leftarrow
    \Adv_2^{\ket{\Gamma(
        f,\red{\Ora_{(C_b,r)\mapsto\widehat C}},\varphi)}}
    (\mathsf{st},\widehat C)$.
    \item Output $1$ if $b'=b$, and output $0$ otherwise.
\end{enumerate}
}
\end{ExperimentBox}

In set expressions involving $D$, we identify the ordered
tuple with its underlying set.

\begin{MyClaim}\zcsetup{reftype=Claim}
\label{claim:io-hyb-1-2}
The joint distributions of the challenge bit, the
pre-challenge output state, the challenge, and the
post-challenge oracle are identical in $H_1$ and $H_2$.
Consequently,
\[
    \Pr[H_1\text{ wins}]
    =
    \Pr[H_2\text{ wins}].
\]
\end{MyClaim}

\begin{proof}
Condition first on
\begin{equation}\zcsetup{reftype=Expression}
\label{eq:io-hyb-1-2-conditioning}
    f,\varphi,\Ora_{-n},b,r,D,\Ora_{T\mapsto D}.
\end{equation}
These variables determine the entire oracle supplied to
$\Adv_1$. Its classical outputs and conditional residual
density operator are therefore independent of the erased
values $\Ora_n|_T$.

Condition further on the classical outputs $C_0,C_1$, and
write $u=(C_b,r)\in T$.
Fix all entries $\Ora_n(v)$ for $v\neq u$, and define
\[
    V
    \coloneqq
    \bits^{5n}\setminus
    \left(
        \{\Ora_n(v):v\in\bits^{2n},\,v\neq u\}
        \cup D
    \right).
\]
Under the original sampling, the remaining entry $\Ora_n(u)$
is uniform over $V$.
Indeed, every injective completion compatible with the
conditioned data has the same probability: $\Ora_n$ is
uniform, and the number of possible ordered choices of $D$
outside its image depends only on the image size.

In $H_1$, the challenge is $\widehat C=\Ora_n(u)$ and is
therefore uniform over $V$.
In $H_2$, first $a=\Ora_n(u)$ is uniform over $V$, and then
$\widehat C$ is uniform over $V\setminus\{a\}$.
For every $h\in V$,
\[
    \Pr_{H_2}[\widehat C=h]
    =
    \sum_{a\in V\setminus\{h\}}
    \frac{1}{|V|}\frac{1}{|V|-1}
    =
    \frac{1}{|V|}.
\]
Thus the challenge is uniform over $V$ in both hybrids.

In either experiment, the post-challenge encoding oracle
agrees with the fixed table away from $u$ and maps $u$ to
$\widehat C$. Its evaluation and decision interfaces are
determined by this table, the fixed $f,\varphi$, and the
fixed family $\mathfrak S_\Psi$.
The conditional residual state of $\Adv_1$ is the same and
independent of the hidden entry $a$.
Hence the relevant joint classical--quantum distributions
are identical.

Averaging over the conditioned variables completes the proof
of \zcref{claim:io-hyb-1-2}.

\end{proof}

\para{Hybrid $H_3$.}
We restore the original pre-challenge oracle.
The challenge sampling and post-challenge reprogramming
remain unchanged. Its difference from $H_2$ is highlighted in \red{red color}.

\begin{ExperimentBox}[label={hybrid:h33}]{Hybrid $H_3$}
{
\begin{enumerate}
    \item $\Ora_n\leftarrow\mathbf I_{2n}^{5n}$.
    \item $b\picks\bits$ and $r\picks\bits^n$.
    \item Sample $D$ uniformly as an ordered subset of size
    $2^n$ of
    $\bits^{5n}\setminus\operatorname{Image}(\Ora_n)$.
    \item
    $(C_0,C_1,\mathsf{st})
    \leftarrow
    \Adv_1^{\ket{\Gamma(f,\red{\Ora},\varphi)}}(1^n)$.
    \item
    $\widehat C\picks
    \bits^{5n}\setminus
    \bigl(\operatorname{Image}(\Ora_n)\cup D\bigr)$.
    \item
    $b'\leftarrow
    \Adv_2^{\ket{\Gamma(
        f,\Ora_{(C_b,r)\mapsto\widehat C},\varphi)}}
    (\mathsf{st},\widehat C)$.
    \item Output $1$ if $b'=b$, and output $0$ otherwise.
\end{enumerate}
}
\end{ExperimentBox}

\begin{MyClaim}\zcsetup{reftype=Claim}
\label{claim:io-hyb-2-3}
It holds that
\[
    \left|
        \Pr[H_3\text{ wins}]
        -
        \Pr[H_2\text{ wins}]
    \right|
    \leq \Delta_{\mathrm{pre}}.
\]
\end{MyClaim}

\begin{proof}
Fix the full reference oracle $\Ora$.
The pre-challenge states are exactly
$\rho_{r,D}$ and $\rho_{\mathrm{ref}}$ from the proof of
\zcref{claim:io-h0-h1}. Hence
\[
    \mathop{\mathbb E}_{r,D}
    \left[
        \operatorname{TD}(\rho_{r,D},\rho_{\mathrm{ref}})
    \right]
    \leq \Delta_{\mathrm{pre}}.
\]

For each fixed $(r,D,b)$, both experiments use the same
continuation: sample $\widehat C$ uniformly outside
$\operatorname{Image}(\Ora_n)\cup D$, and run $\Adv_2$
with oracle
$\Gamma(f,\Ora_{(C_b,r)\mapsto\widehat C},\varphi)$.
This is a common channel on the complete pre-challenge
output state, including the classical circuit descriptions.
Contractivity followed by averaging over $(r,D,b)$ and
$\Ora_n$ proves the claim.

This completes the proof of \zcref{claim:io-hyb-2-3}.

\end{proof}

\para{Hybrid $H_4$.}
We remove the post-challenge encoding reprogramming.
Instead, we set the backup circuit at $\widehat C$ to $C_0$.
Since $\widehat C$ is outside the image of $\Ora_n$, the
challenge evaluates to $C_0^f$ in this experiment. Its difference from $H_3$ is highlighted in \red{red color}.

\begin{ExperimentBox}[label={hybrid:h34}]{Hybrid $H_4$}
{
\begin{enumerate}
    \item $\Ora_n\leftarrow\mathbf I_{2n}^{5n}$.
    \item $b\picks\bits$ and $r\picks\bits^n$.
    \item Sample $D$ uniformly as an ordered subset of size
    $2^n$ of
    $\bits^{5n}\setminus\operatorname{Image}(\Ora_n)$.
    \item
    $(C_0,C_1,\mathsf{st})
    \leftarrow
    \Adv_1^{\ket{\Gamma(f,\Ora,\varphi)}}(1^n)$.
    \item
    $\widehat C\picks
    \bits^{5n}\setminus
    \bigl(\operatorname{Image}(\Ora_n)\cup D\bigr)$.
    \item
    $b'\leftarrow
    \Adv_2^{\ket{\Gamma(
        f,\red{\Ora,\varphi_{\widehat C\mapsto C_0}})}}
    (\mathsf{st},\widehat C)$.
    \item Output $1$ if $b'=b$, and output $0$ otherwise.
\end{enumerate}
}
\end{ExperimentBox}

The next estimate corresponds to
\cite[Claim~4.22]{C:BitDegVai17}.
Its reference oracle uses the modified backup map, which is
why $\mathfrak S_\Psi$ must include all backup maps.

\begin{lemma}[Challenge-Reprogramming Disagreement]
\label{point-wise-4.22}
Fix $f,\Ora,\varphi$, a bit $b$, circuits
$C_0,C_1\in\bits^n$ with $C_0^f\equiv C_1^f$, and
\[
    \widehat C\in
    \bits^{5n}\setminus\operatorname{Image}(\Ora_n).
\]
For every fixed allowed merged-oracle input $w$,
\[
    \Pr_{r\picks\bits^n}
    \left[
        \mathsf G_{\Ora_{(C_b,r)\mapsto\widehat C},\varphi}(w)
        \neq
        \mathsf G_{\Ora,\varphi_{\widehat C\mapsto C_0}}(w)
    \right]
    \leq 2q\,2^{-n}.
\]
\end{lemma}

\begin{proof}
Define the reference tuple
\[
    \Psi_2
    \coloneqq
    \Psi(f,\Ora,\varphi_{\widehat C\mapsto C_0})
\]
and, for each $r\in\bits^n$, define
\[
    \Psi_1^{(r)}
    \coloneqq
    \Psi(f,\Ora_{(C_b,r)\mapsto\widehat C},\varphi).
\]
The reference tuple belongs to $\mathfrak S_\Psi$ because
the family permits arbitrary backup maps.
Since $\widehat C\notin\operatorname{Image}(\Ora_n)$,
the reprogrammed encoding oracle remains injective, so
$\Psi_1^{(r)}\in\mathfrak S_\Psi$ for every $r$.

Fix $w$ and consider its query type.
\begin{enumerate}
    \item \textbf{Queries to $f$.}
    Both tuples use the same $f$, so the disagreement
    probability is zero.

    \item \textbf{Queries to $\Ora$.}
    At length $n$, a fixed input $(C,s)$ changes only if
    $(C,s)=(C_b,r)$. Its disagreement probability is
    therefore at most $2^{-n}$.
    Other lengths are unchanged.

    \item \textbf{Queries to $\mathsf{Eval}$.}
    At the challenge encoding $\widehat C$, the first tuple
    returns $C_b^f(z)$ and the reference tuple returns
    $C_0^f(z)$. These answers agree because
    $C_0^f\equiv C_1^f$.

    At any other encoding $\widetilde C$, the evaluation
    answers can differ only if
    \[
        \widetilde C=\Ora_n(C_b,r).
    \]
    This is the encoding removed by the reprogramming.
    Injectivity implies that a fixed $\widetilde C$ equals
    this value for at most one choice of $r$.
    Thus every fixed evaluation query has disagreement
    probability at most $2^{-n}$.
    Encodings at other lengths are unchanged.

    \item \textbf{Queries to $\msf{Decide}$.}
    Invalid verifier pairs receive $\bot$ in both tuples.
    For a valid query $Q=(V_0,V_1,z)$, choose the critical
    set at the reference tuple,
    \(
        C_Q=C(\Psi_2,V_0,V_1,z)
    \). 
    This set is independent of $r$ and has size at most
    $2q$.
    By \Cref{lem:critical-queries}, disagreement of the
    decision answers requires disagreement at some point
    of $C_Q$. Hence
    \begin{align*}
        \Pr_r\left[
            \msf{Decide}_{\mathfrak S_\Psi}^{\Psi_1^{(r)}}(Q)
            \neq
            \msf{Decide}_{\mathfrak S_\Psi}^{\Psi_2}(Q)
        \right]
        &\leq
        \sum_{u\in C_Q}
        \Pr_r[\Psi_1^{(r)}(u)\neq\Psi_2(u)]\\
        &\leq |C_Q|\,2^{-n}\\
        &\leq 2q\,2^{-n}.
    \end{align*}
\end{enumerate}
Malformed inputs and inputs outside the allowed domain
again have disagreement probability zero.
This completes the proof of \Cref{point-wise-4.22}.

\end{proof}

Set
\[
    \gamma_{\mathrm{post}}
    \coloneqq
    2q\,2^{-n},
    \qquad
    \Delta_{\mathrm{post}}
    \coloneqq
    6q\sqrt{\gamma_{\mathrm{post}}}.
\]

\begin{MyClaim}\zcsetup{reftype=Claim}
\label{claim:io-hyb-3-4}
It holds that
\[
    \left|
        \Pr[H_4\text{ wins}]
        -
        \Pr[H_3\text{ wins}]
    \right|
    \leq \Delta_{\mathrm{post}}.
\]
\end{MyClaim}

\begin{proof}
The pre-challenge oracle is the original oracle in both
experiments. Condition on the classical data
\[
    f,\Ora,\varphi,b,D,C_0,C_1,\widehat C,
\]
and let $\rho_{\mathsf{st}}$ denote the resulting conditional
residual density operator.

The sampling of $D$ and $\widehat C$, and the entire
pre-challenge execution, are independent of $r$.
Consequently, after this conditioning, $r$ remains uniform
over $\bits^n$ and $\rho_{\mathsf{st}}$ is independent of $r$.
Moreover, admissibility in the real experiment implies
$C_0^f\equiv C_1^f$ with probability one here, since the
pre-challenge execution uses the same original oracle.

Couple the post-challenge merged oracles by sampling this
uniform $r$ and taking
\[
    (F_1,F_2)
    =
    \left(
        \mathsf G_{\Ora_{(C_b,r)\mapsto\widehat C},\varphi},
        \mathsf G_{\Ora,\varphi_{\widehat C\mapsto C_0}}
    \right).
\]
The reference oracle $F_2$ is fixed under the conditioning.
By \Cref{point-wise-4.22}, the coupling satisfies the
hypothesis of \Cref{lem:coupled-distributions} with
$\gamma=\gamma_{\mathrm{post}}$.

Apply that lemma to $\Adv_2$ with initial state
\(
    \rho_{\mathsf{st}}
    \otimes
    \ket{\widehat C}\!\bra{\widehat C}
\). 
This state is independent of the sampled oracle pair.
Since $\Adv_2$ makes at most $3q$ merged-oracle queries,
the difference in its probability of outputting the fixed
bit $b$ is at most
\(
    6q\sqrt{\gamma_{\mathrm{post}}}
    =
    \Delta_{\mathrm{post}}
\). 
Averaging over the common distribution of the conditioned
classical data proves the claim.

This completes the proof of \zcref{claim:io-hyb-3-4}.
\end{proof}

\para{Completing the proof of \Cref{thm:full-oracle-iO}.}
In $H_4$, the pre-challenge state, challenge, and
post-challenge oracle are all independent of $b$.
The decision-oracle validity predicate also uses the same
fixed family $\mathfrak S_\Psi$, independently of $b$.
Thus
\[
    \Pr[H_4\text{ wins}]=\frac12.
\]
Combining
\zcref{claim:io-h0-h1,claim:io-hyb-1-2,claim:io-hyb-2-3,claim:io-hyb-3-4},
we obtain
\begin{align*}
    \left|
        \Pr[H_0\text{ wins}]-\frac12
    \right|
    &\leq
    2\Delta_{\mathrm{pre}}+\Delta_{\mathrm{post}}\\
    &\leq
    12q\sqrt{4q\,2^{-n}}
    +
    6q\sqrt{2q\,2^{-n}}\\
    &=
    O\left(q^{3/2}2^{-n/2}\right).
\end{align*}
For $q(n)=O(2^{n/6})$, this is $O(2^{-n/4})$.

The bound is uniform over the fixed $f$ and $\Ora_{-n}$.
Averaging over their independent sampling gives the claimed
security bound for $\Gamma_\varphi$.
The advantage here is the absolute bias after averaging over
the oracle distribution, as in \Cref{PQ-iO}; we do not
interchange this quantity with the expected absolute bias
of individual oracle realizations.

This completes the proof of \Cref{thm:full-oracle-iO}.

\end{proof}

\subsubsection{Consequence for Efficient Black-Box Reductions}

We now connect the two source-security theorems to the
black-box separation.
The implementation convention specified in
\Cref{iO-Oracle} is essential: source circuits have only
$f$-gates, but returned obfuscated programs execute relative
to the ambient oracle tuple.

\begin{corollary}\label{cor:iO-injOWF-black-box-separation}
Under this ambient-oracle implementation convention, there
is no efficient fully black-box quantum construction of a
worst-case quantum-hard language in $\NP\cap\coNP$ from
post-quantum injOWFs and post-quantum $\iO$ for classical
circuits with injOWF-gates.
\end{corollary}

\begin{proof}
Suppose an efficient fully black-box construction
$(V_0,V_1,\mathcal R)$ exists.
Fix a backup map $\varphi$, and instantiate the source
primitives using the independently sampled families
$f,\Ora$ and the obfuscator from \Cref{iO-Oracle}.

\para{Compiling the construction's interfaces.}
Compile the construction's verifiers into classical
$\Psi$-aided verifiers $(\widetilde V_0,\widetilde V_1)$.
An obfuscation query on $(C,r)$ is implemented by querying
$\Ora(C,r)$ and forming the canonical wrapper.
Executing such a wrapper is implemented by the corresponding
$\mathsf{Eval}$-query.
The compilation has polynomial overhead and preserves the
verifiers' behavior.

Every tuple in $\mathfrak S_\Psi$ gives an injective function
and a functionally correct obfuscation implementation.
The universal structure requirement therefore implies that
$(\widetilde V_0,\widetilde V_1)$ is valid relative to
$\mathfrak S_\Psi$.
Consequently, the deterministic adversary
\[
    \mathcal A_{\mathrm{dec}}(z)
    \coloneqq
    \msf{Decide}_{\mathfrak S_\Psi}^{\Psi}
    (\widetilde V_0,\widetilde V_1,z)
\]
decides the constructed language exactly on every input.
At each input length, the verifier machines are compiled
into oracle-aided circuits.
Reversibly forming this decision query and uncomputing its
description supplies the coherent function-evaluation
interface required by the reduction.

\para{Polynomial resource inclusion.}
Replace each call by $\mathcal R$ to
$\mathcal A_{\mathrm{dec}}$ with this decision-oracle
implementation, and compile the source interfaces as above.
Since the reduction is efficient, it uses polynomially many
oracle calls and polynomial-length query registers.
The construction's verifier circuits also have polynomial
size, including after interface compilation.

It follows that a single polynomial $p(n)$ bounds:
\begin{enumerate}
    \item the number of queries to each augmented-oracle
    interface;
    \item the lengths of all evaluation encodings queried
    by the composed reduction; and
    \item the corresponding query counts and evaluation
    lengths inside every verifier submitted to
    $\msf{Decide}_{\mathfrak S_\Psi}$, for every witness
    and every underlying realization.
\end{enumerate}
These are syntactic bounds on the composed algorithms and
circuits, so they also hold in the hybrid executions.
Increasing $p$ if necessary, the induced source adversaries
are $p(n)$-query in the sense of
\Cref{def:full-q-query}.
There is no requirement that $p(n)$ equal the reduction's
original oracle-query bound.

\para{Contradiction with source security.}
By \Cref{thm:full-oracle-injOWF,thm:full-oracle-iO}, the
induced inverter and admissible obfuscation distinguisher
have, respectively, success and advantage at most
\[
    O\left(p(n)^3 2^{-n/2}\right)
    \qquad\text{and}\qquad
    O\left(p(n)^{3/2}2^{-n/2}\right).
\]
Both are negligible; in particular, both are
$O(2^{-n/4})$ for sufficiently large $n$.

On the other hand, $\mathcal A_{\mathrm{dec}}$ decides the
constructed language exactly in the worst case.
The black-box security requirement therefore makes
$\mathcal R$ break one of the source primitives with
inverse-polynomial success or advantage for infinitely many
input lengths. This contradicts the preceding bounds.

This completes the proof of
\Cref{cor:iO-injOWF-black-box-separation}.

\end{proof}

The polynomial-resource inclusion argument is sufficient for
the efficient-reduction impossibility.
The quantitative source-security theorems also apply to
inefficient adversaries satisfying
\Cref{def:full-q-query}, but they do not by themselves give
an exponential lower bound on the bare number of oracle
queries made by an arbitrary inefficient reduction.

\section{Acknowledgments and AI Disclosure} 

We thank Yuval Ishai and Alexis Korb for helpful discussions about their work~\cite{STOC:GIKKLS23}.

GPT-5.6 and GPT-6 were used to improve the clarity and quality of the writing, correct spelling and grammatical errors, and proofread the manuscript.  No AI tools were used to generate any of the ideas or techniques presented in this work, except for minor fixes to bugs identified by AI tools.  

\clearpage

\newpage
{\phantomsection
	\addcontentsline{toc}{section}{References}
}
\bibliographystyle{alphaurl}
\newcommand{\etalchar}[1]{$^{#1}$}

\newpage

{
    \hypersetup{hypertexnames=false}  
    \appendix

	\section{More Basic Preliminaries}
\label{sec:prelim-basic}

\subsection{Probability Inequalities}

\begin{lemma}[Chernoff Bound]\label{chernoff}
    Let $X_1,\ldots,X_n$ be independent random variables taking values in $\{0,1\}$, and let $X=\sum_{i=1}^{n}X_i$. Then for any $\delta\geq0$ and $\mu \geq \mathbb{E}[X]$, we have
    \[
        \Pr[X\geq(1+\delta)\mu]\leq\left(\frac{e^{\delta}}{(1+\delta)^{(1+\delta)}}\right)^\mu
    \]
\end{lemma}
\begin{proof}
    Proof for this version of Chernoff bound can be found in standard literature, e.g., \cite[Theorem 4]{FOCS:GooLiZuc24}.

\end{proof}

\begin{corollary}\label{chernoff_2}
    Let $X_1,\ldots,X_n$ be independent random variables taking values in $\{0,1\}$, and let $X=\sum_{i=1}^{n}X_i$. Then for any $t\geq\mu$ and $\mu \geq \mathbb{E}[X]$, we have
    \[
        \Pr[X\geq t]\leq e^{-\mu}\left(\frac{e\mu}{t}\right)^t
    \]
\end{corollary}
\begin{proof}
    It directly follows from \Cref{chernoff} by setting $\delta=\frac{t}{\mu}-1$.

\end{proof}
\begin{lemma}[Cauchy Schwarz Inequality]\label{cauchy_schwarz}
    For any Random variables $X$ and $Y$, we have
    \[
        \mathbb{E}[XY]\leq \sqrt{\mathbb{E}[X^2]\cdot\mathbb{E}[Y^2]}
    \]
\end{lemma}
\begin{corollary}\label{EX2}
    For any Random variable $X$, we have
    \[
        \mathbb{E}[X]\leq \sqrt{\mathbb{E}[X^2]}
    \]
\end{corollary}
\begin{proof}
    
    Let $Y=1$ in \Cref{cauchy_schwarz}, we have
    \[
        \mathbb{E}[X]\leq \sqrt{\mathbb{E}[X^2]\cdot\mathbb{E}[1^2]}=\sqrt{\mathbb{E}[X^2]}
    \]
\end{proof}

\begin{lemma}[McDiarmid's Inequality]
\label{lem:mcdiarmid}
Let $X=(X_1,\ldots,X_m)$ be a vector of independent random
variables, where each $X_i$ takes values in a finite set
$\mathcal{X}_i$.
Suppose that $F:\prod_{i=1}^{m}\mathcal{X}_i\to\mathbb{R}$
satisfies
\[
    |F(\mathbf{x})-F(\mathbf{x}')|\leq c_i
\]
for some constants $c_1,\ldots,c_m>0$, whenever
$\mathbf{x}$ and $\mathbf{x}'$ differ only in coordinate $i$.
Then, for every $t\geq 0$,
\[
    \Pr\bigl[|F(X)-\mathbb{E}[F(X)]|\geq t\bigr]
    \leq
    2\exp\left(-\frac{2t^2}{\sum_{i=1}^{m}c_i^2}\right).
\]
\end{lemma}

\subsection{Statistical Distance and Properties}

For distributions $P,Q$ on a finite set $\Omega$, denote the \emph{statistical distance} between $P$ and $Q$ as
\[
    \Delta(P,Q)\coloneqq\frac12\sum_{x\in\Omega}|P(x)-Q(x)|.
\]

\begin{lemma}[{\cite[Lemma 3.15]{STOC:GIKKLS23}}]\label{MaxCoupling}
If $A = (A_1,\ldots,A_n)$ and $B = (B_1,\ldots,B_n)$
are sequences of independent random variables such that
\[
    \forall i \in [n],
    \qquad
    \Delta(A_i,B_i) \leq \varepsilon,
\]
then
\[
    \Delta\bigl(
        (A_1,\ldots,A_n),
        (B_1,\ldots,B_n)
    \bigr)
    \leq n\varepsilon.
\]
\end{lemma}

Also recall that statistical distance does not increase under post-processing:
\begin{lemma}
\label{lem:h01-statistical-distance}
Let $P$ and $Q$ be probability distributions on a finite set
$\mathcal X$, and let $\mathsf A$ be a randomized transformation
from $\mathcal X$ to a finite set $\mathcal Y$.
Write $\mathsf A(P)$ and $\mathsf A(Q)$ for the output distributions
obtained by applying $\mathsf A$ to an input drawn from $P$ and $Q$,
respectively. Then
\[
    \Delta(\mathsf A(P),\mathsf A(Q))
    \leq \Delta(P,Q).
\]
In particular, discarding coordinates cannot increase statistical
distance.
\end{lemma}

\begin{proof}
For each $x\in\mathcal X$ and $y\in\mathcal Y$, define
\[
    A(y\mid x)\coloneqq\Pr[\mathsf A(x)=y],
\]
where the probability is over the internal randomness of $\mathsf A$.
For every $x\in\mathcal X$, we have $A(y\mid x)\geq 0$ for all
$y\in\mathcal Y$ and
$\sum_{y\in\mathcal Y}A(y\mid x)=1$.
Writing $P(x)$ and $Q(x)$ for the corresponding probability masses,
the definition of statistical distance and the triangle inequality give
\begin{align*}
    \Delta(\mathsf A(P),\mathsf A(Q))
    &=
    \frac12\sum_{y\in\mathcal Y}
    \left|
        \sum_{x\in\mathcal X}
        (P(x)-Q(x))A(y\mid x)
    \right|\\
    &\leq
    \frac12\sum_{y\in\mathcal Y}\sum_{x\in\mathcal X}
    |P(x)-Q(x)|A(y\mid x)\\
    &=
    \frac12\sum_{x\in\mathcal X}|P(x)-Q(x)|
    \sum_{y\in\mathcal Y}A(y\mid x)\\
    &=
    \frac12\sum_{x\in\mathcal X}|P(x)-Q(x)|\\
    &= \Delta(P,Q).
\end{align*}
The final assertion follows by taking $\mathsf A$ to be the
deterministic projection onto the coordinates being retained.

\end{proof}

\subsection{Efficient Approximability of {$\frac{1}{e}$} and {$1-\frac{1}{e}$}}\label{sec:eff-approx-e}

\begin{lemma}[Efficient Approximability of {$\frac{1}{e}$}]
\label{lem:efficient-approx-inv-e}

There exist a deterministic polynomial-time algorithms
$\msf{Approx}_{\frac{1}{e}}$  output a rational number $\gamma_n=\frac{a_n}{b_n}$ for some $a_n,b_n \in \Naturals$
such that, for every $n\geq 2$ satisfying
\[
    \left|
        \frac{1}{e} -\frac{a_n}{b_n}
    \right|
    \leq 2^{-n}
\]
Consequently, $\frac{1}{e}$ is efficiently approximable.
\end{lemma}

\begin{proof}
On input $1^n$, the algorithm $\msf{Approx}_{1/e}$ outputs
$\gamma_n$ which is a positive rational number and can be written as
\[
    \gamma_n
    =
    \frac{a_n}{b_n},
    \qquad
    a_n
    :=
    \sum_{k=0}^{n}(-1)^k\frac{n!}{k!},
    \qquad
    b_n:=n!,
\]
where $a_n,b_n\in\Naturals$ follows.

Consider the Taylor expansion of $f(x)=e^{-x}$ around $x=0$,
evaluated at $x=1$. Taylor's theorem gives
\[
    \frac{1}{e}
    =
    \sum_{k=0}^{n}\frac{(-1)^k}{k!}
    +R_{n},
\]
where, for some $\xi\in(0,1)$,
\[
    R_{n}
    =
    \frac{f^{(n+1)}(\xi)}{(n+1)!}.
\]
Since
\[
    \left|f^{(n+1)}(\xi)\right|
    =e^{-\xi}
    \leq 1,
\]
we obtain
\[
    \left|
        \frac{1}{e}-\gamma_n
    \right|
    =
    |R_{n}|
    \leq \frac{1}{(n+1)!}.
\]
Since $(n+1)!\geq 2^n$ for every $n\geq 1$, we then obtain
\[
    \left|
        \frac{1}{e}-\frac{a_n}{b_n}
    \right|
    \leq \frac{1}{(n+1)!}
    \leq 2^{-n}.
\]

Finally, the algorithm performs $O(n)$ arithmetic operations on integers
of bit length bounded by $O(\log(n!))=O(n\log n)$. Therefore,
$\msf{Approx}_{1/e}$ runs in deterministic polynomial time.
\end{proof}
\begin{corollary}[Efficient Approximability of {$1-\frac{1}{e}$}]
\label{cor:efficient-approx-one-minus-inv-e}
There exists a deterministic polynomial-time algorithm
$\msf{Approx}_{1-1/e}$ that, on input $1^n$, output a rational number $\gamma_n=\frac{a_n}{b_n}$ for some $a_n,b_n \in \Naturals$
such that, for every $n\geq 1$ satisfying
\[
    \left|
        \left(1-\frac{1}{e}\right)-\frac{a_n}{b_n}
    \right|
    \leq 2^{-n}.
\]
Consequently, $1-\frac{1}{e}$ is efficiently approximable.
\end{corollary}

\begin{proof}
On input $1^n$, the algorithm $\msf{Approx}_{1-1/e}$ invokes
$\msf{Approx}_{1/e}(1^n)$ from
\Cref{lem:efficient-approx-inv-e}. If
\[
    \msf{Approx}_{1/e}(1^n)=\frac{a^{'}_n}{b^{'}_n},
\]
for some $a^{'}_n,b^{'}_n\in \Naturals$

then it outputs
\[
    \gamma_n:=\frac{b^{'}_n-a^{'}_n}{b^{'}_n}
\]
By \Cref{lem:efficient-approx-inv-e},
\[
    \left|\frac{1}{e}-\frac{a^{'}_n}{b^{'}_n}\right|\leq 2^{-n}.
\]
Therefore,
\begin{align*}
    \left|
        \left(1-\frac{1}{e}\right)-\gamma_n
    \right|
    &=
    \left|
        \left(1-\frac{1}{e}\right)-\frac{b^{'}_n-a^{'}_n}{b^{'}_n}
    \right| \\
    &=
    \left|
        \frac{a^{'}_n}{b^{'}_n}-\frac{1}{e}
    \right| \\
    &\leq 2^{-n}.
\end{align*}

The invocation of $\msf{Approx}_{1/e}$ takes polynomial time, and the
additional subtraction $b-a$ can be performed in time polynomial in
the bit lengths of $a$ and $b$. Hence,
$\msf{Approx}_{1-1/e}$ is a deterministic polynomial-time algorithm.
\end{proof}

\section{Length-Tripling Random Oracles as Post-Quantum injOWFs}

The following two lemma justifying the assumption \(\UP\nsubseteq\BQP\) underlying our subsequent oracle-separation results. In particular, we show that a length-tripling random oracle heuristically instantiates a post-quantum injective one-way function with overwhelming probability. Notice that a length-tripling random oracle can be generated independently by setting \(m(n)=3n\) in our interpretation of random oracles by \Cref{lem:random-oracle-decomposition}.
\begin{lemma}[Overwhelming Injectivity of length-tripling random oracle]
    Fix any $n\in \Naturals$, for any length-tripling random oracle $O_{\msf{inj}}:\bits^n\to \bits^{3n}$, we have
    \[\Pr_{O\pick \Ora}[\exists x_1,x_2\in \bits^n \text{ such that } O_{\msf{inj}}(x_1)=O_{\msf{inj}}(x_2) \land x_1\neq x_2]\leq \negl(n)\]\
\end{lemma}
\begin{proof}
    Fix any two distinct inputs $x_1,x_2\in\bits^n$, since $O_{\msf{inj}}$ is a random oracle and its outputs are independent and uniformly distributed over
    $\bits^{3n}$, we have
    \[
        \Pr_{O\pick\Ora}\bigl[O_{\msf{inj}}(x_1)=O_\msf{inj}(x_2)\bigr]
        =\frac{2^{3n}}{2^{6n}}
        = \frac{1}{2^{3n}}.
    \]
    
    There are
    \[
        \binom{2^n}{2}
        < 2^{2n}
    \]
    pairs of distinct elements of $\bits^n$. Therefore, 
        \begin{align*}
    &\Pr_{O\pick\Ora}\left[
        \exists x_1,x_2\in\bits^n :
        x_1\neq x_2
        \land
        O_{\msf{inj}}(x_1)=O_{\msf{inj}}(x_2)
    \right] \\
    &\leq\binom{2^n}{2}\cdot 2^{-3n}  \tag{Union Bound}\\
    &<2^{2n}\cdot 2^{-3n} \\
    &=2^{-n}=\negl(n)
    \end{align*}
\end{proof}
\begin{lemma}[Post-Quantum Oneway-ness of length-tripling random oracle]\label{lem:injOWF-Oracle}
    Fix any $n\in \Naturals$, for any length-tripling random oracle $O_{\msf{inj}}:\bits^n\to \bits^{3n}$ and quantum oracle-aided adversary $\Adv$ making at most $q=\poly(n)$ queries, we have
    \[\Pr_{O\pick\Ora,x\pick\bits^n}\left[
    \Adv^{\ket{O_{\msf{inj}}}}(O_{\msf{inj}}(x))=x
\right]
\leq \negl(n).
\]
\end{lemma}
\begin{proof}
Assume that there exists a quantum adversary $\mcal A$
making at most $q$ queries that inverts a uniformly random
length-tripling oracle $O_{\msf{inj}}:\bits^n\to\bits^{3n}$
with success probability
\[
    \varepsilon
    :=\Pr_{O_{\msf{inj}},x}\left[
        \mcal A^{\ket{O_{\msf{inj}}}}
        (O_{\msf{inj}}(x))=x
    \right].
\]
We construct an adversary $\mcal B$ for the length-preserving
preimage-search problem in \Cref{preimage_search}, with
parameter $m=3n$.

Adversary $\mcal B$ is given quantum oracle access to a
uniformly random function $H:\bits^{3n}\to\bits^{3n}$ and
a challenge $y=H(w)$, where $w\pick\bits^{3n}$ is hidden.
It samples $r\pick\bits^{2n}$ once and runs $\mcal A$ on $y$,
simulating its oracle by$O_r(z):=H(z\mathbin\|r)$

Each quantum query to $O_r$ can be simulated using one
quantum query to $H$ by appending the fixed string $r$
to the query input. If $\mcal A$ outputs $z$, then
$\mcal B$ outputs $z\mathbin\|r$.

For the analysis, write $w=x\mathbin\|s$, where
$x\in\bits^n$ and $s\in\bits^{2n}$.
The event $r=s$ occurs with probability $2^{-2n}$.
Conditioned on this event, $O_r$ is a uniformly random
function from $\bits^n$ to $\bits^{3n}$, $x$ is uniform
and independent of $O_r$, and $y=O_r(x)$.
Hence $\mcal A$ outputs $x$ with probability $\varepsilon$.
When $r\ne s$, $\mcal B$ cannot output $w$. Therefore,
\[
    \Pr_{H,w}\left[
        \mcal B^{\ket H}(H(w))=w
    \right]
    =2^{-2n}\varepsilon.
\]
Applying \Cref{preimage_search} at input length $3n$, we obtain
\[
\begin{aligned}
    2^{-2n}\varepsilon
    &\le
    \Pr_{H,w}\left[
        H\bigl(\mcal B^{\ket H}(H(w))\bigr)=H(w)
    \right]\\
    &\le \frac{6n(2q+1)^2+1}{2^{3n}}.
\end{aligned}
\]
Consequently,
\[
    \varepsilon
    \le \frac{6n(2q+1)^2+1}{2^n}
    =\negl(n),
\]
since $q=\poly(n)$.

\end{proof}

}

\end{document}